\documentclass[12pt]{yalephd}
\usepackage{graphicx}
\usepackage{times}
\usepackage{float}
\usepackage{amssymb}
\usepackage{afterpage}
\usepackage{epsfig}
\usepackage{longtable}
\usepackage[pagebackref=true]{hyperref}
\usepackage{hyperref}
\usepackage{color}
\usepackage{setspace}
\usepackage{titlesec}
\definecolor{linkColor}{rgb}{0.,0.11,0.22}
\definecolor{YaleBlue}{rgb}{0.0,0.22,0.444}
\usepackage{braket}
\usepackage{amsmath}
\renewcommand{\subsubsection}[1]{%
\paragraph{\large\textbullet\ \large #1}\mbox{}\\[1em]}
\usepackage{mathtools}

\DeclarePairedDelimiter\floor{\lfloor}{\rfloor}
\usepackage{setspace}
\usepackage{fancyhdr} 
\usepackage[sort,compress]{cite}
\usepackage{fancyhdr,lipsum}

\usepackage{tikz}
\usetikzlibrary{shapes.geometric, arrows, positioning}
\usepackage{hyperref}
\hypersetup{
    colorlinks,
    linkcolor={blue!80!black},
    citecolor={blue!100!black},
    urlcolor={blue!80!black}
}

\tikzstyle{startstop} = [rectangle, rounded corners, minimum width=3cm, minimum height=1cm, text centered, draw=black, fill=red!30]
\tikzstyle{process} = [rectangle, minimum width=3cm, minimum height=1cm, text centered, draw=black, fill=blue!30]
\tikzstyle{decision} = [diamond, minimum width=3cm, minimum height=1cm, text centered, draw=black, fill=green!30]
\tikzstyle{arrow} = [thick,->,>=stealth]
\usepackage{ifthen}
\usepackage{amssymb}
\usepackage{xcolor}

\newboolean{showcomments}
\setboolean{showcomments}{true}

\makeatletter
\newcommand{\mynote}[3]{%
  \ifthenelse{\boolean{showcomments}}{%
   \fbox{\bfseries\sffamily\scriptsize#1}%
   {\small$\blacktriangleright$\textsf{\emph{\color{#3}{#2}}}$\blacktriangleleft$}}%
  {%
   \@bsphack
   \@esphack
  }%
}
\makeatother

\definecolor{asparagus}{rgb}{0.53, 0.66, 0.42}
\usepackage[xcolor]{mdframed}
\usepackage{changes}[draft]
\begin{document}

% Not to exceed 750 words
\begin{center}
{Abstract}\\
{Quantum Computing in Discrete- and Continuous-Variable Architectures}\\
	% Full Name of Author (as it should appear on the diploma)
	{Shraddha Singh}\\
	% Month and Year Degree is to be awarded
    % (not the month where the dissertation is submitted)
	{2025}
\end{center}

This dissertation develops a theoretical framework for hybrid discrete-variable (DV) and continuous-variable (CV) quantum systems, focusing on control, state preparation, and error correction. Quantum computing holds the potential to surpass classical computation in tasks such as factorization, secure communication, and quantum simulation. Hybrid CV-DV systems offer a promising path by combining the stability and long coherence times of oscillators with the fast gate operations of qubits.  

A central contribution of this work is the development of ``non-abelian quantum signal processing" (NA-QSP), a generalization of quantum signal processing (QSP)~\cite{singh2024towards} where the control parameters are non-commuting quantum operators, such as oscillator position and momentum. We introduce the ``Gaussian-Controlled-Rotation" (GCR) technique, the first non-abelian composite pulse sequence that enables precise control of CV states using DV ancillae. GCR outperforms traditional composite pulse sequences in terms of gate fidelity and robustness to control errors. This framework can be extended to quantum singular value transformation (QSVT). In light of understanding the CV instruction set, we also propose the Gaussian hierarchy for CV operations, a classification of CV operations, analogous to the Clifford hierarchy for qubits, and raise open questions about the comparison and mapping between the two hierarchies.

With the help of GCR, we address deterministic state preparation in oscillators, including squeezed states, two-legged and four-legged cat states, and Gottesman-Kitaev-Preskill (GKP) states. The non-abelian technique enables high-fidelity preparation of these states, which are essential for quantum simulation and error correction, without the need for numerical optimizers~\cite{singh2024towards}. The benefits of our analytical preparation schemes are benchmarked against previous schemes in the literature. Notably, our schemes present the first deterministic preparation circuits for squeezed and GKP states which perform on par with all numerically optimized schemes.

A key challenge in oscillator-based architectures is photon loss, which degrades state coherence. This work~\cite{singh2024gkp_qec} gives the first analysis of probabilistic error correction for photon loss in finite-energy GKP codes, introducing the concept of `probabilistic distance' to quantify error correction performance of the recent GKP experiments showing promising realizations of beyond break-even error correction for qudits~\cite{sivak2023real,brock2024quantum}.   

The dissertation further explores high-fidelity universal control of error-corrected qubits encoded in oscillators. It introduces protocols for high-fidelity logical readout in the presence of residual errors and a pieceable error-corrected gate teleportation. A key finding is that logical operations on GKP qubits using our scheme can achieve high fidelity using GCR, even in the presence of errors, with a biased-noise ancilla. The extension of GCR to multi-mode systems enables efficient entangling gates and error-corrected two-qubit rotations. Our schemes are generalizable to arbitrary qubit as well as qudit GKP lattices.

We also explore how oscillator codes can reduce resource overheads in fault-tolerant quantum computing, alongside potential applications of a hybrid CV-DV architecture. To this end, we also present a quantum phase estimation compilation using an ancillary oscillator and a non-abelian QSP-based circuit, demonstrating the utility of the thesis framework for hybrid CV-DV algorithms. The dissertation establishes NA-QSP as a foundation for hybrid CV-DV quantum control, state preparation, and GKP-based error correction, laying the groundwork for scalable fault-tolerant quantum computation in CV-DV architectures.

	% Dissertation Title: Subtitle
	\font\myfont=cmr12 at 15pt
    \title{\singlespacing\normalfont\myfont Quantum Computing in Discrete- and Continuous-Variable Architectures}
	% Full Name of Author (as it should appear on the diploma)
	\author{Shraddha Singh}
	%
	% where there is an advisory committee, only the chairperson is listed
	\advisor{Steven M. Girvin and Shruti Puri}
	% Month and Year Degree is to be awarded
    % (not the month where the dissertation is submitted)
	\date{May 2025}

%	\singlespace

    \maketitle
    \makecopyright
\titleformat{\chapter}[display]
    {\Huge\bfseries}{\thechapter}{0pt}{\singlespacing\raggedleft\Huge\hrule}% NEW
\titlespacing*{\chapter}{0pt}{50pt}{50pt}% NEW

\newmdenv[backgroundcolor=yellow!20,
            leftline=false,
            rightline=false,
            bottomline=false,
            linewidth=3pt,
            linecolor=black]{myframe}

    \begin{center}
{\bf \large Acknowledgments}
\end{center}
There are countless people to thank for this thesis, but first and foremost is my primary advisor, Steven M. Girvin, without whom this work would not be what it is today. Steve’s rigorous questioning of my understanding has shaped my growth throughout my PhD. He is not only an exceptional scientist, known to all, but also one of the rare researchers who are equally — if not more — gifted as a teacher. His ability to explain complex physics to someone fresh out of undergrad, with only a basic grasp of physics and mathematics, transformed me from someone interested in quantum information into someone who can confidently call herself a quantum physicist.  

Good physicists are rare, but humble physicists are even rarer. I can’t count the number of times someone, upon learning I was Steve’s student, asked, “How is Steve so nice?” Under Steve’s guidance, I didn’t just build research expertise — I became a better communicator, writer, and teacher. Few can combine such research excellence with the ability to instill these skills in their students. Steve not only possesses these qualities but also passes them on effortlessly.  

I still remember my first meeting with Steve. He greeted me with a warm smile — a kind of welcome I didn’t expect from a renowned physicist. I had mentioned in my Statement of Purpose that I wanted to work on GKP codes. In that first technical meeting, Steve asked, “We don’t quite understand how the GKP experiments demonstrated at Yale correct photon loss!” It’s deeply gratifying to say that my dissertation answers this question. It feels like I’ve fulfilled the purpose of my PhD — something that would have been impossible without Steve’s incredible guidance.  

Steve’s mentorship style is hands-off when it comes to problem selection and project ideas, but he’s always eager to engage with any equation you bring to him. His feedback pushes you to be rigorous and precise. The hard part comes next — writing. Steve has written some of the most widely used tutorials in circuit-QED, setting a high bar for clarity and pedagogy. I have never seen a more thorough proofreading process than what Steve applies to documents, including this dissertation and the papers I’ve written with him — some spanning 50 to 150 pages, double-column and single-spaced. He reviews the equations, refines the wording, and pushes you to sharpen the intuition behind the text.  

My motivation for pursuing a physics degree was purely my interest in quantum computing. I have always focused more on equations than intuition — one of the reasons I preferred mathematics over physics in high school and undergrad. Steve has an extraordinary ability to translate mathematical solutions into intuitive pictures, even in quantum mechanics — a task far from trivial. His interpretations of my work helped me develop a deeper intuition for physics, transforming me from someone drawn to the equations into someone who appreciates the underlying physical insights. I never thought this shift was possible — I believed intuition was innate — but Steve showed me otherwise. I am deeply grateful to Steve for not only guiding me through my PhD but also for turning me into a physicist. There are countless other ways Steve’s mentorship has shaped me as a researcher, but I must move on.  

My co-advisor, Shruti Puri, is next among the people I cannot thank enough for being part of my PhD journey. I joined Yale when Shruti was Steve’s postdoc, and she became a professor the following summer. I thank my stars that one day Shruti asked if I’d be interested in working with her on surface codes. I agreed — and the rest is covered in the final chapter of this thesis!  

Shruti is the reason I got to work on Pauli algebra, discrete-variable (DV) codes, and so much fun mathematics and coding — topics I found intuitive and enjoyable even as a second-year student (something I can’t say about GKP codes). Shruti and I worked through the virtual times of 2020, and I can confidently say she was the best collaborator I’ve ever had. Working with Shruti felt like working with a close colleague — it was fun and exciting, and it never really felt like work. I don't think I worked even $40\%$ as hard on this project as I did on others — it was that smooth. As an advisor, Shruti has pushed me to think about the applications of my research. Her constant “But why do I care about this?” questions forced me to consider the practical relevance of my ideas and come up with more concrete solutions. I used to take this ability for granted early in my PhD, but as I approach the stage of advising students myself, I realize how hard it is to not only develop cool ideas but also ensure they have meaningful practical implications. In my first year, Shruti inspired me to become the kind of researcher who can talk to abstract theorists and experimentalists with equal ease — to explain why results make sense (or don’t). The path toward this goal is far from trivial, but thanks to Shruti, I’ve become someone who isn’t afraid to engage with experimentalists anymore.  

I would like to thank my committee members, Robert J. Schoelkopf and Michel H. Devoret, for providing an experimental lab environment where amazing experimentalists thrive. I’ve had the privilege of collaborating with a few of them who need special mention: Volodymyr (Vlad) Sivak, Benjamin J. Brock, Akshay Koottandavida, Vidul Joshi, Andy Z. Ding, Alec Eickbusch, Sumeru Hazra, and Takahiro Tsunoda. My external reader, Liang Jiang, holds a special place in my academic journey. I have referenced so many of his papers while trying to replicate or adapt his ideas during my PhD. I am grateful he agreed to be on my committee — I cannot think of a more suitable external reader.  

Over the course of my PhD at Yale, I have learned not only from my advisors but also from some incredible postdocs who worked with Steve. Baptiste Royer was my postdoctoral supervisor in the early years of my PhD, teaching me everything from high-performance computing to master equation simulations. Most of the early work quoted in this thesis stems from conversations with Baptiste. I had the privilege of collaborating with Kevin C. Smith on our massive ISA paper. Kevin's ability to present complex ideas pedagogically is something I aspire to emulate. Luigi Frunzio generously proofread my thesis despite it being a theoretical work. His meticulous attention to detail — down to misplaced commas and references — was incredible.

I cannot imagine my time at Yale without my partner, Akshay Koottandavida. His experimental insights from Michel’s group helped me climb the steep learning curve of quantum optics, mesoscopic physics, and noise during the early years of my PhD. Akshay and I first connected over quantum information during our undergraduate studies, and it brings me immense joy that we’ve carried that shared passion into Yale — to the point where we are now officially collaborating on projects.

The final years of a PhD can be grueling — trying to find the next step for your career, finishing papers, scheduling a defense date, and, of course, writing the thesis. I didn’t socialize much at Yale until this last year, but afternoons at Good Life Centre and countless dinners with Akshay, Andy, Vidul, and Sumeru played a huge role in keeping me grounded. Andy and I also bonded over late nights working through the never-ending mesoscopic physics (meso) problem sets with Yanni and Tom in our first year.

When I arrived at YQI, it was practically empty due to a leadership transition. In 2021–22, several students joined Shruti’s group, turning YQI into a truly vibrant and supportive place to work. I had the chance to formally supervise two incredible students, Kathleen Chang (Katie) and Harshvardhan Babla (Harsh). Postdocs in Shruti’s group always seem to have an extraordinary ability to express their ideas so clearly that it makes me wonder if I’ve reached that level of clarity yet. I learned a lot about Floquet simulations from Daniel K. Weiss (Danny). Later, Tom Smith joined the group, and we had the chance to collaborate on a major project involving almost the entire QEC team, including Kaavya Sahay, Pei-Kai Tsai, and Qile Su. Pei-Kai and I share a desk wall at YQI. My desk corner was deliberately set up for a quiet, low-profile existence, so the person I talk to most often is Pei-Kai. He’s so much fun to brainstorm with — whether it's about project ideas or random quantum information questions. The few conversations we have had are so engaging that I’m pretty sure I’ve kept him at his desk longer than he intended more than once! 

I would like to thank a few more people who started as professional connections but have become some of my closest friends and greatest motivators. Subhayan Roy Moulik introduced me to quantum cryptography in 2014 and has cheered me on at every major milestone in my career. Whenever I’m weighed down by self-doubt as a researcher, I know that a call with Subhayan will leave me ready to take on whatever comes next. Another person who has been a constant source of motivation is Vijay Singh, my professor during undergraduate studies. VS has always encouraged me to keep producing good work, asking me to send him the next paper I’ve written and sharing his latest updates on devising questions for physics Olympiads. He embodies the kind of teacher and physicist I deeply admire — thoughtful, passionate, and always excited about physics. I cherish the fact that I still get to discuss both physics and life with someone of his caliber. This list could go on — many more names, though not mentioned here, are deeply acknowledged for their role in my professional journey.

Some people deserve gratitude on a more personal level for being there in a non-professional capacity. My aunt's family — Ashish Rai, Priya Rai, and Noor A. Rai — have been my home away from home in New York. Their unwavering support has been invaluable during the intense years of this PhD. NAR and AR are researchers themselves, in addition to being doctors, and spending time with them has taught me a lot about maintaining a healthy work-life balance. My niece, Asha, feels as crucial to this thesis as the projects themselves. Asha is, in fact, as old as the number of years I’ve spent on this PhD. Watching her grow up has been a constant source of much-needed joy and perspective throughout this journey.

At last, I arrive at the most cherished part of this acknowledgment—expressing my gratitude to the people who are the foundation of everything I do: my parents and siblings. My father has been my first and fiercest believer, nurturing my curiosity from the earliest days and urging my siblings and me to be both grounded and fearless—to keep our feet on the ground even as we reached for the stars. My mother is perhaps the sweetest and the most understanding person I know. She has a gift for filtering out the noise in
life and helping us maintain a positive outlook, even during difficult times. Her positivity
radiates through all our lives, helping us move past even the hardest moments. My younger brother, Deep, is the person for whom I’ve always wanted to make things work. My elder sister, Smriti, took on an enormous burden during our father’s illness, so that I could devote myself wholly to this dissertation. I cannot express how important both
of them have been in helping me reach this point. My brother-in-law, Prateek, is one of the most pragmatic people I have met. Amusingly—and quite inspiringly—his words of wisdom, “Intelligence lies in realizing that you’re not going to solve all the world’s problems, so you must choose a few that matter to you and work toward solving them well,” helped me come to terms with narrowing down the number of papers I aimed to finish before graduating. Finally, I must thank my partner, Akshay Koottandavida, who has been my unwavering support system. Whether it was patiently listening to my ramblings or making sure I ate during sleepless deadline-driven nights, Akshay stood beside me every step of the way. I simply could not have done this without him.

    \newpage
\thispagestyle{empty} %no page number
\addtocounter{page}{-1} %ignore this page when counting
\begin{flushright}
\topskip0pt
\vspace*{\fill}
This dissertation is dedicated to my father, Yatindra Singh Chaudhary.
\vspace*{\fill}
\end{flushright}

    \tableofcontents % 
    \listoffigures   % The standard LaTeX form works well.
    \listoftables    % 
    
    \mainmatter      % Reset the page numbering options

\chapter{Introduction} \label{chapter:introduction}
Quantum computing is at the forefront of 21st-century scientific and technological advancements, with promising advantages over classical computing in tasks such as factorization, universally secure communication, and quantum simulation (e.g., the Bose-Hubbard model). Various platforms are being explored to realize these machines, including superconducting circuits~\cite{Blais2004,Blais2020,BlaiscQEDReviewRMP2020}, trapped ions~\cite{bruzewicz2019trapped}, neutral atoms~\cite{Schlosser2001,sortais2007,Kaufman2012}, NV centers~\cite{pezzagna2021quantum}, photonics~\cite{abughanem2024photonic,slussarenko2019photonic,duan2004scalable,o2009photonic}, and acoustics~\cite{barnes2000quantum,satzinger2018quantum,hann2019hardware,yang2024mechanical}. This thesis examines the practical limitations of these architectures in building a fully functional quantum computer that outperforms classical systems. To do this, we need to explore the various types of quantum systems, their fundamental principles, and their unique advantages. To develop an understanding of quantum computation, the first question one needs to ask is, \textit{what is a unit of a (quantum) computer?} 

The fundamental unit of a quantum computer falls into two categories: discrete variable (DV) and continuous variable (CV). A classical computer operates on bits, which exist in one of two states: ``off" ($0$) or ``on" ($1$). In contrast, quantum systems are built on units that exist in superpositions of multiple states. These states are not just binary numbers but vector-like objects, manipulated through matrix operations.
When a quantum system is described by two states, its fundamental unit is a \emph{qubit}. For a system with $d$ levels, the unit is called a \emph{qudit}. These constitute discrete variable quantum systems. More exotic systems, with infinitely many levels, are known as \emph{oscillators} and possess a quantum phase space where operators like position and momentum are defined. Because the position and momentum operators have continuous eigenvalues, such systems are referred to as continuous variable quantum systems.

Naturally, the next question is: \textit{How can we engineer such systems?} Despite rapid advancements across various platforms, it remains unclear which one will ultimately enable a scalable quantum computer. \textit{What would the final machine look like?}
It could be a standalone system built on a single platform or a hybrid design integrating the best components from multiple technologies. For example, superconducting qubits, implemented using Josephson junctions, inductors, and capacitors~\cite{BlaiscQEDReviewRMP2020}, offer fast gate operations but suffer from limited coherence times ($\sim \mathcal{O}(100) \mu\textrm{s}$) and primarily local connectivity~\cite{acharya2024quantum}. Trapped ions, on the other hand, provide long coherence times but have slower gates, leading to comparable error rates over time. However, their architecture allows qubit swapping, enabling more flexible connectivity~\cite{ryan2021realization}. Neutral atoms, confined by optical tweezers, can be physically moved to achieve long-range connectivity, though this process is slow and vulnerable to decoherence and atom loss~\cite{bluvstein2022quantum}. In contrast, photonic platforms face challenges in realizing nonlinear operations—an essential ingredient for universal quantum computation~\cite{aghaee2025scaling}.

In practice, we approximate oscillators using finite-energy systems. Details on how such truncation of the Hilbert space~\footnote{A complete vector space equipped with an inner product that allows the measurement of angles and distances between vectors.} affects oscillator physics is discussed in Chapter~\ref{chapter: paper0}. One way to realize an oscillator is through the electromagnetic modes of a 3D superconducting cavity. These cavities can achieve lifetimes up to a second, though the necessity to control them by coupling to a non-linear element (e.g., a superconducting qubit) can reduce the lifetimes to $\sim 1$–$35$ ms \cite{Reagor_memory_2016,milul2023superconducting}, still significantly longer than typical superconducting qubits. This low loss stems from the 3D cavity design, and efforts are underway to replicate this feature in quasi-2D platforms~\cite{ganjam2024surpassing} to support planar quantum computing architectures. The primary limitation of these systems is photon loss. In ion-based platforms, oscillators are realized in the motional modes of trapped ions, though they suffer from motional heating~\cite{bruzewicz2019trapped,de2022error,fluhmann2019encoding}. Other experimental platforms~\cite{ISA,brown2022,scholl2023} for realizing oscillators remain in early development. Among all platforms, superconducting circuits offer the greatest flexibility in engineering different types of qubits, each with unique advantages. Consequently, this thesis benchmarks its results using parameters from superconducting circuits. A brief discussion of CV-DV systems in superconducting systems is provided in  App.~\ref{App:Phys_Imp}.

%Types of quantum architectures
The next question that comes to mind is: \textit{What does a large-scale quantum computing architecture look like?} A quantum computing architecture can be built using DV systems (qubits), CV systems (oscillators), or a hybrid system (coupled oscillator-qubits). A computer should be reliable despite environmental noise, commonly referred to as ``errors'' or ``noise.'' The goal of reliable quantum computers is to perform useful computation tolerant to such noise in a practical manner, that is, enable fault-tolerant quantum computing (FTQC). Qubit-based architectures are the most experimentally developed due to their relative simplicity, showing significant progress toward scalable, fault-tolerant quantum computation~\cite{acharya2024quantum,ryan2021realization,bluvstein2022quantum,aghaee2025scaling}. More recently, qudit-based systems have gained attention for their potential to encode quantum information more efficiently and protect against errors~\cite{brock2024quantum,yu2025schrodinger,omanakuttan2024fault,wetherbee2024mathematical}.
Oscillator-based architectures, while promising for simulating spin-boson dynamics, lattice gauge theories, and other complex quantum phenomena, face significant experimental challenges. Controlling an oscillator’s infinite-level structure with high fidelity remains difficult, and no theoretical proposal has yet demonstrated a scalable, fault-tolerant oscillator-based quantum computer~\cite{noh2020encoding,hanggli2021oscillator}.
This dissertation explores the third, more unconventional approach: a hybrid CV-DV architecture~\cite{ISA}. 

\section{Pedagogical Outline}
We now turn to the critical questions of errors and control for this esoteric architecture: \textit{What are the different use cases of hybrid CV-DV systems? How do we protect the quantum computing units in such an architecture from noise with affordable overhead? Can we control them while maintaining the protection?}
While answering these questions requires considerable effort, this thesis aims to address some of them. Our focus is on achieving control of the CV-DV architecture, described in detail in Chapter~\ref{chapter: paper0}. The use cases of such an architecture can be classified by focusing on the unit of computation visible to an abstract user or algorithm. We will focus on two classes of applications classified under user-visible abstract machine models (AMMs)~\cite{ISA} for hybrid CV-DV quantum computing. The first use case is the oscillator-centric AMM, where oscillators serve as the primary computing units. In this model, oscillators are used \emph{qua} oscillators to run CV quantum algorithms or quantum simulations, with CV-DV operations enabling control. Techniques for controlling oscillators via DV ancillae are discussed in Chapters~\ref{chapter:na-qsp},~\ref{chapter:state-prep}, and~\ref{AOM}. The second use case is the qubit-centric AMM, where DV systems are the primary computing units. For this thesis, this AMM corresponds to abstracting oscillators as DV units through bosonic encoding~\cite{albert2018performance,gottesman_thesis_1997stabilizer,mirrahimi2014dynamically,michael2016new} to reduce the space-time overhead for practical FTQC~\cite{sivak2023real}. In this model, the goal of the CV-DV system is to engineer a low-error subspace using error correction, requiring high-fidelity design and control. These techniques are explored in Chapters~\ref{chapter:GKP-qec},~\ref{chapter:qec-control}, and~\ref{AQM}. The foundation of this dissertation is a novel theoretical framework, ``non-abelian quantum signal processing (NA-QSP)", designed to orchestrate control in hybrid CV-DV architectures. By leveraging NA-QSP for error correction and control of ``Gottesman-Kitaev-Preskill (GKP)" codes, we outline a path toward high-fidelity qudit-based quantum computing.

The fundamentals of CV systems are subtle, as their continuous nature is non-intuitive compared to qudits or DV systems. In Chapter~\ref{chapter: paper0}, we review the foundations of state space, representation, and operations in a CV-DV architecture. We compare DV and CV systems through the description of states and operators. Here, we move beyond known results and introduce the \emph{CV hierarchy of operations} for continuous variable operations, which we contrast with the existing Clifford hierarchy for DV (qubit) operations. Finally, we introduce the instruction set formalism~\cite{ISA} for CV-DV architectures—finite sets of parametrized operations sufficient for universality in these hybrid systems. These operations have been used in experiments for error correction and quantum simulation using oscillators but have not been formalized as instruction sets. Among the available instruction sets, we focus primarily on the phase-space instruction set, which captures the continuously variable nature of the oscillator in its simplest form. We will demonstrate how unconventional operators in this instruction set can be used to control CV systems with DV ancillae.

In Chapter~\ref{chapter:na-qsp}, we explore the control theory of quantum systems using quantum signal processing (QSP) and composite pulse sequences~\cite{martyn2021grand,low2016methodology,low2017optimal,liu2016power}. QSP generalizes composite pulse sequences to reduce errors in control parameters ($\theta$) for qubit rotations ($R_\phi(\theta)$). This chapter extends the discussion to our theory of NA-QSP~\cite{singh2024towards}, a novel class of quantum signal processing where the control parameters are non-commuting quantum operators in the oscillator phase space, specifically position, and momentum ($\hat \theta = f(\hat x, \hat p)$). The non-commuting nature of these control parameters in hybrid operations, such as position or momentum-controlled qubit rotations $R_\phi(\hat \theta)$, makes them more efficient than traditional composite pulse sequences like BB1(90)\cite{wimperis1994broadband,singh2024towards}. As a key contribution, this chapter introduces the first composite pulse sequence within the non-abelian QSP class, which we developed in Ref.\cite{singh2024towards}. In Chapter~\ref{chapter:na-qsp}, we compare the performance of this sequence, called the ``Gaussian-controlled-rotation (GCR),'' to the more traditional abelian composite pulse sequence, BB1(90) in circuit depth with on-par efficiency in canceling systematic errors.

In Chapter~\ref{chapter:state-prep}, we demonstrate applications of our control sequence GCR, introduced in the previous chapter, for oscillators. We show how to deterministically prepare oscillator states essential for quantum simulations and error correction. Using GCR, we analytically derive schemes to prepare squeezed states, two-legged cat states, GKP states, four-legged cat states, and Fock state $\ket{1}$. We define and explain the significance of each of these states. Techniques developed in this chapter can be used for either AMM. Since the preparation is deterministic (in the absence of errors), ancilla measurements can be used to detect oscillator and qubit errors.  We also show that our state preparation schemes perform on par with the state-of-the-art numerical methods in the literature, without the need for any numerical optimizer, thus, reducing hardware requirements and classical processing costs for control. Finally, we explore the generalization to universal oscillator state preparation. While this may not be critical for high-fidelity control of oscillators, it is important for proving the universality of the phase-space instruction set. 

In Chapter~\ref{chapter:GKP-qec}, we explore the primitives for the FTQC stage and introduce a novel error correction strategy that differs from the conventional stabilizer-based approach. In the stabilizer formalism, corrections are deterministically applied based on error syndrome information. In contrast, we present a new concept of probabilistic error correction based on Ref.~\cite{singh2024gkp_qec}. This probabilistic error correction scheme is effectively understood within the framework of non-abelian QSP. We quantify a `probabilistic distance' for this error correction across different GKP lattices and average photon numbers used in the code space design. Additionally, our methods can be generalized to examine how this error correction distance changes as we encode qudits. While GKP codes were initially designed to correct displacement errors in oscillators, our work offers the first analytical explanation for how the stabilization scheme used in beyond-break-even GKP experiments addresses the photon loss channel from an error correction perspective. This provides a foundation for beyond-break-even quantum error correction (QEC) in systems with dimensions $d=2$\cite{campagne2020quantum, sivak2023real} and ${3,4}$\cite{brock2024quantum}.

In Chapter~\ref{chapter:qec-control}, we explore methods for controlling an error-corrected qudit encoded in an oscillator. Before this thesis, there has been considerable research to engineer gates that are transparent to errors~\cite{ReinholdErrorCorrectedGates,ma2020path,ma2022algebraic}, but the `error-corrected' control of such qudits had remained largely unexplored. The original proposal for GKP codes~\cite{gottesman2001encoding} suggested methods for performing operations on the ideal infinite-energy GKP code states. Subsequent works~\cite{hastrup2021unsuitability, rojkov2023two} showed that these operations have low fidelity for practical realizations that are bounded in energy, even in the absence of errors. Moreover, methods for preparing these complicated states were not discussed in the original proposal. While there have been advancements in optics, the main approach for superconducting circuits and trapped ions has been low-fidelity state preparation through logical measurements of stabilized codewords~\cite{campagne2020quantum, fluhmann2019encoding} or large circuit-depth numerical optimizations~\cite{eickbusch2022fast, sivak2023real, brock2024quantum, hastrup2021measurement}. This chapter uses an analytical preparation that allows for high-fidelity state preparation via post-selection, even in the presence of errors. We also introduce error-corrected single- and two-qubit rotations, which we predict will achieve extremely high fidelity and improve upon previously developed schemes~\cite{rojkov2023two}. We prove how these operations are protected against oscillator errors in CV systems analytically, and confirm our results using numerics. To safeguard against errors in the ancillary DV systems, we designed a pieceable circuit that serves as one of the core findings of this chapter. We also present a framework that offers an analytical understanding of stabilization and readout circuits. Finally, we use traditional QSP pulses to improve the readout fidelity of GKP qubits at the end of a circuit, even in the presence of residual correctable errors. All techniques discussed in this chapter apply to arbitrary qubits and qudit lattices in the GKP encoding.

In Chapter~\ref{chapter:conc}, we explore the applications and prospects of CV-DV architectures. In particular, we describe our work~\cite{singh2022high} where we achieve a significant reduction in the space-time overhead of fault-tolerant quantum computing using bosonic codes. In addition, we offer some concluding insights into open problems related to the concatenation of GKP codes with a scalable DV code, quantum phase estimation, and quantum random walks.  

The structure of this thesis follows a pedagogical approach, with an open question posed at the end of each chapter. We hope this thesis serves not only as a guide to hybrid CV-DV quantum computing, control of oscillator-based qubit and qudit systems, and non-Clifford operations for scalable codes but also as a roadmap for the various options ahead. It aims to highlight the key questions that must be addressed in determining the path forward for enabling useful quantum computations.
\section{Reader's Guide and Author Contributions}
\begin{figure}[htb]
    \centering
    \includegraphics[width=\linewidth]{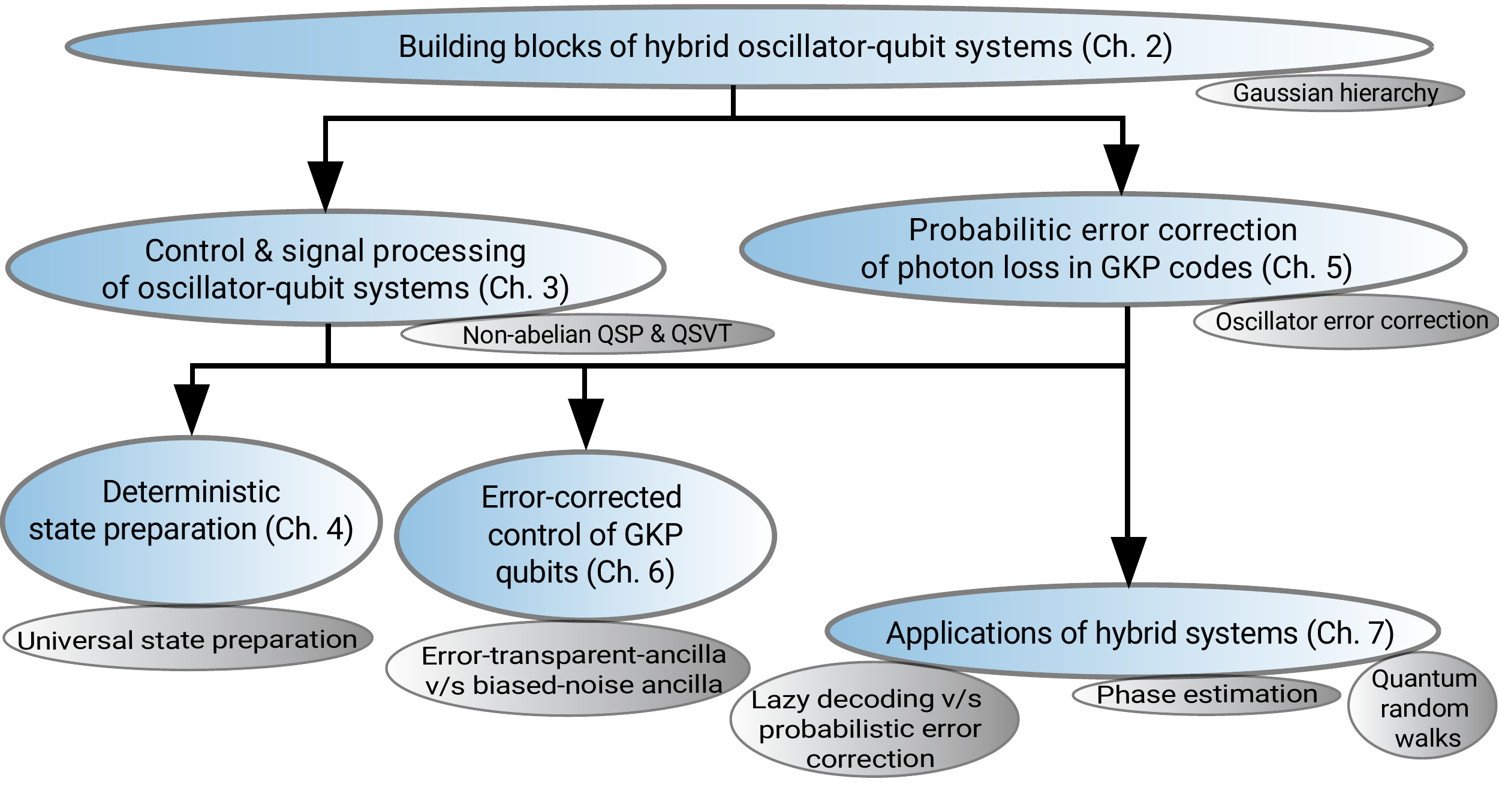}
    \caption[Thesis flowchart]{\textbf{Thesis flowchart.} The blue bubbles in the flowchart represent the key topics covered in this thesis and their interconnections. The gray bubbles attached to each chapter highlight intriguing discussion points related to the corresponding topic.}
    \label{fig:flowchart}
\end{figure}
Chapter~\ref{chapter: paper0} introduces the building blocks of hybrid systems, covering DV and CV state spaces, quantum channels, and operations. It highlights the absence of a structured hierarchy for CV operations comparable to the Clifford hierarchy for qubits and proposes an open problem to define such a framework for potential classification, which we call `Gaussian Hierarchy'. The chapter concludes with an overview of hybrid CV-DV systems, outlining their potential advantages for fault-tolerant quantum computation. Some contents of this chapter and App.~\ref{App:Phys_Imp} are based on,

\singlespacing
\begin{itemize}
    \item Y. Liu$^*$, \textbf{S. Singh}$^*$, K. C. Smith$^*$, E. Crane, J. M. Martyn, A. Eickbusch, A. Schuckert, R. D. Li, J. Sinanan-Singh, M. B. Soley, et al., Hybrid oscillator-qubit quantum processors: Instruction set architectures, abstract machine models, and applications, arXiv preprint arXiv:2407.10381 (2024). \textit{`*' marked authors contributed equally and arranged alphabetically.}
\end{itemize}
\doublespacing

Chapter~\ref{chapter:na-qsp} develops a framework for controlling CV systems using DV ancillae, introducing Gaussian-Controlled-Rotation (GCR), a non-abelian composite pulse technique. GCR extends quantum signal processing (QSP) to non-commuting control parameters, enabling more robust and precise quantum gates. An open problem is posed regarding the generalization of QSP and quantum singular value transformation (QSVT) to non-abelian settings. Chapter~\ref{chapter:state-prep} addresses deterministic oscillator state preparation, including squeezed states, two-legged cat states, and Gottesman-Kitaev-Preskill (GKP) states. It examines the challenges in preparing arbitrary oscillator states and poses the open problem of achieving universal state preparation in hybrid systems. Chapter~\ref{chapter:qec-control} explores high-fidelity control of an error-corrected qudit in an oscillator. It extends GKP-based protocols to non-abelian QSP frameworks and develops logical readout strategies with residual error mitigation. The chapter introduces pieceable gate teleportation for universal qubit rotations and extends GCR to multi-modal operations, presenting an open problem on ancilla-error-transparent conditional displacement gates. All contents in these chapters and Apps.~\ref{app:comp_err},~\ref{app:state-prep} and~\ref{app:qec-control} are based on,

\singlespacing
\begin{itemize}
    \item \textbf{S. Singh}, B. Royer, S. M. Girvin, Towards Non-Abelian Quantum Signal Processing: Efficient Control of Hybrid Continuous- and Discrete-Variable Architectures, arXiv:2504.19992 [quant-ph] (Apr. 2025).
\end{itemize}

\doublespacing
Chapter~\ref{chapter:GKP-qec} focuses on probabilistic error correction of photon loss using finite-energy GKP codes. It introduces the concept of probabilistic distance for quantifying error correction efficiency and explores the autonomous stabilization of GKP codes. Numerical comparisons of different error correction protocols are presented, with an open problem of designing protected qubits and oscillators.
All contents in this chapter and App.~\ref{app:GKp-qec} are based on,

\singlespacing
\begin{itemize}
\item \textbf{S. Singh}, S. Girvin, B. Royer, Error correction of photon loss using GKP states, In preparation (2024). (unpublished)
\item B. Royer, \textbf{S. Singh}, S. M. Girvin, Stabilization of Finite-Energy Gottesman-Kitaev-Preskill States, Phys. Rev. Lett. 125 (2020) 260509.
\item V. Sivak, A. Eickbusch, B. Royer, \textbf{S. Singh}, I. Tsioutsios, S. Ganjam, A. Miano,
B. Brock, A. Ding, L. Frunzio, et al., Real-time quantum error correction beyond break-even, Nature 616 (7955) (2023) 50–55.
\item B. L. Brock, \textbf{S. Singh}, A. Eickbusch, V. V. Sivak, A. Z. Ding, L. Frunzio, S. M.
Girvin, M. H. Devoret, Quantum error correction of qudits beyond break-even, arXiv preprint arXiv:2409.15065 (Accepted in Nature 2025)
%\item B. Royer, \textbf{S. Singh}, S. M. Girvin, Encoding qubits in multimode grid states, PRX Quantum 3 (2022) 010335.
\item A. J. Brady, A. Eickbusch, \textbf{S. Singh}, J. Wu, Q. Zhuang, Advances in bosonic quantum error correction with Gottesman–Kitaev-Preskill codes: Theory, engineering, and applications, Progress in Quantum Electronics (2024) 100496. \textit{(authors arranged alphabetically)}
\end{itemize}

\doublespacing
Chapter~\ref{chapter:conc} discusses applications of oscillators in resource-overhead reduction for fault-tolerance using CV-DV concatenation and gives some future prospects. Some contents in this chapter are based on,

\singlespacing
\begin{itemize}
    \item \textbf{S. Singh}, A. S. Darmawan, B. J. Brown, S. Puri, High-fidelity magic-state preparation with a biased-noise architecture, Physical Review A 105 (5) (2022) 052410.
    \item \textbf{S. Singh}, B. Royer, S. M. Girvin, Towards Non-Abelian Quantum Signal Processing: Efficient Control of Hybrid Continuous- and Discrete-Variable Architectures, arXiv:2504.19992 [quant-ph] (Apr. 2025).
\end{itemize}

\doublespacing
Finally, Chapter~\ref{sec:summary} gives the list of all the open questions we pose throughout this thesis. 

	\chapter{Building Blocks of Hybrid Discrete- and Continuous- Variable Quantum Systems} \label{chapter: paper0}

\begin{myframe}
\singlespacing
\begin{quote}
     \textit{Why do we care about hybrid CV-DV architectures?} The continuous variable (CV) formalism based on position and momentum is an alternative to the Fock space description of oscillators in terms of its countable infinity of integer excitation numbers. Such systems can be truly essential in simulating certain quantum phenomena. However, modifying these systems in any desirable manner, collectively known as ``universal control," is extremely hard to engineer with minimal faults natively. A hybrid CV-DV architecture paves the path to utilize CV systems in their full capacity with the help of qubits or qudits as potential ancillary sources of control.
\end{quote}
\end{myframe}

\doublespacing

In this chapter, we describe the basics of different units of a hybrid CV-DV quantum computing architecture. We start from the simplest and explain what a DV system is, in Sec.~\ref{DV}, and we then move on to CV systems in Sec.~\ref{CV}. We have a special Sec.~\ref{open-hierarchy} on the classification of CV operations, which is an open problem. We highlight why this might be interesting by drawing an analogy to its DV counterpart. Finally, in Sec.~\ref{hybrid}, we discuss the architecture of this hybrid quantum computing platform via the available operations, instruction sets, and its corresponding quantum computing architecture stack. This section has some overlap with our publicly available work on hybrid CV-DV systems~\cite{ISA}.

\section{Discrete Variable (DV) Systems}\label{DV}
In a hybrid CV-DV architecture, discrete-variable (DV) systems are nonlinear systems with typically lower coherence times that can act as controllers to unlock quantum advantage~\cite{brenner2024factoring} within the CV framework. For example, in the superconducting platform, the shorter-lived transmon qudits have proven to be a convenient source of fast universal control of microwave resonators~\cite{eickbusch2022fast,ReinholdErrorCorrectedGates}. 
\begin{figure}[htb]
        \centering
        \includegraphics[width=\linewidth]{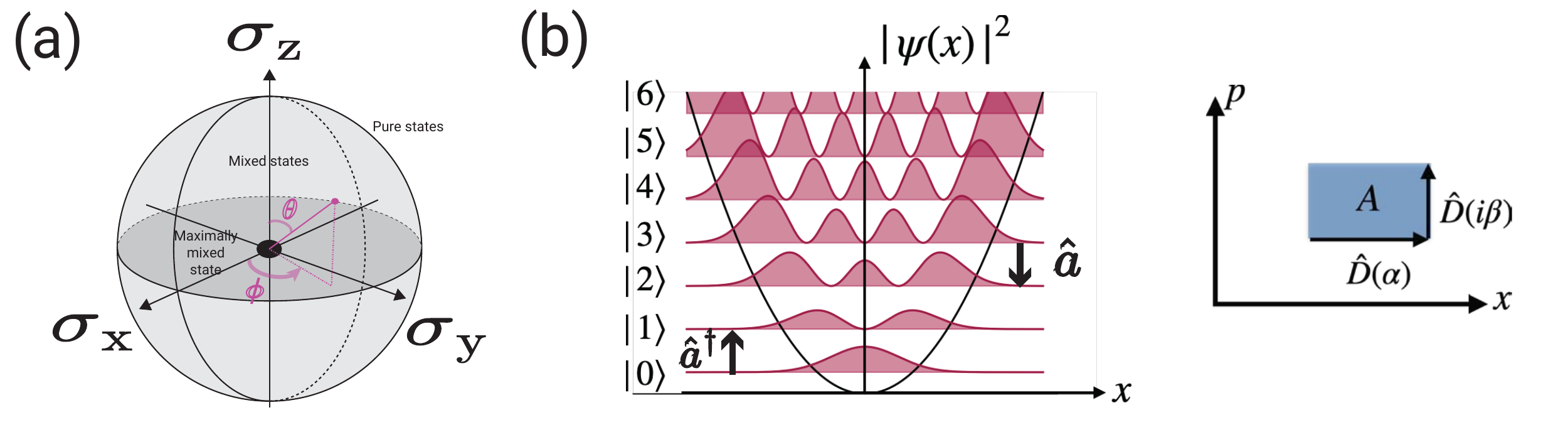}
        \caption[Pictorial representation of hybrid CV-DV space.]{\textbf{The hybrid CV-DV space.} (a) Qubit Bloch sphere (see Sec.~\ref{DV}). The Bloch sphere represents the state space of qubits. Pure quantum states lie on the surface of the 3D object (i.e. on the unit 2-sphere) while the mixed states lie inside it (i.e., the solid ball). The three axes denote the eigenstates (or `basis states') of the Pauli operators which lie on the anti-nodal points. For example, the eigenstates of the $\sigma_\mathrm{z}$ Pauli operator lie on the north and south poles, while the eigenstates of $\sigma_\mathrm{x},\sigma_\mathrm{y}$ Pauli operators lie on the equator. The maximally mixed states lie at the center and are universal across all bases. See Sec.~\ref{DV_gates} for definitions of Pauli operators. (b) Harmonic oscillator. The various levels in the quadratic potential denote the Fock states $\ket{n}$ (see Sec.~\ref{sec:fock_basis}). The lowest state is the vacuum state $\ket{0}$. The x-axis denotes the position of the oscillator. The wave functions in pink denote the state representation in the position basis $|\psi(x)|^2$. On the right, we present the phase space of the oscillator where $p,x$ denote its position and momentum, respectively. The displacements in these two directions do not commute, and their group commutator is given by A, $\hat D(\alpha)\hat D(i\beta)=e^{2iA}\hat D(i\beta)\hat D(\alpha)$. Displacement of a vacuum state in phase space yields a coherent state. See Sec.~\ref{CV} for details.}
        \label{fig:bloch-sphere}
    \end{figure}
\subsection{DV State Space}\label{DV_states}
DV quantum systems can be represented by a finite set of states, known as the basis states. The number of independent basis states $d$ gives the dimension of the DV system, or qu$d$it. Quantum states are described using mathematical objects called spinors represented using a column vector with complex elements. For example, the computational states are given by,
\begin{equation}
\begin{pmatrix} 1 \\ 0 \\ 0 \\ \vdots \\ 0 \end{pmatrix} = \ket{0}, \begin{pmatrix} 0 \\ 1 \\ 0 \\ \vdots \\ 0 \end{pmatrix} = \ket{1}, \begin{pmatrix} 0 \\ 0 \\ 1 \\ \vdots \\ 0 \end{pmatrix} = \ket{2}, \ldots, \begin{pmatrix} 0 \\ 0 \\ 0 \\ \vdots \\ 1 \end{pmatrix} = \ket{d-1},
\end{equation}
where $\ket{i}$ is the shorthand Dirac-ket notation. The dual vectors are written as the corresponding row vectors represented by the Dirac-bra notation $\bra{i}$. The simplest example of a DV system is a qubit, which is a two-level quantum system. In reality, no physical qubit is truly a two-level system. Details of various physical implementations are given in App.~\ref{App:Phys_Imp}. 

\paragraph{Completeness relation: } The outer product of two basis vectors is denoted by $\ket{i}\bra{j}$, which is a $d$-by-$d$ matrix with all elements zero except the $(i,j)$ element which is $1$. It is important for any set of complete orthonormal basis vectors $\{\ket{i}\}$ defined over a $d$-dimensional space to satisfy,
\begin{equation}
    \sum_{i=0}^{d-1}\ket{i}\bra{i} = I_d,\label{eq:resolution-identity}
\end{equation}
where $I_d$ is the $d$-by-$d$ identity matrix. This equation is known as the resolution of identity or completeness relation. It is a fundamental property used to express any quantum state in terms of the basis vectors as $\ket{\psi} = \sum_{i=0}^{d-1}c_i\ket{i}$, where $c_i = \braket{i|\psi}$ are the coefficients of the state $\ket{\psi}$ in the basis $\{\ket{i}\}$. The resolution of identity is also used to express any quantum operator in terms of the basis vectors as $\hat{O} = \sum_{i,j=0}^{d-1}O_{ij}\ket{i}\bra{j}$, where $O_{ij} = \braket{i|\hat{O}|j}$ are the matrix elements of the operator $\hat{O}$ in the basis $\{\ket{i}\}$.

\paragraph{Orthonormal basis vectors: } The inner product of two basis vectors is denoted by $\braket{i|j}$  and satisfies the orthonormality relation, $\braket{i|j} = \delta_{ij}$, where $\delta_{ij}$ is the Kronecker delta function. The norm of an arbitrary quantum state $\ket{\psi}$ is given by $\braket{\psi|\psi}=\sum_i |c_i|^2=1$. The inner product represents the overlap between any two states.  The modulus squared of the overlap between a given approximate state and a target state gives the fidelity of the approximate state $\ket{\psi}$ to the target state $\ket{\phi}$.
\begin{equation}
\mathcal{F}=|\braket{\psi|\phi}|^2, \label{eq:dv-fidelity}
\end{equation}
a measure of the closeness of two states. $\mathcal{F}$ is a dimensionless number between $0$ and $1$, with $0$ indicating orthogonal states while $1$ indicates maximum overlap, i.e., the two states are the same and contain the same quantum information. This definition of the inner product renders a unit modulus complex number $e^{i\phi}$ (called the global phase) multiplied to any quantum state completely insignificant.

\paragraph{Density matrix representation and purity:} For the representation of an ensemble of quantum systems, a density matrix representation of quantum states comes in handy. Somewhat confusingly, density matrices are often referred to as quantum states even though they represent ensembles of systems. In this representation, a quantum state is given by $\rho = \sum_{i,j=0}^{d-1}\rho_{ij}\ket{i}\bra{j}$, where $\rho_{ij} = \braket{i|\rho|j}$ are the matrix elements of the density matrix $\rho$ in the basis $\{\ket{i}\}$. The density matrix is a Hermitian, positive semi-definite matrix with a unit trace that can be used to find the ensemble average of any observable $A$ via $\langle A\rangle=\mathrm{Tr}\left[A\rho\right]$. The diagonal elements of the density matrix are the probabilities of measuring the state in the corresponding basis state, while the off-diagonal elements are the coherences between the basis states. In the density matrix representation, the fidelity between two states (say, density matrices $\rho,\nu$) is given by the Ulhmann's theorem~\cite{jozsa_fidelity_1994}, 
\begin{equation}
\mathcal{F}_d=\big[\mathrm{Tr}(\sqrt{\sqrt{\rho}\nu\sqrt{\rho}})\big]^2
\end{equation}
The \emph{purity} of $\rho$ is $\mathrm{Tr}(\rho^2)$. This quantity is a dimensionless number between $0$ and $1$, with $0$ representing a completely mixed state and $1$ representing a pure state. This definition comes in handy when discussing \emph{entangled} quantum systems, an important quantum mechanical feature. For computation, the superposition of multiple levels in a single qudit system and entanglement of $n$ $m$-dimensional quantum systems are, in fact, synonymous in representing $m^n=d$-dimensional Hilbert space. However, in the representation of multiple quantum systems, an allowed feature is considering a single system irrespective of the other quantum systems (which can be `traced out') in the Hilbert space. If such a standalone representation has a purity of $1$ it is called a pure state. However, if the purity is less than $1$, it is called a mixed state. Now we will discuss what it means to be in a mixed state.

\paragraph{Partial trace and entanglement:} Let us consider a quantum state of two disjoint qubit systems represented using a tensor product as,
\begin{equation}
    \ket{\psi}=\ket{+}\otimes\ket{+}= \frac{(\ket{0}+\ket{1})}{\sqrt{2}}\otimes\frac{(\ket{0}+\ket{1})}{\sqrt{2}}=\frac{1}{2}(\ket{00}+\ket{01}+\ket{10}+\ket{11}).
\end{equation}
This state can be represented as the superposition of four computational basis states in the Hilbert space of a two-qubit system, and alternatively can be factorized into a tensor product of two pure states $\ket{+}$ with unit purity. Thus, the two systems are unentangled. Now, let us consider the two-qubit state,
\begin{equation}
   \ket{\psi}=\frac{1}{\sqrt{2}}(\ket{00}+\ket{11}).\label{eq:bell_state}
\end{equation}
This state cannot be factorized into a tensor product of states in two quantum systems. This is an example of an entangled state. This is a special case of maximally entangled states, called the Bell states. A maximally entangled state collapses onto a known quantum state when one of the systems is measured. To quantify this notion, we use the concept of partial trace. The partial trace of a quantum state $\rho_{AB}$ over one of the subsystems is given by,  
\begin{equation}
   \rho_A = \mathrm{Tr}_B(\rho_{AB})=\sum_i{}_B\bra{i}\rho_{AB}\ket{i}_B,
\end{equation}  
where $\{\ket{i}_B\}$ is any orthonormal basis in the Hilbert space of subsystem $B$. Here, $\rho_A$ is called the reduced density matrix of the subsystem $A$ which is obtained by ``tracing out" the subsystem $B$ from the density matrix $\rho_{AB}$ of the composite system $AB$. The purity of the reduced density matrix $\rho_A=\mathrm{Tr}(\rho^2)$ and the von Neumann entropy is given by $\rho_A=\mathrm{Tr}(\rho\log\rho)$. These quantities are popular measures of the entanglement between the two subsystems. 

The two-qubit entangled state given in Eq.~(\ref{eq:bell_state}) given is also known as the Bell pair. These states have maximal entanglement in that, partial trace on either subsystems yields a maximally mixed state $\rho_\mathrm{max}=\frac{\mathrm{I}}{2}$. Such states cannot be constructed using only rotations on the Bloch sphere. The maximally mixed state, for a $d$-level quantum system, is,
\begin{equation}
    \rho_{\text{max}}=\frac{1}{d}\mathrm{I}_d,
   \end{equation}
is called so because it has the same representation in all bases. 
\paragraph{Visual representation:} The cases $d=2,3,4,..$ are referred to as qubits, qutrits, ququarts, etc., respectively. Qubits have a graphical representation, called the Bloch sphere  (see Fig.~\ref{fig:bloch-sphere}(a)). Note that, mathematically, the term `sphere' denotes the surface of the three-dimensional figure while the term `ball' denotes the inside of it. Thus, while the figure looks like a 3D sphere in space, it is, in fact, a unit 2-sphere (i.e. surface of the 3D object in the figure with a unit radius). The anti-nodal points on the Bloch sphere represent orthogonal spinors or quantum states. The pure quantum states reside on the `Bloch sphere' while the mixed states reside in the `Bloch ball.' For example, the maximally mixed state state lies at the center of the Bloch ball. Collectively, we can write the density matrix of an arbitrary quantum state as,
\begin{align}
    \rho=\frac{\mathrm{I}}{2}+\vec n\cdot\vec{\sigma},\label{density_matrix}
\end{align}
where $\vec{n}=\{n_\mathrm{x},n_\mathrm{y},n_\mathrm{z}\}$ is a Bloch vector whose endpoint within the sphere denotes a quantum state. The $\vec{\sigma}=\{\sigma_\mathrm{x},\sigma_\mathrm{y},\sigma_\mathrm{z}\}$ is the vector form of Pauli matrices (see Sec.~\ref{DV_gates}). If $|n|=1$, that is $\vec{n}\equiv\hat n$ is a unit vector, then $\rho$ is a pure state while $|n|<1$ indicates that $\rho$ is a mixed state. For a pure quantum state, the unit vector $\hat n$ is represented by the polar and azimuthal angles $(\theta,\phi)$ such that,
\begin{align}
    n_\mathrm{x}=\sin{\theta}\cos{\phi},
    n_\mathrm{y}=\sin{\theta}\sin{\phi},
    n_\mathrm{z}=\cos{\theta}.
\end{align}
Thus, an arbitrary pure quantum state takes the form,
\begin{equation}
    \ket{\psi}=\cos(\theta/2)\ket{0}+e^{i\phi}\sin(\theta/2)\ket{1}.
\end{equation}
Note that, $\ket{0},\ket{1}$ are the eigenstates of $\sigma_\mathrm{z}$, forming the computational basis\footnote{ due to its similarity to the binary representation used for classical computation}. The North Pole and South Pole represent these states, respectively. The four cardinal points on the equator represent the four superposition states $\ket{+}=\frac{\ket{0}+\ket{1}}{\sqrt{2}},\ket{+i}=\frac{\ket{0}+i\ket{1}}{\sqrt{2}},\ket{-}=\frac{\ket{0}-\ket{1}}{\sqrt{2}},\ket{-i}=\frac{\ket{0}-i\ket{1}}{\sqrt{2}}$. Here, states $\ket{\pm},\ket{\pm i}$ are eigenstates of $\sigma_\mathrm{x},\sigma_\mathrm{y}$ Pauli operators, respectively. The $\sigma_\textrm{x}$ and $\sigma_\textrm{y}$ bases are known as the Hadamard basis and the $\sigma_\textrm{y}$ basis, respectively. 
A pure quantum state can be transformed into another quantum state via rotations about an axis on the Bloch sphere. We will discuss these operations in detail in the next section. 
 
\subsection{DV Quantum Channels}\label{DV_gates}
A quantum channel describes any physical process that transforms a quantum state (either pure or mixed). It is represented by a completely positive trace-preserving (CPTP) map. The CPTP map is a linear map that preserves the trace of the density matrix and maps positive operators to positive operators. The action of a quantum channel on a quantum state is given by the Kraus representation,
\begin{equation}
   \rho\rightarrow \sum_{i=1}^{N}K_i\rho K_i^{\dagger},\label{eq:kraus}
\end{equation}
where $K_i$ are the not necessarily unitary Kraus operators that satisfy the completeness relation $\sum_{i=1}^{N}K_i^{\dagger}K_i = I$. The Kraus operators are the generalization of the Pauli matrices for quantum channels. The Kraus decomposition may not be unique.

\subsubsection{Unitary Channel}
For pure states $\ket{\psi}=\sum_i c_i\ket{i}$ written in an orthonormal basis $\{\ket{i}\}$, CPTP maps assert the following. The complete positivity imposes that the probabilities $|\braket{i|\psi}|^2=|c_i|^2$ are positive and the trace-preserving property imposes that the sum of these probabilities, $\sum_i |c_i|^2$, is equal to 1. This is achieved by unitary operations which are reversible quantum channels and are represented by a unitary matrix. The action of a unitary operation on a quantum state is given by the unitary transformation,  $\rho\rightarrow U\rho U^{\dagger}$, where $U$ is a unitary operator such that $U^\dagger U=UU^\dagger=I$.

Formally, $N$-qubit quantum states belong to the class of projective complex spaces $\mathbb{CP}^n$ (complex spaces modulo multiplication by complex scalars, since global phase does not matter). For this class of states, the projective unitary group $PU(n)$ (equivalence classes of unitary matrices under multiplication by a constant phase) is exactly equal to the projective special unitary group $\mathrm{PSU}(n)$. Thus, quantum operations on a single-qubit can be spanned using the generators of the $\mathrm{SU}(2)$ group, a special unitary group of $2$-by-$2$ matrices with unit determinant and unit trace. Rotations about two orthogonal axes of the Bloch sphere $R_k(\theta)=e^{-i\frac{\theta}{2}\sigma_k}, \sigma_k\in\{\sigma_\mathrm{x},\sigma_\mathrm{y},\sigma_\mathrm{z}\}$ can generate any $\mathrm{SU}(2)$ rotation of the Bloch sphere. Spinors (quantum states) are considered the fundamental representations of $\mathrm{SU}(2)$.

Mathematically, the relationship between $\mathrm{SU}(2)$ operations and rotations about arbitrary axes of the Bloch sphere is subtle. Even though visual rotations in a Bloch sphere look like rotations in a 3D space, these operations are not equivalent. Rotational symmetries in a 3D sphere are represented by the special orthogonal group $\mathrm{SO}(3)$. There exists a two-to-one homomorphic mapping of the group $\mathrm{SU}(2)$ onto the group $\mathrm{SO}(3)$. If $A\in \mathrm{SU}(2)$ maps onto $R(A)\in \mathrm{SO}(3)$, then $R(A)=R(-A)$. Thus, $\mathrm{SU}(2)$ is a `double cover' of $\mathrm{SO}(3)$, meaning that every rotation in $\mathrm{SO}(3)$ maps to two corresponding transformations in $\mathrm{SU}(2)$. This implies that representations of $\mathrm{SO}(3)$ are also representations of $\mathrm{SU}(2)$ but there are representations in $\mathrm{SU}(2)$ that have no analog in $\mathrm{SO}(3)$. Transformations in $\mathrm{SU}(2)$ act on spinors while rotations in $\mathrm{SO}(3)$ rotate a physical 3D-vector. To conclude the discussion on this difference, rotations on a 3D sphere are represented by the $\mathrm{SO}(3)$ group while rotations on the Bloch sphere are represented by the $\mathrm{SU}(2)$ group.

Now, let us define the different qubit gates to be frequently used in this thesis.
\paragraph{Pauli matrices:} are qubit operators denoted by $\sigma_\textrm{x}$, $\sigma_\textrm{y}$, and $\sigma_\textrm{z}$, which are $2$-by-$2$ traceless matrices that satisfy the commutation relations $[\sigma_i, \sigma_j] = 2i\epsilon_{ijk}\sigma_k$. Here, the cyclic order for determining the Levi-Civita symbol $\epsilon_{ijk}$ is $\sigma_\textrm{x}\rightarrow\sigma_\textrm{y}\rightarrow \sigma_\textrm{z}$. If the indices follow this cyclic order then the value of Levi-Civita symbol is $1$, if they follow the acyclic order the value of the Levi-Civita symbol is $-1$. The Pauli matrices are given by,
\begin{equation}
   \sigma_\textrm{x}=\begin{pmatrix} 0 & 1 \\ 1 & 0 \end{pmatrix}, \sigma_\textrm{y}=\begin{pmatrix} 0 & -i \\ i & 0 \end{pmatrix}, \sigma_\textrm{z}=\begin{pmatrix} 1 & 0 \\ 0 & -1 \end{pmatrix}.
\end{equation}
These operations plus the identity operation
\begin{align}
    \mathrm{I}=\sigma_0=\begin{pmatrix} 1 & 0 \\ 0 & 1 \end{pmatrix}
\end{align}
form a group (taking into account that the matrices may need to be multiplied by $\pm 1$ or $\pm i$) known as the Pauli group which not only serves as the cardinal basis of quantum states (as seen from Eq.~(\ref{density_matrix})) but also as a complete operator basis. That is, all $\mathrm{SU}(2)$ operators can be represented as a sum of these operators,
\begin{equation}
A=\sum_i c_i \mathrm{P}_i.
\end{equation}
where $P_i\in\{\sigma_0,\sigma_\mathrm{x},\sigma_\mathrm{y},\sigma_\mathrm{z}\}$.

\paragraph{Clifford operations:} interchange the Pauli operators (i.e., map the Pauli group onto itself).  The Clifford group is defined to be the normalizer of the Pauli group, that is the set $\mathcal{C}$ of unitaries $U$ obeying $U\mathrm{P}U^\dagger\in \mathrm{P}$, where $\mathrm P$ is the Pauli group. The single-qubit Clifford operations are the Pauli operations themselves plus the Hadamard (H) gate and the Phase (S) gate given by,
\begin{equation}
   \mathrm H=\frac{1}{\sqrt{2}}\begin{pmatrix} 1 & 1 \\ 1 & -1 \end{pmatrix}, \mathrm S=\begin{pmatrix} 1 & 0 \\ 0 & i \end{pmatrix}.
\end{equation}
The Hadamard gate acts as a reflection of the $\sigma_\textrm{x}-\sigma_\textrm{z}$ bases, the $\mathrm{S}$ gate acts as a reflection of the $\sigma_\textrm{x}-\sigma_\textrm{y}$ bases and the combined $\mathrm{HS}$ gate acts as a reflection of the $\sigma_\textrm{z}-\sigma_\textrm{y}$ bases. 

The two-qubit Clifford operations include the $\mathrm{CNOT}$ and $\mathrm{CZ}$ gates,
\begin{equation}
   \mathrm{CNOT}=\begin{pmatrix} 1 & 0 & 0 & 0 \\ 0 & 1 & 0 & 0 \\ 0 & 0 & 0 & 1 \\ 0 & 0 & 1 & 0 \end{pmatrix}, \mathrm{CZ}=\begin{pmatrix} 1 & 0 & 0 & 0 \\ 0 & 1 & 0 & 0 \\ 0 & 0 & 1 & 0 \\ 0 & 0 & 0 & -1 \end{pmatrix}.
\end{equation}
These gates can entangle two-qubit systems, for example,
\begin{equation}
   \mathrm{CNOT}\ket{+}\ket{0}=\frac{1}{\sqrt{2}}(\ket{00}+\ket{11}).
\end{equation}
The Clifford operations for arbitrary numbers of qubits form a group under matrix multiplication and are spanned by the generators $\{\mathrm{CNOT},\mathrm{H},\mathrm S\}$. Circuits composed of these gates are known as Clifford circuits. All Clifford operations can be constructed using Clifford circuits. For example, the $\mathrm{SWAP}$ gate, which swaps the states of two qubits, can be decomposed into three $\mathrm{CNOT}$ gates. 

\paragraph{Clifford circuits are classically simulable:} Even though Clifford circuits can generate large superposition states such as maximally entangled Bell and GHZ states (using the CNOT gate), the Gottesman-Knill theorem   \cite{Gottesman-Knill-Theorem,AaronsonGottesmanCliffordSims} tells us that they are easy to simulate classically.   To see this, consider an $N$-qubit starting state $|000\ldots 000\rangle$.  This state is `stabilized' by (i.e., is a +1 eigenstate of) the set of $N$ single-qubit Pauli $\sigma_\mathrm{z}$ operators, $\{\sigma_{\mathrm{z}_N},\sigma_{\mathrm{z}_{N-1}},\ldots,\sigma_{\mathrm{z}_1}\}$, and is uniquely defined by this list of stabilizers.  Under the action of an arbitrary Clifford circuit, this set of stabilizers is mapped (in the Heisenberg picture) by conjugation to new Pauli strings (generally of weight higher than one under the action of CNOT gates that create entanglement) and the list of transformed stabilizers continues to uniquely define the quantum state at the output of the circuit.  Consider, for example, the Bell state generation circuit 
\begin{align}
   \mathrm{CNOT}_{10} \left [\rule{0pt}{2.4ex} \mathrm{H}\otimes \mathrm{I} \right]|00\rangle = \frac{1}{\sqrt{2}} \left [ \rule{0pt}{2.4ex} |00\rangle+|11\rangle\right ],
\end{align} where the notation $\mathrm{CNOT}_{10}$ is used to indicate that qubit 1 is the control and qubit 0 is the target (and the qubits are numbered right-to-left starting with ordinal number zero).  This circuit transforms the initial stabilizer set $\{\sigma_{\mathrm{z}_1},\sigma_{\mathrm{z}_0}\}$ to the set of weight-two stabilizers $\{\sigma_{\mathrm{x}_1}\sigma_{\mathrm{x}_0},\sigma_{\mathrm{z}_1}\sigma_{\mathrm{z}_0}\}$ that uniquely defines this Bell state.    

The efficiency advantage of the stabilizer formalism~\cite{gottesman_thesis_1997stabilizer} can be seen by noting that if we had a set of $N=2M$ qubits and a Clifford circuit that produced $M$ randomly selected Bell pairs, the quantum state would be described by a large superposition of $2^N$ quantum amplitudes, whereas the set of stabilizers would still only be of size $N$.  Simple classical algorithms exist \cite{AaronsonGottesmanCliffordSims} to update the list of stabilizers of the state according to the Clifford transformations and thus Clifford circuits can be efficiently simulated classically and therefore do not represent the full power of quantum computation.  We define a `stabilizer state' as any state that can be produced from the all-zero state using a Clifford circuit.  Equivalently, a stabilizer state for $N$ qubits is a +1 eigenstate of $N$ independent generators (Pauli strings) of the stabilizer group. We will use these arguments to find analogies of classicality in CV systems in Sec.~\ref{open-hierarchy} as well.

Thus, the advantage of quantum computation arises from the non-Clifford operations, which are not efficiently simulable classically.  One hint that supports this is the following. With Clifford gates, it is possible to create entangled Bell pairs, but with only computational basis measurements, it is not possible to violate the Bell inequalities.  This requires a non-Clifford rotation of the measurement axis (e.g., by pre-pending the computational basis measurement by an HTH gate sequence, where T is a non-Clifford $\pi/8$ phase gate,
\begin{align}
    T=e^{-i\frac{\pi}{8}\sigma_\mathrm{z}}\label{eq:Tgate}.
\end{align}
Popular non-Clifford operations are the T gate and the Toffoli gate, which is a three-qubit controlled-controlled-NOT gate.

\paragraph{Universality:} A universal quantum gate set for $n$ qubits is a discrete set of gates that can approximately generate all possible unitary operations in SU($2^n$) with error
$\mathcal{O}(\epsilon)$ for any non-zero small error $\epsilon$. For \emph{practical} universality, a universal gate set should approximate arbitrary operations in polynomial time. For qubit systems, the Clifford group and one non-Clifford gate are sufficient to achieve universality. It has been shown by Solovay and Kitaev independently that the Clifford $+$ T gate set can achieve any unitary up to $\mathcal{O}(\epsilon)$ in a polynomial number of operations~\cite{SK}.

\subsubsection{Non-Unitary Channels: Measurement}
An example of a non-unitary quantum channel is the measurement of an observable $A$ ensemble averaged over all possible measurement results (i.e., ignoring the particular measurement result obtained).  The Kraus operators for this channel are given by
\begin{align}
    K_i=P_{i}(A),
    \label{eq:Krausformeasurement}
\end{align}
where $P_{i}(A)$ is the projector onto the subspace spanned by the eigenvector(s) of $A$ corresponding to the $i$th eigenvalue of $A$ (i.e., $i$th possible measurement result).  If the measurement results are not ignored, then the density matrix after observing the $j$th possible measurement result is
\begin{align}
    \rho_j=\frac{P_{j}(A)\rho P_{j}(A)}{p_j},
\end{align}
where $p_j=\mathrm{Tr}\left[P_{j}(A)\rho\right]$ is the probability of obtaining the $j$th possible measurement result.  From this we recover the ensemble-averaged result
\begin{align}
    \rho^\prime=\sum_j p_j\rho_j=\sum_j P_{j}(A)\rho P_{j}(A),
\end{align}
consistent with the claim in Eq.~(\ref{eq:Krausformeasurement}).

The measurement process is non-unitary and irreversible, which is a fundamental property of quantum mechanics. Physical implementations yield a classical outcome of the measurement, which is the eigenvalue of the quantum operator corresponding to the eigenvector representing the state after measurement. Any quantum state as mentioned before can be represented in terms of a chosen set of orthonormal basis vectors $\{\ket{i}\}$, with $i = 0, 1, 2, ..., d-1$,
\begin{equation}
   \ket{\psi} = \sum_{i=0}^{d-1}c_i\ket{i}.
\end{equation} 
An arbitrary quantum state can be measured in any basis. The probability of measuring the state $\ket{\psi}$ in the basis state $\ket{i}$ is given by the Born rule~\cite{hallquantum} as
\begin{equation}
   P(i) = |\langle i|\psi\rangle|^2 = |c_i|^2.
\end{equation}
For qubits, measuring the state in one of the Pauli bases $\sigma_\textrm{x},\sigma_\textrm{y},\sigma_\textrm{z}$ collapses the state onto one of the six cardinal states: $\ket{\pm}$ if measuring $\sigma_\textrm{x}$, $\ket{\pm i}$ if measuring $\sigma_\textrm{y}$, $\ket{0}$ or $\ket{1}$ if measuring $\sigma_\textrm{z}$. The $\sigma_\textrm{x}$ and $\sigma_\textrm{y}$ bases are used to measure the state in the Hadamard and $\sigma_\textrm{y}$ bases, respectively. The $\sigma_\textrm{z}$ basis is used to measure the state in the computational basis. Aside from the Pauli bases of measurement, one can measure in any arbitrary basis defined by the orthonormal eigenvectors of any Hermitian quantum operator. The post-measurement state is one of the eigenvectors of the respective quantum operator with the probability of measuring the state in that eigenvector given by the Born rule. Any qubit measurement can be decomposed into a combination of a rotation and a computational basis measurement. For example, measuring in the $\sigma_\textrm{x}$ basis is equivalent to rotating the state by $\pi/2$ about the $\sigma_\textrm{y}$ axis and then measuring in the computational basis. Similarly, measuring in the $\sigma_\textrm{y}$ basis is equivalent to rotating the state by $\pi/2$ about the $\sigma_\textrm{x}$ axis and then measuring in the computational basis. Measuring in the $\sigma_\textrm{z}$ basis is equivalent to measuring in the computational basis directly.
     
\subsubsection{Non-Unitary Channels: Error Channels}
For theoretical purposes, there are traditional non-unitary quantum channels known as error channels used for simulations. For modeling purposes, these error channels generally come under the giant umbrella of the Pauli error channel. Examples of physically important errors that do not belong to this class are amplitude-damping channels and leakage. For simplicity, we will abstain from these examples and only discuss Pauli error channels.

\paragraph{Bit-flip channel:} The bit-flip channel is a quantum channel that acts on a qubit by flipping the state of the qubit with some probability $p$ and leaving the state unchanged with probability $1-p$. The action of the bit-flip channel on a qubit state is given by the map,
\begin{equation}
    \rho\rightarrow p\,\sigma_\textrm{x}\rho\,\sigma_\textrm{x}+(1-p)\rho,
\end{equation}
where $\sigma_\textrm{x}$ is the Pauli matrix. For superconducting qubits, the true error channel typically only takes $\ket{e}\rightarrow \ket{g}$ while the errors like $\ket{g}\rightarrow \ket{e}$ are suppressed due to negligible thermal effects.

\paragraph{Phase-flip channel:} The phase-flip channel is a quantum channel that acts on a qubit by flipping the phase of the qubit with some probability $p$ and leaving the state unchanged with probability $1-p$. The action of the phase-flip channel on a qubit state is given by the map,
\begin{equation}
    \rho\rightarrow p\sigma_\textrm{z}\rho\sigma_\textrm{z}+(1-p)\rho,
\end{equation}
where $\sigma_\textrm{z}$ is the Pauli matrix. 

\paragraph{Depolarizing channel:} The depolarizing channel is a generalization of the bit-flip, phase-flip, and bit-phase-flip channels for qubits. The depolarizing channel is a common noise model in quantum error correction. It is a CPTP map that is represented by the Kraus operators,
\begin{equation}
    K_1 = \sqrt{1-\frac{3p}{4}}I, K_2 = \sqrt{\frac{p}{4}}\sigma_\textrm{x}, K_3 = \sqrt{\frac{p}{4}}\sigma_\textrm{y}, K_4 = \sqrt{\frac{p}{4}}\sigma_\textrm{z},
\end{equation}
where $I$ is the identity operator and $\sigma_\textrm{x}$, $\sigma_\textrm{y}$, and $\sigma_\textrm{z}$ are the Pauli matrices. Here $\frac{p}{4}$ is the probability of depolarization for each of the Pauli matrices. Thus, the action of the depolarizing channel on a quantum state is given by the map,
\begin{align}
\mathcal{E}(\rho)&=\sum_{i\in\{1,2,3,4\}} K_i\rho K_i^\dagger\\
&=\Bigg(1-\frac{3p}{4}\Bigg)\rho+\frac{p}{4}X\rho X+\frac{p}{4}Y\rho Y+\frac{p}{4}Z\rho Z
\end{align}
Now, using Eq.~(\ref{density_matrix}) we can write,
\begin{align}
    \mathcal{E}(\rho)&=\Bigg(1-\frac{3p}{4}\Bigg)\Bigg(\frac{I}{2}+n_x\sigma_\mathrm{x}+n_y\sigma_\mathrm{y}+n_z\sigma_\mathrm{z}\Bigg)+\frac{p}{4}\Bigg(\frac{I}{2}+n_x\sigma_\mathrm{x}-n_y\sigma_\mathrm{y}-n_z\sigma_\mathrm{z}\Bigg)\nonumber\\&\quad+\frac{p}{4}\Bigg(\frac{I}{2}-n_x\sigma_\mathrm{x}+n_y\sigma_\mathrm{y}-n_z\sigma_\mathrm{z}\Bigg)+\frac{p}{4}\Bigg(\frac{I}{2}-n_x\sigma_\mathrm{x}-n_y\sigma_\mathrm{y}+n_z\sigma_\mathrm{z}\Bigg)\\
    &=\Bigg(1-\frac{3p}{4}\Bigg)\Bigg(\frac{I}{2}+\vec{n}\cdot\vec{\sigma}\Bigg)+\frac{p}{4}\Bigg(\frac{3I}{2}-\vec{n}\cdot\vec{\sigma}\Bigg)\\
    &=\Bigg(1-\frac{3p}{4}\Bigg)\rho+\frac{p}{4}(2I-\rho)=p\frac{I}{2}+(1-p)\rho.
\end{align}
 That is, this channel acts on a quantum state by replacing it with a completely mixed state with probability $p$ and the original state with probability $1-p$.
\subsection{Error Correction and Stabilizer Formalism}\label{sec:EC}
Error correction requires us to engineer a mechanism where an environmental defect can be flagged and corrected. For the case of qubits, the most general strategy is to redundantly encode logical qubits into multiple physical qubits. Such redundancy allows us to perform measurements that can flag an error in the logical subspace. We will review the nomenclature of stabilizer formalism~\cite{gottesman_thesis_1997stabilizer} and Knill-Laflamme conditions~\cite{knill2000theory} necessary for error correction.

Stabilizers are a group of commuting operators. For a stabilizer code, the logical codespace is defined as the co-eigenstates for all stabilizers. For a correctable error $E$, $[E,S]=0$ such that,
\begin{align}
    S(E\ket{\psi})=-E(S\ket{\psi})=-E\ket{\psi}
\end{align}
 Thus, the error states are $-1$ eigenstates of these eigenstates which trigger a flagged ($-1$) outcome upon the measurement of a stabilizer. The stabilizer generators are the minimal set of operators required to generate the stabilizer group under matrix multiplication. Thus, the product of two stabilizers is another stabilizer of the code. The logical operations of the code should manipulate the codespace without revealing any information during stabilizer measurements. Thus, it is required that all logical operations commute with stabilizers of the codespace. In other words, logical operators belong to the normalizer of the stabilizer group.
 
 For practical purposes, the codes with Pauli stabilizers work best since the measurement of Pauli stabilizer circuits can be performed using Clifford circuits. For example, measurement of Pauli $X$ operator requires one to apply $CX$ on the required state with the control on an ancilla in the $\ket{+}$ state, followed by measurement in the $X$ (or, Hadamard) basis. This choice is justified since the Pauli operator is a complete basis and represents any channel. Thus, a circuit capable of correcting Pauli errors can correct any error. This brings us to the question, what type of errors can a quantum code correct? The set of errors $\mathcal{E}=\{E_i\}$ are correctable on a code $C$ if,
 \begin{align}
   \braket{\psi_i|E_aE_b|\psi_j}=c_{ab}\braket{\psi_i|\psi_j}=c_{ab}\delta_{ij},
 \end{align}
 where $c_{ab}$ is a constant, $\delta_{ij}$ is the Kronecker delta and $\ket{\psi_i},\ket{\psi_j}$ are the orthogonal logical codewords of $C$~\cite{knill2000theory}. The errors $E_a, E_b\in\mathcal{E}$ are two errors and these conditions give the efficiency with which the two errors can be distinguished in the given codespace. This condition is called the Knill-Laflamme condition and claims that if the above condition is satisfied then there exists a recovery operation that can map the error space $\mathcal{E}$ to the codespace $C$, without learning anything about the quantum information in the system. The last bit of this sentence is extremely important to preserve the information in the superposition of quantum systems.

\paragraph{Code distance:} 
This quantity defines how many errors $E\in\mathcal{E}$ are correctable in a code $C$. For example, the distance of a $d$-qubit repetition code is $d$ and hence it can only correct $t=\frac{d-1}{2}$ errors.

Now that we have defined the well-known DV quantum systems, let us contrast these ideas with the case of CV quantum systems.
\section{Continuous Variable (CV) Systems}\label{CV}
CV systems are modelled using a truncated Hilbert space with a finite cutoff in the number of photons to approximate oscillators. It was recently shown that such truncation does not affect the efficiency of CV simulations in theory, leading to an infinite-dimensional Solovay-Kitaev theorem~\cite{arzani2025can}. They can be constructed using propagating photons or harmonic oscillators, such as superconducting microwave resonators or the mechanical oscillations of trapped ions, which have large, formally infinite, Hilbert space dimensions. In this thesis, we will focus on harmonic oscillators. These systems have long coherence lifetimes but require auxiliary sources of non-linearity to achieve universal control~\cite{braunstein2005quantum,liu2016power,ISA}. We will use hats `$\hat{.}$' for CV operators to distinguish them from the DV operators.

\subsection{CV State Space}
CV states are typically defined on the continuum of oscillator position and momentum and have a (countably infinite but) discrete as well as a continuous variable description. The discrete version is known as the Fock space basis, while the continuous versions are the position or momentum basis. See Fig.~\ref{fig:bloch-sphere}(b).  We will discuss both of these spaces in the following sections to describe the states as well as operators in the Hilbert space of an oscillator. Our focus, however, will very quickly shift to the unusual basis of oscillator phase space, which, like its classical counterpart, is a two-dimensional continuum describing both position and momentum.

\subsubsection{Fock basis}\label{sec:fock_basis}
A harmonic oscillator has a Hamiltonian 
\begin{align}
    H=\hbar\omega(\hat n+1/2).
\end{align}
Here the first term represents the number operator $\hat n$, whose eigenvalues are non-negative integers corresponding to the excitation number of the oscillator.  The corresponding eigenstates (also known as Fock states) are thus equally spaced in energy. For the remainder of this thesis, we shall use the units $\hbar=1$, and ignore the vacuum energy $\hbar\omega/2$ (that is, the energy of the eigenstate of $\hat n$  with the lowest eigenvalue $\ket{0}$) so that the Hamiltonian becomes 
\begin{align}
H=\omega \hat a^\dagger \hat a,    
\end{align}
where $\hat{a}$ and $\hat{a}^\dagger$ are the excitation annihilation and creation operators of the oscillator.
See Fig.~\ref{fig:bloch-sphere}(b) for a pictorial representation.
\paragraph{Fock states:} The basis states of the CV system, or Fock states, are 
\begin{align}
    \ket{n}=\frac{(\hat{a}^\dagger)^n}{\sqrt{n!}}\ket{0},
\end{align}
where $\ket{0}$ is the vacuum state of the oscillator. The Fock states are orthonormal, $\braket{m|n}=\delta_{mn}$, and form a complete basis set, $\sum_{n=0}^\infty\ket{n}\bra{n}=\hat {\mathrm{I}}$. They are the eigenstates of the number operator, $\hat{n}\ket{n}=n\ket{n}$, with eigenvalues $n=0,1,2,...$. The number operator is the (dimensionless) quantum analog of the corresponding classical Hamiltonian. 

Note that the annihilation and creation operators form a closed algebra as $\left[\hat a,\hat a^\dagger \right] = 1$. These are also known as raising and lowering operators or ladder operators since they move the excitation number up and down the ladder of Fock states. 
\begin{align}
     \hat a^\dagger |m\rangle &= \sqrt{m+1}|m+1\rangle\label{eq_sqrt1}\\
    \hat a|m+1\rangle &= \sqrt{m+1}|m\rangle\label{eq_sqrt2}
\end{align}

\paragraph{Coherent states:} Coherent states are continuous basis states of the CV system. The coherent states are the eigenstates of the annihilation operator, $\hat{a}\ket{\alpha}=\alpha\ket{\alpha}$, with eigenvalues $\alpha\in\mathbb{C}$. The coherent states are given by,
\begin{align}
    \ket{\alpha}&= e^{\alpha \hat a^\dagger - \alpha^* \hat a}|0\rangle=e^{-\frac{|\alpha|^2}{2}}e^{\alpha \hat a^\dagger}|0\rangle\label{eq:coherentstateviatranslation}\\
    &=e^{-\frac{|\alpha|^2}{2}}\sum_{n=0}^\infty \frac{\alpha^n}{\sqrt{n!}}|n\rangle,\label{eq:coherentasFock}
\end{align}
where $\alpha$ is a complex number. The coherent states are over-complete, $\frac{1}{\pi}\int d^2\alpha\ket{\alpha}\bra{\alpha}=\hat {\mathrm{I}}$, and form a continuous basis set. Coherent states are the closest quantum states to classical states, and they have the minimum uncertainty in position and momentum. It is important to note that the operators are (1) $\hat a,\hat a^\dagger$ non-Hermitian, and (2) defective in the sense that $\hat a^\dagger$ has no right eigenstates and $\hat a$ has no left eigenstates. This is a non-trivial feature of the Hilbert space of an oscillator in which there exist states from which one can remove a photon and still end up with the same state. The same is not true, however, for adding a photon.

\subsubsection{Phase-space basis}
First quantization in quantum mechanics defines the position and momentum in the discrete space of oscillators using the non-hermitian operators $\hat a$ and $\hat a^\dagger$. These definitions exist in various units, and we will use the following dimensionless `Wigner units,'
\begin{align}
    \hat x=\frac{\hat a +\hat a^\dagger}{2},\quad\hat p=\frac{\hat a -\hat a^\dagger}{2i}.
\end{align}
From this definition it follows that the commutation relation is $[\hat x,\hat p]=\frac{i}{2}\implies \hat p=-\frac{i}{2}\frac{\partial}{\partial \hat x}$. The generator of momentum boosts is thus $2\hat x$ and the generator of displacements is $2\hat p$.

\paragraph{Vacuum state:} In the position basis, vacuum, is given by the derivation,
\begin{align}
    \braket{x|\hat a|0}_\mathrm{vac}&=0\implies \braket{x|(\hat x+i\hat p )|0}_\mathrm{vac}=0
    \\&\implies (x+\frac{1}{2}\frac{\partial}{\partial  x})e^{-\frac{x^2}{\sigma}}=0
    \\&\implies x-\frac{x}{\sigma}=0\implies \sigma=1.\nonumber\\&\implies \braket{x|0}_\mathrm{vac}=\Big(\frac{2}{\pi}\Big)^{1/4}e^{-x^2}.
\end{align}
The minimum position uncertainty for this state is
\begin{align}
    \delta x=\sqrt{\braket{\hat x^2}-\braket{\hat x}^2}=\Big(\frac{2}{\pi}\Big)^{1/4}\sqrt{\int_{-\infty}^\infty \ dx \ x^2e^{-2x^2}}=\frac{1}{2}
\end{align}
Thus, we have $\delta x=\delta p=\frac{1}{2}$ for this state, and the minimum uncertainty principle is satisfied $\delta x\delta p=\frac{1}{4}$. Ignoring the normalization, we can define states with arbitrary position uncertainty as $\psi(x)=e^{-\frac{x^2}{\Delta^2}}$ such that $\Delta^2=4\delta x^2=1$ denotes the vacuum state. 
\paragraph{Position and momentum states: }\label{app:phase-space}
The position and momentum states are continuous basis states of the CV system. These states are the eigenstates of the position and momentum operators, $\hat{x}\ket{x}=x\ket{x}$ and $\hat{p}\ket{p}=p\ket{p}$, with eigenvalues $x,p\in\mathbb{R}$. The position and momentum states are related by (with $\hbar=1$),
\begin{align}
    \ket{x}=\int dp\,e^{i2px}\ket{p},\quad \ket{p}=\int dx\,e^{-i2px}\ket{x},
\end{align}
where $\ket{x}$ and $\ket{p}$ are the position and momentum states, respectively. In the position (momentum) basis, $\ket{x}$ \big($\ket{p}$\big) is represented by a sharply peaked wave function as distribution whose square is a Dirac-delta function. These states are squeezed coherent states with $\Delta\rightarrow 0$, as described below.

\paragraph{Squeezed coherent state:} In these units, if we follow the definitions of a coherent state $\ket{\alpha}$ as $\hat a\ket{\alpha}=\alpha\ket{\alpha}$, the position representation of a squeezed coherent state is given by,
\begin{align}
|\braket{x|\alpha_\Delta}|^2=\Big(\frac{2}{\pi}\Big)^{1/4}e^{-\frac{2(x-\alpha)^2}{\Delta^2}}\label{eq:squeezed}
\end{align}

\paragraph{Completeness and orthogonality:} The Fock basis is a complete orthonormal basis composed of discrete states. The position and momentum states are orthonormal, $\braket{x^\prime|x}=\delta(x-x^\prime)$ and $\braket{p^\prime|p}=\delta(p-p^\prime)$, and form a complete basis set, $\int dx\,\ket{x}\bra{x}=\hat {\mathrm{I}}$ and $\int dp\,\ket{p}\bra{p}=\hat {\mathrm{I}}$. In addition, we also possess the over-complete coherent state basis, as described before. Coherent states are part of a two-dimensional continuum (from the real and imaginary parts of $\alpha$). That is the source of the over-completeness.
 \paragraph{Visual representation: }
   The CV states can be visualized in the oscillator phase space by plotting various types of probability distributions. Below we list the ones used in this thesis.
\begin{figure}[htb]
    \centering
    \includegraphics[width=\linewidth]{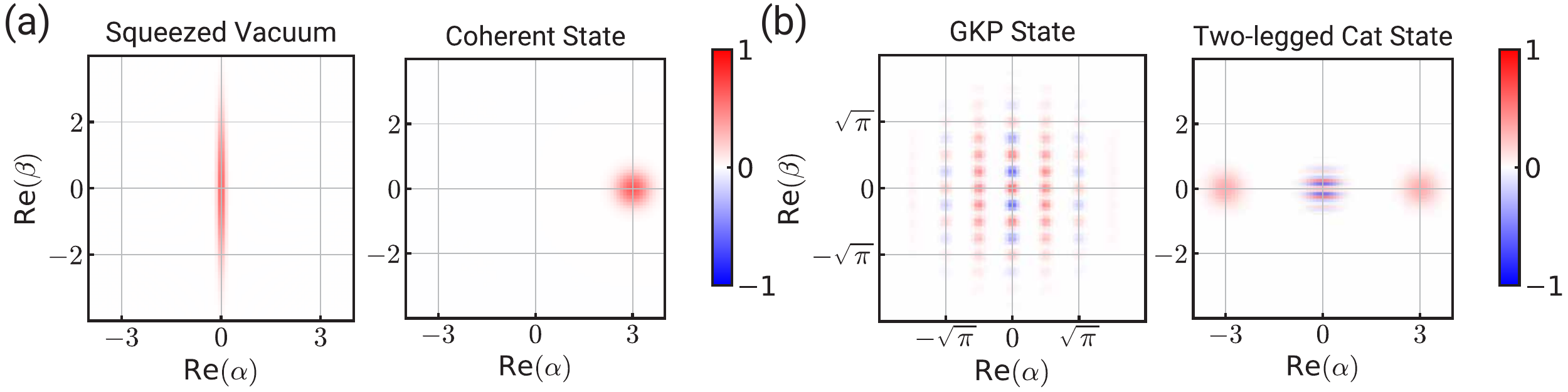}
    \caption[Visualization of CV states]{\textbf{Visualization of CV states.} Wigner function plots for (a) Gaussian and (b) non-Gaussian states.}
    \label{Wigner}
\end{figure}
\begin{itemize}
    \item \textbf{Wave function (marginal probability):} The wave function squared $|\psi(x)|^2=|\braket{x|\psi}|^2$ gives the marginal probability distribution of the state along the position ($x$) of the oscillator. A similar distribution can be obtained along the momentum ($p$) of the oscillator by taking the Fourier transform of the wave function, $|\tilde{\psi}(p)|^2=|\braket{p|\psi}|^2$. The wave function of various Fock states is given by Hermite polynomials (multiplied by the usual Gaussian envelope) in both the position as well as momentum basis. See Fig.~\ref{fig:bloch-sphere}(b).
  \item \textbf{Wigner function (quasi-probability):} The Wigner function for a CV state given by the density matrix $\rho$ is defined as,
\begin{equation}
    W(x,p)=\frac{2}{\pi}\mathrm{Tr}[D(-\alpha)\rho D(\alpha)e^{i\pi\hat a^\dagger \hat a}].
\end{equation}
The density matrix $\rho$ of a system contains the full information needed to predict the (statistical) outcomes of any measurement on that system. The Wigner function also contains the full information and is analogous to the phase space density distribution of classical statistical mechanics.  There is one key difference, however, due to the fact that position and momentum are non-commuting operators.  The Wigner function therefore is a quasi-probability distribution because it can take negative values, which is a signature of non-classicality in quantum states. For example, squeezed states and coherent states which are very similar to classical states do not show Wigner negativity in their Wigner function representation. States that do possess Wigner negativity are known as non-Gaussian states. A few examples of such states are Fock states, cat states, and GKP states. Fig.~\ref{Wigner}(a) shows the Wigner function plots for two different Gaussian states, a squeezed vacuum state and a coherent state $\ket{\alpha}$, respectively. We also contrast this with the plots for some non-Gaussian states, a GKP state and a Two-legged cat state, respectively, in Fig.~\ref{Wigner}(b). Notice the absence of any Wigner negativity in Fig.~\ref{Wigner}(a). Since fully mixed states have no `quantumness' there should be no Wigner negativity in their representation.  

We will use wave functions as well as Wigner representations to visualize CV states in this thesis.
 \end{itemize}
 \paragraph{Partial trace and entanglement: } 
 Partial trace in a CV system is defined as the trace over one of the subsystems of a bipartite system. The partial trace of a CV state $\rho_{AB}$ over one of the subsystems is given by,
    \begin{equation}
        \rho_A = \mathrm{Tr}_B(\rho)=\int dx_B\,{}_B\braket{x|\rho_{AB}|x}_B,
    \end{equation}
    where $\rho_A$ is the reduced density matrix of the subsystem $A$ obtained by tracing out the subsystem $B$ from the density matrix $\rho_{AB}$ of the composite system $AB$. Here, $\{\ket{x}_B\}$ is an arbitrary continuous basis of the subsystem $B$. If the two blobs of cat states in phase space $\ket{\pm\alpha}$ are entangled with another system, then upon tracing out that system, we should not see any interference (or, Wigner negativity) in the Wigner representation of this figure but only two blobs, in contrast with the Fig.~\ref{Wigner}(b). This is a signature of maximal entanglement in CV systems. 
 
 \subsection{CV Quantum Channels}
 CV quantum channels are naturally more complicated than the DV systems due to the unbounded nature of the oscillator operator space. The CV quantum channels are also represented by the completely positive trace-preserving (CPTP) maps. The action of a CV quantum channel on a CV state is given by the Kraus representation as described before in Eq.~(\ref{eq:kraus}). The CV quantum channels can be classified into the following categories based on the type of operations they perform on the CV states.
 
 \subsubsection{Unitary Operations} These are the quantum channels that are reversible and act on a CV state $\rho$ such that, $\rho\rightarrow U\rho U^\dagger$, where $U$ is a unitary operator such that $U^\dagger U=UU^\dagger=I$. These unitary operators act on the states in the (formally infinite) Hilbert space. The finite-dimensional operations that act on the positions and momenta of oscillators allowed in quantum mechanics belong to the symplectic group, preserving the canonical commutation relations for a system with $n$ degrees of freedom (Heisenberg-Weyl algebra). A simple example of such a symplectic transformation for a single oscillator is given below in Eqs.~\ref{eq:FPR1}-\ref{eq:FPR2}. The symplectic groups $\mathrm{Sp}(2,\mathbb{R})$ have an exact correspondence between classical mechanics and quantum mechanics. The Lie algebra of this group $\mathrm{sp}(2,\mathbb{R})$ is isomorphic to the Lie algebras $\mathrm{su}(1,1)$ and $\mathrm{so}(2,1)$ of the groups $\mathrm{SU}(1,1)$ and $SO(2,1)$, respectively. Analogous to the case of $\mathrm{SU}(2)$, the two-dimensional representation of $\mathrm{Sp}(2,\mathbb{R})$ can be obtained from the three-dimensional representation of $\mathrm{SO}(2,1)$ by the stereographic projection of a two-sheet unit hyperboloid from one of its poles to a horizontal plane~\cite{rossi2022multivariable}. In contrast to the discrete variable qubit state space, $\mathrm{SU}(1,1)$ is a non-compact Lie group such that a unitary operation, M, needs to satisfy  $M\omega M^\dagger=\omega$, where $\omega$ is the symmetric bilinear form. The Lie algebra $H$ of $\mathrm{SU}(1,1)$ satisfies the relation, $H^\dagger \omega+\omega H=0$, implying, $U=e^{iHt}$ for $U\in \mathrm{SU}(1,1)$, where $H$ is hermitian. 

 Let us now discuss the parametrized set of CV operations to be used in this thesis.
 \paragraph{Displacement:} The displacement operations on CV states are generated by the Heisenberg-Weyl group of operators, which are the continuous analog of the Pauli group of operators. The Heisenberg-Weyl group is generated by the position and momentum operators of the oscillator, $\hat{x}$ and $\hat{p}$, which satisfy the commutation relation $[\hat{x},\hat{p}]=\frac{i}{2}$. In simpler terms, these operations represent the unitary whose Hamiltonians are linear polynomials in $\hat x, \hat p$ or correspondingly $\hat a, \hat a^\dagger$. The name of these operations owes to the fact that they change the position or momentum of the oscillator. In other words, these operations displace the quantum state in the phase space of the oscillator. The displacement operation is given by,  
    \begin{align}
        D(\alpha)=e^{\alpha\hat{a}^\dagger-\alpha^*\hat{a}}=e^{2i(\mathrm{Im}(\alpha)\hat x-\mathrm{Re}(\alpha)\hat p)},
    \end{align}
    for a complex number $\alpha$.
This is a state displaced by $\mathrm{Re}(\alpha)$ along the position axis and $\mathrm{Im}(\alpha)$ along the momentum axis. Thus, as pointed out before, the generator of displacement in the position basis is $2\hat p$. Analogous statements can be made about momentum boosts and $2\hat x$. We will often use the following commutation relations associated with displacement operators,
\begin{eqnarray}
    D(\beta) D(\alpha) &=& D(\alpha+\beta)e^{+\frac{1}{2}[\beta\alpha^*-\beta^*\alpha]},\label{eq:UCDcompos1}\\
    D(\alpha) D(\beta) &=& D(\alpha+\beta)e^{-\frac{1}{2}[\beta\alpha^*-\beta^*\alpha]},\label{eq:UCDcompos2}\\
    &=&   D(\beta) D(\alpha)e^{-[\beta\alpha^*-\beta^*\alpha]},\label{eq:UCDcompos3}\\
    &=&D(\beta) D(\alpha)e^{-2iA(\alpha,\beta)},\label{eq:UCDcompos4}
\end{eqnarray} 
where $A(\alpha,\beta)$ is the oriented area of the parallelogram formed by the displacements $\alpha,\beta$ (see Fig.~\ref{fig:bloch-sphere}(b)). Displacements in phase space form a complete operator basis and can represent the operators on oscillator Hilbert space. Any CV channel can be generated using only displacements for the Kraus operator. 

\paragraph{Gaussian operations:} The Gaussian unitary operations are exponentials of quadratic polynomials in $\hat x, \hat p$ or correspondingly $\hat a, \hat a^\dagger$, in addition to displacement operations. For example, the list of essential Gaussian operations includes the phase-space rotation operation, the beam-splitter operation, and the squeezing operation: \begin{itemize}
\item the phase-space rotation operation is given by,
\begin{align}
    \hat{\mathrm{P}}(\theta)=e^{-i\theta\hat{a}^\dagger\hat{a}}=e^{-i\theta\hat{n}}=e^{-i\theta\frac{(x^2+p^2)}{4}}.\label{eq:Ptheta}
\end{align}
This operation rotates the quantum state in the phase space of the oscillator by an angle $\theta$. The well-known quantum Fourier transform gate ($\hat {\mathrm{F}}$) is a special case of $\theta=\pi/2$, such that,
\begin{align}
    \hat {\mathrm{F}}^\dagger \hat x \hat {\mathrm{F}} &=+\hat p\label{eq:FPR1}\\
    \hat {\mathrm{F}}^\dagger \hat p \hat {\mathrm{F}} &=-\hat x.
    \label{eq:FPR2}
\end{align}
Notice that this is a simple example of a symplectic transformation that preserves the commutation relation $[\hat x,\hat p]=\frac{i}{2}$.
\item the beam-splitter operation is given by, 
\begin{align}
    \hat{\mathrm{BS}}(\theta,\varphi)=e^{-i\frac{\theta}{2}\left[e^{i\varphi} a^\dag b+e^{-i\varphi} ab^\dag\right]}.\label{eq:BSgateunitary}
\end{align}
where $\hat{a}_1$ and $\hat{a}_2$ are the annihilation operators of two oscillators. The beam-splitter operation entangles the two oscillators by creating a superposition of the two oscillators.   
\item the squeezing operation is given by, 
\begin{align}
	\hat{S}(r) &= e^{\frac{1}{2}r(a^2 - {a^\dagger}^2)}=e^{ir(\hat x\hat p+\hat p\hat x)}.
\end{align}
where $r$ is the squeezing parameter. The squeezing operation squeezes the quantum state in the phase space of the oscillator along one of the quadratures.
\item the two-mode squeezing operation is given by,
\begin{align}
    \hat{\mathrm{TMS}}(r)=e^{r(\hat{a}_1\hat{a}_2-\hat{a}_1^\dagger\hat{a}_2^\dagger)}=e^{i\frac{r}{2}(\hat p_1\hat x_2+\hat x_1\hat p_2)},
\end{align}
where $\hat{a}_1$ and $\hat{a}_2$ are the annihilation operators of two oscillators. The two-mode squeezing operation squeezes the two oscillators along one of the joint quadratures in the 4-dimensional phase space.

Note that the operations listed above are sufficient to synthesize any Gaussian operation~\cite{bartlett2002efficient}. For example, the circuit in Fig.~\ref{fig:SUM}(a) realizes the Bloch-Messiah decomposition \cite{Braunstein2005squeezing,Weedbrook2012,PhysRevA.94.062109} of the two-mode squeezing operation using a pair of beam-splitters and single-mode squeezers as follows,
\begin{equation}
   \hat{\mathrm{TMS}}(r,\pi/2)=\hat{\mathrm{BS}}(\pi/2,0)[\hat{S}(r)\otimes \hat{S}(r)]\hat{\mathrm{BS}}(\pi/2,\pi).
   \label{eq:BMdecompTMS}
\end{equation}
Here the gate symbols represent the full unitary operators acting on the Hilbert space, not the symplectic matrix representation of the gates. The direct product $\hat S(r)\otimes \hat S(r)$ represents single-mode squeezing applied to each arm of the interferometer in Fig.~\ref{fig:SUM}(a).   
The derivation of this circuit is given in Ref.~\cite{ISA} as is the symplectic transformation of the quadrature coordinates. 

\item the two-mode SUM gate (another popular gate as required by the GKP logical encoding~\cite{gottesman2001encoding}) is given by, 
\begin{align}
    \mathrm{SUM}(\lambda) &= e^{i2\lambda x_1\hat p_2}
\end{align}
where $\lambda\in\mathbb{R}$. This operator displaces one oscillator mode indexed $2$ by an amount proportional to the position of the oscillator mode indexed $1$. The Bloch-Messiah decomposition~\cite{ISA} for the SUM gate has the advantage that two-mode squeezing can be replaced by simpler single-mode squeezing,
\begin{align}
   \mathrm{SUM}(\lambda)
   &=\mathrm{BS}(\pi+2\theta,-\pi/2)[\hat S(r)\otimes \hat S(-r)]\mathrm{BS}(2\theta,-\pi/2)\label{eq:SUMBlochMessiah}\\
   \sinh{r}&=\frac{\lambda}{2},\label{eq:SUMBM2}\\
   \cos(2\theta)&=\tanh(r),\label{eq:SUMBM3}\\
   \sin(2\theta)&=-\mathrm{sech}(r),\label{eq:SUMBM4}
\end{align}
where we use the tensor product ordering convention that $B\otimes A$ means that $A$ is applied to the upper arm of the interferometer and $B$ is applied to the lower arm.  Thus, $\hat S(-r)$ is applied to the upper arm and $\hat S(+r)$ is applied to the lower arm of the interferometer in Fig.~\ref{fig:SUM}(b). We again direct readers to the tutorial~\cite{ISA} for derivation of this decomposition. 

\begin{figure}[tb]
    \centering
    \includegraphics[width=\textwidth]{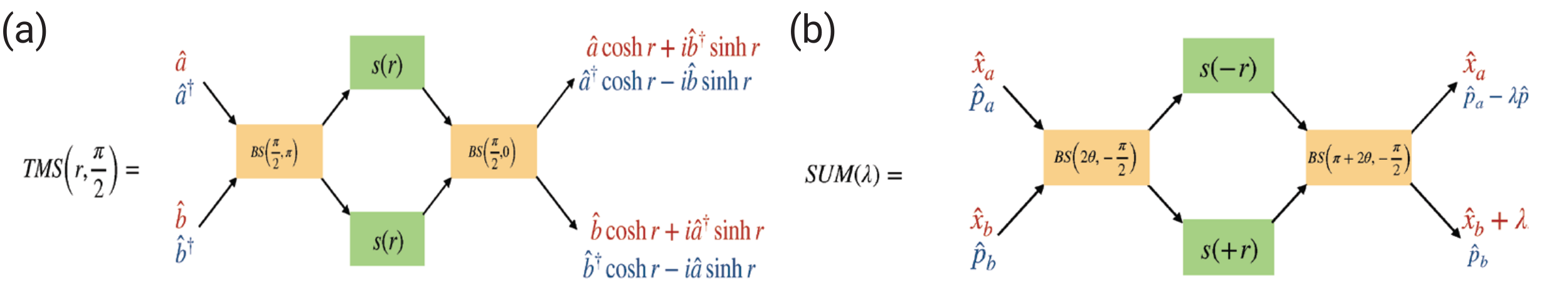}
    \caption[Bloch-Messiah decomposition of Gaussian operations]{\textbf{Bloch-Messiah decomposition of Gaussian operations.} Decomposition of (a) TMS$\Big(r,\frac{\pi}{2}\Big)$, and (b) SUM$(\lambda)$ gate using photon-number preserving beam-splitter gates and single-mode squeezing gates. The transformation from the initial mode quadrature operators on the left to the final quadratures on the right follows the SUM gate transformation. Here $a$ and $b$ denote different oscillators. The effect of this operation is described by how the position and momentum operators are transformed. The red (blue) operators on the left are transformed into red (blue) operators on the right for each oscillator. Note that the terms in Eq.~(\ref{eq:SUMBlochMessiah})(Eq.~(\ref{eq:BMdecompTMS})) are applied right to left while the circuit diagram should be read left to right.}
    \label{fig:SUM}
\end{figure}

\end{itemize}     
\paragraph{Non-Gaussian operations:}
All non-Gaussian operations are represented by unitaries with Hamiltonians $f(\hat x,\hat p)$, or correspondingly $g(\hat a,\hat a^\dagger)$, that are polynomials of degree three or higher. Non-Gaussian operations are non-linear operations that cannot be decomposed into a sequence of Gaussian operations. It has been shown that sequences consisting only of Gaussian operations are efficiently simulable on classical hardware \cite{bartlett2002efficient}. Thus, non-Gaussianity via non-Gaussian operations or non-Gaussian states is essential for quantum advantage in CV quantum computation. 
\paragraph{Universality:}
We remind the readers that there are additional universal instruction sets in a hybrid CV-DV architecture (discussed in Sec.~\ref{hybrid}). These CV-DV instruction sets are the focus of our work. However, let us look into the CV-only options first. Since CV systems cannot achieve non-classicality with only Gaussian operations, this leads us to the following two alternatives for an architecture composed only of oscillators. 
\begin{itemize}
\item\textbf{Cubic Instruction Set} This instruction set requires a cubic Hamiltonian in $\hat x, \hat p$ or correspondingly $\hat a, \hat a^\dagger$. This is a generator of non-Gaussian operations and is sufficient for universality, in addition to the Gaussian operations described above. To understand why a single term in the Hamiltonian higher than quadratic is sufficient for universality, we can examine the commutation relations for different degrees of polynomial (using standard dimensionless units for which $[\hat x,\hat p]=+i$)
    \begin{align}
        [\hat{x},\hat{x}^m\hat{p}^n]&= in\hat{x}^m\hat{p}^{n-1},\\
        [\hat{x}^2,\hat{x}^m\hat{p}^n]&= \hat{x}[\hat{x},\hat{x}^m\hat{p}^n]+[\hat{x},\hat{x}^m\hat{p}^n]\hat x\nonumber\\
        &=in(\hat{x}^{m+1}\hat{p}^{n-1}+\hat{x}^m\hat{p}^{n-1}\hat{x}),\\
        [\hat{x}^3,\hat{x}^m\hat{p}^n]&= \hat{x}[\hat{x}^2,\hat{x}^m\hat{p}^n]+[\hat{x^2},\hat{x}^m\hat{p}^n]\hat x,\\
        &=in(\hat{x}^{m+2}\hat{p}^{n-1}+\hat{x}^{m+1}\hat{p}^{n-1}\hat{x}\nonumber\\
        &\quad+\hat{x}^{m+1}\hat{p}^{n-1}\hat{x}+\hat{x}^{m}\hat{p}^{n-1}\hat{x}^2).
    \end{align}
    As is evident, the degree of a polynomial in $\hat x,\hat p$ is preserved by its commutator with a quadratic term $\hat x^2$. In contrast, the commutator of a degree $m+n$ polynomial with a cubic term $\hat x^3$ yields a polynomial of degree $m+n+1$. Similarly, we can prove the same properties for $\hat p, \hat p^2, \hat p^3$ and $\hat x^m \hat p^n$. Thus, the algebra generated by the Lie brackets of these terms is infinite in the presence of a cubic or higher-order polynomial. As a specific example, consider
    \begin{align}
       [\hat x^3,\hat p^2]=i3(\hat p\hat x^2+\hat x^2\hat p),\\
       [\hat x^3,[\hat x^3,\hat p^2]]=-18\hat x^4.
    \end{align}
Using this inductive proof we have shown that, the algebra generated by the control Hamiltonian set which includes quadratic Hamiltonians with only one cubic Hamiltonian can generate arbitrary order polynomials enabling universal control. Note that engineering a native cubic interaction is rather hard experimentally. Thus, resorting to hybrid CV-DV platforms for universal control of oscillators can be useful, as will be shown in this thesis.

\item Another alternative requires non-Gaussian states in addition to Gaussian operations~\cite{Baragiola}. Non-Gaussian states cannot be prepared without non-Gaussian resources and hence are a great resource for non-classicality as well. However, whether a specific non-Gaussian state is sufficient for universality requires further mathematical analysis. We employ hybrid CV-DV architectures for deterministic preparation of useful oscillator states, including non-Gaussian states, like Cat states, GKP states, Four-legged cat states, and Fock $\ket{1}$, in Chapter~\ref{chapter:state-prep}. The Wigner representation of some non-Gaussian states is shown in Fig.~\ref{Wigner}(b)
\end{itemize}
\paragraph{Truncation and operator fidelity:} As mentioned before, in practice CV systems use a truncated Hilbert space with a finite cutoff in the number of photons to approximate oscillators~\cite{arzani2025can}. In a truncated Hilbert space, the closeness between two CV operators, say $U$ and $V$, can be defined using the Hilbert-Schmidt product: 
\begin{align}
    \Bigg|\frac{1}{d}\mathrm{Tr}(PU^\dagger V)\Bigg|^2.
\end{align}
This distance measure is called the operator fidelity and is computed on the oscillator-qubit subspace with projector $P=\sum_{\ell=0}^{d-1} \ket{\ell}\bra{\ell}$. This is the projector on the space of the truncated oscillator with $d$ levels. 
\clearpage
\subsubsection{Measurement}
 The Fock basis measurement also known as the photon number measurement (PNM) reveals whether the cavity is in state $|m\rangle$ with the help of the projector, $\hat P_m=|m\rangle\langle m|$. This measurement is rather non-trivial and requires the help of hybrid CV-DV architecture in itself. Thus, we do not get into the basics of this measurement operation. We shine some light on the Homodyne detection, available easily in photonic architectures, in App.~\ref{App:Phys_Imp}. This is also not a measurement process that is easily available on other platforms. Measuring the quantum information contained in an oscillator is one of the problems of CV control where a DV ancilla has often proved helpful. We will talk about some of these strategies in Chapters~\ref{chapter:na-qsp} and~\ref{chapter:GKP-qec}.

\subsubsection{Error Channels}
Damping and decoherence of quantum systems coupled to a bath are typically described using a master equation (ME) for the density matrix. The ME is derived by making the Born-Markov approximation on the assumption that the coupling to the bath is weak and the bath is memoryless.  The Lindblad form of the master equation guarantees that the time-evolution of the density matrix corresponds to a completely positive trace-preserving (CPTP) map
\begin{equation}
    \frac{d\rho}{dt} = -i[H,\rho] + \sum_j \mathcal{D}(E_j)\rho,
\end{equation}
where $\mathcal D(E_j)$ is a `superoperator', also called the Lindbladian, whose action on the density matrix is given by
\begin{equation}
    \mathcal{D}(E_j)\rho = E_j\rho E_j^\dagger - \frac{1}{2}\{E_j^\dagger E_j,\rho\},
\end{equation}
where $\{A,B\}=AB+BA$ and the $E_j$ are `jump' operators acting on the Hilbert space of the oscillator, describing the effects of coupling to the bath. We use master equations to simulate a noisy CV error channel in this thesis with the help of QuTiP~\cite{Johansson2013}. Below we give the Kraus operators (see Eq.~(\ref{eq:kraus})) or the jump operators for the Lindbladian for various oscillator error channels
\paragraph{Amplitude damping:}\label{app:ampdamp}  
Following Ref.~\cite{michael2016new} we label the Kraus operators by the total number of quanta (photons or phonons) $\ell$ lost during time $\tau$
\begin{equation}
    \hat K_\ell =\sqrt{\frac{(1-e^{-\kappa \tau})^\ell}{\ell !}} e^{-\frac{\kappa}{2}\tau \hat n}\hat a^\ell;\,\,\ell=0,1,2,3,\ldots
\end{equation}
where $\hat n$ is the photon number operator. Here, $\kappa$ is the decay rate for photon loss. We see that the $\hat a^\ell$ term in $\hat K_\ell$ destroys $\ell$ photons, however, even if by chance, no photons are lost, the $\ell=0$ Kraus operator $\hat K_0=e^{-\frac{\kappa}{2} \tau \hat a^\dagger \hat a}$ enacts a `no-jump backaction,' causing the probability amplitude for Fock states (all but $\ket{0}$) to be reduced. Remarkably, the no-jump evolution dissipates any state towards the phase-space origin (or, the vacuum state) \emph{even though no photons are leaving the cavity}.  To first order in the time interval $\tau$, we can neglect all but the lowest two Kraus operators
\begin{align}
   \hat K_0 &\approx I - \frac{\kappa}{2} \tau \hat a^\dagger \hat a,\\
   \hat K_1 &\approx \sqrt{\kappa\tau}\,\hat a.
\end{align}
Keeping terms only that are linear in $\tau$, these two operators satisfy the completeness relation for Kraus maps. From the action of these Kraus operators for small time intervals, we can derive the master equation for the continuous evolution of the density matrix under the amplitude damping channel,
\begin{align}
    \frac{d\rho}{dt}=\kappa\Bigg(\hat{a}\rho\hat{a}^\dagger-\frac{1}{2}\{\hat{a}^\dagger\hat{a},\rho\}\Bigg).
\end{align}
We would like to clarify that $[.,.]$ indicates commutators while $\{.,.\}$ indicates anti-commutators. Amplitude damping is the dominant source of error in superconducting and photonic architectures and has thus been used for analysis in Chapters~\ref{chapter:GKP-qec} and~\ref{chapter:qec-control}. 
\paragraph{Dephasing channel:} This error channel is the dominant source of error in trapped-ion architectures. The dephasing channel is a rotation in phase space  under the action of $\hat P(\theta)=e^{i\theta \hat a^\dagger \hat a}$ (defined in Eq.~(\ref{eq:Ptheta})) by a random angle $\theta$,
\begin{equation}
    \mathcal{E}(\rho)=\int_{-\pi}^{\pi} d\theta\ p(\theta) e^{-i\theta a^\dagger a}\rho e^{i\theta a^\dagger a} \,.
\end{equation}
The channel randomizes the phase according to the probability distribution $p(\theta)$. When $\rho$ is expressed in the photon number basis
\begin{equation}
   \mathcal{E}(\rho)=\sum_{m,n}\bra{n}\rho\ket{m}\int_{-\pi}^{\pi} d\theta\  p(\theta) e^{-i\theta(n-m)}\ket{n}\bra{m} \,, 
\end{equation}
we see that the channel preserves the diagonal elements and the unit trace of the channel (as a CPTP map) but reduces the magnitude of the off-diagonal elements, indicating decoherence. 
\paragraph{Heating:} Heating is the process of adding energy to the system. The error jump operator, in this case, is $\sqrt{\gamma} \hat a^\dagger$. Compared to photon loss,  the rate for this error is significantly lower in superconducting circuits and photonics. Hence, we do not discuss this error channel in detail.
\paragraph{Displacement error channel: } The displacement error channel is a quantum channel that acts on a quantum state by displacing the state in the phase space of the oscillator according to some probability distribution, that fades away for large displacements. Taking this probability distribution to be a Gaussian, the action of the displacement channel on a quantum state is given by the map,
\begin{align}
    \rho\rightarrow \int \ d^2\alpha \ p(\alpha)D(\alpha)\rho D^\dagger(\alpha),
\end{align}
where $D(\alpha)$ is the displacement operator and $p(\alpha)=\frac{1}{\pi\sigma^2}e^{-|\alpha|^2/\sigma^2}$. Here, $\sigma$ denotes the noise strength. For GKP codes, whose stabilizers are displacement operators, measurement-based QEC twirls the photon loss channel into the displacement error channel. This result is the premise of the Gottesman-Kitaev-Preskill codes for CV quantum error correction in photonic architectures where homodyne measurements are considered a free resource.
\subsection{Error Correcting Codes in CV systems}
 The multi-level system of a truncated oscillator can be used to redundantly encode a subspace representing a smaller number of levels (qudits) where errors can be detected or corrected. The stabilizer formalism discussed in Sec.~\ref{sec:EC} can be used to define the two broad classes of stabilizer codes encoding a qubit (or even a qudit) in a single oscillator, namely Gottesman-Kitaev-Preskill (GKP) codes~\cite{gottesman2001encoding} and the rotationally symmetric (RS) codes~\cite{grimsmo2020quantum}. The GKP codes, named after their inventors, were among the first codes discovered in the class of bosonic codes or oscillator codes. These codes are the quantum analog of the lattice codes~\cite{royer2022encoding,conrad2024lattices}. On the other hand, the much newer class of rotationally symmetric codes are the quantum analog of the spherical codes~\cite{jain2024quantum}. In this thesis we will cover the GKP codes in detail in Chapters~\ref{chapter:GKP-qec} and~\ref{chapter:qec-control}, discussing novel gates and error correction processes. We discuss the CV-DV architecture methods used to stabilize (or error-correct) such codespaces in Sec.~\ref{hybrid}. Here, we lay out a brief overview of both classes of codes for the discussions in the thesis that precede these chapters.
\clearpage
\subsubsection{The GKP codes}
The ideal GKP codes are infinite-energy (un-normalizable) translationally invariant states that were designed to protect against displacement noise in phase space. These codes yield optimal protection under a Gaussian displacement channel~\cite{albert2018performance}. The photon loss operator, which is the physically realistic noise channel, can be expressed in terms of the displacement channel. This was the foundation for the proposal in the original design~\cite{gottesman2001encoding} of the codes. 

The ideal (or, infinite-energy) GKP codewords are superpositions of the eigenstates of position and momentum operators. These codes can be realized using an arbitrary lattice in the phase space with certain constraints, such that the logical operators,
\begin{align}
    \mathrm{Z}_L=D(i\sqrt{\pi /d}), \ \mathrm{X}_L=D(\sqrt{\pi /d})
\end{align}
cover an area of $\pi/d$ in phase space. From Eq.~(\ref{eq:UCDcompos4}), we note that this condition satisfies, $\mathrm{Z}_L\mathrm{X}_L=-\mathrm{X}_L\mathrm{Z}_L=$ for $d=2$ as required for Pauli matrices. Also, the distance of the code against displacement errors is $(1/2)\sqrt{\pi/d}$. That is, under a displacement error of magnitude less than $\sqrt{\pi/d}$ along position or momentum, the erroneous state can be mapped back to the codespace without a logical error. Since there are two independent bases in which displacement errors could take place, position, and momentum, the code requires two stabilizers to stabilize the codespace given by,
\begin{align}
   S_ \mathrm{Z}=D(i\sqrt{\pi d}), \ S_\mathrm{X}=D(\sqrt{\pi d}).
\end{align}
These operators (and thus, the code) are invariant under displacement by $\sqrt{\pi d}$. Note that the length of stabilizer displacements increases while the length of displacements associated with the logical operators decreases with an increase in dimension $d$ of the code space. Thus, for encoding a qudit, the distance of the code (defined here as the minimum length of displacement that would cause a logical error) decreases. In addition, the error correction procedure would take a longer time when measuring the stabilizers corresponding to longer displacements. Longer-time circuits are prone to more errors in a CV-DV architecture. Thus, increasing $d$ yields lower performance for stabilization since ancilla errors present the limiting bottleneck to the experimental implementations~\cite{sivak2023real}. However, we are also protecting a larger subspace with these longer stabilizers, which may likely increase the efficiency of quantum computation~\cite{brock2024quantum,jordan2012quantum}.

In realistic systems, the photon number distribution in the code words needs to be bounded and hence they are not fully translationally invariant. For example, Fig.\ref{Wigner}(b) shows the Wigner function of the logical $\ket{0}$ codeword of the GKP code whose lattice is cut off with a Gaussian envelope. The definition of these finite-energy code words is given in the next Chapter for a discussion of deterministic state preparation. We will not elaborate on the effects of these bounded systems until Chapter~\ref{chapter:GKP-qec}. Ref.~\cite{albert2018performance} showed that there exists a recovery map (or an error correction map) under which these codes are optimal for the correction of photon loss. To justify this result, the authors also gave a mapping of the photon loss channel to a displacement channel, when composed with the amplification noise channel. The authors note this in Ref.~\cite{albert2018performance}, and it was also later shown that the associated amplification noise could worsen the performance of GKP error correction~\cite{hastrup2023analysis} meaning that this is not the optimal decoder. A theoretically optimal recovery map was derived in Ref.~\cite{zheng2024performance}, however, it is not clear how these circuits can be realized in practical circuits. Multi-mode extensions of these codes~\cite{harrington2001achievable,conrad2022gottesman,conrad2024lattices,royer2022encoding} which use lattices in a $2N$-dimensional symplectic space to encode qudits in $N$ oscillators. These codes could be used for encoding an oscillator into many oscillators~\cite{noh2020encoding}. Note that, however, such an encoding does not have a threshold for the random displacement channel~\cite{hanggli2021oscillator}. We pose the open question: \textit{Is this true for the case of photon loss channel also?}
\subsubsection{Rotationally Symmetric (RS) codes}
These codes use $N-$ legged cat states, which are superpositions of $N$ blobs of coherent states, located on a circle of some radius. See Fig.~\ref{Wigner}(b) for the example of a two-legged cat state. The rotationally symmetric codewords are symmetric under rotations by $2\pi/N$ in phase space if using $N-$ legged cat codes as the basis states for the encoding.
Thus, the stabilizer $S_\mathrm{z}$ and logical operator $\mathrm{Z}_L$ are defined as,
\begin{align}
    S_\mathrm{z}=\hat P(2\pi/N), \  \mathrm{Z}_L=\hat P(\pi/N)
\end{align}
where $\hat P(\theta)$ is the phase space rotation defined in Eq.~(\ref{eq:Ptheta}). The distance of the code against rotation errors is $\pi/N$. That is, under a rotation error of magnitude less than $\sqrt{\pi/N}$ in phase space, the erroneous state can be mapped back to the codespace without a logical error. The $X$-basis stabilizer and logical operators depend on the phase operator ($\hat \theta$) which follows, $[\hat \theta,\hat n]=i$ where $\hat n$ is the number operator. The quantum phase operator is difficult to rigorously define mathematically~\cite{barnett1986phase}, and has a complicated practical realization in realistic systems. Hence RS codes are proposed~\cite{grimsmo2020quantum} to be fitting for the measurement-based quantum computation formalism where the universal set replaces $X$-type operations with $X$-basis measurements.
\section{Open Problem: Hierarchy of CV Operations}\label{open-hierarchy}
A classification of DV quantum operations, known as the Clifford hierarchy, has proven extremely useful in developing the foundations for practical and universal fault tolerance. The Clifford hierarchy was introduced in~\cite{gottesman1999demonstrating} in connection with a generalized technique for gate teleportation to reduce the resource cost of fault-tolerant quantum computation. Given a known unitary $U$, gate teleportation takes an input quantum state $\ket{\psi}$ to $U\ket{\psi}$ with some corrections using a Clifford circuit. The corrections related to teleportation, corresponding to any unitary $U$, belong to the set described by $UPU^\dagger$, where $P$ is the set of Pauli gates $\{X,Y,Z\}$. This relation yields the following hierarchy,
\begin{equation}
    CPC^\dagger \in \mathcal{C}_{n-1}\quad \text{if}\quad C\in \mathcal{C}_n\quad \text{for }n>1
\end{equation}
Here $\{\mathcal{C}_n,n=1,2,3,\ldots\}$ represents an infinite sequence of finite sets, where $\mathcal{C}_1$ is the set of Pauli gates. The so-called `Clifford group' is $\mathcal{C}_2$, the second level of the Clifford hierarchy which conjugates a Pauli into another Pauli. As a result, circuits comprising only Clifford group operations are efficiently simulable classically. We have that
\begin{equation}
    \mathcal{C}_n\subset \mathcal{C}_{n+1},\quad\forall n.
\end{equation}
Any gate that does not belong to $\mathcal{C}_2$ is a non-Clifford gate and can be used for universality along with Clifford group generators. The teleportation of the non-Clifford gates in $\mathcal{C}_3$ requires only a Clifford teleportation circuit with Clifford correction. Thus, teleportation of a non-Clifford gate is a viable means to perform universal fault-tolerant quantum computation for any quantum computing architecture with fault-tolerant Clifford operations and a resource `magic' state (created with a non-Clifford gate, typically the $T$ gate defined in Eq.~(\ref{eq:Tgate})). Thus, this method of applying non-Clifford operations has proven essential for fault tolerance. 

Additionally, the hierarchy was later used in Ref.~\cite{bravyi2013classification} to show the relationship between the transversality of logical operations and the locality of quantum error correcting codes. Transversality is the easiest form of logical operation that is naturally fault-tolerant. A transversal logical gate has constant support on the codeword of a scalable code, irrespective of the distance of the code. This constant support introduces a constant amount of error on the codewords which can be corrected better with increasing distance (and consequently, size) of the code. The existence of such constructions proves useful for FTQC. Thus, the Clifford hierarchy is extremely useful in studying quantum error correction codes for fault tolerance.
\begin{table}
    \centering
    \begin{tabular}{|c|c|}
    \hline
         Discrete (finite) & Continuous (infinite) \\
         \hline
         Qubit & Oscillator \\
         Pauli group generated by $\{X,Z\}$ & Displacements generated by $\{e^{i\hat{x}},e^{i\hat{p}}\}$\\
         Stabilizer states & Gaussian states \\
         Clifford group & Gaussian operations\\
         Pauli/Clifford Channels & Gaussian Channels\\
         Pauli measurements & Homodyne measurements\\
         State Tomography & Wigner function\\
         Stabilizer states/Clifford operations 2-design & Gaussian states/operations are not 2-design\\
         \hline
    \end{tabular}
    \caption[CV-DV analogy]{Rough analogies between discrete and continuous variable quantum information~\cite{iosue2022continuous}}
    \label{tab:disc_cont}
\end{table}

\subsubsection{The Gaussian Hierarchy} 
 Given the usefulness of the Clifford hierarchy in quantum error correction, we would like to ask if such a classification is possible for CV operations as well. Note that, however, the set of CV unitaries is a continuous dense set. Thus, it might be better to define this hierarchy on the set of parameterized gates, i.e., the Hamiltonians of the CV unitary operations. This approach was used in Ref.~\cite{bartlett2002efficient} to prove that Gaussian operations are classically simulable. Thus, we propose the \emph{Gaussian hierarchy}, a classification of CV operations.
\begin{align}
    \mathcal{G}_n=\{U|[H(D),H(U)]=H(U^\prime)\implies U^\prime\in \mathcal{G}_{n-1}\}: D\in\mathcal{G}_1\quad\forall\quad n>1,
\end{align}
where $H(U)$ denotes the algebra that generates the corresponding parametrized unitary operation (that is, the Hamiltonian of the unitary, ignoring its overall scale). Here, $D$ refers to displacement operations, and hence this hierarchy can be studied using the commutators of Hamiltonians $H(U)$ with the phase space basis vectors $\hat x$ and $\hat p$.  
\subsubsection{$\mathcal{G}_1:$ Displacements -- Analogous to $\mathcal{C}_1$}
For CV quantum computation, in~\cite{bartlett2002efficient} the analog of the Pauli group for oscillators is the Heisenberg-Weyl group $HW(1)$ which is a continuous Lie group composed of displacements. The algebra that generates this group $hw(1)$ is spanned by the canonical operators $\{\hat x, \hat p,iI\}$ satisfying the required commutation relations. We know that the product of two displacements is (up to a global phase) another displacement and that every displacement has an inverse. Similar to the Pauli group, the displacement operators form a complete basis for CV operations. Thus, the displacements in phase space are equivalent to the Pauli group.
\subsubsection{$\mathcal{G}_2:$ Gaussian Operations -- Analogous to $\mathcal{C}_2$}
 Any unitary generated by quadratic Hamiltonian in $x,p$, also known as Gaussian operations, preserves displacements under conjugation and hence, is the analogous class of gates corresponding to $\mathcal{C}_2$ in the Clifford hierarchy. These operations simply induce a linear transformation (rotation, translation, and symplectic rescaling) on the phase-space coordinates. It follows from this that a phase-space displacement is mapped onto a different displacement under conjugation by any Gaussian operation. Thus, Gaussian operations are the Clifford group ($\mathcal{C}_2$) analog of the CV hierarchy.

We can draw an analogy \cite{bartlett2002efficient} between the simplicity of computing time evolution under quadratic bosonic Hamiltonians (Gaussian operations) and the Gottesman-Knill theorem \cite{Gottesman-Knill-Theorem,AaronsonGottesmanCliffordSims} that a quantum computer based on qubits and using only Clifford group operations is easy to simulate classically.  
The fact that DV Clifford circuits can be efficiently simulated classically is related not just to the ability to do stabilizer updates but also to the fact that the full non-classical correlations inside Bell states cannot be revealed without making non-Clifford (e.g., T-gate 45-degree) rotations on the Bloch sphere (or rotating the measurement axis by 45 degrees) to violate the Bell inequalities. Such non-Clifford rotations are sometimes described as introducing `magic' (or `non-stabilizerness') to a state \cite{Emerson_Magic_Resource,PhysRevLett.115.070501,doi:10.1098/rspa.2019.0251,Bravyi2019simulationofquantum,PRXQuantum.4.010301}.  Similarly, CV transformations by Gaussian operations only update the mean and variance of Gaussian states (analogous to stabilizer updates), and fail to introduce any negativity in the Wigner function (defined in Sec.~\ref{CV}). If we start with a Gaussian state (without any Wigner negativity), then that property is preserved under all Gaussian operations. Hence the expectation value of bosonic observables can be readily obtained by classical importance sampling of the wave function, without suffering from any sign problems \cite{PhysRevLett.109.230503,Veitch_2013}. This is not generically the case for non-Gaussian states and thus Wigner negativity for bosonic systems is akin to `magic' in qubit states. Ref.~\cite{Jaffe_doi:10.1073/pnas.2304589120} highlights the similarity between Gaussian CV states and stabilizer DV states.

\subsubsection{Non-Gaussian Operations -- Beyond $\mathcal{C}_2$ and $\mathcal{G}_2$}
Note that each level in this hierarchy contains operations,
\begin{align}
    \mathcal{G}_n=\{U|U=\exp\{i\hat f(\hat x,\hat p)t\}\quad \mathrm{s.t.}\quad \textrm{deg}(f)=n\},
\end{align}
and is an infinite continuous set of unitaries. Can we say something special about the operations for $n>2$ in the Gaussian hierarchy? The first question to ask here is, if we restrict ourselves to a discrete encoding, say a GKP qubit encoding, what is the correspondence between $\mathcal{C}$ and $\mathcal{G}$. We will give a specific example to narrate why this question might be of interest. Even though we use jargon from the GKP codes literature in this section, the reader does not need to know the details of the encoding to follow the arguments laid out in this section. For details on this encoding, we direct the readers to Chapters~\ref{chapter:GKP-qec} and~\ref{chapter:qec-control}.

Let us assume a square GKP encoding, where displacements execute Pauli operations. In this case, an operation $U=e^{ig(\hat a^\dagger \hat a)^2}\in\mathcal{G}_4$ yields a square root of logical Hadamard $\sqrt{H}$ on the codespace~\cite{royer2022encoding}. Where does this gate lie in the Clifford Hierarchy? We prove that $\sqrt{H}\notin \cup_n\mathcal{C}_n$, that is, this gate is one of the uncountably many gates that lie outside the Clifford Hierarchy!

Our proof uses repeated conjugation of Pauli operators. Upon first conjugation, we get,
\begin{align}
    \sqrt{H}\sigma_\textrm{x}\sqrt{H}^\dagger=\frac{H+\sigma_\mathrm{y}}{\sqrt{2}}=H_\mathrm{x}^{(1)},\quad \sqrt{H}\sigma_\textrm{z}\sqrt{H}^\dagger=\frac{H-\sigma_\mathrm{y}}{\sqrt{2}}=H_\mathrm{z}^{(1)}\quad \quad (1).  
\end{align}
The Pauli $\sigma_\textrm{y}$ operator conjugation is straightforward hence we do not discuss it. Now we further conjugate Paulis with the resulting gates, 
\begin{align}
    H_\mathrm{x}^{(1)}\sigma_\textrm{x}(H_\mathrm{x}^{(1)})^\dagger,\quad H_\mathrm{x}^{(1)}\sigma_\textrm{z}(H_\mathrm{x}^{(1)})^\dagger,\quad H_\mathrm{z}^{(1)}\sigma_\textrm{x}(H_\mathrm{z}^{(1)})^\dagger, \quad H_\mathrm{z}^{(1)}\sigma_\textrm{z}(H_\mathrm{z}^{(1)})^\dagger \quad\quad (2),
\end{align}
and recursively repeat this process for $N$ times. We call the set of gates we collect after $N$ recursions a conjugacy set.
\begin{align}
    \mathrm{Conj}_H=\{H_\mathrm{x}^{(1)},H_\mathrm{z}^{(1)},...,H_\mathrm{x}^{(N)},H_\mathrm{z}^{(N)}\}&=\Big\{\frac{H\pm \sigma_\mathrm{y}}{\sqrt{2}},..\Big\}\not\supset P,\\ G(\mathrm{Conj}_H)&=\textrm{const.}\quad\forall\quad N,
\end{align}
where $G(\mathrm{Conj}_H)$ is the cardinality of $\mathrm{Conj}_H$.
A gate in the Clifford hierarchy through such successive conjugations of the Pauli operator goes up the ladder in the Clifford hierarchy, eventually yielding a Pauli operation at some point. In contrast, after only $N=4$, (1) we realize that the cardinality $G$ of the conjugacy set $\mathrm{Conj}_H$ is constant for increasing $N$, and (2) this set does not include the Pauli gates. That is, conjugation under $\sqrt{H}$ will never end up in $\mathcal{C}_1$ under recursive conjugations of Pauli gates. Hence, our conjecture that, this gate lies outside the Clifford Hierarchy, is confirmed. 

Formally, we present the following insights and questions. 
\begin{itemize}
    \item If a gate belongs to $\mathcal{C}_n$, what is the lowest degree of polynomial for the Hamiltonian which corresponds to a logical gate in a bosonic error correcting code? The example that triggered this question was the fact that $\sqrt{H}\in\mathcal{C}_\infty$ can be obtained for GKP qubits using only the Kerr non-linearity ${\hat a}^{\dagger^2} \hat a^2$, a generator of unitaries in $\mathcal{G}_4$. This result indicates that there is no one-to-one correspondence between the complexity of Hamiltonians used to engineer logical gates and the Clifford hierarchy (for GKP logical qubit codes in an oscillator) as one might wrongly perceive from the analogy between Gaussian operations (obtained from quadratic or lower-order Hamiltonians) and Clifford operations. The aforementioned example indicates a structure in CV systems for quantum computation or quantum error correction in terms of feasibility or ease of logical operations which, in turn, could yield high-fidelity and low-overhead logical quantum operations.
    \item The above question relates to the hierarchy of CV gates in terms of the degree of polynomial for the Hamiltonians used to describe logical operations of various bosonic error correction codes. This is the hierarchy we suggest to obtain a classification of bosonic error correcting codes analogous to Pauli and non-Pauli stabilizer codes/linear and non-linear codes. In this structure, the lowest hierarchy (Pauli-stabilizer) QEC code is the Gottesman-Kitaev-Preskill  (GKP) code. Rotation Symmetric (RS) Codes, on the other hand, have a highly complicated position in this hierarchy. See Tables~\ref{tab:gkp_compare_codes} and~\ref{tab:rs_compare_codes}. Such classification could give better insights into the feasibility of error correction and fault tolerance for CV error-correcting codes.
    \begin{table}
        \centering
        \begin{tabular}{||c|c|c||}
            \hline
            \hline
            Logical Space &CV operations&$\mathcal{G}_n$\\
            \hline
            Stabilizers & $\{D(\sqrt{2\pi}),D(i\sqrt{2\pi})\}$ &$n=1$\\
            Pauli Operators & $\{D(\sqrt{\pi/2}),D(i\sqrt{\pi/2})\}$ &$n=1$\\
            Clifford Operators & $\{\mathrm{CX}=e^{2i\hat x\otimes\hat p},\mathrm{H}=\hat P(\pi/4),\mathrm{S}=e^{i4\hat x^2}\}$ &$n=2$\\
            Non-Clifford Operators &$\mathrm{T}=f(\hat x)$~\cite{hastrup2021unsuitability,gottesman2001encoding}&$n= 3$\\
            \hline
            \hline
        \end{tabular}
        \caption[Gottesman-Kitaev-Preskill codes in light of the Gaussian hierarchy.]{Gottesman-Kitaev-Preskill codes in  light of the Gaussian hierarchy. Here $n$ denotes the lowest Gaussian hierarchy level each set of operations belongs.}
        \label{tab:gkp_compare_codes}
    \end{table}
     \begin{table}
        \centering
        \begin{tabular}{||c|c|c||}
            \hline
            \hline
            Logical Space &CV operations&$\mathcal{G}_n$\\
            \hline
            Stabilizers &$S_\mathrm{z}=\hat P(2\pi/N)$ &$n=2$\\
            Pauli Operators & $\mathrm{Z}_L=\hat P(\pi/N)$ &$n=2$\\
            Clifford Operators & $\{\textrm{CZ}:e^{i\frac{\pi}{N^2}(\hat N\otimes\hat N)},\quad \textrm{S}:e^{i\frac{\pi}{2N^2}\hat N^2}\}$ &$n=4$\\
            Non-Clifford Operators &$\mathrm{T}=e^{i\frac{\pi}{4N^4}\hat N^4}$&$n=8$\\
        \hline
        \hline
            \end{tabular}
        \caption[Rotationally symmetric codes in the light of the Gaussian hierarchy.]{Rotationally symmetric codes in the light of Gaussian hierarchy. Note that we only give $Z$-type gates since the code is proposed for a computing architecture with only diagonal gates and $X$-basis measurements. Here $N\in2\mathbb{Z}$ defines the encoding used for the rotationally symmetric codes. Same as Table.~\ref{tab:gkp_compare_codes} $n$ denotes the lowest Gaussian hierarchy level each set of operations belongs to.}
        \label{tab:rs_compare_codes}
    \end{table}
    \item The gate $\sqrt{H}$ can be decomposed into a short depth circuit using the Clifford + T set as follows,
    \begin{align}
        \sqrt{H}=iSHTH SHT^\dagger H S^\dagger.
    \end{align}
    An open question is: Is it possible to find a gate that has a longer circuit depth or does not have an exact decomposition in Clifford + T but lies in a specific level of the Gaussian hierarchy? 
\end{itemize}

\begin{myframe}
\singlespacing\begin{quote}
To summarize, it is interesting that, for the GKP encoding of qubit in an oscillator, $\mathcal{C}_n\subseteq\mathcal{G}_n$ for $n\ge 4$ since there is a gate in $\mathcal{G}_4$ which is not in $\mathcal{C}_4$ and $\mathcal{G}_{n-1}\subset\mathcal{G}_n,\mathcal{C}_{n-1}\subset\mathcal{C}_n$. The questions that arise as the next steps in studying this hierarchy are, \textit{Is $\mathcal{C}_n\subset\mathcal{G}_n$ for $n\ge 4$? Is this result true for $n=3$?} Such a result will establish genuine non-correspondence between the non-Clifford and non-Gaussian operations. A follow-up big-picture question is, \textit{What are the repercussions of such analogies on a CV analog of the Solovay-Kitaev theorem~\cite{SK} and ideas of transversality~\cite{bravyi2013classification} in multi-mode CV codes?}
\end{quote}
\end{myframe}

\doublespacing
\section{Hybrid CV-DV Systems}\label{hybrid}

 The hardware efficiency and power of hybrid oscillator-qubit systems has been recently demonstrated with quantum error correction for memory close to or beyond the break-even point using a variety of CV quantum error correcting codes in microwave resonators: the cat code~\cite{ofek2016extending}, the binomial code~\cite{ni2023beating}, the truncated 4-component cat code \cite{gertler2021protecting}, and the Gottesman-Kitaev-Preskill (GKP) code~\cite{sivak2023real,campagne2020quantum}. Error correction with the GKP code has also been demonstrated in a trapped-ion system~\cite{fluhmann2019encoding, de2022error}. Another natural application for these hybrid systems is the quantum simulation of physical models containing bosons, for example, lattice gauge theories~\cite{crane2409hybrid}, simulating the physics of spin-boson systems~\cite{ISA}, etc. Several recent experiments have explored the measurement of Franck-Condon factors in molecular photo-electron spectroscopy using efficient boson sampling in both the optical~\cite{quesada2019franck} and the microwave domain~\cite{hu2018simulation,wang2020efficient} and also explored non-adiabatic dynamics near conical intersections in molecular energy surfaces~\cite{wang2023observation}. These microwave boson-based simulations used quite modest hardware and achieved results that would have required circuit depths far beyond the capabilities of any currently existing qubit-only hardware systems. %usefulness of CV-DV in error correction and simulations

Thus, the convergence of CV and DV systems in hybrid architectures opens new frontiers in quantum information processing, both theoretically and practically. However, a dearth of efficient bosonic control methods including state preparation and measurements poses a challenge for any useful computation from these quantum systems. Note that an \emph{efficient circuit}, here, refers not only to the circuit with the shortest depth but also one that is robust in the presence of errors. Errors in the auxiliary control qubits during a long circuit can inhibit the advantages achieved by the hybrid architecture. Recent progress improving fault tolerance to ancilla errors in bosonic quantum error correction has been made through novel bosonic code designs \cite{royer2022encoding} and use of multi-level ancillae \cite{rosenblum_fault-tolerant_2018, ReinholdErrorCorrectedGates, tsunoda2023error,TeohPNAS_DualRail, khaneja2005optimal, heeres2015cavity,eickbusch2022fast}. 

In a hybrid superconducting-atom architecture, Rydberg atoms can be used to control a superconducting microwave resonator via the Jaynes-Cummings Hamiltonian \cite{haroche2013nobel,kumar2023quantum,hogan2012driving,garcia2019single}. In trapped-ion systems, the mechanical oscillation of the ions is controlled via lasers supplying forces that depend on the spin state of the individual ions~\cite{bruzewicz2019trapped,de2022error}. Mechanical oscillations of cold atoms in trapping potentials are only just beginning to be explored as a quantum control/computation platform~\cite{ISA,brown2022,scholl2023}. There have also been demonstrations where quantum opto- and electro-mechanical systems interface mechanical motion with the electromagnetic modes of optical resonators and microwave circuits~\cite{chu2020perspective}. The hybrid state space is composed of the states in the joint CV-DV Hilbert space. 

We will discuss the operations available for universal control in this architecture, and then give an interpretation of the CV-DV architecture stack shown in Ref.~\cite{ISA} in the light of this thesis.
\subsection{Hybrid Operations and Control}
These operations include joint maps on the space of oscillators and qubits. In Ref.~\cite{ISA} we develop instruction set architectures for the hybrid CV-DV processor. This architecture includes various options for universal hybrid control. Our work is primarily focused on the phase space instruction set, so we will only discuss this specific instruction set in this section. 
\subsubsection{Phase-Space Instruction Set}\label{phase-space}
The set comprises two parameterized gates: arbitrary qubit rotations ($\mathrm{R}$) (about any axis in the equatorial plane of the Bloch sphere) and conditional displacements ($\mathrm{CD}$), defined as,
\begin{equation}
    \mathrm{R}_\phi(\theta)=e^{-i\frac{\theta}{2}\sigma_\phi},\quad \mathrm{CD}(\beta,\sigma_\phi)=e^{(\beta \hat a^\dagger -\beta^* \hat a)\otimes\sigma_\phi}.
    \label{eq:phasespaceIS}
\end{equation}
Here $\sigma_\phi=\cos{\phi}\,\sigma_\mathrm{x}+\sin{\phi}\,\sigma_\mathrm{y}$, while $\hat a,\hat a^\dagger$ are the annihilation and creation operators on the oscillator 
subspace, respectively, and $\sigma_\mathrm{x},\sigma_\mathrm{y},\sigma_\mathrm{z}$ are the Pauli operators on the qubit subspace. The $\mathrm{CD}$s are hybrid operations that displace the oscillator by $\pm\beta$ in phase space depending on the qubit state through the eigenvalue of $\sigma_\phi$. Here $\beta=\Delta x+i\Delta p$ is a complex number parameterizing the phase space displacement. These gates have been realized in various superconducting circuits and trapped-ion experiments by means of dispersive interaction~\cite{leghtas_hardware-efficient_2013,campagne2020quantum,eickbusch2022fast,ding2024quantum} and sideband interaction~\cite{fluhmann2019encoding,de2022error}, respectively. The dispersive coupling and sideband interaction are given by $e^{i\chi\hat a^\dagger \hat a\sigma_\mathrm{z}}$ and $e^{\hat a\sigma_{-}+\hat a^\dagger\sigma_{+}}$, respectively. The sideband interaction, a native interaction available in the trapped-ion platform, can be easily represented as the product of two conditional displacements. The dispersive coupling, while not a native interaction to superconducting circuits, is achievable using Schrieffer Wolff tranformation in the dispersive regime. See Ref.~\cite{BlaiscQEDReviewRMP2020} for details. However, these gates are not trivially related to conditional displacement gates, Ref.~\cite{leghtas_hardware-efficient_2013} introduced a method to perform conditional displacement in a displaced oscillator frame with constant (or, strong) dispersive coupling. Such strong dispersive coupling could induce large unwanted Kerr nonlinearities. In Refs.~\cite{campagne2020quantum,eickbusch2022fast}, authors extended this idea to the regime of weak dispersive coupling where the effects due to Kerr nonlinearities are suppressed. Finally, in Ref.~\cite{ding2024quantum}, a Kerr-cat biased noise ancilla (see App.~\ref{App:Phys_Imp}) was used where the native coupling between Kerr-cat and oscillator is trivially a conditional displacement gate. 

For this thesis, we use so-called Wigner units \cite{ISA} in which the oscillator quadrature operators are $
\hat x=\frac{a+a^\dagger}{2},\hat p=\frac{\hat a-\hat a^\dagger}{2i}$. For these units, we have $[\hat x,\hat p]=\frac{i}{2}$ and the wave function of the minimum uncertainty vacuum state is given by, $\psi(x)=\Big(\frac{2}{\pi}\Big)^{1/4}e^{-x^2}$. See Sec.~\ref{app:phase-space} for details on these units. This instruction set is useful in the control of non-overlapping superpositions of Gaussian wave functions, such as squeezed states, cat states, and GKP states (see Chapter~\ref{chapter:state-prep}). 

We note that while the qubit rotations $\mathrm{R}_\phi(\theta)$ are only for axes lying on the equator of the Bloch sphere, they provide universal single-qubit control. As an aside, we also note that even if the natively available conditional displacement gate is controlled on $\sigma_\mathrm{z}$, conjugation with qubit rotations allows easy synthesis of $\mathrm{CD}(\beta,\sigma_\phi)$. In this convention, for purely real $\beta$, we have, 
\begin{align}
\braket{x|C\mathrm{D}(\beta,\sigma_\phi)|0_\Delta,+\phi}=\braket{x|e^{-i2\beta\hat p\otimes \sigma_\phi}|0_\Delta,+\phi}=\Big(\frac{2}{\pi}\Big)^{1/4}e^{-\frac{(x-\beta)^2}{\Delta^2}}\otimes\ket{+\phi},
\end{align}
where $\ket{+\phi}$ is the eigenstate of $\sigma_\phi$ corresponding to the $+1$ eigenvalue. This is a state displaced along the position axis by $|\beta|$ to the left or right conditioned on the qubit state. Alternatively, we can interpret $\mathrm{CD}$, not as a qubit-controlled displacement of the oscillator, but rather as an oscillator-controlled rotation of the qubit
\begin{align}
\mathrm{CD}(\beta,\sigma_\phi)&=e^{i2\hat v\otimes \sigma_\phi}=\mathrm{R}_\phi(-4\hat v(\beta)),\\
\text{where} \ \hat v(\beta)&=\mathrm{Im}(\beta)\hat x-\mathrm{Re}(\beta)\hat p,
\label{eq:hatvbetadef}
\end{align}
and where the qubit rotation angle $\hat v$ is now a quantum operator acting on the oscillator Hilbert space. We will denote $\mathrm{CD}$s with $\mathrm{Re}(\beta)=0$ as conditional momentum boosts. In addition, $\sigma_\phi$ for the case of $\phi=0,\pi/2$ will be denoted by $\sigma_\textrm{x},\sigma_\mathrm{y}$, respectively. To distinguish states of the qubit from oscillator Fock states, we will use $\ket{g},\ket{e}$ to represent the qubit ground $\ket{0}$ and excited $\ket{1}$ states.

\paragraph{Higher-order nonlinearities:} As we have mentioned before, DV systems such as transmons are nonlinear systems, while CV systems are (very close to) linear. For quantum control, when these systems are coupled, the CV system inherits unwanted non-linearities from the DV system. A common example is the Kerr nonlinearity (a perturbation term $\propto {b^\dagger}^2 b^2$ for the mode whose Hamiltonian is $\propto b^\dagger b$). Such terms affect the higher Fock states much more than the lower Fock states, deforming the CV states. Ignoring the effect of such non-linearities while designing control pulses can limit the performance of a hybrid CV-DV architecture. 

\paragraph{Dissipative channels:} The hybrid architecture can be used to engineer dissipative channels, that have proven useful for state preparation. Let us say that the target state, $\ket{\psi}$, has a dissipator $\hat d$ such that $\hat d\ket{\psi}=0$. In this case, we can use the DV ancillae to engineer dissipation under this operator using the Hamiltonian,
\begin{align}
    \mathcal{D}(\hat d)=\hat d\hat\sigma_{+}+\hat d^\dagger \hat\sigma_{-}
\end{align}
For example, the dissipator for a vacuum is $\hat d=\hat a=\hat x+i\hat p$ which is the dissipator that exists naturally in nature. The dissipators for squeezed states (represented by Eq.~(\ref{eq:squeezed})) and coherent states, shown in Fig.~\ref{Wigner}(a), are given by, $\hat d=\hat x+i\Delta^2\hat p$ and $\hat d=\hat a-\alpha$, respectively.

The dissipator for more exotic (non-Gaussian) states like $N$-legged cat states is given by $\hat d=\hat a^n-\alpha^n$ (compare the case of $n=2$ with two-legged cat states Fig.~\ref{Wigner}(a) for intuition).  Importantly, for states like vacuum, and squeezed states, $\hat d=f(\hat x)$ where the degree of $f$ is $1$. In this case, the above dissipation can be engineered using trotterized circuits composed of conditional displacements and conditional momentum boosts. These channels have been useful for the stabilization of bosonic codes. This aspect will be discussed in more detail in Chapter~\ref{chapter:GKP-qec}.
\paragraph{Operator fidelity:} For hybrid systems we extend the definition of operator fidelity between two operators $U,V$ given in Chapter~\ref{CV} as, 
\begin{align}
    \Bigg|\frac{1}{2d}\mathrm{Tr}(PU^\dagger V)\Bigg|^2.
\end{align}
This operator fidelity is computed on the oscillator-qubit subspace with projector $P=\sum_{\ell=0}^{d-1} \ket{\ell}\bra{\ell}\otimes \sum_{q=0}^1 \ket{q}\bra{q}$. This is the projector on the joint subspace of a truncated oscillator with dimension $d$ and a qubit. 
\subsection{Hybrid Architecture}
\begin{figure}[htb]
\centering
    \includegraphics[width=\linewidth]{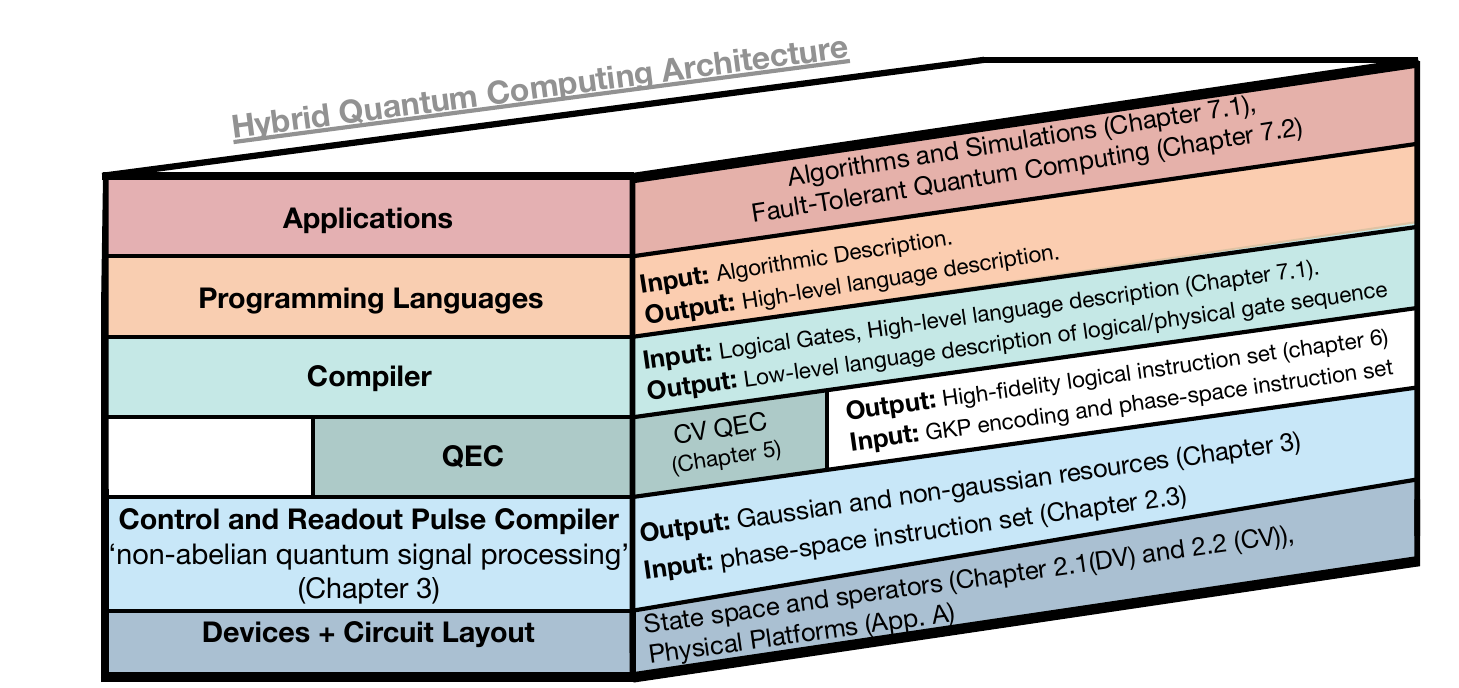}
  \caption[Thesis structure concerning hybrid CV-DV quantum computing architecture.]{\textbf{The structure of this thesis in light of a bottom-up architecture for hybrid CV-DV quantum processors.} See text for details.}
    \label{fig:hybrid-stack}
\end{figure}

A hybrid architecture comprises various layers as shown in the stack in Fig.~\ref{fig:hybrid-stack}. 
The various layers of an architecture stack include the physical layer which constitutes the building block or the hardware layer. For each layer in the stack presented by the square face of the cuboid, we give the input it uses from the stack below it and the output it gives to the stack above it. We have discussed details regarding the physical layer in an abstract sense here in the current chapter. The next layer is a control layer designed to process readout and gate sequences for the higher layers in the stack using the physical operations available. Such control sequences include the non-abelian composite sequence described in Chapter ~\ref{chapter:na-qsp}. The next layer is the compiler layer which may or may not contain a QEC encoding. For the case of no QEC, we give Gaussian and non-Gaussian resource state preparation in Chapter~\ref{chapter:state-prep}. However, these resource states can be easily used for error correction. For this thesis, we give special focus to the case of including a bosonic QEC layer, discussed in Chapter~\ref{chapter:GKP-qec}. Thus, for the compiler layer, with QEC, we give high-fidelity logical gate sequences using the control architecture in Chapter~\ref{chapter:qec-control}. Finally, as an application to this architecture in the NISQ era, without QEC, Chapter~\ref{AOM} gives compilation schemes for phase estimation and insights into quantum random walks in phase space. Chapter~\ref{AQM} on the other hand discusses its use case in a fully fault-tolerant quantum computer using multiple modes when a single-mode or two-mode bosonic encoding is not enough to achieve quantum advantage. As mentioned before, each chapter includes its open problem, or as we may put it, a few of the many unanswered questions for each stack in the hybrid CV-DV architecture.
    \chapter{Control of CV Systems using DV Ancillae: Towards Non-Abelian Quantum Signal Processing} \label{chapter:na-qsp}
\begin{myframe}
\singlespacing
\begin{quote}
\textit{How can we control a CV system using a DV ancilla efficiently?} The continuous-variable (CV) quantum information resources available in hybrid oscillator-qubit systems enable us to harness quantum advantage at a lower overhead relative to discrete-variable (DV) systems that rely solely on qubits. In this chapter, we present new techniques to harness this quantum advantage by extending the concept of quantum signal processing (QSP)~\cite{low2017optimal,martyn2021grand,motlagh2023generalized,rossi2022multivariable}
from the DV domain with classically controlled qubit rotations to the CV domain where the qubit rotations are controlled by the non-commuting position and momentum coordinates of quantum oscillators. We utilize the rich commutator algebra of such hybrid systems to build several experimentally practical and useful circuit `gadgets,' thereby taking some first steps towards a full theory of \emph{non-abelian quantum signal processing}. 
\end{quote}
\end{myframe}

\doublespacing

Robustness against systematic errors in quantum control is essential for reliable quantum operations below the quantum error correction threshold. Quantum signal processing (QSP) is a technique for transforming a unitary operation parameterized by a variable $\theta$ into a unitary parameterized by a polynomial function $f(\theta)$.  This technique underlies many important quantum algorithms and is a descendant of composite pulse techniques developed in NMR spectroscopy which were designed to make spin rotations robust against systematic fluctuations in the value of the classical control parameter $\theta$.  There has been some prior advancement in the direction of CV-DV control using QSP~\cite{rossi2022multivariable,sinanan2023single}. 
However, these works correspond to the case of commuting variables. In this chapter, we extend the concept of quantum signal processing to the case of multiple control parameters $\hat\theta_1,\hat\theta_2,\ldots$ which are themselves non-commuting quantum operators—namely the positions and momenta of quantum harmonic oscillators. The non-commutativity of the control parameters implies that they unavoidably suffer intrinsic quantum fluctuations. Still, the richer commutator algebra also significantly enhances the power of QSP and reduces circuit depths. 

We show this by introducing a composite pulse sequence using non-commuting quantum control variables which we dub the Gaussian-Controlled-Rotation ($\mathrm{GCR}$).  
The $\mathrm{GCR}$ sequence is designed to produce a well-defined rotation of the ancilla qubit that is robust against errors due to quantum fluctuations in the position and momentum of the oscillator, and we show that it achieves a minimum of $4.5\times$ reduction in circuit depth compared to the best-known QSP pulses with commuting variables, such as $\mathrm{BB1}(90)$ that produces a 90-degree qubit rotation, an important example task for applications discussed in this chapter.  
Throughout this thesis, we present several analytical primitives for efficient optimal control of bosonic systems using DV systems. All our primitives belong to the class of non-abelian QSP, a term we introduce for the class of pulses discussed in this chapter which will serve as the key to optimal universal control of  CV systems in the presence of errors. Our analytical understanding of error cancellation in non-abelian composite pulses suggests the outlines of a prospective hierarchy of non-abelian QSP, delineating the challenges and framework necessary for efficient control of hybrid CV-DV quantum computing.

\begin{figure*}[ht]
    \centering
    \includegraphics[width=\textwidth]{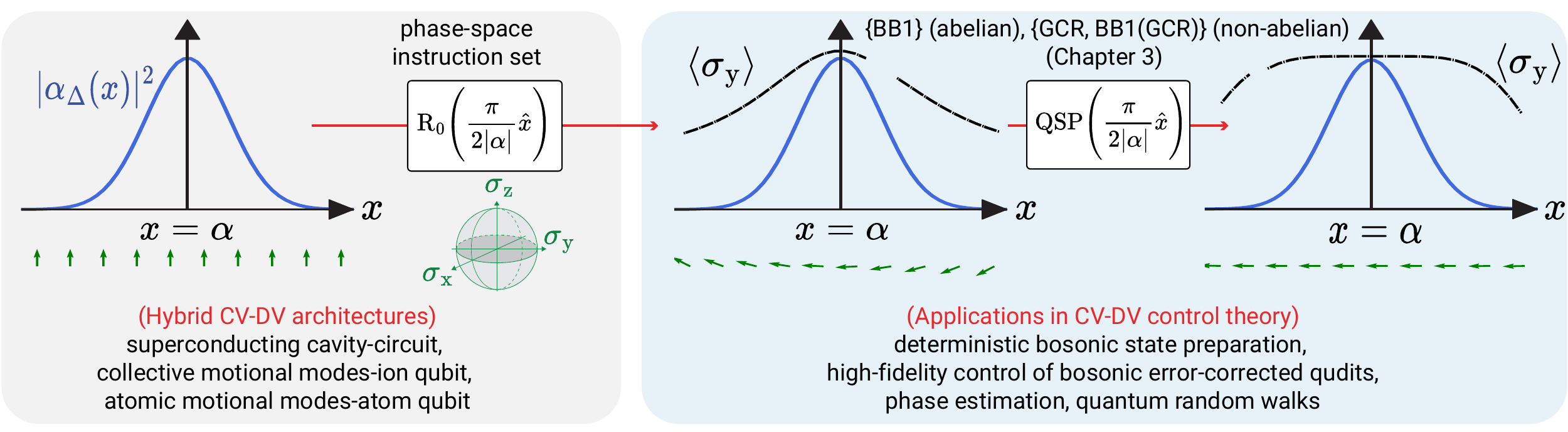}
    \caption[Framework of composite pulses in phase space and its applications.]{\textbf{Framework of composite pulses in phase space and its applications.} The blue curves show the Gaussian probability distribution $|\braket{x|\psi}|^2$ of the oscillator position, and the green arrows indicate the spin orientation of the ancilla qubit for the state $\ket{g}\otimes\ket{\alpha_\Delta}$ (see Eq.(\ref{eq:Gaussian})). The objective is to use only oscillator-controlled rotations to uniformly rotate the spin from $\ket{g}$ to $\ket{\mp i}$ depending on $\braket{x}=\pm\alpha$, independent of the oscillator’s position uncertainty. Quantum signal processing (QSP) pulses are used to achieve this. The black dashed curves show the QSP response function, plotted as the expectation value $\braket{\sigma_\mathrm{y}}$; flatter curves indicate better QSP performance. The gray panel depicts the hybrid CV-DV system in a product state, manipulated by a phase-space instruction set consisting of conditional displacements ($\mathrm{CD}$) and arbitrary qubit rotations ($\mathrm{R}$) (see Eq.(\ref{eq:phasespaceIS})). Conditional displacements $\mathrm{CD}(\gamma,\sigma_\phi)$ produce oscillator-controlled qubit rotations $\mathrm{R}_\phi(\gamma \hat{x})$.
The blue panel introduces the central idea of this chapter: applying composite pulses in oscillator phase space to control hybrid systems. QSP corrects spin rotation errors caused by the position uncertainty of the Gaussian state, as illustrated in the left figure of the blue panel.
We develop a non-abelian composite pulse sequence, $\mathrm{GCR}$, and compare it with traditional composite pulses like $\mathrm{BB1}$\cite{wimperis1994broadband} in Sec.~\ref{sec:GCR}. A concatenated pulse BB1(GCR), combining BB1 and GCR, is also introduced (recipe in Sec.~\ref{GCR-BB1}). The right figure in the blue panel highlights the improvement in rotation fidelity and summarizes several applications discussed throughout this chapter.
    }
    \label{fig:Gaussian-qubit}
\end{figure*} 
\section{Technical Background and Preliminaries}\label{sec:preliminaries}

In this section, we first give preliminary information required for the construction presented in this thesis. We introduce various novel constructions including the notion of composite pulses in phase space, rotation gadgets, and quantum operator valued control parameters in Sec.~\ref{sec:comp}. We then give an overview of the main results with the help of a QSP hierarchy for the control of CV-DV architectures in Sec.~\ref{sec:techsummary}.

\subsection{Framework of Composite Pulses in Phase Space}\label{sec:comp}

Quantum states in an oscillator cannot have fixed position and momentum eigenvalues with infinite precision, as per the uncertainty principle. The simplest quantum state, a coherent state $\ket{\alpha}$, is given by a Gaussian wave function where the width of the Gaussian $\Delta$ gives the uncertainty in determining the position or momentum of the state. For example, in the position basis, 
\begin{align}
    \braket{x|\alpha_\Delta}=\alpha_\Delta(x)=\Big(\frac{2}{\pi\Delta^2}\Big)^{1/4}e^{-\frac{(x-\alpha)^2}{\Delta^2}}\label{eq:Gaussian};
\end{align}
is a state whose mean position is $\alpha\in\mathbb{R}$ with an uncertainty of $\delta x = \frac{\Delta}{2}$. This uncertainty is often just the natural uncertainty associated with the zero-point fluctuations of the oscillator ground state or vacuum (for which $\Delta=1$) but may be smaller or larger in squeezed states (see Sec.~\ref{sec:squeezing}). For coherent states without squeezing, we will drop $\Delta$ from the notation and use $\ket{\alpha}$.  For general squeezed coherent states,  the most useful information related to $\ket{\alpha_\Delta}$ is (for our purposes) in the mean position and momentum determined by $\alpha$, and (secondarily) in the squeezing $\Delta$. We will use composite pulses inscribed in phase space using oscillator-controlled qubit rotations to access this phase-space information. As we will show, this technique is especially handy for the control of states represented by sums of non-overlapping Gaussian wave functions.

\paragraph{Rotation gadgets:} 
Let us first consider the task of extracting a single bit of information about a CV state, the sign of the mean value of the position operator (i.e., distinguish between $\ket{+\alpha_\Delta}, \ket{-\alpha_\Delta}$), using a DV auxiliary qubit. For convenience, we will focus on the case where $\alpha$ is a real number, with a straightforward generalization to arbitrary vector $\hat v$ in phase space with complex $\alpha$ discussed in App.~\ref{ssec:pnot0}. That is, given the knowledge of $|\alpha|$, the final DV qubit state should be independent of $\Delta$. For large enough $|\alpha|/\Delta$ where $\ket{+\alpha_\Delta}$ and $\ket{-\alpha_\Delta}$ are nearly orthogonal, this task can be achieved if we manage to rotate the qubit in state $\ket{g}$ by an angle $\mathrm{R}_0\Big(\frac{\pi}{2}\frac{\alpha}{|\alpha|}\Big)$, or equivalently $\mathrm{R}_0\Big(\frac{\pi}{2}\mathrm{sign}(\hat x)\Big)$, and measure the qubit in the Y-basis of the auxiliary Bloch sphere. In short, we aim to find a unitary transformation $\hat U$ such that \begin{equation}
    \hat U \ket{g}\otimes \ket{\pm\alpha_\Delta} \approx \ket{\mp i}\otimes \ket{\pm \alpha_\Delta}.
\end{equation} 

To address this problem, we derive a rotation gadget that executes a good approximation to $\mathrm{R}_0\Big(2\theta\frac{\alpha}{|\alpha|}\Big)$ for arbitrary $\theta$ in the large $|\alpha|/\Delta$ limit\footnote{The case of small $|\alpha|/\Delta$ is discussed in Sec.~\ref{sec:squeezing} and App.~\ref{app:small_cats}}. 
With this goal in mind, let us study the effect of a conditional momentum boost
\begin{align}
 \mathrm{R}_\phi\Bigg(-\frac{\theta}{|\alpha|}\hat x\Bigg)=  e^{i\frac{\theta}{2|\alpha|}\hat x\otimes \sigma_\phi} 
\end{align}
applied using the phase space instruction set in Eq.~(\ref{eq:phasespaceIS}). In the position basis, the position-controlled rotation by $\hat \theta(\hat x)=\frac{\theta}{|\alpha|}\hat x$ applies a qubit rotation that is linear in the position $\hat x$ of the oscillator. However, this operation suffers fluctuations due to uncertainty in the position of the oscillator, yielding a distribution of the spin polarization on the oscillator position $x$ as depicted by Fig.~\ref{fig:Gaussian-qubit} for $\theta=\pi/2,\phi=0$. Our goal is to develop a QSP sequence that (approximately) converts $\hat x$ to $f(\hat x)=\mathrm{sign}(\hat x)$ (or more precisely to a periodic square-wave function of $\hat x$) that is antisymmetric in $\hat x$ and has period $2\alpha$ so that the flat tops are centered on $\pm \alpha$.

If we define  $\hat \epsilon=\frac{\hat x-\alpha}{\alpha}$ then the qubit is erroneously rotated by,
\begin{align}
\mathrm{R}_\phi\Bigg(-\frac{\theta}{|\alpha|}\alpha(1+\hat \epsilon)\Bigg) &= \mathrm{R}_\phi\Bigg(-\frac{\theta}{|\alpha|}\alpha\,\hat \epsilon\Bigg)\mathrm{R}_\phi\Bigg(-\frac{\theta}{|\alpha|}\alpha\Bigg),
\end{align}
since $\epsilon(x)\equiv \langle x|\hat\epsilon|x\rangle=0$ only at position $x=\alpha$. Because of the Gaussian envelope of the wave function, the probability of finding the oscillator in a region of large ${\epsilon(x)}$ is small. This reduces the effect of over- and under-rotations at ${x}\neq \alpha$ for large $\alpha/|\Delta|$. 
This idea has been used in several works~\cite{hastrup2021unconditional,hastrup2021measurement} to achieve the disentanglement of qubit and oscillator after a hybrid operation. 

The errors associated with the quantum fluctuations in position need to be corrected in a QND manner, to learn information about the oscillator accurately. In this chapter, we show that these errors can be strongly reduced using composite pulse QSP sequences. Previous works~\cite{eickbusch2022fast,hastrup2021improved,hastrup2021measurement,hastrup2021unconditional} have achieved similar improvements using numerical optimization on circuits composed of $\mathrm{CD}$s to prepare oscillator states using DV systems. Our construction, on the other hand, is completely analytical and intuitive. This intuition gives rise to novel protocols discussed in Chapters~\ref{chapter:state-prep}-\ref{chapter:conc}. But first, we ask if one can employ traditional pulses from the classical NMR spin control literature to achieve such error cancellations, removing the over- and under-rotation errors due to quantum fluctuations $(\hat \epsilon)$ in the position.
\paragraph{Composite pulses:} Note that the above-defined task amounts to executing a pulse sequence after which the spin-polarization (more importantly, the expectation value $\braket{\sigma_\textrm{y}}$) resembles a square waveform as a function of the position of the oscillator. This waveform should have a flat top near the peak of the Gaussian function representing the oscillator state (see Fig.~\ref{fig:Gaussian-qubit}). The NMR community has developed classical error-canceling pulse sequences for DV-only architectures that can produce a corrective rotation to compensate for the presence of systematic errors in control variables, say  $\theta_\epsilon=\frac{\theta}{|\alpha|}\alpha(1+\epsilon)$.  In the absence of any correction, the infidelity in achieving the target rotation $\mathrm{R}_\phi(\theta)$ by such an erroneous rotation $\mathrm{R}_\phi(\theta_\epsilon)$ is given by $\sim \epsilon^2$. To reduce this infidelity, the composite pulses $f_\phi(\theta_\epsilon)$ apply,
\begin{align}
 f(\theta_\epsilon)&=\mathrm{R}_\phi\Bigg(-\frac{\theta}{|\alpha|}\alpha\epsilon\Bigg)+\mathcal{O}(\epsilon^n),
\end{align}
such that $f(\theta_\epsilon)$ cancels all $\mathcal{O}(\epsilon^{n-1})$ error terms, improving the fidelity of rotation to $\mathcal{O}(\epsilon^{2n})$. Some of the best-known composite pulse sequences, for classical (scalar) control variables $\theta_\epsilon$, include BB1~\cite{wimperis1994broadband}, SCROFULOUS~\cite{cummins2003tackling}, TYCKO~\cite{tycko1985composite}, etc. Among these, the BB1 pulse sequence demonstrates the best error cancellation (or the flattest square waveform), and so, we shall use this scheme as the standard for comparisons.

\paragraph{Quantum control variables:} Remarkably, for hybrid control, we can replace the classical control variables $\theta_\epsilon$ in traditional QSP pulses with quantum control variables $\hat \theta=(\theta/|\alpha|)\hat x$, provided that all the quantum arguments for all rotations in a composite pulse commute. This replacement can be used conveniently with any signal designed using ideas of univariate QSP with commuting variables~\cite{low2016methodology,low2017quantum}. For example, the $\mathrm{BB1}$ pulse to achieve a target rotation of $\mathrm{R}_0(\theta)$ is given by,
\begin{align}
    \mathrm{BB1}(\theta)&=\mathrm{R}_{\phi_1}\Bigg(\frac{\pi}{\theta}\theta_\epsilon\Bigg)\mathrm{R}_{3\phi_1}\Bigg(\frac{2\pi}{\theta}\theta_\epsilon\Bigg)\mathrm{R}_{\phi_1}\Bigg(\frac{\pi}{\theta}\theta_\epsilon\Bigg) \mathrm{R}_0(\theta_\epsilon),\quad \phi_1=\cos^{-1}\Bigg(-\frac{\theta}{4\pi}\Bigg).
\end{align}
We can adapt the sequence to achieve the corresponding correction of the oscillator-controlled qubit rotation $\mathrm{R}_0\Big(\frac{\theta}{|\alpha|}\hat x\Big)$ using the following sequence,
\allowdisplaybreaks{
\begin{align}
     \mathrm{BB1}\Bigg(\frac{\theta}{|\alpha|} \hat x\Bigg)&= \mathrm{R}_{\phi_1}\Bigg(\frac{\pi}{|\alpha|}\hat x\Bigg)\mathrm{R}_{3\phi_1}\Bigg(\frac{2\pi}{|\alpha|}\hat x\Bigg)\mathrm{R}_{\phi_1}\Bigg(\frac{\pi}{|\alpha|}\hat x\Bigg)\times \mathrm{R}_0\Bigg(\frac{\theta}{|\alpha|}\hat x\Bigg).\\\label{eq:BB1}
    &= \mathrm{CD}\Bigg(-\frac{i\pi}{4|\alpha|},\sigma_{\phi_1}\Bigg)\mathrm{CD}\Bigg(-\frac{i\pi}{2|\alpha|},\sigma_{3\phi_1}\Bigg)\mathrm{CD}\Bigg(-\frac{i\pi}{4|\alpha|},\sigma_{\phi_1}\Bigg) \nonumber\\ &\quad\times\mathrm{CD}\Bigg(-\frac{i\theta}{4|\alpha|},\sigma_\textrm{x}\Bigg).
\end{align}
}
In the last equation, we have used $\sigma_{\phi=0}=\sigma_\textrm{x}$ for clarity. By using these extra conditional momentum boosts to create a composite pulse sequence, we can boost the fidelity of the target rotation against the quantum error $\hat \epsilon$, and more accurately relay information about the mean value $\pm\alpha$ of the oscillator position distribution to the control qubit. We discuss the detailed performance metrics for this sequence in Sec.~\ref{sec:GCR}.

\subsection{Non-Abelian QSP for Quantum Control of Hybrid Systems}
\label{sec:techsummary}
In this section, we formalize the requirements for QSP to achieve universal state preparation and control, and summarize helpful results for each class in the hierarchy and their applicability towards our goal of universal oscillator control. 
\paragraph{Univariate QSP:} Here $[\hat v_i,\hat v_j]=0$, that is, all the allowed vectors in phase space used for $\mathrm{CD}$ are parallel. Thus even though the rotation angles $\hat v_j$, are quantum operators, all the angles in the QSP sequence commute with each other, allowing us to directly utilize the univariate classical QSP methods introduced in \cite{low2016methodology}. Our $\mathrm{BB1}$ analog used for comparison above is an adaptation of this QSP class to quantum control variables. This formalism was also used for single-shot interferometry in~\cite{sinanan2023single} which introduced `Bosonic QSP' independent of the present work. 

We combine ideas and formalisms presented in ~\cite{ISA,rossi2022multivariable,nemeth2023variants,sinanan2023single,martyn2021grand} and this chapter to obtain the following general non-abelian QSP  sequence for hybrid CV-DV architectures,
\begin{align}
U_{\vec{\phi}}(\hat v_1,\hat v_2,..)&=e^{i\phi_0\sigma_\mathrm{z}}\prod^k_{j=1} \mathrm{CD}(\beta_j,\sigma_0)e^{i\phi_j\sigma_\mathrm{z}},\quad (\vec{\phi}=\{\phi_0,\phi_1,...\}),\\
&\equiv e^{i\phi_0\sigma_{\phi_0}}\prod^k_{j=1}\mathrm{CD}(\beta_j,\sigma_{\phi_j}),\\
&=\mathrm{R}_{Z}\Big(\phi_0\Big)\prod^k_{j=1} \mathrm{R}_{\sigma_{\phi_j}}(\hat v_j),
\end{align}
where $\hat v_j = \hat v(\beta_j)$ as defined in Eq.~(\ref{eq:hatvbetadef}). Note that, in traditional QSP where $\hat v_j\equiv\theta$ would correspond to a fixed rotation (`quantum signal') about the $x$ axis of the Bloch sphere. In contrast, here the rotation angle also depends on the index $j$ and is an operator on the oscillator Hilbert space that may not commute with other operators $\hat v(\beta_k)$. 
It is useful to note that the QSP sequence $U_{\vec\phi}$ defined above can be written in the form of a $2\times 2$ operator acting on the DV qubit,
\begin{align}
    U_{\vec{\phi}}(\hat v_1,\hat v_2,..) &= \left( \begin{array}{cc} W_{gg}&W_{ge}\\W_{eg}&W_{ee}   \end{array}  \right),
\end{align}
where each of the $W$ blocks is a CV operator acting only on the oscillator, for example, $W_{ge}=\langle g|  U_{\vec{\phi}}(\hat v_1,\hat v_2,..) |e\rangle$, etc.

For each problem below, the goal is that the qubit should be completely unentangled from the oscillator after $U_{\vec{\phi}}$ is applied to the starting state $|g,0\rangle$ (qubit in $|g\rangle$ and cavity in vacuum $|0\rangle$). That is, we want $U_{\vec{\phi}}$ to be block diagonal and $W_{gg}$ to perform a specified target (unitary) operation $U_\mathrm{t}$ on the oscillator. If these conditions are not perfectly satisfied, then we have several important measures of fidelity.  First, how close is $W_{gg}$ to $U_\mathrm{t}$?  This is the fidelity, $F_\mathrm{ps}$, of the operation post-selected on measuring the qubit to be in $|g\rangle$.  Second, it is useful to know the success probability for the post-selection
\begin{align}
    P_g &= 1-P_e = \langle 0|W^\dagger_{gg}W_{gg}|0\rangle = 1-\langle 0|W^\dagger_{eg}W_{eg}|0\rangle.\label{eq:prob}
\end{align}
If $P_e$ is small relative to errors in other operations of the system, then we can completely ignore the qubit outcome or use it to detect ancilla errors. In this case, we care about a third quantity, the hybrid fidelity $F_\mathrm{H}$,
\begin{align}
    F_\mathrm{H}=|\braket{\psi|W_{gg}|0}|^2,\label{eq:fid}
\end{align}
where $\ket{\psi}$ is the target oscillator state, in case of state preparation. For universal control, this quantity will correspond to oscillator fidelity with the target operation $V$. There are additional fault-tolerance metrics one can consider in the case that the ancilla qubit can raise a flag indicating a leakage error has occurred~\cite{ReinholdErrorCorrectedGates,ma2020path,ma2022algebraic} but this is beyond the scope of the present work.

If we allow $\mathcal{O}(\epsilon)$ upper bound on a qubit-oscillator entanglement error, then the problem statements are framed as follows:

        \textbf{(Problem 1) Universal State Preparation} to realize an arbitrary oscillator state $\ket{\psi}$ starting from vacuum. We require a hybrid unitary $U_{\vec{\phi}}(\hat v_1,\hat v_2,..)$ such that:
    \begin{itemize}
            \item $1-|\braket{\psi|W_{gg}|0}|^2=1-F_\mathrm{H}=\mathcal{O}(\epsilon)$, and
        \item $||W_{eg}\ket{0}||=\sqrt{P_e}=\mathcal{O}(\epsilon)$.
    \end{itemize}
    
    \textbf{(Problem 2) Universal Control} to synthesize a polynomial Hamiltonian $\hat H (\hat x,\hat p)$ that realizes an arbitrary oscillator unitary ($e^{-iHt}$):  Defining $||\hat A^\dagger \cdot \hat B||=\frac{1}{d}\sqrt{\mathrm{Tr}(\hat A^\dagger \hat B)})$ as the operator fidelity between operators $\hat A,\hat B$ on a $d$-dimensional space, we need $U_{\vec{\phi}}(\hat v_1,\hat v_2,..)$ to obey:     
   \begin{itemize}
    \item $1-||W_{gg}^\dagger \cdot e^{-i\hat H(\hat x,\hat p)t}||=1-F_\mathrm{H} =\mathcal{O}(\epsilon)$, and
    \item $||W_{eg}||=\mathcal{O}(\epsilon)$.
    \end{itemize}
    
General QSP techniques for single-qubit rotations were introduced in the context of classical $\theta$ rotation angle variables in~\cite{low2016methodology}. These were extended to multi-variable QSP schemes~\cite{nemeth2023variants,rossi2022multivariable} where the polynomial is a function of more than one variable $\theta_1,\theta_2,...$. Note that, multi-variate or multi-variable does not necessarily refer to having many oscillators each with their own $\hat x, \hat p$, but rather (for the case of a single oscillator at least) to having multiple directions in phase space along which displacements can be made. The class of QSP techniques that use only conditional displacements and conditional momentum boosts (or any other orthogonal displacement generators in phase space) along with qubit rotations is universal. However, the availability of displacements using an arbitrary number of generators $\hat v_i=\alpha_i\hat x+\beta_i \hat p$ can yield more efficient circuits for the universal control of oscillators. An important addition to the present Chapter is the attempt to generalize these schemes towards universal control of bosonic systems. Here the target polynomials, $\hat f(\hat x, \hat p)$, are in general a function of two non-commuting variables. While a full constructive theory of this generalization remains an open problem, we suggest a hierarchy of QSP schemes in the following discussion, summarized in Table~\ref{tab:qsp_summary}, that can yield insights into the development of novel techniques with readily available QSP methods. 
\begin{table*}[htb]
    \centering
    \begin{tabular}{||p{0.275\textwidth}|p{0.155\textwidth}|p{0.365\textwidth}|p{0.115\textwidth}||}
    \hline
    \hline
    \textbf{Types of CV-DV QSP} & \textbf{Conditions} & \textbf{Use-case} & \textbf{Refs.} \\ \hline
    Univariate QSP & $[\hat v_i,\hat v_j]=0$ &  \begin{tabular}[c]{@{}l@{}}Traditional QSP methods~\cite{low2016methodology,low2017optimal,martyn2021grand}\\ w/ quantum control variables \end{tabular} & \cite{sinanan2023single} \\ \hline
    \begin{tabular}[c]{@{}l@{}}Multivariate QSP \\ w/ commuting variables\end{tabular} &  \begin{tabular}[c]{@{}l@{}}$v_1,v_2,v_3,.. \ \mathrm{s.t.}$ \\ $[\hat v_i,\hat v_j]=0$\end{tabular} &  Control of multiple oscillators & \cite{rossi2022multivariable,nemeth2023variants} \\ \hline
    \begin{tabular}[c]{@{}l@{}}Bivariate QSP w/\\ non-commuting variables\end{tabular} & $[v_i,v_j]\neq 0$ &  \begin{tabular}[c]{@{}l@{}}High-fidelity control of single \\oscillator with low circuit-depth \end{tabular} & This Chapter w/~\cite{ISA}\\ \hline
    \begin{tabular}[c]{@{}l@{}}Multivariate QSP w/\\  non-commuting variables\end{tabular} &  \begin{tabular}[c]{@{}l@{}}$[\hat v_i,\hat v_j]\neq 0$  \\ (for some $i,j$)\end{tabular} &  \begin{tabular}[c]{@{}l@{}}High-fidelity control of multiple\\oscillators with low circuit-depth \end{tabular}  & N/A \\ \hline
    \hline
    \end{tabular}
    \caption[ Hierarchy of CV-DV QSP Framework.]{\textbf{Hierarchy of CV-DV QSP Framework.}  Overview of different types of QSP techniques, with commuting and non-commuting quantum variables, found in literature, developed towards universal oscillator control. The bottom two rows belong to the largely unexplored territory of non-abelian quantum signal processing (or NA-QSP; see Ref.~\cite{ISA} for a formal introduction to NA-QSP). 
    Its applications in various arenas of CV-DV control theory are listed in Fig.~\ref{fig:Gaussian-qubit}.}
    \label{tab:qsp_summary}
\end{table*}

\paragraph{Multivariate QSP with commuting variables:}
This scheme is the primitive version of multivariate QSP introduced in~\cite{rossi2022multivariable,nemeth2023variants}.
\begin{itemize}
    \item \textit{Bivariate QSP with commuting variables.} Here $\hat v_i\in\{\hat v_1,\hat v_2\}\quad \mathrm{s.t.}\quad [\hat v_1,\hat v_2]=0$.  In~\cite{nemeth2023variants} the authors prove that it is possible to construct $U_{\vec{\phi}}$ for an arbitrary target Hamiltonian $\hat H(\hat v_1,\hat v_2)$, if $\mathrm{deg}_{\hat v_1}(\hat H)\le 1\quad \mathrm{or}\quad \mathrm{deg}_{\hat v_1}(\hat H)\le 1$. For state preparation defined above, $\braket{x|U_{\vec{\phi}}|0}=\psi(x)$, that is, it already satisfies the condition used in this theorem. The theorem can be  put to use towards oscillator state preparation if we restrict ourselves to the regime where $|[\hat v_1,\hat v_2]|$ is sufficiently small. The efficiency for universal control in the commuting variable regime is likely to be similar to Suzuki-Trotter decompositions.

    \item \textit{Multivariate QSP ($\mathrm{m-qsp}$) with commuting variables.} The adaptation of any composite pulse using quantum control variables is an example of a composite pulse that belongs to this QSP class if it allows $|v_i|\neq|v_j|\quad\text{if}\quad i\neq j$. In~\cite{rossi2022multivariable} the authors introduce this QSP class towards control of single or multiple oscillators using operators which commute, for example, $\hat x_1=\hat x\otimes I, \hat x_2=I\otimes \hat x$ on the joint space of two oscillators.
    
    As before, if we restrict that pairwise products of the magnitudes obey $|v_i||v_j|=\mathcal{O}(\lambda),\,\, \forall\, i, j$, we can use $\mathrm{CD}$s along arbitrary vectors in phase space to achieve state preparation using m-qsp even if $[v_i,v_j]\neq 0$. Such inputs will make QSP sequences more efficient to prepare states that have low mean photon numbers and are rotationally symmetric in phase space, for example, Fock state $\ket{1}$. However, unlike the abelian bivariate case, there is no constructive proof of an algorithm to generate $U_{\vec{\phi}}$ in this case.
\end{itemize}

\paragraph{Multivariate QSP with non-commuting variables:}
The scheme is lightly touched upon by~\cite{nemeth2023variants}.
\begin{itemize}
    \item \textit{Bivariate QSP.} Our scheme is the first example of a composite pulse that belongs to this QSP class. We have already seen the advantages of such schemes in achieving high-fidelity outputs for low circuit depth via $\mathrm{GCR}$. In addition, this class is necessary to achieve universal control which beats the efficiency of methods like Suzuki-Trotter.
    \item \textit{Multivariate QSP.} This class lies at the top of the hierarchy and is the most efficient resource for optimal universal control of oscillators. The resource of multiple non-commuting variables was used with gradient-descent-based techniques to achieve highly efficient circuits for state preparation of various non-Gaussian states in~\cite{eickbusch2022fast}. The final goal towards universal control of CV-DV architectures will be to understand analytical constructions for this QSP class.
\end{itemize}

Non-abelian QSP offers a powerful resource for hybrid quantum systems that will be a broadly useful tool for the realization of quantum advantage in continuous-variable quantum computing. The composite Gaussian Controlled Rotation (GCR) pulse scheme introduced in the next section is a first step in this direction.

\section{Gaussian-Controlled-Rotation (GCR): A Non-Abelian Composite Pulse} \label{sec:GCR}

In this section, we introduce an analytic non-abelian QSP composite pulse sequence, the Gaussian-controlled-rotation, $\mathrm{GCR(\theta)}$. This is a non-abelian instance of a rotation gadget for the control problem defined in Sec.~\ref{sec:comp}.
We prove here that  GCR($90$) achieves a target rotation with similar error cancellation as the abelian QSP protocol BB1($90$) but with a reduction of circuit depth by a factor of at least $4.5$. We shine light upon the usefulness of GCR sequences in Chapters~\ref{chapter:state-prep}-\ref{chapter:conc}.

Using the additional freedom afforded by NA-QSP, we define a Gaussian-Controlled-Rotation, 
\begin{align}
  \mathrm{GCR}(\theta)\ket{g}\otimes\ket{\alpha_\Delta}&\equiv e^{i\frac{\theta}{2|\alpha|} \hat x\sigma_\textrm{x}}e^{i\frac{\theta\Delta^2}{2|\alpha|} \hat p\sigma_\textrm{y}}\ket{g}\otimes\ket{\alpha_\Delta}\label{eq:GCR}\\
  &=\mathrm{R}_{0}\Bigg(-\frac{\theta}{|\alpha|}\hat x\Bigg)\mathrm{R}_{\frac{\pi}{2}}\Bigg(-\frac{\theta\Delta^2}{|\alpha|}\hat p\Bigg)\ket{g}\otimes\ket{\alpha_\Delta},\\
  &\approx \mathrm{R}_0\Bigg(-\theta\frac{\alpha}{|\alpha|}\Bigg)\ket{g}\otimes\ket{\alpha_\Delta}.
 \end{align}

The above sequence works equally well if the qubit operators are rotated by angle $\phi$ about the $z$ axis such that $\sigma_\textrm{x}\rightarrow\sigma_\phi, \sigma_\textrm{y}\rightarrow\sigma_{\phi+\pi/2}$. Similarly, the sequence is also generalizable to accommodate rotations conditioned on generators of arbitrary (but perpendicular) displacements in phase space such that $\hat x\rightarrow \hat v$ and $\hat p\rightarrow \hat v_{\perp}$ in the CV phase space.

For a CV quantum state $\ket{\alpha_\Delta}$ with wave function given in Eq.~(\ref{eq:Gaussian}), the composite pulse $\mathrm{GCR}(\theta)$ performs a rotation of the qubit state $\ket{g}$ about an arbitrary axis on the equator of the qubit Bloch sphere by a fixed angle $\pm\theta$ whose sign is determined by the sign of $\alpha$. Here, the momentum-controlled rotation (or, conditional displacement) applies a pre-correction to the first order in the uncertainty of $\hat \theta(\hat x)$. Below we give a proof of correctness and an error analysis for this construction, computing the quantities $P_e$ and $F_\mathrm{H}$ (see Eqs.~\ref{eq:prob}-\ref{eq:fid}) for GCR and compare it against the case no QSP correction and BB1. The post-selected fidelity $F_\mathrm{ps}$ is also computed in App.~\ref{app:error-analysis}.

 \subsection{Proof of Correctness} To understand the effect of a conditional displacement in Eq.~(\ref{eq:GCR}) in the position basis, we note that the momentum operator acts as the derivative operator ($p=-\frac{i}{2}\frac{d}{dx}$) on $\alpha_\Delta(x)$ yielding powers of $(x-\alpha)$. This observation indicates that this operation could be used to correct for rotation errors proportional to $(x-\alpha)$ as follows,
 \allowdisplaybreaks{
\begin{align}
\ket{\psi}=e^{i\frac{\lambda}{2}\hat p\otimes \sigma_\textrm{y}}\ket{g}\otimes\ket{\alpha_\Delta}
=&[\cos{(\lambda\hat p/2)} I+i\sin{(\lambda\hat p/2)}\sigma_\mathrm{y}]\ket{g}\otimes\ket{\alpha_\Delta},\\
\approx &[I+i\frac{\lambda}{2}\hat p\sigma_\mathrm{y}]\ket{g}\otimes\ket{\alpha_\Delta},\quad\lambda\rightarrow 0,\\
\therefore\braket{x|\psi}\approx&[I+\frac{\lambda}{4}\frac{d}{dx} \sigma_\textrm{y}]e^{-\frac{(x-\alpha)^2}{\Delta^2}}\ket{g}\label{eq:pos_basis},\\
=&[I-\frac{\lambda}{2}\frac{x-\alpha}{\Delta^2} \sigma_\textrm{y}]e^{-\frac{(x-\alpha)^2}{\Delta^2}}\ket{g},\\
= &[I-i\frac{\lambda}{2}\frac{x-\alpha}{\Delta^2} \sigma_\textrm{x}]e^{-\frac{(x-\alpha)^2}{\Delta^2}}\ket{g},\label{eq:GCR-corr}\\
|\psi\rangle\approx & \mathrm{R}_0\Bigg(\frac{\lambda}{\Delta^2}(\hat x-\alpha)\Bigg)\ket{g}\otimes\ket{\alpha_\Delta}.
\end{align}
}
Through these steps, we have converted a momentum-controlled qubit rotation, about the $\mathrm{y}$ axis of the ancillary Bloch sphere, to a position-controlled qubit rotation, about the $\mathrm{x}$ axis of the ancillary Bloch sphere. The key step of this derivation is based on the second to last equation where we use $\sigma_\mathrm{y}\ket{g}=i\sigma_\mathrm{x}\sigma_\mathrm{z}\ket{g}=i\sigma_\mathrm{x}\ket{g}$.  Therefore, this scheme only works if the initial qubit state is $\ket{g}$.

To first order in $\lambda\hat p\sim \frac{\lambda}{\Delta^2}$, this equality changes the expression into a unitary rotation gate. Thus, a small conditional displacement can be seen as a rotation on $\ket{g}\otimes\ket{\alpha_\Delta}$ which cancels the erroneous rotation $R\Big(-\frac{\theta}{|\alpha|}\alpha\hat \epsilon\Big)$ up to first order in $\epsilon(x)$, provided
\begin{align}
\lambda=\frac{\theta\Delta^2}{|\alpha|}.
\end{align}
This quantity also decides the back-action on the oscillator due to the pre-correction. If the initial qubit state were instead $\ket{\psi}\otimes\ket{e}$, the pre-correction requires reversing the sign of $\lambda$. Thus, the momentum-controlled rotation cannot yield the desired cancellation if applied to a qubit state that is not an eigenstate of $\sigma_\mathrm{z}$. 
 
While this is a case of bivariate QSP, studied in~\cite{rossi2022multivariable}, it is different in that the two variables under consideration are non-commuting ($[f(\hat x),g(\hat p)]\neq 0$) and this feature has favorable implications on reducing circuit depth for error cancellations, as we will prove below. QSP for non-commuting variables is briefly outlined in~\cite{nemeth2023variants} but no explicit instance of a pulse sequence is presented. Our composite pulse GCR($\theta$) is an example of bivariate QSP using non-commuting variables which we will refer to as `non-abelian QSP.' We will now quantify the advantages of our scheme. Our main goal in this analysis is to justify the reduced circuit depth by bounding the error of the scheme and its back action on the oscillator state for the circuit depth reduction achieved here.

\subsection{Error Bounds}\label{sec:composite_error} 

Here, we will talk about the correctness of our scheme for $\ket{\alpha_\Delta}$. Since we use non-commuting control operators, the back action on the conjugate basis should be studied to check the validity of the framework. To calculate the error in the process we will, at first, only consider a single basis state $\ket{\alpha_\Delta}$. Defining, $U=e^{i\frac{\theta}{2|\alpha|}(\hat x-\alpha)\sigma_\mathrm{x}}$, $
V=e^{i\frac{\lambda}{2}\hat p\sigma_\mathrm{y}}, \lambda=\frac{\theta\Delta^2}{|\alpha|}$, Eq.~(\ref{eq:GCR}) can be rewritten as 
\begin{align}
    \mathrm{GCR}(\theta)\ket{g}\ket{\alpha_\Delta}=\mathrm{R}_0\Bigg(-\theta\frac{\alpha}{|\alpha|}\Bigg)UV\ket{g}\ket{\alpha_\Delta}.\label{eq:U}
\end{align}
In the position basis, the action of $U$ and $V$ is given by the following equations.
\allowdisplaybreaks{
\begin{align}
\braket{x|U|\alpha_\Delta}\ket{g}&=\sum_{m=0}^{\infty}\frac{[i\theta(x-\alpha)\sigma_\mathrm{x}]^m}{2^m|\alpha|^mm!}\alpha_\Delta(x)\ket{g}=\sum_{m=0}^{\infty} r_m\ket{g}\\ 
\braket{x|V|\alpha_\Delta}\ket{g}&=\braket{x|\sum_{n=0}^\infty\frac{(i\lambda\hat p\sigma_\mathrm{y})^n}{2^n n!}|\alpha_\Delta}\ket{g}\nonumber\\&=\sum_{n=0}^\infty \Big(-\frac{\lambda\sigma_\mathrm{y}}{4\Delta}\Big)^n \frac{1}{n!}H_n\Big(\frac{x-\alpha}{\Delta}\Big)\alpha_\Delta(x)\ket{g}=\sum_{n=0}^{\infty} s_n\ket{g}\label{eq:V}
\end{align}
where $H_n(x)$ denotes the $n^\mathrm{th}$ physicist's Hermite polynomial. Note that, in the absence of the corrective operation (`pre-rotation') $V$, the qubit rotation error caused by $U$ is small when the uncertainty of $\frac{\theta}{4|\alpha|}\hat x$ is small. For conciseness, we will give expressions in terms of twice this uncertainty,
\begin{align}
    \chi=\frac{\theta\Delta}{2|\alpha|}.
\end{align}
 The corrective operation $V$ cancels the rotation errors to the first order in $\chi$, and so the fidelity of the process is proportional to $\chi^4$. We need to compute the distance between our approximate correction $V$ and the exact cancellation operator $U^\dagger$ for the initial state $\ket{g}\otimes\ket{\alpha_\Delta}=\ket{g,\alpha_\Delta}$. 

\begin{figure*}[th]
     \centering
     \includegraphics[width=\textwidth]{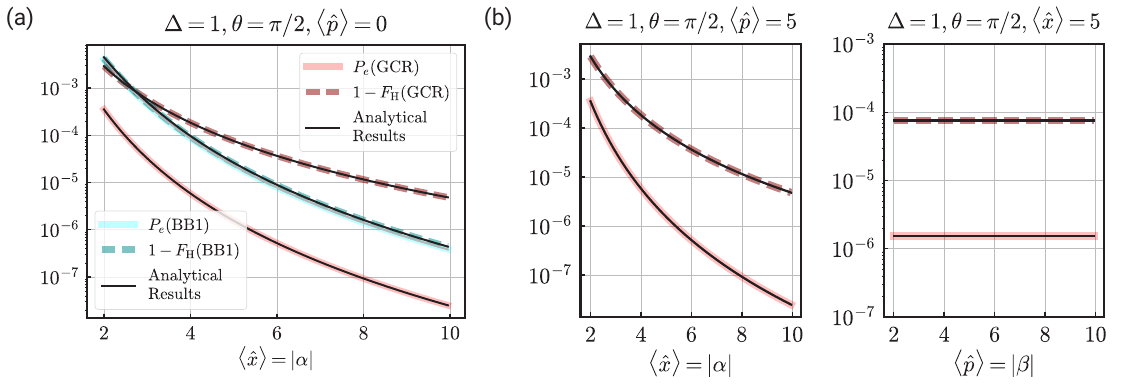}
     \caption[Performance of non-abelian composite pulse sequence $\mathrm{GCR}(\theta)$ in quantum phase space.]{\textbf{Performance of non-abelian composite pulse sequence $\mathrm{GCR}(\theta)$ in quantum phase space.} (a) Comparison against $\mathrm{BB1}(\theta)$ for the case of $\theta=\pi/2$ and $\Delta=1$ so that $\chi=\frac{\pi}{4|\alpha|}$. The colored lines denote the various merits of correctness (failure probability: solid, infidelity: dashed) obtained from simulations using QuTiP \cite{Johansson2013} and the black lines denote the corresponding analytical expressions quoted in Eqs.~\ref{eq:fail_gcr}-\ref{eq:reset_fid_gcr}. See App.~\ref{app:comp_err} for a derivation. The infidelities of $\mathrm{GCR}(\theta)$ scale as $\chi^4$ while the failure probability of both schemes scales as $\chi^6$. (b) Performance of $\mathrm{GCR}(\theta)$ for the coherent basis $\{\ket{\alpha_\Delta+i\beta_\Delta}\}$ where $\alpha\neq0,\beta\neq 0$. (Left) For varying $|\alpha|$ and fixed $|\beta|(=5)$, the simulated failure probability (solid) and infidelity (dashed) show that this variation of $\mathrm{GCR}(\theta)$ also improves upon the rotation errors with the same efficiency as confirmed by the black lines, again plotting Eqs.~\ref{eq:fail_gcr}-\ref{eq:reset_fid_gcr}. (Right) For varying $|\beta|$ and a fixed $|\alpha|=5$, we show that this improvement does not depend on $|\beta|$ as suggested by Eq.~(\ref{eq:pnot0}) since it just requires a simple rotation to keep the anomaly coming from this state with center at $\braket{\hat x}\neq 0$ and $\braket{\hat p}\neq 0$ in check.}
     \label{fig:Correctness}
 \end{figure*}

 Note that, throughout the analysis below we are only interested in the deviation of GCR($\theta$) from the desired operation. To do this, we focus on the deviation of $UV$ from the identity operation on the hybrid oscillator-qubit space. The hybrid infidelity $1-F_\mathrm{H}$ increases as the implemented operation deviates further from the identity on the oscillator-qubit space. The failure probability $P_e$ is non-zero iff the operation is not an identity on the qubit subspace. See Fig.~\ref{fig:Correctness}. The details of the analytical calculations quoted below can be found in App.~\ref{app:error-analysis}. }

\textit{Failure Probability.} %A failure event is when the qubit is rotated incorrectly to $\ket{e}$ at the end of the $UV$. The probability of this event is given by, 
%{\allowdisplaybreaks
%\begin{align}
%\mathcal{P}_{\mathrm{fail}}(e)&=1-\mathcal{P}_{\mathrm{success}}(g)\\&=1-\int_{-\infty}^\infty  dx\quad |\braket{g,x|UV|\alpha_\Delta,g}|^2
%\end{align} 
The probability of incorrectly rotating the qubit (i.e., ending up in $|e\rangle$) is only affected by $\mathcal{O}(\chi^3)$ and $\mathcal{O}(\chi^5)$ terms in the expansion of $UV$, and hence,
\begin{align}
P_e(\mathrm{GCR})\sim 0.1\chi^6+\mathcal{O}(\chi^{8}), \ \chi\ll 1.\label{eq:fail_gcr}
\end{align}
%}
 Thus, the probability of making an erroneous rotation has been proved to scale as $\chi^6$. As $\chi\rightarrow 0$, the probability goes to 1, that is, the delta-function limit $\Delta\rightarrow 0$ or zero-rotation limit $\alpha\rightarrow \infty$ yields a unit probability of success, as expected. The approximation is not well-suited for $\alpha\rightarrow 0$ since higher-order terms come into play while in $\Delta\rightarrow \infty$ limit the momentum-basis is more suitable for the peak-dependent rotation of the qubits.
 
 In the case of no QSP correction, the failure probability in the asymptotic limit of large $\alpha$ is given by, $P_e(\mathrm{no-QSP})=0.25\chi^2$ (see App.~\ref{app:no_corr}). Improving upon which, the failure probability for BB1 is given by,
\begin{align}
    P_e(\mathrm{BB1})&=1.85\chi^6.
\end{align}
See App.~\ref{app:err_BB1} for details of these calculations and contrast this with the case no QSP correction and GCR. Note that this success probability scales with the same power of $\chi^6$ and a $10\times$ worse prefactor compared to GCR (see App.~\ref{app:err_BB1}).

\textit{Hybrid Infidelity.} If the failure probability is low enough, we can afford to ignore the outcome of the qubit and let it reset. In this case, the hybrid state fidelity is important.
\begin{align}
  1-F_\mathrm{H}(\mathrm{GCR})&=| \braket{\alpha,g|UV|\alpha, g}_\Delta|^2\\
  &=\chi^4/8+\mathcal{O}(\chi^6),\label{eq:reset_fid_gcr}
\end{align}
 We see that hybrid state infidelity has a lower scaling of $\mathcal{O}(\chi^4)$ for our scheme due to unwanted back action from the conditional displacement $e^{i\frac{\lambda}{2}\hat p\sigma_\mathrm{y}}$. The BB1 correction, on the other hand, has the same scaling as failure probability, and thus, a smaller back action ($1-F_\mathrm{H}(\mathrm{BB1})\sim P_e(\mathrm{BB1})=1.85\chi^6$), because the scheme is composed of only conditional momentum boosts.

\textit{Circuit-depth:} In terms of general gate counting methods to quantify circuit complexity, $\mathrm{BB1}$ uses four gates while $\mathrm{GCR}$ uses only two gates. However, the duration of the hybrid gates scales with its displacement amplitude ($T_{\mathrm{CD}(\alpha,\sigma_\phi)}\propto\alpha$), and hence, we believe that the correct way to quantify the circuit depth is by comparing the total magnitude of conditional displacements and conditional momentum boosts used in Eqs.~(\ref{eq:BB1}) and~(\ref{eq:GCR}). Using this figure of merit, the circuit duration for our non-abelian scheme is proportional to $T_\mathrm{GCR}\propto \frac{\pi}{4|\alpha|}(1+\Delta^2)$. For the case of BB1 correction, the total duration of conditional momentum boosts that we apply is $T_\mathrm{BB1}\propto \frac{2\pi}{|\alpha|}+\frac{\pi}{4|\alpha|}=\frac{9\pi}{4|\alpha|}$ (see Eq.~(\ref{eq:BB1})). The duration of our scheme decreases for squeezed states ($\Delta<1$) whereas it does not have any effect on the circuit duration of $\mathrm{BB1}$. For the worst-case scenario of $\Delta=1$ for our scheme\footnote{Note that for $\Delta<1$, the correction pulse is smaller than the case of $\Delta=1$. For the case of $\Delta>1$, the position fluctuations are anti-squeezed and the momentum fluctuations are squeezed.  Hence we should use the GCR sequence with $\hat x\rightarrow \hat p$, $\hat p\rightarrow -\hat x$ in which case $\Delta$ is replaced by $1/\Delta$. Hence, in this case as well the correction is smaller than the case of $\Delta=1$. A larger amplitude for correction yields a larger back action on the oscillator, and thus, $\Delta=1$ is the worst-case scenario for $\mathrm{GCR}$.}, our circuit depth is still shorter than $\mathrm{BB1}(90)$ by a factor approaching
\begin{align}
    \frac{T_\mathrm{GCR}}{T_\mathrm{BB1}}=4.5
\end{align}
in the limit of large $|\alpha|$. This is an appreciable improvement when it comes to high-fidelity performance in the presence of non-deterministic (random) errors, such as, DV ancilla decay. This is the dominant source of error in hybrid CV-DV architectures, where a longer circuit would induce more errors in the system and hence would be less suitable for high-fidelity outcomes. Thus, in the presence of such errors, our scheme's shorter circuit depth would take precedence if the failure probability $P_e$ and hybrid infidelity $1-F_H$ scaling are comparable. 

We confirm our analytical results using numerics. In Fig.~\ref{fig:Correctness}(a), we plot $1-P_e(\mathrm{GCR})$ and $1-F_{H}(\mathrm{GCR})$ against $|\alpha|$ for the case $\Delta=1$. We find that the analytical results match with the numerical results for both $\mathrm{GCR}$ and $\mathrm{BB1}$. The figure (along with detailed expressions in App.~\ref{app:comp_err}) also implies that for the case of coherent states without squeezing ($\Delta=1$) and $\theta=\pi/2$, we need $\alpha>2$ to obtain any advantage from $\mathrm{GCR}$ or $\mathrm{BB1}$. Thus, we have justified the correctness and validity of our scheme. Importantly, we have shown that the failure probability $P_e(\mathrm{GCR})$ is low enough for most $|\alpha|$ values to be negligible. Hence, absent qubit or cavity decay errors, the scheme is effectively deterministic and does not need to rely on ancilla measurements. As a result, ancilla measurements can be used to herald the failure of the gate due to extrinsic factors such as detect qubits and cavity decay errors.

Our scheme has comparable performance to one of the best-known composite pulse sequences $\mathrm{BB1}(90)$. Comparing the plots in Fig.~\ref{fig:Correctness}(a), we note the following: The failure probability of both schemes scales as $\chi^6$ but the prefactor of $\mathrm{BB1}$ is an order of magnitude worse. The reset fidelity, on the other hand, scales as $\chi^6$ for $\mathrm{BB1}$ while it scales as $\chi^4$ for $\mathrm{GCR}$. Thus, the back action of $\mathrm{GCR}$ on the oscillator is worse than BB1 for large $\alpha$. However, it is important to note that our scheme achieves this performance despite being shorter by a factor of at least $4.5$ (more if $\Delta\neq 1$) in circuit duration. Thus, in the presence of loss as well in terms of time cost, our non-abelian-QSP-inspired sequence can be a better alternative to $\mathrm{BB1}$ type correction for CV-DV control.

\section{Composing Abelian and Non-Abelian QSP:BB1(GCR)}\label{GCR-BB1}

\begin{figure*}[!htb]
     \centering
     \includegraphics[width=\textwidth]{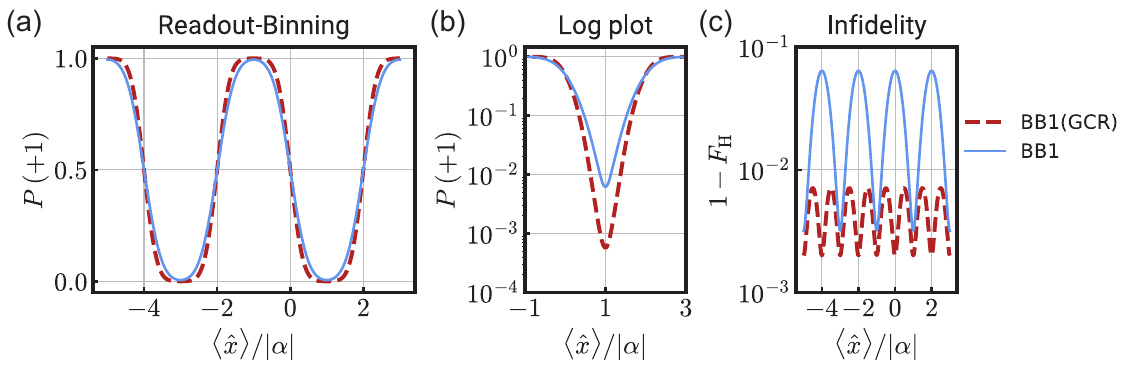}
     \caption[Readout binning using Gaussian-controlled-$\mathrm{BB1}$ pulse sequence, $\mathrm{BB1}(\mathrm{GCR})$.]{\textbf{Readout binning using Gaussian-controlled-$\mathrm{BB1}$ pulse sequence, $\mathrm{BB1}(\mathrm{GCR})$.}  All plots follow the same legend. \textbf{(a)} Readout binning for the case of $|\alpha|=\sqrt{\pi}/2$. The plot gives the probability to measure a $+1$ outcome upon $\sigma_\textrm{y}$ measurement, post the $\mathrm{BB1}(\mathrm{GCR})$ and $\mathrm{BB1}$ in red and blue respectively. The x-axis represents the initial oscillator state with mean modular position value $\braket{x}/|\alpha|$ on which this pulse was applied. Note that a binning readout tells us whether the oscillator is in a specific bin (of size $2|\alpha|$) or not using qubit measurement outcome. Note that $\mathrm{BB1}(\mathrm{GCR})$ yields a flatter response function compared to $\mathrm{BB1}$. \textbf{(b)} Logarithmic scale for plot (a) to quantify the advantage of $\mathrm{BB1}(\mathrm{GCR})$ precisely, for bins which support opposite qubit measurement outcomes. Note the order of magnitude improvement in BB1(GCR) compared to BB1 at the peaks of the target square wave response. \textbf{(c)} The hybrid fidelity $F_\mathrm{H}$ after each pulse. Interestingly, the $\mathrm{BB1}(\mathrm{GCR}(90))$ pulse has better fidelity as well.
     }
     \label{fig:GCR_BB1}
 \end{figure*}
 While the non-abelian QSP sequence GCR corrects for errors due to Gaussian uncertainty, it will be rendered less efficient if the state were to experience a small displacement error such that $\braket{x}\neq \pm\alpha$. We can solve this problem by concatenating GCR with BB1(90) which is designed to be resilient to such displacement errors. The net result for this QSP response function will be an approximate square wave that corresponds to a modular position measurement that is encoded into the ancilla qubit. Let us call this pulse BB1(GCR). The incorporation of GCR brings the response function somewhat closer to an ideal square wave, relative to just using BB1.
 
 We can achieve the desired concatenation by converting each rotation in the $\mathrm{BB1}$ (see Eq.~(\ref{eq:BB1})) into a Gaussian-controlled rotation.
\begin{align}
    \mathrm{BB1}(\mathrm{GCR}(\theta))&: \mathrm{GCR}_0\Bigg(\frac{\theta}{|\alpha|}\hat x\Bigg)\mathrm{GCR}_{\phi_1}\Bigg(\frac{\pi}{|\alpha|}\hat x\Bigg) \mathrm{GCR}_{3\phi_1}\Bigg(\frac{2\pi}{|\alpha|}\hat x\Bigg)\mathrm{GCR}_{\phi_1}\Bigg(\frac{\pi}{|\alpha|}\hat x\Bigg)\label{eq:GCR-BB1},
\end{align}
with the same expression for $\phi_1$ as given in Eq.~(\ref{eq:BB1}). We have presented a Gaussian-controlled version of the reversed BB1 sequence\footnote{The BB1 correction can be run backward with the same performance. This sequence is equivalent to pre-pending the three corrective rotations. Note that BB1 correction can be appended at the beginning, end, or even in the middle of the target rotation.}. This order was chosen to match the order of pre-correction required for $\mathrm{GCR}$. Here, for example, 
\begin{align}
     \mathrm{GCR}_{\phi_1}\Bigg(\frac{\pi}{|\alpha|}\Bigg)=e^{i\frac{\pi}{2|\alpha|}\hat x\sigma_\phi}e^{i\frac{\pi\Delta^2}{2|\alpha|}\hat p\sigma_\gamma}.
\end{align}
It is important to note that the pre-correction for GCR$_{\phi_1}(\pi/|\alpha|)$ is conditioned on the qubit Bloch sphere axis $\sigma_\gamma$, to be determined as follows. After the previous rotation GCR$_{3\phi_1}(2\pi/|\alpha|)$, we compute the state to which qubit is rotated in the ideal case of no errors. Let us call this state $\ket{\zeta}$. Then,
\begin{align}
    \sigma_\gamma\ket{\zeta}=i\sigma_\phi\ket{\zeta}.
\end{align}
In the reverse BB1 case, $\sigma_\gamma$ for $\mathrm{GCR}_{0}$ depends on the state after the BB1 correction, which in the ideal case of no errors is the same as the starting qubit state $\ket{g}$. This makes the pre-corrections less intrusive and more efficient.

Remember that the goal here is not just to distinguish between $\braket{x}=\pm\alpha$ anymore. We would like to take advantage of the $\mathrm{BB1}(\mathrm{GCR}(90))$ and extract the following bit-wise information about oscillator position~\cite{de2024modular},
\begin{align}
    \frac{\braket{x}}{|\alpha|} \mod 2.
\end{align}
\begin{itemize}
    \item We start in the hybrid state $\ket{g}\otimes\ket{\pm\alpha'_\Delta}$ where $\alpha'=m|\alpha|, m\in \mathbb{Z}$.
    \item If $m\in +\mathbb{Z}$, odd (even)  $m$ will yield $\ket{-i}$ ($\ket{+i}$) outcome.
    \item Else if $m\in -\mathbb{Z}$, odd (even)  $m$ will yield $\ket{+i}$ ($\ket{-i}$) outcome.
\end{itemize}
See Fig.~\ref{fig:GCR_BB1} for numerical results for this protocol. For a coherent state with $\Delta=1$, this sequence doubles the pulse length but also gives a better response. In addition, the more squeezed the state is, the shorter the additional pre-corrections are. From Fig.~\ref{fig:GCR_BB1}, we find that for $|\alpha|=\sqrt{\pi}/2,\Delta=0.34$ the BB1(GCR) performs better in both metrics, defined in App.~\ref{performance metrics}, terms of failure probability $P_e$ as well as fidelity $F_\mathrm{H}$. The improvement is same for both qubit measurement outcomes ($\pm 1$). This is an important requirement for measurement pulses, otherwise, it is not straightforward to say that the measurement fidelity (see App.~\ref{performance metrics}) has been improved. Note that improvements for both bins are identical. Our choice of $|\alpha|,\Delta$ will be useful in Chapter~\ref{chapter:qec-control} when discussing efficient end-of-the-line readout of logical GKP codewords. This composition is generalizable to all existing composite pulse sequences in the literature designed for qubit-only architectures~\cite{cummins2003tackling,tycko1985composite}.

\section{Open Problem: Non-Abelian QSP and QSVT}
The ultimate goal of this chapter is two-fold. The first goal, as described in various aspects of this chapter, is to achieve high-fidelity control of hybrid oscillator-qubit architecture. In this context, we employ the techniques developed in this chapter for state preparation, error correction, and control in the following chapters of this thesis. The second goal is to raise an open question about the formalism of non-abelian quantum signal processing. \vspace{1em}

\begin{myframe}
\singlespacing
\begin{quote}
Can we extend the theory of the quantum singular value transformation for quantum algorithms to the class of non-abelian QSP? The idea is to extend this formalism to a a constructive complete constructive theory of non-abelian QSP. This generalization could pave the way for non-abelian quantum singular value transformation (QSVT), in analogy to the abelian QSVT that unifies quantum algorithms on qubit-only platforms. We anticipate that non-abelian QSVT may similarly unify hybrid CV-DV quantum algorithms. Moreover, the principles of non-abelian QSP may extend beyond hybrid systems to multi-qubit gate synthesis, broadening the impact of this framework. We believe that our work offers a foundational step toward this vision.
\end{quote}
\end{myframe}

\doublespacing
% etc
    \chapter{Deterministic Oscillator State Preparation} \label{chapter:state-prep}
\begin{myframe}
\singlespacing
\begin{quote}
    \textit{How can non-abelian QSP sequences enable efficient control of CV systems?} Non-abelian QSP lies at the pinnacle of the hierarchy of QSP variants tailored for CV-DV architectures, offering a potent resource for hybrid quantum systems poised to realize quantum advantage in continuous-variable quantum computing. To demonstrate this, we present applications of $\mathrm{GCR}$ in the control of hybrid continuous-variable (CV) and discrete-variable (DV) architectures. With the help of $\mathrm{GCR}$ we design analytical schemes for high-fidelity preparation of several CV states such as squeezed states, cat states, Fock states, and GKP states. The fidelity and circuit depth of our analytical schemes are comparable to numerically optimized methods. Moreover, the analytical approach gives a sound method to track error propagation and its mitigation. The unique feature of our result lies in the identification of a structure in these circuits which makes the state preparation and control circuits more fault-tolerant to ancilla errors. Such a structure is difficult to achieve or tailor via numerical optimization. 

\end{quote}
\end{myframe}

\doublespacing

The phase-space instruction tool is a very powerful resource as Ref.~\cite{eickbusch2022fast} shows that these operations can be done extremely quickly even in the weak dispersive regime, which is not the case for other qubit-based universal instruction sets. The weak-dispersive regime is key to reducing errors from higher-order terms such as Kerr effects. Conditional displacements and single-qubit rotations, as discussed in Chapter~\ref{chapter: paper0}, are a universal set of instructions that can map an oscillator in vacuum to an arbitrary CV state. This set can also implement arbitrary quantum channels. This instruction set is particularly useful in the preparation of non-overlapping superpositions of Gaussian wave functions like those in two-legged cat states and GKP states. Now, we will demonstrate how the composite pulse sequence designed in Sec.~\ref{sec:GCR} can achieve deterministic preparation of states that are superpositions of non-overlapping Gaussian wave functions. Towards this direction, using our non-abelian QSP sequence $\mathrm{GCR}$, we first give gadgets to squeeze an oscillator in Sec.~\ref{sec:squeezing}, as well as entangle and unentangle a qubit from oscillator states which are represented as non-overlapping Gaussian wave functions in Sec.~\ref{ssec:Cat_States}. 

With the help of these gadgets, we design preparation schemes of these simplest non-Gaussian states which can be represented as a superposition of non-overlapping finite-energy basis states $\{\ket{\alpha}_\Delta\}$ in the phase-space representation. This includes squeezed vacuum (Sec.~\ref{sec:squeezing}), two-legged cat states~\cite{mirrahimi2014dynamically} (Sec.~\ref{ssec:Cat_States}), and GKP codewords~\cite{gottesman2001encoding} (Sec.~\ref{ssec:GKP-States}). Then, we discuss the preparation of rotationally symmetric states in Sec.~\ref{ssec:universal}. We explain why it might be better to use an abelian sequence like $\mathrm{BB1}$ for $N$-legged cat states with high rotation symmetry (i.e., $N>2$). We also present an amplification gadget to prepare the rotationally symmetric Fock states, setting the floor for future works to pursue universal state preparation for completeness. 

For this section, we use the tensor product ordering $\ket{\mathrm{osc}}\otimes\ket{\mathrm{qubit}}$ for the joint Hilbert space. We will the total amplitude of $\mathrm{CD}$s as the circuit duration (quoted in $\mu$s). The exact conversion into the runtime of circuit is given in App.~\ref{app:squeezing}. 
\section{Squeezed States}\label{sec:squeezing}
\begin{figure*}[ht]
    \centering
    \includegraphics[width=\linewidth]{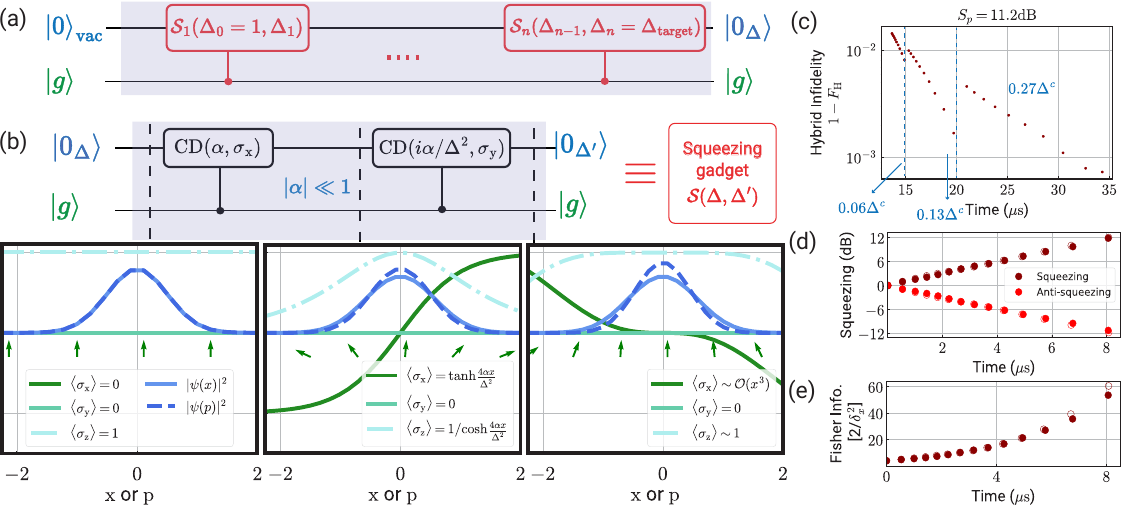}
    \caption[Deterministic preparation of squeezed states.]{\textbf{Deterministic preparation of squeezed states.} (a) Deterministic squeezing protocol with incremental $\mathrm{GCR}$. (b) Narration of $\mathrm{GCR}$ as a squeezing gadget $\mathrm{S}(\Delta,\Delta^\prime)$. The plots show how this sequence introduces a small amount of squeezing while unentangling the qubit from the final state for $\Delta=1,|\alpha|=0.25$. (c) Variation in fidelity and circuit duration with varying squeezing rate $|\alpha|_{k+1}=a\Delta_{k}^c$ where $c\in[-3,0]$ and $a\in\{0.06,0.13,0.27\}$ for a target squeezing of $11.2 \ \mathrm{dB}$. (d) Squeezing (maroon) and anti-squeezing (red) are shown as a function of the circuit duration for the faster protocol with $c=2$. See App.~\ref{app:squeezing} for definitions of $S_x,S_p$ in terms of $\Delta$. (e) Fisher information for the faster protocol. The empty circles in (d,e) represent a plot of the results for the case when post-selection is activated. 
    }
    \label{fig:squeezing}
\end{figure*}

We present a novel approach for generating squeezed states, marking a significant advancement in oscillator control.  This result is comparable to state-of-the-art schemes in Refs.~\cite{hastrup2021unconditional,eickbusch2022fast} without the need for any numerical optimization tools. In addition, the structure identified by our scheme gives a more versatile approach to optimize the fidelity with respect to circuit duration. Our protocol has a basic unit composed of $\mathrm{GCR}$. This basic unit follows an alternate explanation of $\mathrm{GCR}(\theta)$ as a deterministic small-even-cat preparation circuit for small $\theta$. The squeezing gadget uses the back action of GCR on the oscillator state to yield a squeezing gadget. Let us understand this effect in more detail for the case of momentum squeezing.

\paragraph{Squeezing gadget $\mathcal{S}(\Delta,\Delta^\prime)$:} We pose the first step of this problem as modifying the uncertainties ($\delta x=\Delta/2,\delta p=1/(2\Delta)$) of an oscillator state,
\begin{align}
\psi(x)=e^{-\frac{(x-\beta)^2}{\Delta^2}}.
\end{align}
 Our protocol begins with the oscillator in vacuum, that is, $\Delta=2\delta x=1,\beta=0$. After a conditional displacement $\mathrm{CD}(\alpha)=e^{i\alpha\hat p\otimes\sigma_\mathrm{x}}$ is applied to the joint state $\ket{0_\Delta}\otimes\ket{g}$, the expectation values of the qubit operators conditioned on the position of the oscillator are given by,
\begin{align}
 \braket{\sigma_\textrm{x}}_x&=\sin(\theta)=\tanh{\frac{4\alpha x}{\Delta^2}},\ 
 \braket{\sigma_\textrm{y}}_x=0, \ 
 \braket{\sigma_\textrm{z}}_x=\cos(\theta)=\mathrm{sech}{\frac{4 \alpha x}{\Delta^2} },
\end{align}
where $\Delta/2=\delta x$ is the position uncertainty of the input state for the units used in this chapter (see Sec.~\ref{app:phase-space}). Note that, since the rotation axis was $\sigma_\mathrm{x}$, 
\begin{align}
\int_{-\infty}^\infty \ dx \ \braket{\sigma_\mathrm{x}}=0,
\end{align} 
as should be the case. However, for a specific value of $x$, $\braket{\sigma_\mathrm{x}}\neq 0$ is possible. Maintaining a small slope ensures that $\braket{\sigma_\mathrm{x}}$ is proportional to $x$ as long as $\psi(x)$ has significant amplitude. To unentangle the qubit from the oscillator such that $\braket{\sigma_\mathrm{z}}=1$, we apply a rotation about $\sigma_\textrm{y}$ by an angle $\theta=\sin^{-1}{(\tanh{\frac{4\alpha x}{\Delta^2}})}\approx\frac{4\alpha x}{\Delta^2}$ (if $4|\alpha|x/\Delta^2\ll 1$), i.e., $\mathrm{R}_\mathrm{y}(-4\alpha\hat x/\Delta)=e^{i\frac{2\alpha}{\Delta^2}\hat{x}\sigma_\textrm{y}}=\mathrm{CD}(i\alpha/\Delta^2,\sigma_\textrm{y})$. This corrects the linear part, setting
\begin{align}
\braket{\sigma_\textrm{x}}\rightarrow 0,\braket{\sigma_\textrm{z}}\sim 1,
\end{align}
for the range where $x\ll \Delta^2/4|\alpha|$. This sequence is equivalent to $\mathrm{GCR}$ in the momentum basis. In Sec.\ref{sec:GCR}, we analyzed a conditional momentum boost from the position basis, whereas here we analyze a conditional displacement from the position basis. Exact expressions for $\braket{\sigma_\textrm{x}}$ and $\braket{\sigma_\textrm{z}}$ are given in App.\ref{app:squeezing}.

Thus, to maximize un-entanglement between the oscillator and qubit, $|\alpha|/\Delta$ must be sufficiently small to satisfy the condition. A larger\footnote{but still small enough for the output state to resemble a Gaussian} $|\alpha|$ yields greater squeezing but worse un-entanglement. Nevertheless, after a single application of the squeezing gadget, the state's position uncertainty $\delta x$ increases (see Fig.~\ref{fig:squeezing}(a)), enabling larger $|\alpha|$ values in successive rounds to achieve even greater squeezing. This, in turn, broadens the range over which $\psi(x)$ has significant amplitude. A small even-cat state is simply a slightly squeezed vacuum, thus naturally leading to the preparation of squeezed oscillator states.

\par\textbf{Protocol:} In Fig.~\ref{fig:squeezing}(a) we show that repeated application of the squeezing $\mathrm{GCR}$ circuit ($\mathcal{S}$, see Fig.~\ref{fig:squeezing}(b)) yields the desired target squeezing. Careful selection of $|\alpha|_k$ for successive steps $k$ is crucial for this purpose, as it dictates the convergence of squeezing with each cat step.  For optimal squeezing, $|\alpha|$ should be as high as possible while ensuring that the slope of $\braket{\sigma_\mathrm{y}}$ is linear over the range $|x|\le 2\delta x$. Another important detail is that the (small) even cat state is a sum of two highly overlapping Gaussian functions, posing a very high fidelity to a squeezed vacuum. 

To determine the right parameter for correction in the next round $\frac{\alpha_{k}}{\Delta_{k}^2}$ we need to approximate it to the closest Gaussian function, that is, identify the resulting $\Delta$ after each application of $\mathcal{S}$. This can be computed using various approximations/numerical methods. We derive that for a linear slope over $\mathrm{FWHM}$ of the oscillator state, $|\alpha|_{k+1}\ll 0.13\Delta^{1/2}$, see details in App.~\ref{app:squeezing}. 

To understand the relationship between convergence (which determines circuit duration) and unentanglement (which determines hybrid oscillator-qubit fidelity) for various $|\alpha|_{k+1}=a\Delta^{c}$, we use this protocol to obtain a squeezing of $S_p=11.2 \ \mathrm{dB}$ in Fig.~\ref{fig:squeezing}(c). We note that increasing $a,c$ yields better fidelity but longer circuit duration. This behavior, however, is reversed for $c>0$. 

The upper bound on $c$ is $2$ since, for this value, the slope of $\braket{\sigma_\mathrm{y}}$ is constant for varying $\Delta$. This is the fastest rate of convergence one could choose. However, for the case of momentum-squeezing analyzed here, $\Delta\ge 1$. Thus, with each step $k$, $\braket{\sigma_\mathrm{y}}$ will be now more nonlinear, making un-entanglement harder. So, we approximate corrections to the linear slope $4\alpha/\Delta^2$ numerically (see App.~\ref{app:squeezing}). Using this faster protocol, we obtain Figs.~\ref{fig:squeezing}(d,e). The squeezing efficiency, measured in dB, has a linear dependence on circuit duration\footnote{The conversion from amplitudes of $\mathrm{CD}$ to time is given in App.~\ref{app:squeezing}}, as shown in Fig.~\ref{fig:squeezing}. We note that this protocol results in faster convergence compared to Fig.~\ref{fig:squeezing}(c). 

Squeezed states often find use cases in measuring the net displacement or momentum boost in a state. The sensitivity of this measurement is usually determined via the Fisher information, which for a Gaussian state, like the momentum-squeezed state, is given by~\cite{paris2004quantum} $2/\delta_x^2$, where $\delta_x^2$ is the variance of the position operator. Note that this is not the right formula to compare non-Gaussian states generated using conditional displacements; however, we still use this metric for comparison with results in Refs.~\cite{hastrup2021unconditional,eickbusch2022fast}. In addition, our final state is close to the desired Gaussian squeezed state compared to the state prepared in Ref.~\cite{hastrup2021unconditional} that has more of the unwanted Wigner negativit. That is, the squeezed state prepared using our scheme is more suited to be used with this formula. Thus, quantifying the efficiency of this scheme in accurate position or momentum measurements, the highest Fisher information of $F=2/\delta x^2=53.5$ is reported for the final state with squeezing $S_p=11.2 \ \mathrm{dB}, S_x=-11.9 \ \mathrm{dB}$ at infidelity of $0.008$ in $8.06\mu\textrm{s}$.

We also find that post-selection over qubit being in the state $\ket{g}$ (empty squares), after each application of $\mathcal{S}_k$, does not improve results by much in the absence of errors. This observation indicates that we are achieving optimal un-entanglement with the help of $\mathrm{GCR}$ for our choices of $|\alpha|_k$. In addition, each $\mathcal{S}_k$ is small enough such that we can achieve better performance with the help of this post-selection, that could enable detection of ancilla errors in the middle of the circuit. This is known as mid-circuit error detection. 

\par\textbf{Discussion} Our analytically derived circuit performs on par with the semi-analytical and numerically optimal methods in Refs.~\cite{hastrup2021unconditional,eickbusch2022fast}. We plot results against circuit duration instead of circuit depth, given the speed and errors in a conditional displacement gate depend significantly on the length of the displacement \cite{eickbusch2022fast}. Our protocol achieves squeezing levels, with $S_p=8.5 \ \mathrm{dB}$ of squeezing and $S_x=-8.4 \ \mathrm{dB}$ of anti-squeezing, alongside an infidelity of $\sim \mathcal{O}(10^{-3})$ in $5.8\mu\mathrm{s}$  while Ref.~\cite{hastrup2021unconditional} reports $\delta x=8.5 \textrm{dB}$ and $\delta p=-9.9 \textrm{dB}$ with an infidelity of $\sim \mathcal{O}(10^{-2})$. The performance of our scheme is also on par with numerically optimized schemes~\cite{eickbusch2022fast}, offering improved oscillator control. Details of the comparisons here can be found in App.~\ref{app:squeezing}.

Let us briefly discuss the reason behind our improvement upon the results in Ref.~\cite{hastrup2021unconditional}. In that work, the authors use a large conditional displacement in the first step in contrast to our approach of incrementing the amplitude of conditional displacements with increasing Gaussian width of the oscillator state. Due to this approach, the protocol requires numerical techniques to unentangle the qubit. On the other hand, our protocol is completely analytical and yields squeezed states with better fidelity. Our protocol outputs states with fairly less interference (i.e., Wigner negativity), yielding higher fidelity with a squeezed state (which is a Gaussian state, and hence shows no interference/Wigner negativity).

\section{Two-Legged Cat States}\label{ssec:Cat_States}
\begin{figure*}[htb]
    \centering
    \includegraphics[width=\linewidth]{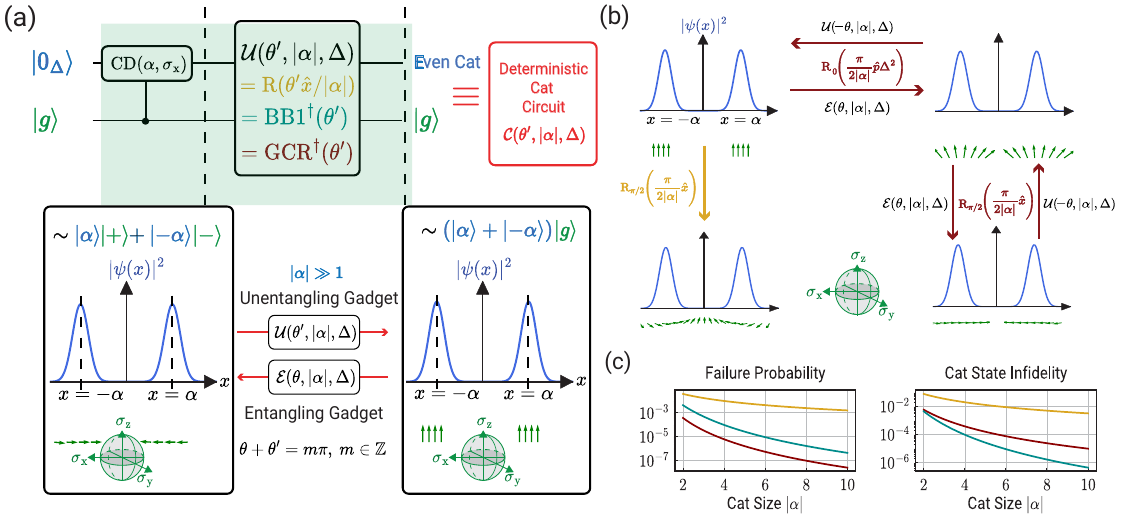}
    \caption[Deterministic preparation of two-legged cat ctates.]{\textbf{Deterministic preparation of two-legged cat states.} (a) Deterministic cat state preparation requires a unentangling sequence given by $\mathcal{U}$. (b) Entangling-unentangling gadgets using $\mathrm{GCR}$. (c) We show numerical results with options of no correction ($\mathcal{U}=\mathrm{R}_\textrm{y}(\theta^\prime\hat x/|\alpha|)$ in yellow), univariate or traditional QSP correction ($\mathcal{U}=\mathrm{BB1}$ in cyan), bivariate non-abelian QSP correction ($\mathcal{U}=\mathrm{GCR}$ in red). (Left) Success probability of ancilla ending in-state $\ket{g}$. (Right) Fidelity of output oscillator state with the desired cat state upon success.}
    \label{fig:Cat_states}
\end{figure*}
The superposition of two coherent states located at diametrically opposite locations in the phase space of an oscillator is known as a two-legged cat state. The interference pattern at the origin is determined by the local phase of this state, also termed the `whiskers of the cat.'
\begin{align}
\ket{C_{+\alpha}}&\propto(\ket{\alpha}+\ket{-\alpha})\quad \text{Even Cats}\\
    \ket{C_{-\alpha}}&\propto(\ket{\alpha}-\ket{-\alpha})\quad\text{Odd cats}.
\end{align}

If the basis states $\ket{\pm\alpha_\Delta}$ with $\Delta\neq 1$ are used in this definition, then $\ket{C_{\pm\alpha}}$ are squeezed cat states. 

A deterministic preparation of cat states will require the qubit to be unentangled from an oscillator after a large (equal to cat size) conditional displacement $\mathrm{CD}(\alpha,\sigma_\textrm{x})$ (see Fig.~\ref{fig:Cat_states}(a)) leading to the state $\ket{C_{+\alpha}}$ if the initial qubit state was in $\ket{g}$. That is, ignoring the normalization constant, we need
\begin{align}
\mathcal{U}(\ket{\alpha}\ket{+}+\ket{-\alpha}\ket{-})\approx\ket{C_{+\alpha}}\ket{g}.
\end{align}
So, the first question we address here is how to entangle or unentangle oscillator states from qubits with minimal back action on the oscillator. This can be done with the help of $\mathrm{GCR}$ which, as we saw in Sec.~\ref{sec:GCR}, will rotate a qubit entangled with states $\ket{\pm\alpha}$ by $\pm \pi/2$.  Thus, we now interpret $\mathrm{GCR}$ as \emph{entangling-untangling gadgets} of the hybrid oscillator-qubit system, to be used repeatedly in the remainder of this chapter.

\subsection*{Entangling Oscillators and Qubits}~\label{ssec:ent-unent}
The primary requirement to use the rotation gadgets ($\mathrm{GCR}$ or $\mathrm{BB1}$) for entangling oscillators and qubits will be that the oscillator state be represented by a sum of non-overlapping Gaussian wave functions. Consider a state (ignoring normalization), 
\begin{align}
    \ket{\psi_\Delta}\propto\ket{+\alpha_\Delta}\pm\ket{-\alpha_\Delta},
\end{align}
with $\braket{\alpha_\Delta|-\alpha_\Delta}\rightarrow 0$, such that $\ket{\psi_{\Delta}}$ is a sum or difference of non-overlapping Gaussian wave functions. The entangling gadget $\mathcal{E}$ and unentangling gadget $\mathcal{U}$ are defined as (up to normalization constants),
 \begin{align}
\mathcal{E}\ket{\psi_\Delta}\ket{g}&=\ket{+\alpha_\Delta}\ket{+}\pm\ket{-\alpha_\Delta}\ket{-}\label{eq:Gaussian1},\\
\ket{\psi_\Delta}\ket{g} &=\mathcal{U}(\ket{+\alpha_\Delta}\ket{+}\pm\ket{-\alpha_\Delta}\ket{-}). \label{eq:ent_unent}  
\end{align}
It can be trivially seen that $\mathcal{U}=\mathcal{E}^{-1}$ works, yet we will see next that this is not the only option available for the unentangling gadget $\mathcal{U}$.

\paragraph{$\mathcal{E}$:} A Gaussian-peak-dependent entangling gadget can be defined as 
\begin{align}
    \mathcal{E}(\theta,|\alpha|,\Delta)&:\mathrm{GCR}(\theta)\big(\ket{\psi_\Delta}\otimes\ket{q}\big), \    \textrm{ or} ~\ \mathrm{BB1}(\theta)\big(\ket{\psi_\Delta}\otimes\ket{q}\big)
\end{align}
 when the initial qubit state is $\ket{q}$. For the case of cat states in Fig.~\ref{fig:Cat_states}(a) or Eqs.~\ref{eq:ent_unent}, $\theta=\pi/2$.
 
\paragraph{$\mathcal{U}$:} If we start in an entangled hybrid oscillator-qubit state, an unentangling circuit ensures that the qubit state at the end of the circuit is fixed (see Fig.~\ref{fig:Cat_states}(a)). Thus, we define the unentangling gadget as any circuit of the form,
\begin{align}
    \mathcal{U}(\theta^\prime,|\alpha|,\Delta)&:\mathrm{GCR}^\dagger(-\theta^\prime)\ket{\phi_\Delta} \ \textrm{or} ~\ \mathrm{BB1}^\dagger(-\theta^\prime)\ket{\phi_\Delta}
\end{align}
where $\ket{\phi}$ represents a hybrid oscillator-qubit entangled state, such as the output of Eq~\ref{eq:Gaussian1}. Note that the parameters of the unentangling gadget can be different from those given by $\mathcal{E}^{-1}$\footnote{We use inverse rather than adjoint because $\mathrm{GCR}$ can be viewed as a non-abelian QSP sequence dependent on the qubit being in a specific state. See discussion around Eq.~(\ref{eq:GCR-corr}}), as noted from the periodicity of single-qubit rotations; they only need to satisfy Eq.~(\ref{eq:ent_unent}). The condition for $\mathcal{E}-\mathcal{U}$ for a specific oscillator state is given by,
\begin{align}
    \theta+\theta^\prime=m\pi\quad m\in\mathbb{Z}.
\end{align}
 For example, for the case of cat states in Fig.~\ref{fig:Cat_states}(a), $(\theta,\theta^\prime)=(-\pi/2,\pi/2)$ is just one of the many choices. Now, we can summarize the cat state preparation protocol. For a pictorial representation of the cancellation of errors at $\ket{\pm\alpha}$ simultaneously, see Fig.~\ref{fig:Cat_states}(b).

\par\textbf{Protocol:} We assume that the Gaussian functions have negligible overlap, that is, $\alpha>1$. We start with the hybrid oscillator-qubit state, $\ket{0}\otimes\ket{g}$ and perform a conditional displacement (up to normalization constants), 
\begin{align}
    e^{-i2\alpha p\sigma_\mathrm{x}}(\ket{0}_\textrm{vac}\otimes\ket{g})&\propto\ket{\alpha}\ket{+}+\ket{-\alpha}\ket{-}
\end{align}  
This leaves the oscillator-qubit in an entangled state. At this point, if we were allowed to use measurements, we could probabilistically prepare even or odd cat states by measuring the qubit in the $\sigma_\mathrm{z}$ basis~\cite{sun2014tracking,wineland2013nobel,haroche2013nobel}. However, for a deterministic process, we need to avoid any measurement, and this is where the unentangling gadget $\mathcal{U}$ can help.
 Therefore, the cat preparation circuit is given by,
\begin{align}
    \mathcal{U}(\pi/2,|\alpha|,1)e^{-i2\alpha \hat p \sigma_\mathrm{x}}(\ket{0}_\textrm{vac}\otimes\ket{g})&\propto  \mathcal{U}(\pi/2,|\alpha|,1)(\ket{\alpha}\ket{+}+\ket{-\alpha}\ket{-})\label{eq:cat-statea}\\
     &=  e^{-i\frac{\pi}{4|\alpha|}\hat p \sigma_\mathrm{x}}e^{i\frac{\pi}{4|\alpha|}\hat x \sigma_\mathrm{y}}(\ket{\alpha}\ket{+}+\ket{-\alpha}\ket{-})\label{eq:cat-stateb}\\
   &\approx (\ket{\alpha}+\ket{-\alpha})\otimes\ket{g}+O(\chi^2)\ket{\psi^\prime}\otimes\ket{g}\nonumber\\
    &\quad +O(\chi^3)\ket{\psi^{\prime\prime}}\otimes\ket{e},
\end{align}
where $\chi=\theta\Delta/2|\alpha|$ is the error parameter for QSP sequences like GCR, as defined in Chapter~\ref{chapter:na-qsp}. This circuit requires no measurement and yields the even cat $\propto(\ket{\alpha}+\ket{-\alpha})$ state in the cavity. Odd cats $\propto(\ket{\alpha}-\ket{-\alpha})$ can similarly be prepared by starting in qubit state $\ket{e}$ or using 
\begin{align}
    \mathcal{U}(\pi/2,|\alpha|,1)e^{-i2\alpha \hat p \sigma_\mathrm{x}}(\ket{0}_\textrm{vac}\otimes\ket{g})&\propto  \mathcal{U}(\pi/2,|\alpha|,1)(\ket{\alpha}\ket{+}+\ket{-\alpha}\ket{-})\label{eq:cat-statec}\\
     &=  e^{-i\frac{\pi}{4|\alpha|}\hat p \sigma_\mathrm{x}}e^{-i\frac{\pi}{4|\alpha|}\hat x \sigma_\mathrm{y}}(\ket{\alpha}\ket{+}+\ket{-\alpha}\ket{-})\label{eq:cat-stated}\\
   &\approx (\ket{\alpha}-\ket{-\alpha})\otimes\ket{e}+O(\chi^2)\ket{\psi^\prime}\otimes\ket{e}\nonumber\\
    &\quad +O(\chi^3)\ket{\psi^{\prime\prime}}\otimes\ket{g}
\end{align}
to end up in $\ket{e}$ with maximum probability. In these scenarios, the final state will have the highest fidelity with an odd cat of size $\alpha$. As described in Sec.~\ref{sec:composite_error}, the success probability and fidelity depend on the value of $\chi=\theta\Delta/2|\alpha|$, so the cat fidelity increases with increasing $\alpha$ and decreasing $\Delta$.

\par\textbf{Discussion} In Figs.~\ref{fig:Cat_states}(c,d) we vary $\alpha$ and plot two quantities for the worst case scenario of $\Delta=1$, (i) $1-P_e$ which is marked by the measurement of ancilla in $\ket{e}$ and (ii) $1-F_\mathrm{H}$, the infidelity of the hybrid output state against the desired even cat state with the qubit in $\ket{g}$. Low $P_e$ for $\mathrm{GCR}$ and $\mathrm{BB1}$ indicate that ancilla measurement is not required for unentangling the qubit here. As shown in Sec.~\ref{sec:composite_error}, the un-entanglement error $P_e$ decreases with increasing $|\alpha|$ and decreasing $\Delta$. We can see the same trend as given by the corresponding correctness metrics for our framework in Fig.~\ref{fig:Correctness}. To study the effectiveness of QSP, we also show the respective curves for performing no correction with $\mathcal{U}=\mathrm{CD}(\theta^\prime/2|\alpha|,\sigma_\textrm{y})$. The analytical expression $P_e=\mathcal{O}(\chi^2)$ for this curve has been derived in App.~\ref{app:cat_I}. The analytical fidelity for the $\mathrm{GCR}$ and $\mathrm{BB1}$ has been computed in App.~\ref{app:comp_err}. Thus, we have shown orders of magnitude improvement achieved from our framework of using composite pulses in phase space when correcting continuous-variable rotation errors on qubits. 

\section{GKP States}\label{ssec:GKP-States}
\begin{figure*}
    \centering
    \includegraphics[width=\linewidth]{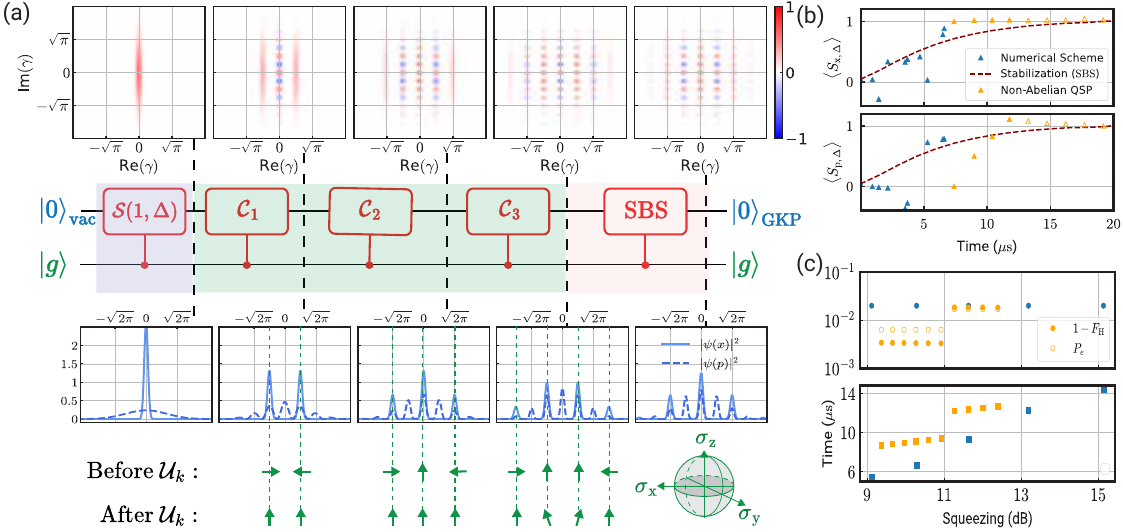}
    \caption[Deterministic GKP logical $\ket{+Z}$ state preparation.]{\textbf{Deterministic GKP logical $\ket{+Z}$ state preparation.} \textbf{(a)} Circuit components for GKP preparation with $\Delta=0.34$ (as used in recent experiments~\cite{eickbusch2022fast,sivak2023real}). $\mathcal{S}$ denotes the squeezing circuit from Fig.\ref{fig:squeezing}, $\mathcal{C}_k = \mathcal{U}_k e^{-i\sqrt{2\pi}\hat{p}\sigma_\mathrm{z}}$ represents the gate sequence given by Eq.~(\ref{GKP-circuit}) in the main text, and $\mathrm{SBS}$ is one round of the small-big-small protocol. See Table~\ref{tab:GKP-prep-circuit-depth} for circuit-depth justification. (Top) Wigner functions $\mathcal{W}(\alpha,\beta)$~\cite{ISA} are shown after each step. Note the change in the state before and after $\mathrm{SBS}$. (Bottom) Wave functions and corresponding spin polarizations are shown before and after each $\mathcal{U}_k$. After $\mathcal{U}_3$, the qubit remains slightly entangled, with $\braket{\sigma_\mathrm{z}} = 0.9937$ and GKP fidelity $\mathcal{F}_\mathrm{GKP} = 0.9989$.
\textbf{(b)} Convergence of various protocols, tracked using the expectation values of finite-energy stabilizers $\braket{S_{\mathrm{x},\Delta}}$ and $\braket{S_{\mathrm{p},\Delta}}$ (see Eqs.~(\ref{eq:GKP-stab-p}-\ref{eq:GKP-stab-x})). The non-abelian QSP protocol using $\mathrm{GCR}$ (via $\mathcal{C}_k$'s) starts at $8.06\mu\textrm{s}$, accounting for the time to achieve $11.2\ \mathrm{dB}$ of squeezing (see Fig.\ref{fig:squeezing} and Sec.\ref{sec:squeezing}). Empty triangles mark the section where SBS is appended. Our protocol matches the performance of the numerical scheme\cite{eickbusch2022fast} while significantly outperforming stabilization-based methods (using only $\mathrm{SBS}$\cite{royer2020stabilization}).
\textbf{(c)} Hybrid infidelity, failure probability (top), and circuit duration (bottom) versus squeezing (or $\Delta$) for the numerical and analytical (non-abelian QSP) circuits. Our protocol achieves fidelity and circuit duration comparable to the numerical scheme, while maintaining a low failure probability. All simulations here neglect physical errors such as damping, heating, and decoherence, which are addressed in Sec.~\ref{chapter:qec-control}.} 
    \label{fig:GKP-prep}
\end{figure*}
GKP codes have applications for quantum sensing and bosonic error correction~\cite{sivak2023real,campagne2020quantum,fluhmann2019encoding,brady2024advances}. In this section, we will only discuss the definition of logical codewords required for preparation, leaving all other details to Sec.~\ref{chapter:GKP-qec} where we discuss these codewords in more detail. While we focus on the square GKP codewords to give explicit constructions for preparation, all our protocols are easily generalizable to arbitrary lattices of hexagonal and rectangular GKP codes. Due to the completely analytical constructions, we note that our preparation schemes are easily generalizable to GKP qudits also. Such generalizations are not accessible to numerically optimized circuits, such as the ones shown in~\cite{eickbusch2022fast}\footnote{For each qudit codeword, a new numerical optimization needs to be run since the circuit constructions are highly non-intuitive and provide no structure for any generalization.}. The computational basis square-GKP codewords are defined as~\cite{gottesman2001encoding},
\begin{align}
    \ket{0}_\mathrm{GKP}&\propto\sum_{m=-\infty}^{\infty} e^{-(m\sqrt{\pi})^2\Delta^2}\mathrm{D}(m\sqrt{2\pi})\ket{0_\Delta},\label{eq:GKP-logical0}\\
    \ket{1}_\mathrm{GKP}&\propto \sum_{m=-\infty}^{\infty} e^{-((2m+1)\sqrt{\pi})^2\Delta^2}\mathrm{D}\big([m+1/2]\sqrt{2\pi}\big)\ket{0_\Delta},\label{eq:GKP-logical1}
\end{align}
where $m\in \mathbb{Z}, \ \mathrm{D}(\alpha)=e^{\alpha \hat a^\dagger-\alpha^*\hat a}=e^{2i(\mathrm{Im}(\alpha)\hat x-\mathrm{Re}(\alpha)\hat p)}$ denotes an unconditional displacement of the oscillator by $|\alpha|$ generated by the operator $\hat v(\alpha)=2\mathrm{Im}(\alpha)\hat x-2\mathrm{Re}(\alpha)\hat p$. Here, $\ket{0_\Delta}=\ket{0_{2\delta x}}$ are the position-squeezed states (see Eq.~(\ref{eq:Gaussian})). Note that, $\ket{\pm}_L=(\ket{0}_\mathrm{GKP}\pm\ket{1}_\mathrm{GKP})/\sqrt{2}$ states are exactly equal to the same superposition of finite-energy momentum-squeezed states $\ket{0_{2\delta p}}$.

There are various definitions of GKP states in the literature, in addition to the above equations, all of which are equivalent~\cite{matsuura2020equivalence}. Note that the states we prepare in this chapter will closely resemble those described by the above equations. The most important task for a preparation routine is to prepare a state close to the GKP manifold such that the stabilization (i.e., subsequent rounds of error correction) can take care of the residual small errors. Thus, in order to remove any non-uniformity in fidelity using various definitions, we will also compute the expectation values of the finite-energy stabilizers~\cite{royer2020stabilization} (see Chapter~\ref{chapter:GKP-qec} and App.~\ref{dissipation-engineering} for further details), 
\begin{align}
    S_\mathrm{x,\Delta}&=e^{i2\sqrt{2\pi}(\cosh{\Delta^2}\hat x-\hat p\sinh{\Delta^2})}\label{eq:GKP-stab-x}\\
    S_\mathrm{p,\Delta}&=e^{i2\sqrt{2\pi}(\cosh{\Delta^2}\hat p-\hat x\sinh{\Delta^2})}\label{eq:GKP-stab-p}
\end{align}

We now present the first analytical measurement-free protocol, derived using non-abelian QSP, for the preparation of GKP codes. We give comparisons to other, numerically optimized schemes in Refs.~\cite{eickbusch2022fast,hastrup2021measurement}. We also give a comparison against using the stabilization scheme which can cool any CV state towards the GKP manifold~\cite{royer2020stabilization,de2022error}. The comparisons in this section discuss circuit depth. Our scheme has an additional advantage towards error correction which will be discussed in Chapter~\ref{chapter:qec-control}.

\par\textbf{Protocol:} GKP states are an extension of two-legged cat codes where the deterministic preparation scheme is now required to create superpositions of multiple squeezed coherent states unentangled from the qubit. A sketch of the preparation scheme is given in Fig.~\ref{fig:GKP-prep}(a). We now give the algorithm described by the circuit construction shown in Fig.~\ref{fig:GKP-prep}(b) to prepare logical Pauli eigenstates of the GKP code. 

We start with a squeezed vacuum (prepared using the protocol in Sec.~\ref{sec:squeezing}), and use the cat-state preparation circuit $\mathcal{C}_1$ (described in~\ref{ssec:Cat_States}) to prepare squeezed cats, leaving the ancilla unentangled. Circuits $\mathcal{C}_k$ producing $k+1$ peaks represent the gate sequence,
\begin{equation}    \mathcal{C}_k:\mathcal{C}\Big(\frac{\pi}{4k},\sqrt{\frac{\pi}{2}},\Delta_{k-1}\Big)=e^{-i\sqrt{2\pi}\hat p\sigma_\mathrm{z}}\mathcal{U}\Big(\frac{\pi}{4k},\sqrt{\frac{\pi}{2}},\Delta_{k-1}\Big),\label{GKP-circuit}
\end{equation}
where we use the definition of the unentangling gadget $\mathcal{U}(\theta,|\alpha|,\Delta)$, conditioned on the input state, from Sec.~\ref{ssec:ent-unent}. In this case, the circuit components $\mathcal{C}_k$ are decided by the lattice spacing ($\sqrt{2\pi}$ in Wigner units for square-GKP codes~\cite{gottesman2001encoding}), the finite-energy parameter $\Delta$ of the target GKP state and the previous state after $k-1$ snippet of the circuit. We first start with a squeezed state which can be prepared using the protocol in Sec.~\ref{sec:squeezing}. Next, we create a squeezed cat state of size $\sqrt{\pi/2}$ where the non-abelian QSP correction is $\sqrt{\pi/2}\Delta^2$. This process is repeated to create a superposition of three Gaussian wave functions. Note that the information regarding integer multiple $m$ is not required while designing the unentangling gadget. It is so because the gap between each peak is $\sqrt{2\pi}$ (in Wigner units), and that means each peak subsequently away from the origin rotates the qubit by an extra angle of $\sqrt{2\pi}|\alpha|=2\pi$. Thus, each peak will rotate the qubit by the same amount (as $2\pi$ is the period of all trigonometric functions).  Note that, in our scheme, expectation values exceeding unity in Fig.~\ref{fig:GKP-prep}(b) are artifacts of the non-abelian pulse. This arises because the final state after $\mathcal{C}_3$ has not yet fully converged to the GKP code space; in particular, the most displaced squeezed states in the superposition lack the Wigner negativity characteristic of true GKP states. At this stage, applying a single round of SBS can effectively refine the state, bringing it closer to the ideal GKP form. Crucially, using SBS at this point almost deterministically prepares the logical $\ket{0}_\mathrm{GKP}$ state.

\paragraph{Fidelity and circuit-depth:} The state prepared using this method yields a state whose Gaussian peaks have amplitudes that are binomial coefficients, whereas the usual definition of GKP states uses a Gaussian envelope. Hence, we use the Newton-Raphson method to find the number of steps $N$ required for a given $\Delta$ where the binomial coefficients reach a Gaussian distribution. This method is highlighted in App.~\ref{app:GKP-prep}.
\begin{table}[htb]
    \centering
    \begin{tabular}{|c|c|c|c|}
    \hline
    $\Delta$ & Squeezing (in dB)& $N$ &$1-\mathcal{F}$\\
    \hline
    $0.10$     & $20$&$31$&$\mathcal{O}(10^{-5})$\\
    $0.20$     & $14$&$7$&$\mathcal{O}(10^{-3})$\\
    $0.30$ &  $10.45$& $3$&$\mathcal{O}(10^{-2})$\\
    $0.4$     & $7.95$&$1$&$\mathcal{O}(10^{-2})$\\
    \hline
    \end{tabular}
    \caption[Finite-energy envelope parameters for equivalence between Gaussian and Binomial GKP amplitude distribution.]{Circuit depth for different squeezing levels starting with the initial state $e^{-\frac{x^2}{\Delta^2}}$. Here, $N$ is the optimal number of large conditional displacements ($\sqrt{\pi}$) involved in preparing a $\ket{0}_\mathrm{GKP}$ circuit for the desired finite-energy parameter, obtained using $N\Delta^2=0.32$ as solved above. Here, a state with $\Delta=0.5$ is achieved with very high fidelity because, for states with such high finite-energy parameter ($\Delta$), a GKP logical $\ket{0}_\mathrm{GKP}$ state is a squeezed vacuum while a GKP $\ket{1}_\mathrm{GKP}$ is the grid state which is similar to (not same as) a squeezed cat state.
    }
    \label{tab:GKP-prep-circuit-depth}
\end{table}
With this circuit depth we compare the fidelity of our GKP logical $\ket{\mu=\{0,1\}}$ states using the definition,
\begin{align}
    \ket{\mu}_\mathrm{GKP}=\mathcal{N}_\mu \sum_{m=-\frac{N+\mu}{2}}^{\frac{N-\mu}{2}+1} b_1\mathrm{D}\big([m+\mu/2]\sqrt{2\pi}\big)\ket{0_\Delta},\label{eq:GKP-bin-logical}
\end{align}
where $m\in \mathbb{Z},b_\mu={N\choose m+\mu+\floor{N/2}}$ and $N=\floor{0.32/\Delta^2}$ (see Table~\ref{tab:GKP-prep-circuit-depth}). These equations are justified by showing the evolution of the prepared states in Fig.~\ref{fig:GKP-prep}(b). Note that we have accounted for the circuit depth of squeezing in Fig.~\ref{fig:GKP-prep}(b), as the first point for non-abelian QSP starts at $8.06\mu\textrm{s}$.

\paragraph{Success probability:} For the circuits $C_k$ we need $\mathcal{U}(\theta^\prime,|\alpha|,\Delta)$ with $\theta^\prime=\pi/4$ for $k<3$. For $k=3$, as can be seen in Fig.~\ref{fig:GKP-prep}(a), the angle required to rotate the qubits at peaks on the farther end is $\pi/12$. Thus, the unentanglement gadget $\mathcal{U}_k$ at this point rotates the qubits by $\theta=\pm\pi/12$ at the two extreme peaks as required. However, it also rotates the qubits entangled with the peaks in the middle, ones that did not require any rotation. While the rotation at the central peaks is not significant, there is a different angle of rotation compared to $\pi/4k$ which could produce better unentanglement for $k\ge 3$. We compute the optimal angle of rotation using the procedure given in App.~\ref{app:GKP-prep}. With this protocol, we obtain Figs.~\ref{fig:GKP-prep}(a,b) for $\Delta=0.34$ (used in recent experiments~\cite{eickbusch2022fast,sivak2023real}). The state after $k=3$ yields $4$ peaks with a fidelity of $F_\mathrm{H}=0.9989$ to the target GKP state, while the success probability was $P_g=0.99$. Thus, we may also reset the ancilla after each $\mathcal{C}_k$. Such high success probability justifies using the measurements to keep ancilla errors in check (see Sec.~\ref{ssec:GKP-errors}). 

\paragraph{Other GKP lattices:}
Finally, to achieve different square and hexagonal GKP lattices, we will only need to change the lattice spacing $l$ and the rotation angle $\theta$ according to the position of the deformed lattice peaks $k\alpha$, in the circuit components $\mathcal{C}_k$,
\begin{align}
    \mathcal{C}_k=\mathcal{U}\Bigg(\frac{\pi}{4k},\frac{l}{2},\Delta_{k-1}\Bigg)e^{-i\sqrt{2}\alpha\hat p\sigma_\mathrm{z}}.
\end{align}

Arbitrary GKP states (other than Pauli eigenstates) can be prepared using the gate-teleportation circuit discussed in Sec.~\ref{ssec:piecewise-teleportation}. 

\par\textbf{Discussion} We show a comparison of our scheme against two different GKP preparation methods~\cite{eickbusch2022fast,royer2020stabilization} in Fig.~\ref{fig:GKP-prep}(b,c). It should be noted that our scheme is different from Ref.~\cite{hastrup2021measurement} where the authors propose to prepare GKP states using the same pattern of alternating conditional displacements and conditional momentum boosts; however, due to the lack of the non-abelian QSP pulse, in this case the qubit needs to be unentangled using numerical schemes. The scheme in Ref.~\cite{hastrup2021measurement} starts with a large cat and then creates multiple superpositions by moving inwards towards the origin. The demerit of this scheme is that one needs additional numerical optimization to adjust the coefficient of each peak, in the absence of which, the GKP state has an external envelope of two Gaussian functions centered at the peaks of the cat state prepared in the first step. In addition, this means that our protocol uses smaller conditional displacements in one step. This is an important distinction since this incurs less error during one step, and so if post-selection upon qubit measurement after each step is used, our protocol will naturally yield a higher success probability. This direction has been discussed in detail in Sec.~\ref{ssec:GKP-errors} as mid-circuit ancilla error detection. 

Another method to prepare the GKP states is by using a code space stabilization scheme followed by measurement of the $Z_L$ operator on the cavity state. We find that our scheme is twice as fast compared to the stabilization scheme \emph{small-big-small}~\cite{royer2020stabilization,de2022error}, described in detail in Chapter~\ref{chapter:GKP-qec} and App.~\ref{dissipation-engineering}. Finally, our circuit depth and fidelity (see table~\ref{tab:GKP-prep-circuit-depth} and Fig.~\ref{fig:GKP-prep}) are on par with the optimized E\textrm{CD} circuits in~\cite{eickbusch2022fast}. Importantly, our scheme gives us a way to make the scheme tolerant to circuit errors and achieve higher fidelity in the presence of faults, impossible for the long numerically optimized circuits in~\cite{eickbusch2022fast}. After each $\mathcal{C}_k$, in the absence of errors, the qubit is in a known pure state untangled from the oscillator with a very high probability ($>0.99$) as indicated by the low failure probability in Fig.~\ref{fig:GKP-prep}(c). At this point, GKP states can be post-selected, given the qubit is found in the desired state. Thus, we can also keep qubit errors in check with this scheme.

\section{Open Problem: Universal State Preparation}\label{ssec:universal}
The phase-space instruction set discussed in this chapter is more suited to oscillator states discussed above in terms of efficient circuits for preparation and control. However, this instruction set is universal, and thus, for completeness, we discuss the preparation of rotationally symmetric states like $N$-legged cat states and Fock states. Finally, we will give insights into applications in the construction of arbitrary superposition of Fock states. We note that such states could be better prepared with the help of Fock-space instruction set~\cite{ISA} using the hybrid SNAP gates and unconditional displacement or momentum boosts of the oscillator.
\begin{figure}
    \centering
    \includegraphics[width=\textwidth]{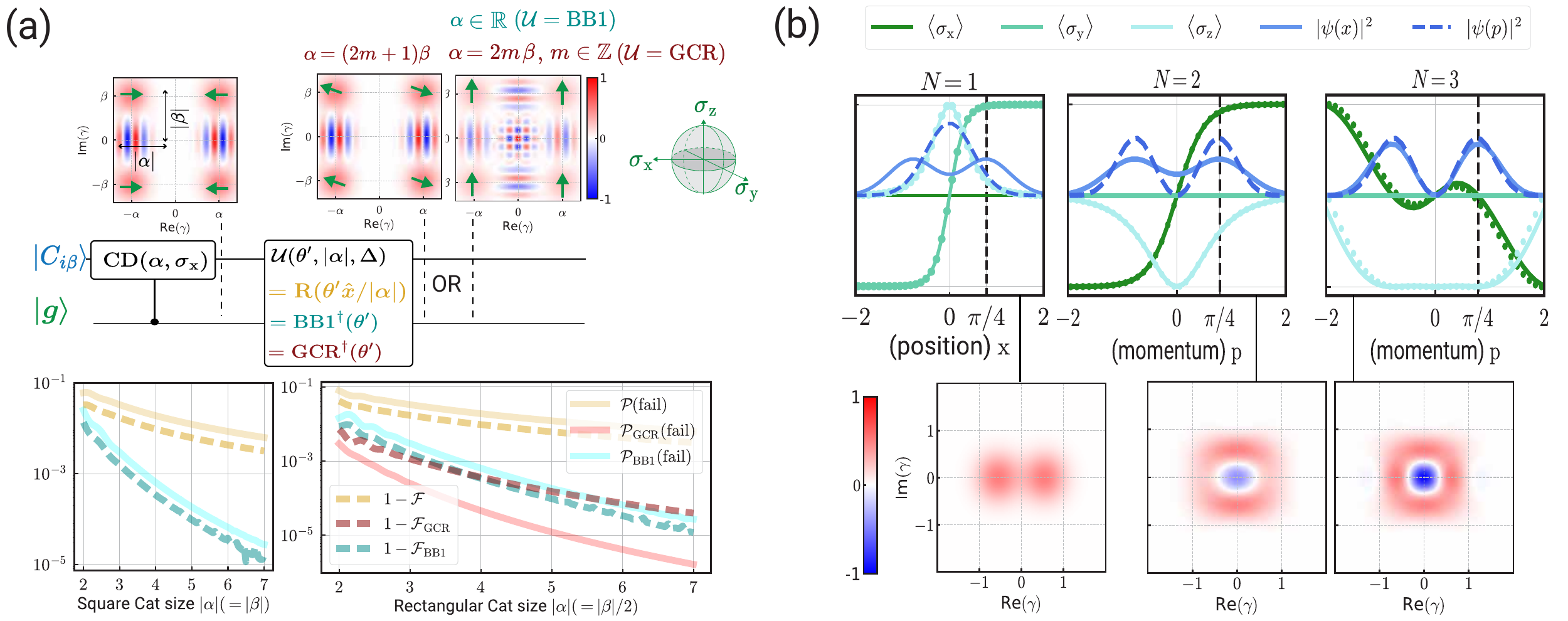}
    \caption[Circuit for preparation of rotationally symmetric states using QSP sequences in phase space instruction set.]{\textbf{Circuit for preparation of rotationally symmetric states using QSP sequences in phase space instruction set.} \textbf{(a)} Four-Legged Cat States. (Top) Two-legged cat preparation using the circuit in Fig.~\ref{fig:Cat_states}(a), starting from a (perfect) momentum-cat state. Green arrows in the Wigner plots indicate the qubit spin polarization at each blob. Note that the protocol does not require an initially unentangled two-legged cat; starting from vacuum with two conditional displacements along $\hat{x}$ and $\hat{p}$ quadratures also suffices. Using a perfect cat here simplifies simulation without underestimating infidelity. At the end of the circuit, alignment of arrows across all blobs is necessary for deterministic preparation; otherwise, the qubit remains entangled with the oscillator. The non-abelian QSP sequence $\mathrm{GCR}$ can only prepare rectangular cat states with an even aspect ratio $|\beta|/|\alpha|$, while the abelian BB1 sequence has no such constraint. (Bottom) Performance of GCR and BB1 for rectangular (square) 4-legged cat states. The yellow curve denotes performance without QSP correction. \textbf{(b)} Fock State $\ket{1}$. Wavefunctions $\psi(x)$ and $\psi(p)$ plotted against position $x$ and momentum $p$ to analyze the oscillator subspace.
(Left) Expectation values $\braket{\sigma_\mathrm{x}}$, $\braket{\sigma_\mathrm{y}}$, and $\braket{\sigma_\mathrm{z}}$ versus oscillator position after applying $\mathrm{CD}(\alpha_1,\sigma_\mathrm{y})$. Solid curves show simulations; dotted curves show small-cat analytical predictions (Sec.\ref{app:squeezing}). The hybrid fidelity is $F_\mathrm{H} \sim 0.58$. The Wigner plot indicates it is a large-cat state centered at $\alpha_1/2$, suggesting a no-QSP correction with $\beta_1 = (\pi/4)(\alpha_1/2)$.
(Middle) Expectation values versus oscillator momentum after $\mathrm{CD}(\beta_1,\sigma_\mathrm{x})$. Fidelity improves to $F_\mathrm{H} \sim 0.84$. Solid and dotted curves match well, validating the small-cat approximation for the next step. (Right) Expectation values after $\mathrm{CD}(\alpha_2,\sigma_\mathrm{y})$. Solid curves match the analytical predictions for $\braket{\sigma_\mathrm{y,z}}^\mathrm{new}$ (Eqs.(\ref{eq:sx_new}, \ref{eq:new_sx}), App.\ref{app:squeezing}). Final hybrid fidelity reaches $F_\mathrm{H} \sim 0.99$. The Wigner plot shows a symmetric state resembling the Fock state $\ket{1}$. Achieved fidelities and circuit durations match the optimized protocols of Ref.\cite{eickbusch2022fast}.}
    \label{fig:4-legged-cat}
\end{figure}

\subsection*{Rotationally Symmetric Codewords}
Here, we will discuss the preparation of four-legged cat states and their extension to $N$-legged cat states. $N$-legged cat states are superposition of $N$ coherent states located at the vertices of an $N$-sided polygon, centered at the phase space origin. Superposition of $N$ basis states requires a minimum of $\log_2{N}$ conditional displacement applications, at the end of which the oscillator and qubit should be completely entangled. We again use QSP corrections to assist in un-entangling the ancilla qubit.

For example, in order to generate a $4$-legged cat state, our first cat preparation circuit generates a two-legged cat state, and then the next circuit in the orthogonal direction generates a four-legged cat. The challenge is to un-entangle the qubit in this case where it is entangled with four oscillator states, all at $x\neq 0$ and $p\neq 0$. 

\par\textbf{Protocol:} From the preparation of even cat states $\ket{C_{i\beta}}=\mathcal{N}(\ket{i\beta}+\ket{-i\beta})$ (where $\mathcal{N}$ is the normalization constant), we proceed as follows, 
\begin{align}
  &\mathcal{U}(\theta,|\alpha|,1)e^{-i2\alpha\hat p\sigma_\mathrm{x}}\ket{C_{i\beta}}\ket{g}\nonumber\\&\quad=\mathcal{U}(\theta,|\alpha|,1)[e^{-i2\alpha\hat p}\ket{C_{i\beta}}\ket{+}+e^{+i2\alpha\hat p}\ket{C_{i\beta}}\ket{-}]
\end{align}
where $\mathcal{U}(\theta,|\alpha|,1)\equiv \mathrm{GCR}$ or $\mathrm{BB1}$. Here, $\theta=\frac{\pi}{2}$. Fig.~\ref{fig:4-legged-cat} summarizes the effects of both abelian and non-abelian QSP pulses discussed in our work. 
\paragraph{$\mathrm{GCR}$:} We now need to simultaneously unentangle two displaced cat states from the qubit, analogous to the case of displaced coherent states (see App.~\ref{ssec:pnot0}). The effect of the unentanglement gadget on the two cat states in superposition is given by,
\begin{align}
  &\mathcal{U}(\theta,|\alpha|,1)e^{-i2\alpha\hat p\sigma_\mathrm{x}}\ket{C_{i\beta}}\ket{g}\nonumber\\
  &\quad\approx[e^{i\alpha\frac{\pi}{4\beta}\sigma_\mathrm{x}}e^{-i2\alpha\hat p}\ket{C_{i\beta}}+e^{-i\alpha\frac{\pi}{4\beta}\sigma_\mathrm{x}}e^{i2\alpha\hat p}\ket{C_{i\beta}}]\ket{g}.
\end{align}
Thus, the un-entanglement fails unless $\frac{\pi\alpha}{4\beta}=\frac{m\pi}{2}\implies\frac{\alpha}{\beta}=2m\quad\text{where}\quad m\in\mathbb{Z}$. Therefore, using the non-abelian composite pulse sequence GCR we could only realize a four-legged cat state $\ket{4\mathcal{C}}$ which is rectangular,
\begin{equation}
  \ket{4\mathcal{C}}\propto (\ket{2\beta,i\beta}-\ket{2\beta,-i\beta})-(\ket{-2\beta,i\beta}+\ket{-2\beta,-i\beta})
\end{equation}
We could squeeze this state back to fix the gaps but that would squeeze the individual blobs and is not recommended. We show that with increasing $\alpha$, our protocol gives an increase in fidelity to the rectangular four-legged cat state $\ket{4\mathcal{C}}$. This specific example gives us some insight into how GKP states are ideal for $\mathrm{GCR}$. For any logical Pauli state, the spacing between each blob in the grid is such that $\frac{\alpha}{\beta}=\frac{2m\sqrt{\pi}}{\sqrt{\pi}}=2m\quad\text{where }m\in\mathbb{Z}$. 
\paragraph{$\mathrm{BB1}$:} The above problem disappears if we use an abelian QSP sequence such as the $\mathrm{BB1}$ scheme for the un-entanglement $\mathcal{U}(\theta,|\alpha|,1)$. Denoting the un-entangling gadget as the QSP sequence which rotates qubits based on their position eigenvalue (with uncertainty $\Delta=1$), we have,
\begin{align}
  &\mathcal{U}(\theta,|\alpha|,1)e^{-i2\alpha\hat p\sigma_\mathrm{x}}\ket{C_{i\beta}}\ket{g}\nonumber\\&\quad=\mathcal{U}(\theta,|\alpha|,1)[e^{-i\alpha\beta/2}\ket{\alpha+i\beta}\ket{+}+e^{i\alpha\beta/2}\ket{-\alpha+i\beta}\ket{-}\nonumber\\&\quad\quad e^{-i\alpha\beta/2}\ket{\alpha-i\beta}\ket{+}+e^{i\alpha\beta/2}\ket{-\alpha-i\beta}\ket{-}]
  \\&\quad\approx[\ket{\alpha+i\beta}+e^{i\alpha\beta}\ket{-\alpha+i\beta}\nonumber\\&\quad\quad+ e^{-i\alpha\beta}\ket{\alpha-i\beta}+\ket{-\alpha-i\beta}]\ket{g}
\end{align}
All operations in $\mathcal{U}\equiv \mathrm{BB1}$ are controlled momentum boosts, whose action depends on the position of the oscillator post $e^{-i2\alpha\hat p\sigma_\mathrm{x}}$. So, the un-entanglement is the same as the two-legged cat state case except for the additional local phase $e^{i\alpha^2}$ on the blobs along $\frac{\hat x-\hat p}{\sqrt{2}}$\footnote{There is no phase if $\alpha^2=2\pi$.}.

\par\textbf{Discussion:} Similar strategies can be applied for other rotationally symmetric codewords, that is, $N$-legged cat states for $N>4$. However, we will not dive into these strategies and move on to discuss the more general rotationally symmetric states, Fock states. Since the Fock basis is a complete orthogonal basis for the oscillator Hilbert space, this discussion takes a step forward towards universal state preparation using non-abelian QSP.
\subsection*{Fock State Preparation}
For any instruction set to be universal, it should be able to generate the Fock basis. Thus, this section is targeted at the generation of Fock states using the phase-space ISA. Although this preparation scheme may be inefficient given conditional displacements are more suited to states with translation symmetry, we give this construction for the sake of completeness. To this end, we first realize that Fock states can be approximately represented as a sum of coherent states as
\begin{align}
    \ket{\psi_n}&=\frac{1}{\mathcal{N}}\sum_{j=0}^{m-1}e^{i\frac{2\pi n}{m}j}\ket{\alpha e^{-i\frac{2\pi j}{m}}},\label{eq:fock}\\
    F_n&=|\braket{n|\psi_n}|^2=m^2\frac{\alpha^{2n}}{n!}\frac{e^{-\alpha^2}}{\mathcal{N}^2}\label{eq:prob_fock},\quad \alpha\in\mathbb{R},
\end{align}
where $F_n$ gives the fidelity of $\ket{\psi_n}$ with respect to the Fock state $\ket{n}$. Here $m,\alpha$ should be chosen such that the coherent states are on a ring completing an angle of $2\pi$. To represent each Fock state $\ket{n}$ there is a minimum requirement on $m=m_n$ and the fidelity to the Fock states increases with increasing $m>m_n$.

Unlike our previous examples, Fock states are not superpositions of non-overlapping Gaussian wave functions. To prepare Fock states, a straightforward recipe is to use a trotterized circuit for the anti-Jaynes-Cummings Hamiltonian as described in App.~\ref{law-eberly}. Evolution under the Anti-JC Hamiltonian,
\begin{align}
    \mathrm{AJC}=\hat a\sigma_-+\hat a^\dagger\sigma_+=2(\hat x\sigma_x-\hat p\sigma_y),
\end{align}
where $\sigma_{\pm}=\sigma_\mathrm{x}\mp i\sigma_\mathrm{y}$. This Hamiltonian allows the simultaneous addition (or removal) of a single photon to (or from) the qubit and the oscillator. Note that R.H.S.\footnote{right hand side} can be approximated using a conditional displacement and a conditional momentum boost. Further trotterization could yield even better approximations. App.~\ref{law-eberly} shows that Fock states prepared using this method are better than numerically optimized circuits~\cite{eickbusch2022fast} in terms of circuit duration for the preparation of Fock state $\ket{1}$. In addition, we can employ the Law-Eberly protocol~\cite{law1996arbitrary} to prepare arbitrary superpositions of Fock states.

%If the initial oscillator state is a vacuum, Anti-JC prepares $\ket{1}$ if the coefficient of the Hamiltonian $\mathrm{AJC}$ is $1.110$. Thus, the approximation of the required unitary evolution under $\mathrm{AJC}$ using conditional displacements is lower-bounded in circuit duration. 
In terms of circuit depth or gate counts (if for any scheme this is a useful quantity) the numerically optimized circuit is still unmatchable. In this section, we 
 develop an alternative analytical scheme that matches the gate count, for the respective fidelity, of the numerically optimal circuits given in Ref.~\cite{eickbusch2022fast}. So, in this section, we derive an analytical protocol which matches the gate count of the numerically optimized circuits.
 
 \textbf{Protocol:} Let us focus on the simplest case of Fock state $\ket{n=1}$. To begin with, we realize that Eq.~(\ref{eq:fock}) gives us the Fock state $\ket{1}$ in the form of an odd small cat with $m$ blobs in phase space. The smaller $\alpha$ is, the better the fidelity to $\ket{1}$. However, using our small cat preparation circuit laid out in Sec.~\ref{sec:squeezing}, the probability of projecting onto small odd cat states is lower than small even cat states. This problem is explained in the context of two-legged cat states in App.~\ref{app:small_cats}. We label the gate count as the number of conditional displacements and denote this quantity as $N$. The smaller the cat size, the smaller is the probability of projecting the oscillator onto an odd cat state. Thus, there is an optimal $\alpha$ that can achieve the preparation of small odd cat states with low failure probability while maintaining a high fidelity with Fock state $\ket{1}$ for the case of $N=1$. 
 \paragraph{$N=1$:} The optimal $\alpha$ for $\mathrm{CD}(\alpha,\sigma_\textrm{y})$, if we start with $\ket{0,g}$, that yields $\ket{1,e}$ is $\alpha=\frac{\pi}{2}$.
 \begin{align}
     \ket{0,g}\xrightarrow{\mathrm{CD}{(\alpha/2,\sigma_\textrm{x}})}&  \ \ket{\alpha}\ket{+}-\ket{-\alpha}\ket{-}\\&= \ \mathcal{N}_\textrm{odd}(\ket{\alpha}-\ket{-\alpha})\ket{e}+\mathcal{N}_\textrm{even}(\ket{\alpha}+\ket{-\alpha})\ket{g}.
 \end{align}
 Here, $\mathcal{N}_\mathrm{odd}$ ( $\mathcal{N}_\mathrm{even}$) are the normalization constants of the odd and even superpositions of $\ket{\pm\alpha}$ states. See App.~\ref{app:small_cats} for details. The probability of projecting the qubit onto $\ket{e}$ is given by $|\mathcal{N}_\textrm{odd}|^2/|\mathcal{N}_\textrm{even}|^2$ (see Eq.~(\ref{Fidelity_small_odd}) in App.~\ref{app:small_cats}). The fidelity of this state with $\ket{1}$ is given by Eq.~(\ref{eq:prob_fock}). The maximum of the product of these quantities lies at $\alpha=\pi/2$. This parameter is the same as the case of JC for $\ket{0,g}\rightarrow\ket{1,e}$ but with a single conditional displacement
\paragraph{$N=2$:} The optimal circuit for two $\mathrm{CD}$s can be directly given by the first-order trotterized circuit for $\mathrm{AJC}$. As explained in  App.~\ref{law-eberly}, this circuit adds a photon to both the oscillator and the qubit, simultaneously, $\ket{0,g}\rightarrow \ket{1,e}$ at $\alpha=\pi/2$. In fact, for arbitrary Fock state $\ket{n}$, we have $\alpha=\pi/2\sqrt{(n+1)}$ for the evolution $\ket{n,g}\rightarrow \ket{n+1,e}$. This preparation is the same as the protocol for a small odd-cat state preparation discussed in Sec.~\ref{sec:squeezing} (also, see Apps.~\ref{app:squeezing},\ref{app:small_cats}) for $|\alpha|=\pi/2$. The fidelity with $\ket{1,e}$ at this stage is $0.70$, slightly less than the numerically optimized circuits~\cite{eickbusch2022fast,ISA}. So, we try an alternative strategy.

 See the Wigner function in Fig.~\ref{fig:4-legged-cat}(b) after the $N=1$ circuit. It represents two well-separated blobs which is a mixed state representing a large cat state (entangled with a qubit, traced out). Thus, in this case, it might be good to check if the large cat preparation circuit works better in this regime. We find that the second $\mathrm{CD}$ after $\mathrm{CD}(\alpha/2,\sigma_\mathrm{y})$, in fact, yields better fidelity with $e^{i\frac{\pi}{4(\alpha/2)\hat x}\otimes\sigma_\mathrm{y}}$. This is the unentanglement sequence for large cat states, without any QSP correction, discussed in Sec.~\ref{ssec:Cat_States} (also, see Sec.~\ref{sec:comp} and App.~\ref{app:cat_I}). The hybrid fidelity after this $N=2$ circuit is $0.84$, the same as the numerical scheme.

The unentanglement\footnote{Note that, since $|\alpha|<2$ we cannot use a QSP correction here (see Fig.~\ref{fig:Correctness}).} in this case involves a favorable back action in the momentum basis\footnote{This feature is favorable for creating rotationally symmetric states such that due to the enhanced non-commutativity of displacement and momentum boosts at small amplitudes, $|\beta|\neq|\alpha|$ creates a perfect square inscribed inside the Wigner distribution of the Fock state}. See the wave function and Wigner plots for the oscillator state in Fig.~\ref{fig:4-legged-cat}(b) after $N=2$ circuit. The peaks in the momentum basis almost coincide with the peaks in the position basis, yielding a state close to Fock $\ket{1}$. The Wigner distribution of the position peaks overlaps with the momentum peaks, yielding a near-circular quasi-probability distribution. Thus, the back action and overlapping peaks in the two bases both aid in the preparation of a Fock state. 

\paragraph{$N=3$:} Now, since $|\alpha|/2>2$ we can resort to QSP corrections for large cat states. We will, thus, use ideas from small cat state preparation. So far, in Sec.~\ref{sec:squeezing}, we have discussed this idea for creating a squeezed vacuum which is just an even cat state. Now, we switch to the preparation of small \textit{odd} cat states. The problem of using the same QSP correction with extremely small cat states is given in App.~\ref{app:small_cats}. Thus, a medium cat state is the best way to approach Fock states, which resemble a small odd cat state. The case of a cat state of size $|\alpha|/2=\pi/4$ belongs to this class of states. 

Given various preparation schemes discussed in previous sections and the JC Hamiltonian approach using this construction, we identify the following sequence (first used in Ref.~\cite{hastrup2022universal}),
\begin{align}
\mathrm{CD}(\alpha_1,\sigma_\mathrm{y})\mathrm{CD}(i\beta_1,\sigma_\mathrm{x})\mathrm{CD}(\alpha_2,\sigma_\mathrm{y})\mathrm{CD}(i\beta_2,\sigma_\mathrm{x})...\label{eq:CD_circuit}    
\end{align}
 as the most general form of the non-abelian unentanglement circuits. This sequence has implications in quantum random walk, which we will discuss in Chapter~\ref{chapter:conc}. We have $\alpha_1,\beta_1$ from the circuit to prepare a large cat state with
 \begin{align}
    \alpha_1=\frac{\pi}{2},\beta_1=\frac{\pi}{4}\Big(\frac{\alpha_1}{2}\Big).
 \end{align}

 Now, for the next gate, we choose the momentum-controlled rotation or $\mathrm{CD}(\alpha_2/2,\sigma_\mathrm{y})$ as follows. We note that the action of this gate on $\braket{\sigma_\mathrm{y}}$, in the position basis, is
\begin{align}
\braket{\sigma_\mathrm{y}}_\mathrm{new}=\braket{\sigma_\mathrm{y}}_\mathrm{old}\cos{(-2\alpha_2 p)}-\braket{\sigma_\mathrm{z}}_\mathrm{old}\sin{(-2\alpha_2 p)}\label{eq:sx_new}.
\end{align}
See Eq.~(\ref{eq:new_sx}) in App.~\ref{app:squeezing} for details. 
Following the small cat protocol, we note that for unentanglement of the qubit, we need, $\braket{\sigma_\mathrm{y}}_\mathrm{new}=0$, which yields the condition,
\begin{align}
    \tan{(-2\alpha_2 p)}=\frac{\braket{\sigma_\mathrm{y}}_\mathrm{old}}{\braket{\sigma_\mathrm{z}}_\mathrm{old}}
\end{align}  
From Fig.~\ref{fig:4-legged-cat}(b), we notice that, the circuit for $N=2$ has created a large cat of size $\beta_1$ along the momentum quadrature, then we have (ignoring the local phase induced by the first gate) 
\begin{align}
\frac{\braket{\sigma_\mathrm{y}}_\mathrm{old}}{\braket{\sigma_\mathrm{z}}_\mathrm{old}}=\frac{\tanh{2\beta_1p}}{\mathrm{sech}{2\beta_1 p}}=\sinh{2\beta_1 p}
\end{align}
Thus, we need to satisfy, 
\begin{align}
    \tan{2\alpha_2 p}=\sinh{2\beta_1 p}
\end{align}  
Since the state is not centered at the origin we cannot use the linearity condition, but we satisfy the equation at the maximas of the wave functions in $p$. Thus, imposing $\braket{\sigma_\mathrm{y}}=0$ at the maximas of $\psi(p)$, that is at  
$p=\alpha_1/2$ (see Fig.~\ref{fig:4-legged-cat}(b)), we get, 
\begin{align}
    \alpha_2=\frac{1}{\alpha_1}\tan^{-1}({\sinh{2\beta_1 \alpha_1}})/\alpha_1=(\tan^{-1}{\sinh{(\pi/2)}})
\end{align}  
Thus, we have our third gate with this choice of $\alpha_2$,
\begin{align}
    \mathrm{CD}(\alpha_2/2,\sigma_\mathrm{y}).
\end{align}
This $N=3$ sequence further amplifies the hybrid state fidelity to $0.99$. The sequence matches the fidelity and circuit duration of the numerically optimized circuits for depth $N=3$ in Ref.~\cite{eickbusch2022fast}. In addition, when compared with the Law-Eberly Hamiltonian, this sequence has a higher operator fidelity to $\mathrm{AJC}$ Hamiltonian evolution compared to the symmetric second-order trotterized circuit $\mathrm{CD}(\alpha_1/2,\sigma_\mathrm{x})\mathrm{CD}(i\alpha_1,\sigma_\mathrm{y})\mathrm{CD}(\alpha_1/2,\sigma_\mathrm{x})$. 

We describe the above calculations pictorially in Fig.~\ref{fig:4-legged-cat}(b). This method can be generalized to obtain algorithms for arbitrary circuit depth $N$ and Fock state $\ket{n}$. However, given the minimal use case of this strategy in practical cases, we leave this discussion for future work addressing universal state preparation with the phase-space ISA.

Our work gives different strategies to approach the problem of universal state preparation. Concluding this section, we point out that a single algorithm in this direction does not seem like the optimal solution, given the possibility of ancilla decay during a conditional displacement or momentum boost gate.
%%%%%%%%%%%%%%%%DISCUSS FINAL REMARKS FOR THIS SECTION HERE%%%%%%%%%%%%%%%%%%%%%%%%%%%%%%%%%%%%%%%%%%%%%%%%%%%

    \chapter{Probabilistic Error Correction of Photon Loss} \label{chapter:GKP-qec}

\begin{myframe}
\singlespacing
\begin{quote}
    \textit{How can we tolerate errors in an oscillator?} Errors in an oscillator arise due to a beam-splitter unitary (see Chapter~\ref{chapter: paper0}) applied on the joint Hilbert space of the user-accessible quantum system and the environment. When the environment is traced out, the user-accessible Hilbert space can be seen under the action of a photon loss channel (again, described in Chapter~\ref{chapter: paper0}). In this chapter, we will answer questions related to the correction of photon loss in an oscillator used to encode a qubit. More specifically, we will look into the correction of photon loss using the Gottesman-Kitaev-Preskill codes~\cite{gottesman2001encoding}. Recently, these codes have been demonstrated to achieve beyond break-even memory for encoding qudits with $d=2$~\cite{sivak2023real} and $d=\{3,4\}$ dimensions~\cite{brock2024quantum} using an autonomous error correction scheme~\cite{royer2020stabilization}. In this chapter, we will see an analytical explanation of how this scheme probabilistically corrects photon loss with the help of a DV ancillary system. We will give comparisons between two different GKP qubit codes, the square GKP and hexagonal GKP. %, qudits with different dimensions, and single-mode versus two-mode GKP code corrections. 
\end{quote}
\end{myframe}

\doublespacing
A discrete variable encoding composed of qubits requires at least a five-qubit encoding to correct for all Pauli errors (four-qubit encoding for amplitude damping)~\cite{leung1997approximate}. However, a single oscillator suffices to correct for all possible errors in an oscillator (photon loss, dephasing, etc.) encoding a qubit. Thus, oscillators could serve as a powerful resource to improve the space overhead of error correction. Error correction for memory using a single superconducting cavity encoding a GKP qubit has already been demonstrated to be on par with a subsequent DV encoding (surface codes) using $\sim 50$ qubits~\cite{sivak2023real,acharya2024quantum}. GKP bosonic codes have also enabled the first demonstration of beyond-break-even error correction with qutrits and ququarts, enabling higher-dimensional qudits as the building block~\cite{brock2024quantum}.

The problem is, however, that the superconducting cavities used for GKP experiments are massive (few centimeters)  compared to the transmons with access for wiring and ground planes used for the surface-code experiments (hundreds of microns). While these numbers are obsolete and improvements have been made in coherence times of transmons ($\sim$ ms) and size of cavities (few mms), we still need to compare the space-time overhead of using either DV only or hybrid CV-DV error correction architectures. That is, to use the GKP encoding, for any reduction in the overhead due to a larger chip size or slower gate time, the logical error of the qubit must compensate for this increase in size with the increase in the already high coherence time. In this context, we will try to understand the bottlenecks of error correction with GKP encoding and find the minimum logical error possible with practical recovery maps.

Encoding a qubit or a qudit, the GKP codes were designed to correct for errors that cause a shift in the values of position and momentum, that is a displacement channel (see Chapter~\ref{chapter: paper0}). Since any quantum channel can be represented in terms of displacements, any error space can be mapped back to the codespace (possibly with a logical error) by the GKP stabilization used in experiments~\cite{sivak2023real,brock2024quantum}. Ref.~\cite{albert2018performance} shows that among the various bosonic codes, the GKP codes achieve optimal error correction against photon loss, the dominant source of errors in oscillators. This analysis used numerical optimization via semi-definite programming over arbitrary recovery maps, optimizing the fidelity of the code space under the action of photon loss. In Ref.~\cite{zheng2024performance} the authors gave a theoretical derivation of this recovery map for optimal correction of photon loss using the GKP codes. However, to date, we do not understand how to implement this optimal recovery map in practical systems. 

In reality, the implementation of recovery maps also entails errors. In fact, in recent experiments such errors are the dominant cause of the current floor in the logical error probability~\cite{sivak2023real,brock2024quantum}. The recovery map used in these experiments is the so-called \emph{small-big-small} scheme~\cite{royer2020stabilization}. This scheme was derived using a dissipation-based method and is equivalent to the phase estimation technique described in~\cite{terhal2020towards}. The phase estimation technique is designed to correct displacement errors and has no intuitive explanation for the correction of photon loss. On the other hand, the dissipation-based method was engineered to stabilize the code to the GKP eigenspace but does not entail any understanding of why such a stabilization protects against logical errors. 

In this chapter, we will exactly explain how this scheme corrects photon loss and protects the finite-energy GKP states against errors. A finite-energy GKP state is identified by two parameters: the lattice constant ($l$) which is equal to the amount of displacement along the stabilizer vectors under which the GKP lattice is invariant, and the finite-energy parameter $\Delta$ which decides the size (or expanse) of finite-energy GKP states in phase space. The small-big-small (SBS) circuit does not require any measurements and hence is an autonomous error correction scheme. We will show how this scheme applies a probabilistic error correction of the logical GKP codewords. Our analytical understanding helps us understand the beyond-break-even experiments using the square GKP qubit lattice~\cite{sivak2023real,brock2024quantum}. It will further help us extend these ideas to hexagonal GKP codes, to help understand the relationship between the finite-energy parameter, the lattice constant, and autonomous error correction using the SBS scheme.

For this chapter, we intend to use the units where $\hat a=\frac{\hat x+i\hat p}{\sqrt{2}}$. We will specifically use the finite-energy states with a Gaussian envelope introduced by the operator $\hat E=e^{-\Delta^2\hat n}$. All equations derived in this chapter are generalizable to arbitrary qubit and qudit lattices, but we will primarily focus on the square GKP code which has the lattice constant $l_\mathrm{sq}=2\sqrt{\pi}$, as described in Chapter~\ref{chapter: paper0}.

\section{Finite-energy GKP codespace and Error space}
We have already seen a definition of the GKP codewords in Chapter~\ref{chapter:state-prep} when discussing the preparation of these states.  Here, we will use the definition which is equivalent~\cite{matsuura2020equivalence} but proves more straightforward for this chapter. We will ignore normalization constants while writing equations in this section for convenience, but they will be accounted for, in the next section. The Pauli $Z$ basis eigenstates are defined as, 
\begin{align}           
\ket{0}_\mathrm{GKP}&\propto\hat E\sum_{n\in 2\mathbb{Z}} \mathrm{D}(n\sqrt{\pi/2})\ket{0}_x\label{eq:GKP0}\\
    \ket{1}_\mathrm{GKP}&\propto\hat E\sum_{m\in 2\mathbb{Z}+1} \mathrm{D}(m\sqrt{\pi/2})\ket{0}_x\label{eq:GKP1}\\
    \text{where}\quad  \ket{0}_x&=S(\infty)\ket{0}, \ \hat E=e^{-\Delta^2\hat n},\\ \mathrm{D}(\alpha)&=e^{\alpha a^\dagger-\alpha^*a}, \ S(r,\phi)=e^{-\frac{r}{2}(e^{i\phi}{a^\dagger}^2-e^{-i\phi}a^2)}
\end{align}
The envelope operator $\hat E$ takes the finite-energy parameter $\Delta$ as a parameter to truncate the Hilbert space from ideal GKP to the realistic, normalizable, finite-energy GKP codes. Here, $\ket{0}_x$ represents the infinitely-squeezed vacuum state with a mean position of $\braket{x}=0$. For the Wigner plot of the logical $\ket{0}_\mathrm{GKP}$ codeword and its probability distribution, see Fig.~\ref{fig:GKP-prep} in Chapter~\ref{chapter:state-prep}. Note that while the logical Z and logical X eigenstates are very nearly orthogonal (for sufficiently small $\Delta$), the logical magic states of the GKP code (eigenstates of logical Hadamard)
\begin{align}
    \ket{+H}_\mathrm{GKP}&=\frac{\ket{0}_\mathrm{GKP}+\ket{+}_\mathrm{GKP}}{\sqrt{2}}\\
    \ket{-H}_\mathrm{GKP}&=\frac{\ket{0}_\mathrm{GKP}-\ket{+}_\mathrm{GKP}}{\sqrt{2}},    
\end{align}
are exactly orthogonal. Below we plot these logical codewords for the square GKP code and give a flow picture that shows the correction of displacement errors on these states after a single round of the stabilization protocol SBS~\cite{royer2020stabilization} used in experiments~\cite{sivak2023real,brock2024quantum}. Each blue dot represents the mean position and momentum values of an erroneously displaced GKP state which is displaced between $0$ to $l_\mathrm{sq}$. The orange dots denote the mean position and momentum values of the state after a single round of stabilization on the corresponding states. Note that the stabilization not only corrects for displacement errors but also refocuses the envelope to the center. Hence, all vectors from the initial point (blue) to the final point (orange) can be seen as moving towards the Voronoi cell~\cite{shaw2022stabilizer}, after one step of stabilization. We will show later how these various displacement errors (shown in blue) are corrected to an error or no-error state in Fig.~\ref{fig:hex_sq}, after several rounds of stabilization. 
\begin{figure}[htb]
    \centering
    \includegraphics[width=\linewidth]{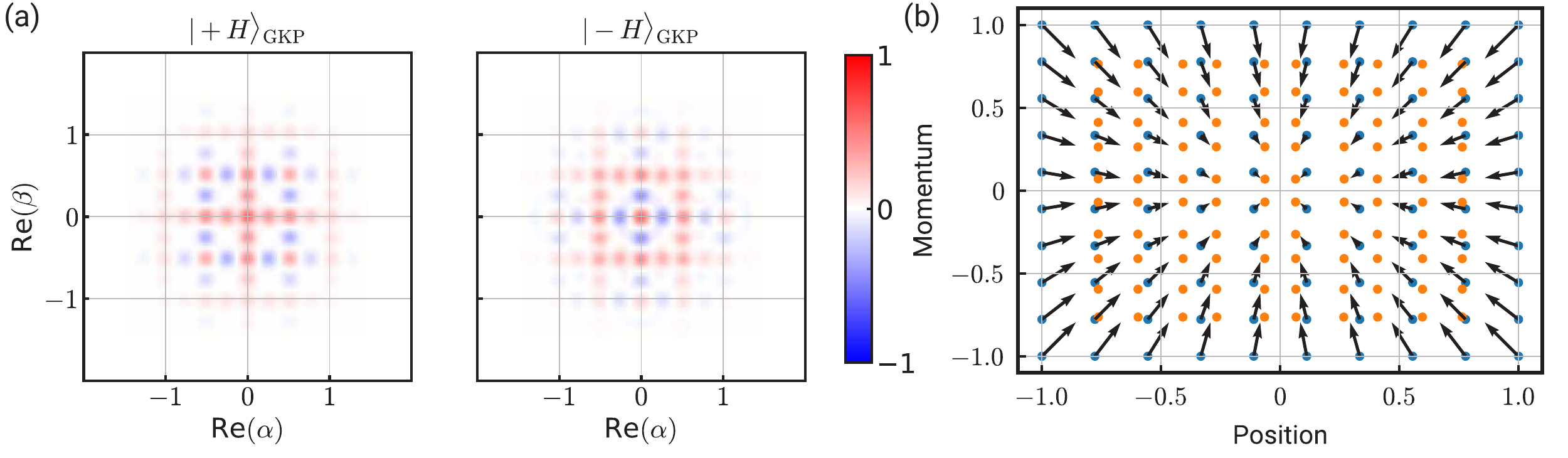}
    \caption[Finite-energy GKP flow for displacement errors with the SBS stabilization scheme.]{\textbf{Finite-energy GKP flow for displacement errors with the SBS stabilization scheme.} All axes are labeled in units of lattice constant $l_\mathrm{sq}$. (a) Logical GKP magic states. (b) The blue dots represent the original mean values of position and momentum of various erroneously displaced GKP states. The orange dots show the mean values of the position and momentum of the corresponding states after a single round of stabilization protocol SBS. The displaced states move towards the central Voronoi cell after a single stabilization round.}
    \label{fig:gkp-flow}
\end{figure}

\begin{figure}
    \centering
    \includegraphics[width=\linewidth]{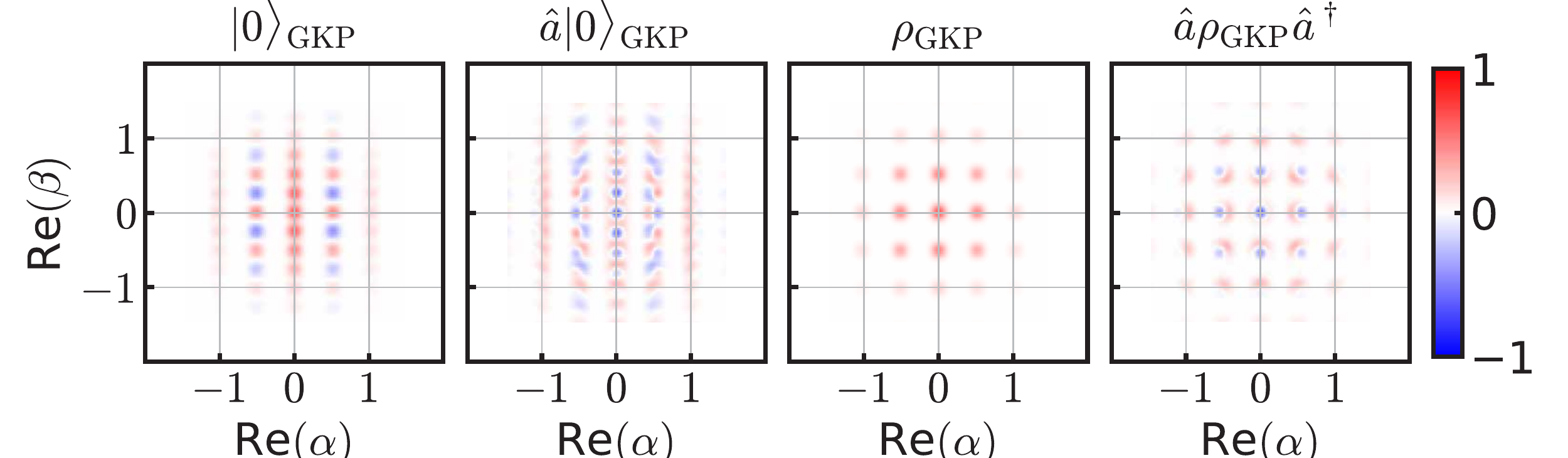}
    \caption[Logical GKP codewords and error words for square lattice encoding]{Logical GKP codeword $\ket{0}_\mathrm{GKP}$ and corresponding error word are shown in the first two plots. The maximally mixed state in the logical codespace $\rho_\mathrm{GKP}=\frac{\ket{0}_\mathrm{GKP}\bra{0}_\mathrm{GKP}+\ket{1}_\mathrm{GKP}\bra{1}_\mathrm{GKP}}{2}$ and the error space $\hat a\rho_\mathrm{GKP}\hat a^\dagger$ are shown in the next two plots. All axes are labeled in units of lattice constant $l_\mathrm{sq}$.}
    \label{fig:GKP-error-words}
\end{figure}

Now, the oscillator error channel, for small time $\tau$, applies a single photon loss with some probability $\kappa\tau$ where $\kappa$ is the photon loss decay rate (see Chapter~\ref{chapter: paper0}). Thus, a single photon loss on these codewords can describe error words~\cite{sivak2023real} for which we give the following mathematical description, using $\hat a=\sum_n \sqrt{n}|n-1\rangle\langle n|$ and ignoring normalization constants,
\begin{align}
    \hat a\ket{0}_\mathrm{GKP}&=e^{-\Delta^2}\hat E\sum_{n\in 2\mathbb{Z}} \mathrm{D}(n\sqrt{\pi/2})(\hat{a}+n\sqrt{\pi})\ket{0}_x\\
    &=e^{-\Delta^2}\hat E\sum_{n\in 2\mathbb{Z}} [n\sqrt{\pi}\mathrm{D}(n\sqrt{\pi/2})\ket{0}_x-\mathrm{D}(n\sqrt{\pi/2})\ket{1}_x ]\label{eq:photon-loss}
\end{align}
where we have used $\hat E^{-1}\hat a\hat E=e^{-\Delta^2}\hat a$ (see App.~\ref{app:aE}) and,
\begin{align}
    \braket{x|\hat a|0}_x=\braket{x|i\hat p|0}_x&=\frac{d}{dx}\lim_{\sigma\rightarrow 0}(2\pi\sigma^2)^{-1/4}e^{-\frac{x^2}{2\sigma^2}}\\
    &=\lim_{\sigma\rightarrow 0}-\frac{(2\pi\sigma^2)^{-\frac{1}{4}}}{\sigma^2}xe^{-\frac{x^2}{2\sigma^2}}\\
    &=-\braket{x|1}_x,
\end{align}
 For the error word, it can be seen that at the phase space origin, the squeezed Fock state $\ket{1}_x$ term will dominate since only the $n=0$ contributes, while at other locations in phase space, each peak is an unequal superposition of displaced squeezed Fock states $\ket{0}_x$ and $\ket{1}_x$. Fig~\ref{fig:GKP-error-words} shows a GKP logical codeword and corresponding error word $\hat a\ket{0}_{GKP}$. To shine a light on the entire GKP codespace we also show the logical maximally mixed state in the GKP codespace and error space (corresponding to single photon loss). A special feature of the GKP codes is that the density matrix of the logically mixed state has no Wigner negativity. See discussion on Wigner functions and non-classicality in Chapter~\ref{chapter: paper0}. On the other hand, bosonic codes like binomial codes or four-legged cat codes have a logical maximally mixed state with some Wigner negativity. That is, they are classical in the sense of being fully mixed within the logical manifold but are not classical in the CV sense.
 
Eq.~(\ref{eq:photon-loss}) shows that photon loss on GKP states changes the coefficients of various squeezed Fock states in the superposition, with increased contribution from higher squeezed Fock states closer to the origin. It is counter-intuitive that photon loss would yield such error states; however, note that there is always a Gaussian envelope of $e^{-\Delta^2\hat n}$ acting on these states, which limits the extent of the state to a circle of radius $\Delta/\sqrt{2}$ in phase space with center at the origin, where the origin corresponds to a negative quasi-probability density. For an arbitrary number of photon loss events, we have,
\begin{align}
    \hat a^m\ket{0}_\mathrm{GKP}&=e^{-\Delta^2}\hat E\sum_{n\in 2\mathbb{Z}}\sum_{r=0}^m{}{m\choose r}[(n\sqrt{\pi})^{m-r}\times \mathrm{D}(n\sqrt{\pi/2})\ket{r}_x].
\end{align}
In the absence of a photon loss event, the so-called `jump' event, the state goes through energy relaxation under the action of $e^{-\frac{\kappa}{2}\tau\hat n}$ (see Chapter~\ref{chapter: paper0}). This no-jump evolution will result in just changing the envelope size of the GKP code. 

\section{Autonomous Stabilization of Finite-energy GKP Codes}
Importantly, the GKP qubit is not defined to have bounded support in the Fock basis like some other bosonic codes, for example, binomial codes~\cite{michael2016new}. Thus, there is no leakage space for a GKP code, which is an error space that cannot be mapped back to the codespace via this type of dissipative stabilization. Ideally, this encoding under some recovery map can bring any oscillator state back to the codespace with or without logical error given a finite amount of time. This is the reason why the experiment in Ref.~\cite{sivak2023real} was run for 800 rounds and at the end of this experiment, the states were still found to be in the logical manifold with high probability. The question is: How can we quantify the logical error in this process? For that, we need to understand exactly how this stabilization scheme corrects errors in the GKP manifold. In particular, we want to answer this question for the GKP stabilization scheme used in experiments~\cite{sivak2023real,brock2024quantum}.

The stabilization scheme (also referred to as the error correction map or the recovery map), small-big-small or SBS was introduced in Ref.~\cite{royer2020stabilization} and has been explicitly laid out in App.~\ref{app:GKp-qec}. This stabilization circuit is given by the combination of the following unitaries,
\begin{align}     
\mathrm{SBS}^\mathrm{X}_\mathrm{GKP}&=e^{i\epsilon_x\hat{x}\hat{\sigma}_y}e^{-i\sqrt{\pi}\hat{p}\hat{\sigma}_x}e^{i\epsilon_x\hat{x}\hat{\sigma}_y}, \textrm{ where } \epsilon_x=\frac{\sqrt{\pi}}{2}\Delta_x^2,\\
 \textrm{ and } \mathrm{SBS}^\mathrm{Z}_\mathrm{GKP}&=e^{-i\epsilon_p\hat{p}\sigma_y}e^{-i\sqrt{\pi}\hat{x}\hat{\sigma}_x}e^{-i\epsilon_p\hat{p}\hat{\sigma}_y}, \textrm{ where } \epsilon_p=\frac{\sqrt{\pi}}{2}\Delta_p^2.\label{eq:SBS}
\end{align}
Here $\mathrm{SBS}^\mathrm{X}_\mathrm{GKP}(\mathrm{SBS}^\mathrm{Z}_\mathrm{GKP}$) gives stabilization for the logical X (Z) basis.
For a rectangular GKP code, we have $\Delta_x\neq \Delta_p$ while for square codes $\Delta_x=\Delta_p=\Delta$. We can also generalize the SBS circuit to arbitrary lattices by using different stabilizer vectors (instead of $\hat x,\hat p$), for example, the Hexagonal lattice discussed in Sec.~\ref{Hex_gkp}. The SBS circuit implements dissipation, but also deterministically applies a logical Pauli operator, a fact we often obscure to simplify the explanation. It should be noted that, before Ref.~\cite{royer2020stabilization}, a similar circuit was obtained for GKP error correction via adaptive phase estimation using single-qubit ancillae in Ref.~\cite{terhal2016encoding} 

\paragraph{No error:} While the SBS circuit was interpreted as another explanation for the adaptive phase estimation protocol, it is different in the sense that in the ideal GKP limit $\Delta\rightarrow 0$, the SBS protocol comes down to only applying the `big' conditional displacement $e^{i\sqrt{\pi}\hat{x}\sigma_x}$ without any correction. That is, for an ideal GKP state with $\Delta=0$,
\begin{align}
\mathrm{SBS}^\mathrm{Z}\equiv\cos{\sqrt{\pi}\hat x}\ket{g}\bra{g}+i\sin{\sqrt{\pi}\hat x}\ket{e}\bra{g}
\end{align}

Thus, for an infinite-energy logical GKP state $\ket{\psi}_\mathrm{GKP}$ we have,
\begin{align}
 \textrm{SBS}^\mathrm{Z} \ \ket{\psi}_\mathrm{GKP}\ket{g}&=  \cos{\sqrt{\pi}\hat x}\ket{\psi}_\mathrm{GKP}\ket{g}+i\sin{\sqrt{\pi}\hat x}\ket{\psi}_\mathrm{GKP}\ket{e}\\&=\cos{\sqrt{\pi}\hat x}\ket{\psi}_\mathrm{GKP}\ket{g}=\ket{\bar \psi}_\mathrm{GKP}\ket{g},
\end{align}
where $\ket{\bar \psi}_\mathrm{GKP}=\cos{\sqrt{\pi}\hat x}\ket{\psi}_\mathrm{GKP}=e^{i\sqrt{\pi}\hat x}\cdot \Big(\frac{\mathrm{I}+\mathrm{S}^\mathrm{Z}}{2}\Big)\ket{\psi}_\mathrm{GKP}=\mathrm{Z}_\mathrm{GKP}\ket{\psi}_\mathrm{GKP}$. Recall from Chapter~\ref{chapter: paper0} that $\mathrm{S}^Z=e^{\pm i2\sqrt{\pi}\hat x}$ stabilizes the logical $Z$ basis and $e^{i\sqrt{\pi}\hat x}$ applies a logical $Z$ operator for the infinite energy ($\Delta=0$) case. Following these equations, we call $\cos{\sqrt{\pi}\hat x}$ the symmetrized version of the logical operator, $\mathrm{Z}_\mathrm{GKP}$. This equation is true because $\sin{\sqrt{\pi}\hat x}\ket{\psi}=0$. We call this operator the \emph{logical annihilator} of the codespace.

Now let us look at the realistic codespace with $\Delta\neq 0$. In this case, the envelope operator $\hat E$ applies a deformation of logical space where the infinite-energy logical operators $\hat A$ and states $\ket{\psi}$ are transformed as,
\begin{align}
\ket{\psi}\rightarrow \ket{\psi_\Delta}=\hat E\ket{\psi}\implies A_\Delta\rightarrow \hat EA\hat E^{-1}.
\end{align}
Thus, we can write $\hat E\hat A\hat E^{-1}\hat E\ket{\psi}_\mathrm{GKP}= \hat E(\hat A\ket{\psi}_\mathrm{GKP})$ which shows that the logical quantum computation will remain the same as the infinite-energy case, but with an envelope operator applied at the end of the circuit, leading to bounded logical operators and normalizable, but (since $\hat E$ is not unitary) not quite orthogonal states. Following this deformation, the exact version of the finite-energy stabilization scheme can be written as,
\begin{align}
    \mathrm{SBS}_\Delta^\mathrm{Z}\equiv \hat E\cos{\sqrt{\pi}\hat x}\hat E^{-1}\ket{g}\bra{g}+i\hat E\sin{\sqrt{\pi}\hat x}\hat E^{-1}\ket{e}\bra{g}.
\end{align}
 This is equivalent to the small-big-small scheme's $\mathrm{SBS}^\mathrm{Z}_\mathrm{GKP}$ up to first order in $\Delta$. Using this equation, we can compute the action of exact finite-energy stabilization on the finite-energy logical GKP state $\ket{\psi_\Delta}_\mathrm{GKP}$,
\begin{align}
\textrm{SBS}^\mathrm{Z}_\Delta\ket{\psi_\Delta}_\mathrm{GKP}\ket{g}&:  E\cos{\sqrt{\pi}\hat x}E^{-1}E\ket{\psi_\Delta}_\mathrm{GKP}\ket{g}+E\sin{\sqrt{\pi}\hat x}E^{-1}E\ket{\psi_\Delta}_\mathrm{GKP}\ket{e}\\&~~=\hat E\ket{\bar \psi}\ket{g}=\ket{\bar \psi_\Delta}\ket{g}.   
\end{align}
Thus, exact `SBS' stabilization for finite-energy codes has the same effect as the `SBS' stabilization for infinite-energy codes in the no error case.

\paragraph{Photon loss:} For the infinite-energy case, we note that single photon loss operators commute with the symmetrized logical operators on the logical states\footnote{Note that $[f(\hat x),\hat a]\ket{\psi}=(1/\sqrt{2})[f(\hat x),i\hat p]\ket{\psi}=(1/\sqrt{2})[f(\hat x),d/d \hat x]\psi(x)=-f'(\hat x)\psi(x)$. Here $f'(\hat x)$ denotes the derivative of $f(\hat x)$ and $\psi(x)$ is the state $\ket{\psi}$ written in the position basis. We have used $\hat p=-id/d \hat x$.}.
\begin{align}
    [\cos{\sqrt{\pi}\hat x},\hat a]|\psi\rangle_{\textrm{GKP}}&=\sqrt{\pi/2}\sin{\sqrt{\pi}\hat x}|\psi\rangle_{\textrm{GKP}}=0.
\end{align}
This is because the peaks of the GKP codes lie at $x=m \ l_\mathrm{sq}/2=m\sqrt{\pi}$ along the position axis where $m\in\mathbb{Z}$. For example, see Eqs.~\ref{eq:GKP0}-\ref{eq:GKP1} and the Wigner function of the state $\ket{0}_\mathrm{GKP}$ in Fig.~\ref{fig:GKP-error-words}. And for the logical annihilator, we simply have,
\begin{align}
    [\sin{\sqrt{\pi}\hat x},\hat a]|\psi\rangle_{\textrm{GKP}}&=-\sqrt{\pi/2}\cos{\sqrt{\pi}\hat x}|\psi\rangle_{\textrm{GKP}}=-\sqrt{\pi/2}|\bar\psi\rangle_{\textrm{GKP}}.
\end{align}

This simple equation tells us that the logical operators of an ideal GKP code are transparent to a single photon loss. 
Now, the action of the SBS circuit in the presence of an error can be described as follows, ignoring normalization constants,
\begin{align}
 \textrm{SBS}^\mathrm{Z} \ \hat a\ket{\psi}_\mathrm{GKP}\ket{g}&:  \cos{\sqrt{\pi}\hat x}\hat a\ket{\psi}_\mathrm{GKP}\ket{g}+i\sin{\sqrt{\pi}\hat x}\hat a\ket{\psi}_\mathrm{GKP}\ket{e}\\&=(\hat a\cos{\sqrt{\pi}\hat x}+\sqrt{\pi/2}\sin{\sqrt{\pi}\hat x})\ket{\psi}_\mathrm{GKP})\ket{g}\nonumber\\&~~~~~+i(\hat a\sin{\sqrt{\pi}\hat x}-\sqrt{\pi/2}\cos{\sqrt{\pi}\hat x})\ket{\psi}_\mathrm{GKP})\ket{e}\\&=\hat a\ket{\bar{\psi}}\ket{g}-i\sqrt{\pi/2}\ket{\bar \psi}\ket{e}.
\end{align}
 This is the basis of our probabilistic error correction perspective on the SBS scheme. The scheme can be seen as a logical operation when the ancilla at the end of the $\mathrm{SBS}^\mathrm{Z}$ is in the ground state ($|g\rangle$); if it is excited ($|e\rangle$), the circuit acts as a corrector of single photon loss error. The measurement outcome does not give any information that can be used for further correction. In the event of a $+1$ outcome ($|g\rangle$), the resulting state could have been the uncorrected state or in a no-error state while a $-1$ outcome indicates that the error has already been corrected or is in the process of being corrected (in the event of higher-order errors like $\hat a^2$). In either case, we cannot straightforwardly use this measurement outcome to further advantage. 
 
The commutation of a finite-energy stabilizer with the finite-energy symmetrized version of $\mathrm{Z}_\mathrm{GKP}$ is given by, 
\begin{align}
    [\hat E\cos{\sqrt{\pi}\hat x}\hat E^{-1},\hat a]\hat E|\psi\rangle_{\textrm{GKP}}&=( \hat E\cos{\sqrt{\pi}\hat x}\hat E^{-1}\hat a-\hat a \hat E\cos{\sqrt{\pi}\hat x}\hat E^{-1}) \hat E\ket{\psi}_\mathrm{GKP}\\
    &=e^{-\Delta^2}E [\cos{\sqrt{\pi}\hat x},\hat a]\hat E^{-1}\hat E\ket{\psi}_\mathrm{GKP}\\&=e^{-\Delta^2}\sqrt{\pi/2}\hat E\sin{\sqrt{\pi}\hat x}\ket{\psi}_\mathrm{GKP}=0,
\end{align}
using $\hat E^{-1}\hat a\hat E=e^{-\Delta^2}\hat a$ (see App.~\ref{app:aE}). Thus, for the error word $\hat a\ket{\psi}_\mathrm{GKP}$, ignoring normalization constants we have,
\begin{align}
  \textrm{SBS}^\mathrm{X}_\mathrm{GKP} \ \hat a\ket{\psi_\Delta}_\mathrm{GKP}\ket{g}&:  \hat E\cos{\sqrt{\pi}\hat x}\hat E^{-1}\hat a\hat E\ket{\psi_\Delta}_\mathrm{GKP}\ket{g}+i\hat E\sin{\sqrt{\pi}\hat x}\hat E^{-1}\hat a \hat E\ket{\psi_\Delta}_\mathrm{GKP}\ket{e}\\&~=\hat a \hat E\ket{\bar \psi}_\mathrm{GKP}\ket{g}-ie^{-\Delta^2}\sqrt{\pi/2}\hat E\ket{\bar \psi}_\mathrm{GKP}\ket{e}\\&~=\hat a \ket{\bar \psi_\Delta}_\mathrm{GKP}\ket{g}-ie^{-\Delta^2}\sqrt{\pi/2}\ket{\bar \psi_\Delta}_\mathrm{GKP}\ket{e}.
\end{align}
Here, $[\hat E\sin{\sqrt{\pi}\hat x}\hat E^{-1},\hat a]\hat E|\psi\rangle_{\textrm{GKP}}=-e^{-\Delta^2}\sqrt{\pi/2}\hat E\cos{\sqrt{\pi}\hat x}\ket{\psi}_\mathrm{GKP}$, and we use the notation $\ket{\bar \psi_\Delta}_\mathrm{GKP}=E\ket{\bar \psi}_\mathrm{GKP}$. Thus, probabilistic correction equations hold in the case of finite-energy as well, with a coefficient of $e^{-\Delta^2}$ ($<1$) that decreases the probability of correction\footnote{Note that this factor was previously seen in the expression for the error-word $\hat a\hat E\ket{\psi}_\mathrm{GKP}$ also in Eq.~(\ref{eq:photon-loss}}). The probability of correction is given by,
\begin{align}
    p_e=\frac{e^{-2\Delta^2}\frac{\pi}{2}||\ket{\bar \psi_\Delta}||}{||\hat a\ket{\bar \psi_\Delta}||+e^{-2\Delta^2}\frac{\pi}{2}||\ket{\bar \psi_\Delta}||}=\frac{\frac{\pi}{2}e^{-2\Delta^2}}{\frac{||\hat a\ket{\bar \psi_\Delta}||}{||\,\ket{\bar \psi_\Delta}||}+\frac{\pi}{2}e^{-2\Delta^2}}.
\end{align}
Here $||\ket{\psi}||=|\braket{\psi|\psi}|$. This equation shows us that the probability of correction depends on the ratio of the normalization constants of the codeword and the error word. In the next section, we will derive analytical expressions for these normalization constants to see how this ratio varies for different lattice sizes, changing the probability of error correction with envelope size. 

\paragraph{Complete stabilization:} The complete stabilization round involves the stabilization of both the logical $X$ and $Z$ Pauli bases. Thus, after one round of stabilization, defining $\ket{\bar{\bar \psi}_\Delta      }_\mathrm{GKP}=\mathrm{X}_\mathrm{GKP}\mathrm{Z}_\mathrm{GKP}\ket{\psi_\Delta}_\mathrm{GKP}$ we have, 
\begin{align}
&\textrm{SBS}^\mathrm{X}_\mathrm{GKP}\textrm{SBS}^\mathrm{Z}_\mathrm{GKP} \ \hat a\ket{\psi_\Delta}_\mathrm{GKP}\ket{gg}\nonumber\\
&=\hat a\hat E\cos{\sqrt{\pi}\hat p}\ket{\bar{\psi}}_\mathrm{GKP}\ket{gg}+ie^{-\Delta^2}\sqrt{\pi/2}\hat E\sin{\sqrt{\pi}\hat p}\ket{\bar{\psi}}_\mathrm{GKP}\ket{gg}\nonumber\\&\quad-ie^{-\Delta^2}\sqrt{\pi/2}\hat E\cos{\sqrt{\pi}\hat p}\ket{\bar{\psi}}_\mathrm{GKP}\ket{ge}\nonumber\\&\quad +i\hat a\hat E\sin{\sqrt{\pi}\hat p}\ket{\bar{\psi}}_\mathrm{GKP}\ket{eg}+e^{-\Delta^2}\sqrt{\pi/2}\hat E\cos{\sqrt{\pi}\hat p}\ket{\bar{\psi}}_\mathrm{GKP}\ket{eg}\nonumber\\&\quad+e^{-\Delta^2}\sqrt{\pi/2}\hat E\sin{\sqrt{\pi}\hat p}\ket{\bar{\psi}}_\mathrm{GKP}\ket{ee}\\
&=\hat a\ket{\bar{\bar\psi}_\Delta}_\mathrm{GKP}\ket{gg}-ie^{-\Delta^2}\sqrt{\pi/2}\ket{\bar{\bar\psi}_\Delta}_\mathrm{GKP}\ket{ge}+e^{-\Delta^2}\sqrt{\pi/2}\ket{\bar{\bar\psi}_\Delta}_\mathrm{GKP}\ket{eg},
\end{align}
where 
\begin{align}
  [\hat E\cos{\sqrt{\pi}\hat p\hat E^{-1},\hat a}]=ie^{-\Delta^2}\sqrt{\pi/2}\sin{\sqrt{\pi}\hat p}, \ [\hat E\sin{\sqrt{\pi}\hat p\hat E^{-1},\hat a}]=-ie^{-\Delta^2}\sqrt{\pi/2}\cos{\sqrt{\pi}\hat p}  
\end{align}
and $\sin{\sqrt{\pi}\hat p}\ket{\bar{\psi}}_\mathrm{GKP}=0$. We also use the notation 
\begin{align}
\ket{\bar{\bar \psi}_\Delta}_\mathrm{GKP}=\hat E \cos{\sqrt{\pi}\hat p}\ket{\bar\psi}_\mathrm{GKP}=\hat E \mathrm{D}(\sqrt{\pi/2})\ket{\bar\psi}_\mathrm{GKP}=\hat E\mathrm{X}_\mathrm{GKP}\ket{\bar\psi}_\mathrm{GKP}.    
\end{align}
Note that the first (second) qubit index from the right belongs to the first or X (second or Z) stabilization. Thus, the total error correction probability $p_\mathrm{ge}+p_{eg}=2p_e$ is doubled after one round of stabilization using the exact stabilization protocol for finite-energy GKP states defined using a Gaussian envelope. 

Now, we can describe the effect of the channel $\mathcal{SBS}$ described by the exact stabilization scheme, where qubits are reset after each round, on a GKP code word $\rho=\ket{\psi_\Delta}\bra{\psi_\Delta}$ as,
\begin{align}
    \mathcal{SBS}(\rho_\mathrm{GKP})=\ket{\bar{\bar\psi}_\Delta}\bra{\bar{\bar\psi}_\Delta}=\bar{\bar\rho}.
\end{align}
On the error word $\hat a\rho\hat a^\dagger=\hat a\ket{\psi_\Delta}\bra{\psi_\Delta}\hat a^\dagger$ we have,
\begin{align}
   \mathcal{SBS}(\hat a\rho\hat a^\dagger)=p_{gg}\hat a\bar{\bar\rho} a^\dagger+(p_{eg}+p_{ge}) \bar{\bar\rho}. 
\end{align}
Here $p_{eg}+p_{ge}=\frac{\pi e^{-2\Delta^2}}{\pi e^{-2\Delta^2}+\frac{||\hat a\rho\hat a^\dagger||}{||\rho||}}$, where $||\rho||=\mathrm{Tr}(\rho)$.
\paragraph{Energy relaxation:} Before we dive into the calculations of the probability of error correction, we should analyze the error word related to the no-jump evolution (at short time intervals, see Chapter~\ref{chapter: paper0}). That is,
\allowdisplaybreaks{
\begin{align}
    \mathrm{SBS}^\mathrm{Z}_\Delta \ \hat n \ \ket{\psi_\Delta}_\mathrm{GKP}\ket{g}&=  \hat E  \cos{\sqrt{\pi}\hat x} \hat E^{-1} \ \hat n \ \hat E\ket{\psi}_\mathrm{GKP}\ket{g}\nonumber\\
    &\quad+i    \hat E \sin{\sqrt{\pi}\hat x} \hat E^{-1} \ \hat n \ \hat E\ket{\psi}_\mathrm{GKP}\ket{e}\\
    &=\hat E  \cos{\sqrt{\pi}\hat x} \ \hat n \ \ket{\psi}_\mathrm{GKP}\ket{g}+i    \hat E \sin{\sqrt{\pi}\hat x} \ \hat n \ \ket{\psi}_\mathrm{GKP}\ket{e}.\label{eq:hatn}
\end{align}
Thus, this error is corrected the same as the case of infinite-energy GKP (in contrast to the factors of $e^{-\Delta^2}$ in the case of photon loss). Now the required commutators are given by,
\begin{align}
    [\cos{\sqrt{\pi}\hat x,\hat n}]\ket{\psi}_\mathrm{GKP}&=[\cos{\sqrt{\pi}\hat x,\hat a^\dagger}]\hat a+\hat a^\dagger [\cos{\sqrt{\pi}\hat x,\hat a}]\\
    &=-\sqrt{\pi/2}\sin{\sqrt{\pi}\hat x}\hat a\ket{\psi}_\mathrm{GKP}+ \sqrt{\pi/2}\hat a^\dagger\sin{\sqrt{\pi}\hat x}\ket{\psi}_\mathrm{GKP}\\
    &=-\sqrt{\pi/2}\sin{\sqrt{\pi}\hat x}\hat a\ket{\psi}_\mathrm{GKP}\\
    &=-\sqrt{\pi/2}\hat a\sin{\sqrt{\pi}\hat x}\ket{\psi}_\mathrm{GKP}+\pi/2 \cos{\sqrt{\pi}\hat x}\ket{\psi}_\mathrm{GKP}\\
    &=-\pi/2 \cos{\sqrt{\pi}\hat x}\ket{\psi}_\mathrm{GKP}=-\pi/2\ket{\bar{\psi}}_\mathrm{GKP}
\end{align}
and,
\begin{align}
    [\sin{\sqrt{\pi}\hat x,\hat n}]\ket{\psi}_\mathrm{GKP}&=[\sin{\sqrt{\pi}\hat x,\hat a^\dagger}]\hat a+\hat a^\dagger [\sin{\sqrt{\pi}\hat x,\hat a}]\\
    &=\sqrt{\pi/2}\cos{\sqrt{\pi}\hat x}\hat a\ket{\psi}_\mathrm{GKP}- \sqrt{\pi/2}\hat a^\dagger\cos{\sqrt{\pi}\hat x}\ket{\psi}_\mathrm{GKP}\\
    &=\sqrt{\pi/2}\cos{\sqrt{\pi}\hat x}\hat a\ket{\psi}_\mathrm{GKP}- \sqrt{\pi/2}\hat a^\dagger\ket{\bar\psi}_\mathrm{GKP}\\
    &=\sqrt{\pi/2}\hat a\cos{\sqrt{\pi}\hat x}\ket{\psi}_\mathrm{GKP}+\pi/2 \sin{\sqrt{\pi}\hat x}\ket{\psi}_\mathrm{GKP}\nonumber\\&\quad- \sqrt{\pi/2}\hat a^\dagger\ket{\bar\psi}_\mathrm{GKP}\\
    &=\sqrt{\pi/2}\hat a\ket{\bar\psi}_\mathrm{GKP}- \sqrt{\pi/2}\hat a^\dagger\ket{\bar\psi}_\mathrm{GKP}=i\sqrt{\pi}\hat p\ket{\bar\psi}_\mathrm{GKP}
\end{align}
Therefore, continuing Eq.~(\ref{eq:hatn}) we have,
\begin{align}
    \mathrm{SBS}^\mathrm{Z}_\Delta \ \hat n \ \ket{\psi_\Delta}_\mathrm{GKP}\ket{g}&=(\hat n-\pi/2)\ \ket{\bar\psi_\Delta}_\mathrm{GKP}\ket{g}-\hat E\sqrt{\pi}\hat p  \ket{\bar\psi}_\mathrm{GKP}\ket{e}.
\end{align}
Now, for the complete round of stabilization, we can write,
\begin{align}
&\mathrm{SBS}^\mathrm{X}_\Delta\mathrm{SBS}^\mathrm{Z}_\Delta \ \hat n \ \ket{\psi_\Delta}_\mathrm{GKP}\ket{gg}\\&=\hat E\cos{\sqrt{\pi}\hat p}(\hat n-\pi/2)\ \ket{\bar\psi}_\mathrm{GKP}\ket{gg}-\hat E\sqrt{\pi}\hat p \cos{\sqrt{\pi}\hat p} \ket{\bar\psi}_\mathrm{GKP}\ket{ge}\\&\quad+i\hat E\sin{\sqrt{\pi}\hat p}(\hat n-\pi/2)\ \ket{\bar\psi}_\mathrm{GKP}\ket{eg}-i\hat E\sqrt{\pi}\hat p\sin{\sqrt{\pi}\hat p}  \ket{\bar\psi}_\mathrm{GKP}\ket{ee}\\
&=(\hat n-\pi/2) \ket{\bar{\bar\psi}_\Delta}_\mathrm{GKP}\ket{gg}+\hat E[\cos{\sqrt{\pi}\hat p},\hat n] \ket{\bar\psi}_\mathrm{GKP}\ket{gg}\nonumber\\&\quad-\hat E\sqrt{\pi}\hat p \ket{\bar{\bar\psi}}_\mathrm{GKP}\ket{ge}+i\hat E[\sin{\sqrt{\pi}\hat p},\hat n]\ \ket{\bar\psi}_\mathrm{GKP}\ket{eg}\\
&=\hat n \ket{\bar{\bar\psi}_\Delta}_\mathrm{GKP}\ket{gg}-\hat E \sqrt{\pi}\hat p \ket{\bar{\bar\psi}_\Delta}_\mathrm{GKP}\ket{ge}+\hat E\sqrt{\pi}\hat x\ \ket{\bar{\bar\psi}}_\mathrm{GKP}\ket{eg}\\
&=\hat n \ket{\bar{\bar\psi}_\Delta}_\mathrm{GKP}\ket{gg}-i \sqrt{\pi/2}(e^{-\Delta^2}\hat a^\dagger-e^{-\Delta^2}\hat a) \ket{\bar{\bar\psi}_\Delta}_\mathrm{GKP}\ket{ge}\nonumber\\&\quad+\sqrt{\pi/2}(e^{-\Delta^2}\hat a^\dagger+e^{-\Delta^2}\hat a)\ \ket{\bar{\bar\psi}_\Delta}_\mathrm{GKP}\ket{eg},
\end{align}
using $[\cos{\sqrt{\pi}\hat p},\hat n]=(\pi/2)\cos{\sqrt{\pi}\hat p}, \ [\sin{\sqrt{\pi}\hat p},\hat n]=-i\sqrt{\pi}\hat x\cos{\sqrt{\pi}\hat p}.$ We have also used $\hat E\hat a\hat E^{-1}=e^{-\Delta^2}\hat a$
Remember that the first (second) qubit index from the right belongs to the first or $Z$ (second or $X$) stabilization. Thus, if the state ends up in one of the error words $\{\hat a\ket{\bar{\bar\psi}_\Delta}_\mathrm{GKP},\hat a^\dagger \ket{\bar{\bar\psi}_\Delta}_\mathrm{GKP}\}$, it will be corrected in the next round of stabilization.
}

How do our predictions compare with the approximate finite-energy stabilization scheme, \emph{small-big-small}? We can analyze this using the Kraus map formalism, detailed in our work~\cite{sivak2023real}. We find that all the predictions from our theory match with the Kraus map formalisms shown in Fig. S13 of Ref.~\cite{sivak2023real}. Importantly, this figure shows that single photon loss can be corrected in one complete stabilization round if the qubit outcomes were $ge$ or $eg$. In addition, it also shows that errors like $\hat n$ cannot be corrected in a single round of stabilization. We give details of the Kraus map formalism in App.~\ref{app:Kraus_GKP}. 

Now that we have seen the analytical evidence of probabilistic correction of single photon loss, we ask what are the probabilities of these corrections and how they vary with $\Delta$?

\section{Probabilistic Distance}
The distance of GKP codes against displacement errors is half the lattice constant, $\sqrt{\pi}$ for square GKP codes. That is, the code cannot correct for any displacement error larger than $\sqrt{\pi}/2$. In the case of autonomous correction of photon loss, the probability of correction depends on the lattice constant as well as $\Delta$, determining the error correction capacity of the SBS scheme in correcting photon loss on a given finite-energy GKP codespace. We will denote this quantity as the `probabilistic distance' of the GKP code against photon loss, and it is equal to $p_d=p_\mathrm{ge}+p_{eg}$ for the case of GKP qubit codes. The probabilistic distance only depends on the finite-energy parameter $\Delta$ and lattice constant $l$. In this section, we will compute the exact dependence of the probabilistic distance on these two quantities for exact stabilization $SBS_\mathrm{\Delta}^\mathrm{X}SBS_\mathrm{\Delta}^\mathrm{Z}$ and compare it against numerical results for the approximate scheme given by $SBS_\mathrm{GKP}^\mathrm{X}SBS_\mathrm{GKP}^\mathrm{Z}$.

The probabilistic distance $p_d$ depends on the ratio $|(\mathcal{N}_0/\mathcal{N}_1)le^{-\Delta^2}|^2$ where $\mathcal{N}_i$ is the normalization constant of $\hat a^i\ket{\psi_\Delta}_\mathrm{GKP}$. We will calculate these quantities analytically and compare it with the numerics in this section. Here, we assume that the normalization constants for orthogonal GKP codewords are the same up to a factor so small that it is insignificant compared to this probability. For completeness, we will discuss the probabilistic distance for $\hat a^\dagger$ errors as well. Note that our derivation can be used with any type of envelope, and is not restricted to a Gaussian envelope. However, this envelope operator has a closed-form expression in the displacement basis which makes the computation of $\mathcal{N}_0,\mathcal{N}_1$ straightforward as given below. 

For the logical $\ket{\psi_\Delta}=\ket{0_\Delta}_\mathrm{GKP}$ (we will drop the subscript GKP from now on) codeword of the square lattice encoding a qubit, the normalization is given by,
\begin{align}
1/\mathcal{N}_0^2&=\braket{\psi_\Delta|\psi_\Delta}=\sum_{\alpha,\alpha^\prime\in 2m\sqrt{\pi}}\braket{\alpha|\hat E^2|\alpha^\prime}=\sum_{\alpha,\alpha^\prime\in 2m\sqrt{\pi}}\braket{\alpha|e^{-2\Delta^2\hat n}|\alpha^\prime}\\
&~~~~~~~~~~~~~=\frac{1}{\pi(1-e^{-2\Delta^2})}\sum_{\alpha,\alpha^\prime\in 2m\sqrt{\pi}}\int d^2\alpha^{\prime\prime} e^{-\frac{|\alpha^{\prime\prime}|^2}{2\tanh{\Delta^2/2}}}\braket{\alpha|\mathrm{D}(\alpha^{\prime\prime})|\alpha^\prime}\\
&~~~~~~~~~~~~~=\frac{1}{\pi(1-e^{-2\Delta^2})}\sum_{\alpha,\alpha^\prime\in 2m\sqrt{\pi}} e^{-\frac{|\alpha-\alpha^\prime|^2}{2\tanh{\Delta^2/2}}}\\
&~~~~~~~~~~~~~=\frac{1}{\pi(1-e^{-2\Delta^2})}\sum_{m,m^\prime\in \mathbb{Z}} e^{-\frac{4\pi}{\Delta^2}|m-m^\prime|^2}\\
&~~~~~~~~~~~~~=\frac{1}{\pi(1-e^{-2\Delta^2})}\sum_{z\in \mathbb{Z}} e^{-\frac{4\pi z^2}{\Delta^2}}\\
&~~~~~~~~~~~~~=\frac{1}{\pi(1-e^{-2\Delta^2})}\vartheta_3{(0,e^{-\frac{4\pi}{\Delta^2}})}.\label{eq:norm0}
\end{align}
Here $\vartheta_3{(0,a)}$ is the third Jacobi-theta function. Now, for the corresponding error word,
\begin{align}
    1/\mathcal{N}_1^2&=\braket{\psi_\Delta|a^\dagger a|\psi_\Delta}=\sum_{\alpha,\alpha^\prime\in 2m\sqrt{\pi}}\braket{\alpha|\hat Ea^\dagger aE|\alpha^\prime}\\
    &=\sum_{\alpha,\alpha^\prime\in 2m\sqrt{\pi}}\braket{\alpha|e^{-2\Delta^2\hat n}a^\dagger a|\alpha^\prime}\\
     &=-\frac{1}{2}\frac{d}{d\Delta^2}\Bigg[\frac{1}{\pi(1-e^{-2\Delta^2})}\vartheta_3{(0,e^{-\frac{4\pi}{\Delta^2}})}\Bigg].\label{eq:norm1}
\end{align}
The derivative of the Jacobi theta function is the Jacobi elliptic theta function. For the case of error $\hat a^\dagger$, this turns out to be,
\begin{align}
    1/{\mathcal{N}^\prime_1}^2&=\braket{\psi_\Delta|a a^\dagger|\psi_\Delta}=\sum_{\alpha,\alpha^\prime\in 2m\sqrt{\pi}}\braket{\alpha|\hat Ea a^\dagger \hat E|\alpha^\prime}\\
    &~~~~~~~~~~~~~~~~~~~~~=\sum_{\alpha,\alpha^\prime\in 2m\sqrt{\pi}}\braket{\alpha|\hat e^{-2\Delta^2\hat n}(a^\dagger a+1)|\alpha^\prime}\\
     &~~~~~~~~~~~~~~~~~~~~=1/\mathcal{N}_1^2+1/\mathcal{N}_0^2.
\end{align}

\begin{figure}[htb]
    %\centering
    \includegraphics[width=1.1\textwidth]{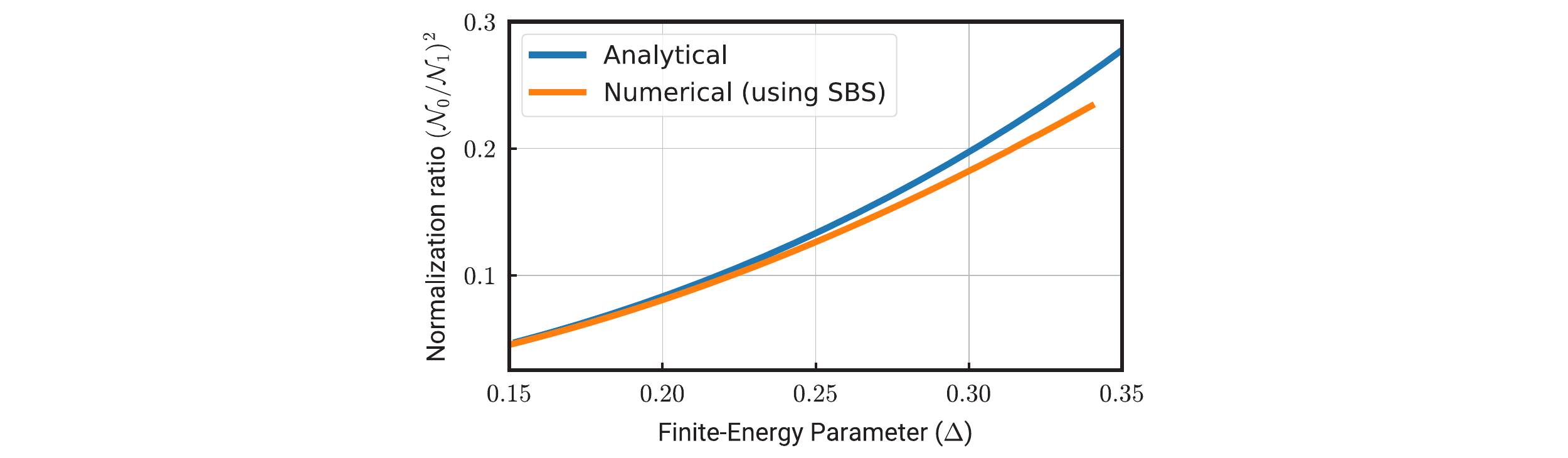}
    \caption[Norms of square GKP codewords and error words in relation to probabilistic distance.]{\textbf{Norms of GKP states and error words in relation to probabilistic distance.} The squared ratio of the norm of codeword $\ket{0}$ ($\mathcal{N}_0$) and the norm of error word $\hat a\ket{0}$ ($\mathcal{N}_1$). The analytical curve shown in blue is obtained using Eqs.~\ref{eq:norm0}-\ref{eq:norm1}. The numerical curve is obtained using the states stabilized by the approximate small-big-small scheme for maximum equivalence to the definition of GKP states used in our analytical calculations in this section. The deviation with increasing $\Delta$ is due to SBS being an approximate only correct to first-order in $\Delta$. Despite this, we see decent agreement in the two curves.}
    \label{fig:norm}
\end{figure}

Thus, we see an asymmetry between correction probabilities for $\hat a, \hat a^\dagger$. Thus, the probabilistic distance for various envelope sizes against $\{a,a^\dagger\}$ errors can be computed using the above expressions with $p_d=\frac{x}{1+x}$, for $x=\{\pi e^{-2\Delta^2}(\mathcal{N}_0/\mathcal{N}_1)^2,\pi e^{2\Delta^2}(\mathcal{N}_0/(\mathcal{N}_1+\mathcal{N}_0))^2\}$, respectively. We find that for small $\Delta$ the probabilistic distance for photon loss is better than the probabilistic distance for photon gain. To compare these expressions with numerics we plot the ratio of importance here, $\mathcal{N}_0/\mathcal{N}_1$ using numerics and our analytical expressions. For numerics, we have used the states stabilized by the small-big-small scheme to get the closest states to the definitions used here. Thus, the values deviate with increasing $\Delta$, however, we see in Fig.~\ref{fig:norm} that our analytical expression is still in agreement with the numerical curve and hence obtained for smaller values of $\Delta$.

\section{Numerical Comparison}
The above calculations were carried out for the exact stabilization of the finite-energy GKP manifold. However, the approximate scheme is a first-order trotterized approximation of this exact version, correct to only first-order in $\Delta$. So, the question is, how does our probability of correction derived for the exact stabilization compare to the approximate stabilization scheme? For this purpose, we now give comparisons between our theory calculations and numerically obtained results using the stabilization circuits. 

We use the logical $\ket{\pm H}_\mathrm{GKP}$ states stabilized by the SBS scheme~\cite{royer2020stabilization} for this analysis. For the numerical analysis, we run a single round of the approximate stabilization scheme and extract the probability of a $ge$ or $eg$ outcome. Note that, this scheme is approximate and gets worse with an increase in $\Delta$. Thus, we also compute the fidelity of the corrected state with the original state as a sanity check. We then compare this with the probabilistic distance computed for the exact stabilization 
\begin{align}
p_d=\frac{\pi e^{-2\Delta^2}}{\pi e^{-2\Delta^2}+\frac{||\hat a\rho\hat a^\dagger||}{||\rho||}}=  \frac{\pi e^{-2\Delta^2}}{\pi e^{-2\Delta^2}+\bar n_\mathrm{GKP}}
\end{align}
After half a round of stabilization, $SBS^\mathrm{Z}_\mathrm{GKP}$, the outcome probability of correction is equal to $p_d/2$, while after the complete round  $SBS^\mathrm{X}_\mathrm{GKP}SBS^\mathrm{Z}_\mathrm{GKP}$, the probability of correction is equal to $p_d$. 

The approximate stabilization scheme that we have derived in Ref.~\cite{royer2020stabilization} is only correct upto first order in $\Delta^2$, so we can modify this formula to,
\begin{align}
p_d=\frac{\pi}{\pi+\bar n_\mathrm{GKP}}.
\end{align}
The $\bar n_{GKP}$ calculated analytically uses the exact GKP description and hence will be the reason for any deviations in the analytical values and the numerically computed values. Since $\bar n_\mathrm{GKP}\rightarrow \infty$ as $\Delta\rightarrow 0$, this equation indicates that the exact autonomous stabilization will yield a lower probability of correction as the expanse of GKP state increases in phase space. That is, a smaller $\Delta$ is preferable for the exact stabilization. However, we just pointed out that the approximate scheme is only valid for small $\Delta$. Thus, there exists an optimal $\Delta$ which should be yield optimal performance with probabilistic error correction. Ref.~\cite{sivak2023real} optimized the value of $\Delta$ to $0.34$ using reinforcement learning without any knowledge of the effect we have pointed out here. Thus, our work also indicates why this optimized $\Delta$ was not any smaller or larger.

\begin{figure}[htb]
    \centering
    \includegraphics[width=\linewidth]{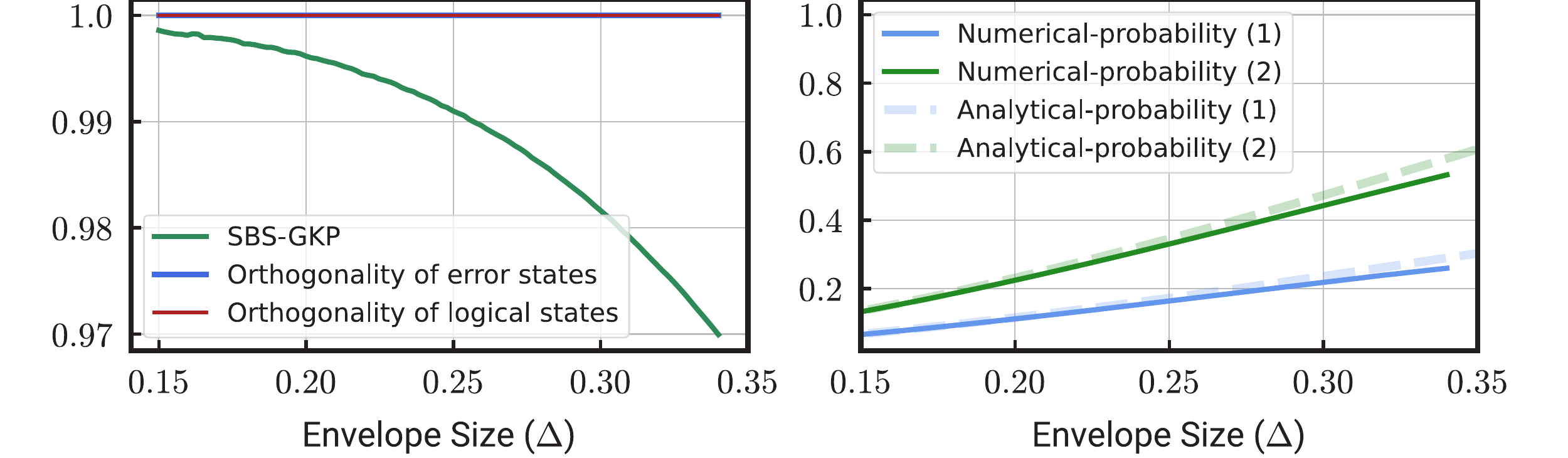}
    \caption[Probabilistic error correction of photon loss on a square GKP state using SBS.]{\textbf{Probabilistic error correction of photon loss on a square GKP state using SBS.} (Left) We plot the orthogonality of the codeword and the error word $(\ket{+H}_\mathrm{GKP},\hat a\ket{+H}_\mathrm{GKP})$ (purple), and the logically orthogonal codewords $(\ket{+H}_\mathrm{GKP}, \ket{-H}_\mathrm{GKP})$ (red). Here, $1$ on the y-axis indicates that the states are completely orthogonal. The green curve in this plot shows the admissibility of using SBS for the analysis by comparing the fidelity between the GKP states stabilized by SBS and the states to which an error state is corrected to, after one round of $\mathrm{SBS}^\mathrm{X}$. Note that SBS is a first-order approximation to the exact finite-energy GKP stabilization, and so there is a decrease in the efficiency of stabilization as $\Delta$ increases. (Right) The probability of correction after $\mathrm{SBS}^\mathrm{X}$ obtained analytically (dotted) and numerically (solid) is given in blue. The probability of correction after $\mathrm{SBS}^\mathrm{Z}\mathrm{SBS}^\mathrm{X}$ obtained analytically (dotted) and numerically (solid) is given in green. These quantities are probabilities of getting the right qubit ($ge$ or $eg$) outcome on an error state $\hat a\ket{\psi}$. While this quantity increases with $\Delta$, the SBS scheme strays further away from stabilizing the GKP codespace as $\Delta$ increases. Thus, there is an optimal $\Delta$ where the probability of achieving a corrected GKP state is maximum. These plots together give the complete picture of the probabilistic error correction of GKP states with the SBS scheme.}
    \label{fig:SBS_sq}
\end{figure}

As can be seen from Fig.~\ref{fig:SBS_sq}, the numerical results match the probabilistic distance to good precision; however, as expected, the admissibility of SBS as a GKP stabilization scheme decreases with an increase in $\Delta$. As pointed out in Ref.~\cite{royer2020stabilization}, the cooling rate, which is related to what we call the probabilistic distance, decreases with an increase in the size of the GKP code. The slight disagreement with increasing $\Delta$ is due to the reasons pointed out in the previous paragraph. Importantly, experiments in Refs.~\cite{sivak2023real,brock2024quantum} do not use the BSB scheme, because it is a longer circuit which will induce more circuit errors before ancilla reset. This is why we do not show the calculations for this circuit, even though it has a better probability of correction.

\subsection{Single-mode Lattices}\label{Hex_gkp}
\begin{figure}
    \centering
    \includegraphics[width=\linewidth]{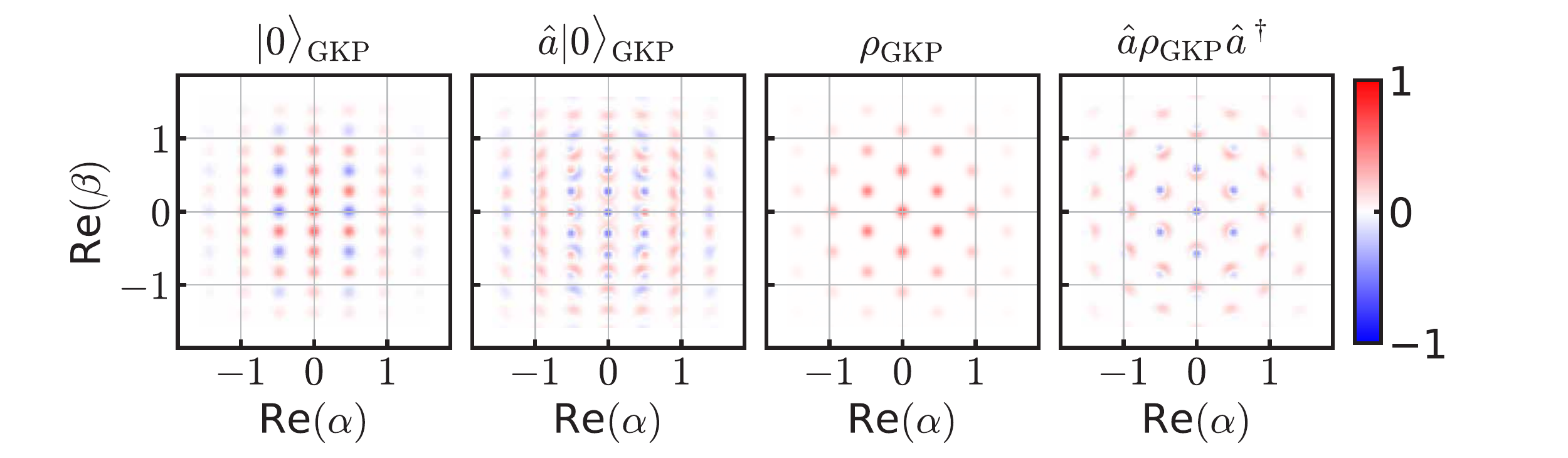}
    \caption[Logical GKP codewords and error words for hexagonal lattice encoding]{Logical GKP codeword $\ket{0}_\mathrm{GKP}$ for the hexagonal code and corresponding error word are shown in the first two plots. The maximally mixed state in the logical codespace $\rho_\mathrm{GKP}=\frac{\ket{0}_\mathrm{GKP}\bra{0}_\mathrm{GKP}+\ket{1}_\mathrm{GKP}\bra{1}_\mathrm{GKP}}{2}$ and the error space $\hat a\rho_\mathrm{GKP}\hat a^\dagger$ are shown in the next two plots. All axes are labeled in units of lattice constant $l_\mathrm{hex}$. Note the hexagonal envelope in these figures in contrast to Fig.~\ref{fig:GKP-error-words}.}
    \label{fig:GKp-hex}
\end{figure}

Other options for single-qubit GKP encoding include the rectangular GKP code and the hexagonal GKP code. A GKP encoding can be designed as an arbitrary lattice in two dimensions using the stabilizer vectors $S^\mathrm{X}$ and $S^\mathrm{Z}$ in phase space, satisfying the required commutation relations~\cite{gottesman2001encoding}. Among these, the hexagonal codes are predicted to be the most efficient for displacement errors due to their resemblance with a closest packed lattice~\cite{harrington2001achievable,gottesman2001encoding}. Thus, we will next analyze this lattice from the perspective of generalizing our results to arbitrary qubit lattices. Here, we use the hexagonal lattice stabilization for which the lattice constant is, 
\begin{align}
    l_\mathrm{hex}=2\sqrt{2\pi}/3^{1/4}.
\end{align}
The logical operators $P\in\{\mathrm{Z},\mathrm{X}\}$ are defined along vectors $\hat v\in\{\hat p,\cos{(\pi/6)}\hat x+\sin{(\pi/6)}\hat p\}$ while the stabilizers of logical $P\in\{\mathrm{Z},\mathrm{X}\}$ bases are along the vectors $\hat v\in\{\hat x,\cos{(2\pi/3)}\hat x+\sin{(2\pi/3)}\hat p\}$. The codewords and error words for the hexagonal code are given in Fig.~\ref{fig:GKp-hex}.

\begin{figure}
    \centering
    \includegraphics[width=\linewidth]{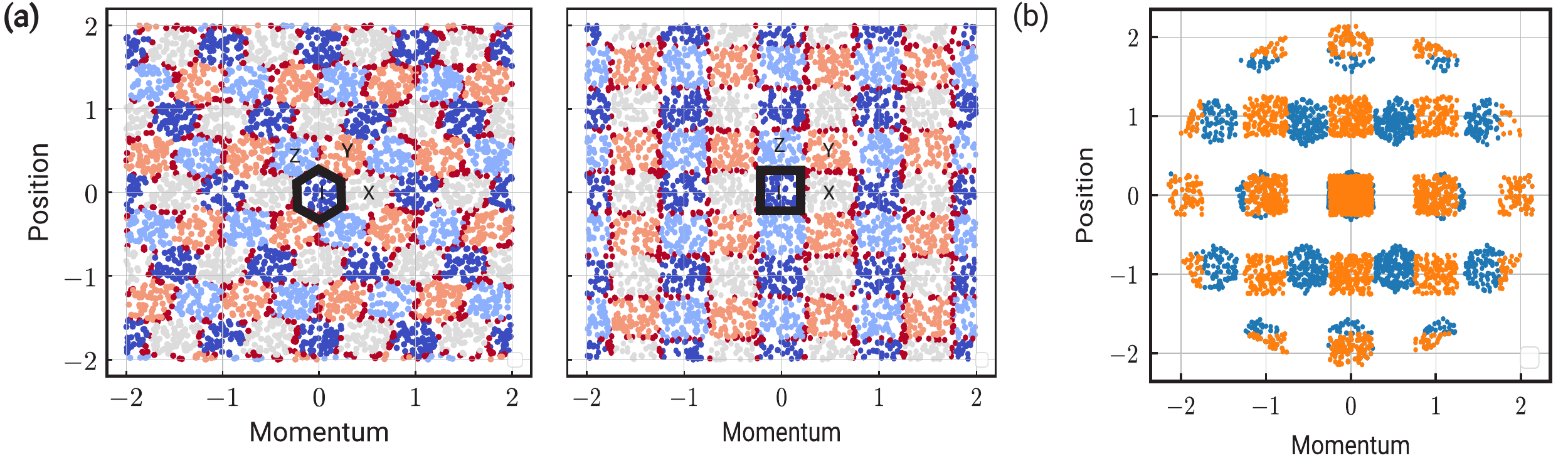}
    \caption[Voronoi cells of square and hexagonal GKP codes]{\textbf{Voronoi cells of square and hexagonal GKP codes.} (a) Regions marked for errors that are safe for Pauli P eigenstates for hexagonal codes (left) and square codes (right) using the SBS stabilization scheme for displacement errors. The dark blue region marks I, that is, all states will be corrected without any logical error. Similarly, the {gray, orange, light blue} regions mark the areas which are safe for logical Pauli ${\mathrm{X}_\mathrm{GKP},\mathrm{Y}_\mathrm{GKP},\mathrm{Z}_\mathrm{GKP}}$ eigenstates. The unit cell in the center (marked in black) is known as the Voronoi cell for the corresponding lattice. The red dots correspond to the ambiguous regions that could fall into either cell it is on the boundary of. (b) An illustration of how the correctable or error-safe regions for hexagonal lattice are more in the area compared to the square code. Here the blue (orange) tiling corresponds to the tilings of only the Voronoi cell (no logical error case) for the hexagonal (square) code. The hexagons (squares) are apart along the hexagonal (square) GKP stabilizers by a length of $l_\mathrm{hex}$ ($l_\mathrm{sq}$).}
    \label{fig:hex_sq}
\end{figure}

Note that the lattice constants follow  $l_\mathrm{sq}<l_\mathrm{hex}$. 
Thus, the displacement errors are more protected or have a larger distance for the hexagonal GKP encoding compared to the square GKP encoding~\cite{harrington2001achievable,gottesman2001encoding}. We give a Voronoi cell interpretation of how errors are corrected in the two different lattices also in Fig.~\ref{fig:hex_sq}. That is, given a certain displacement error, which displacements in phase space can be mapped back to the codespace without error? 

The ratio of the total correctable area is proportional to the ratio of the areas of the Voronoi cell for both lattices. Voronoi cell is the unit cell in the reciprocal lattice (recall Brillouin zone from solid state physics). A Voronoi cell can be constructed by drawing perpendicular bisectors of each logical operator. the region enclosed by these bisectors is the Voronoi cell. Errors in this Voronoi cell will be corrected back to the original state while errors in the cell of the same area and shape connected to the Voronoi cell correspond to one of the three Pauli logical errors. This tiling is repeated throughout phase space to yield the distribution shown in Fig.~\ref{fig:hex_sq}(a,b) for hexagonal and square lattices. In these figures, we confirm via simulation that errors in the specific region are mapped to no error for eigenstates of $I$ (all states, dark blue), Pauli $X$ (gray), Pauli $Y$ (orange), and Pauli $Z$ (light blue) operators. The red dots show the region of ambiguity where the states could go to either region. These regions lie at the mark of $l/4$ for the lattice constant $l$. The area of the Voronoi cell for the square lattice is $\pi/4=0.79$ (square of length $l_\mathrm{sq}/4$). The corresponding area for the hexagonal lattice is $3^{1/4}\frac{\sqrt{\pi}}{2\sqrt{2}}=0.82$ (hexagon of side $\sqrt{l_\mathrm{hex}/6}$). Thus, the hexagonal lattice can correct roughly a factor of $1.05$ more errors compared to the square lattice. This is illustrated in Fig.~\ref{fig:hex_sq}(c) where the hexagons of side $l_\mathrm{hex}$ tiled in phase space contain some squares of side $l_\mathrm{sq}$ tiled in phase space.

\begin{figure}
    \centering
    \includegraphics[width=\linewidth]{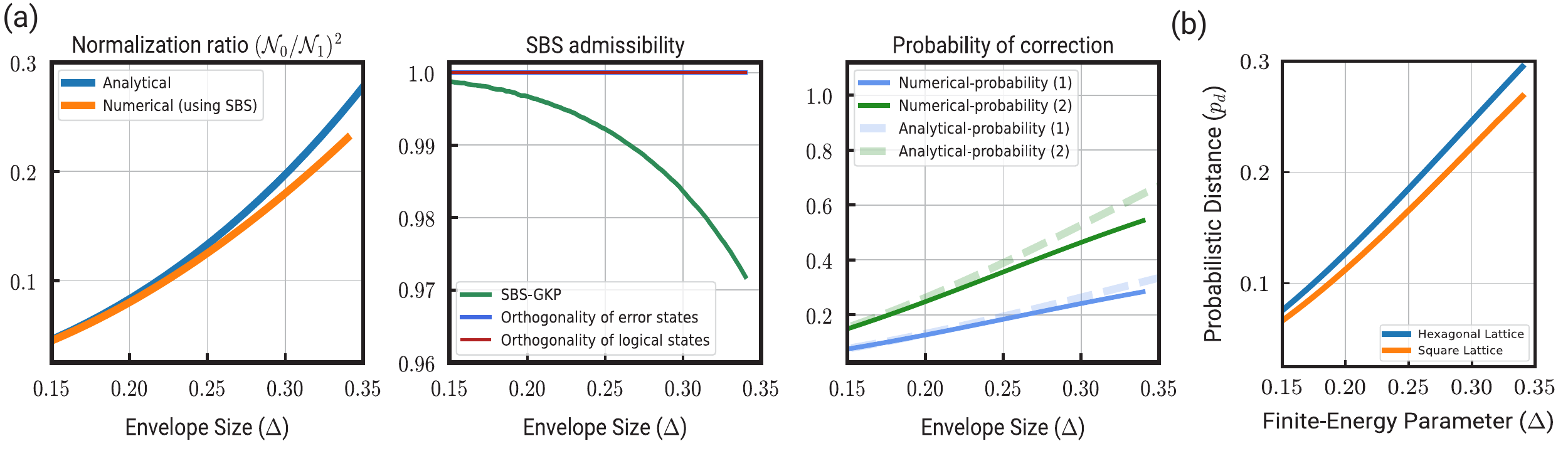}
    \caption[Probabilistic correction of hexagonal GKP lattices and comparison with square GKP codes.]{\textbf{Probabilistic correction of hexagonal GKP lattices and comparison with square GKP codes.} (a) Plots corresponding to Figs.~\ref{fig:SBS_sq}-\ref{fig:norm} for the hexagonal GKP encoding. We can see agreement in numerical and analytical schemes. The performance, in this case, matches the performance of the square GKP code. (b) Comparison of probabilistic distance, that is, the probability of correction of photon loss, after one complete round of stabilization, between hexagonal and square GKP codes. The improvement in hexagonal lattice compared to square GKP agrees with the intuition gained from the dependence of $p_d$ on $l_\mathrm{hex}\ge l_\mathrm{sq}$ in Eq.~(\ref{eq:prob_hex}). }
    \label{fig:SBS_hex}
\end{figure}

Let us talk about the case of photon loss now. The SBS stabilization scheme for the hexagonal codes is given by,
\begin{align}     
\mathrm{SBS}^\mathrm{P}_\mathrm{GKP}&=e^{i\epsilon_v\hat{v}_\perp\hat{\sigma}_y}e^{i(l/2)\hat{v}\hat{\sigma}_x}e^{i\epsilon_v\hat{v}_\perp\hat{\sigma}_y}, \textrm{ where } \epsilon_v=\frac{l}{4}\Delta^2,
\label{eq:SBS_hex}
\end{align}
where $\hat v\in\{\hat x,\cos{(2\pi/3)}\hat x+\sin{(2\pi/3)}\hat p\}$ for logical $P\in\{\mathrm{Z},\mathrm{X}\}$ bases respectively. Calculations similar to the case of square GKP code can be repeated here to deduce the probability of correction after one complete round of stabilization $\mathrm{SBS}^\mathrm{X}_\mathrm{GKP}\mathrm{SBS}^\mathrm{Z}_\mathrm{GKP}$ as, 
\begin{align}
    p_\mathrm{corr}=2p_e=\frac{e^{-2\Delta^2}\frac{l^2}{4}}{\bar n_\mathrm{GKP}+e^{-2\Delta^2}\frac{l^2}{4}}\approx \frac{\frac{l^2}{4}}{\bar n_\mathrm{GKP}+\frac{l^2}{4}}.\label{eq:prob_hex}
\end{align}
The right hand side represents the probability of correction for the approximate stabilization with small-big-small (correct upto first order in $\Delta^2$). The formula will deviate from the 
Here $\bar n_\mathrm{GKP}=||\hat a\ket{\psi}||/||\ket{\psi}||$ is the ratio of norms of the GKP error word and codeword. These equations can further be generalized to understand error correction for qudits as well, where for square GKP codes, the lattice constant is simply $l_\mathrm{sq,d}=\sqrt{2\pi d}$ with lattice vectors $(\hat v=\hat x,\hat v_\perp=\hat p)$. Intuitively the correction probability in this case should be higher, given the normalization constants of the codewords and error words do not enforce a different result. In Fig.~\ref{fig:SBS_hex}, we plot the probability of correction for a complete round of stabilization of the hexagonal GKP codes using the approximate small-big-small scheme. 
Fig.~\ref{fig:SBS_hex} shows that, as expected the probabilistic distance of hexagonal codes is larger than the square code, the same as the relationship between the distance against displacement errors. The hexagonal probabilistic distance is at least a factor of $1.1$ (minimum at $\Delta=0.15$ in Fig.~\ref{fig:SBS_hex}(b)) more than the case of square GKP. This is an interesting outcome of our result that we are able to confirm the analytical results known for displacement errors with respect to photon loss.

Thus, in this chapter, we have given an analytical understanding of how the SBS scheme corrects single photon losses, providing deeper insights into the recent beyond-break-even memory experiments~\cite{sivak2023real,brock2024quantum} with GKP encoding. The question is if this stabilization scheme could be still improved to achieve better performance, as indicated by Refs.~\cite{zheng2024performance,albert2018performance}. This relates to further our calculations and gets an analytical expression for the logical error rate under the photon loss error channel. Note that, in deterministic error correction using measurement-based feedback, the decoding graph uses the probability of errors to decipher the most likely error chain that might have occurred, given the stabilizer measurement outcomes. More on such methods can be found in Chapter~\ref{chapter:conc}. As a concluding step to our analytical understanding, it would be interesting to understand the following. There is a hint from an analysis that Vlad Sivak and Baptiste Royer did which shows a numerical optimization of the stabilization scheme under photon loss is sensitive to the photon loss rate. While we do not have any further analysis to support this argument, this analysis hints that the stabilization scheme could be improved with knowledge about the noise channel, a component missing in the current scheme. How can the probabilistic error correction use the information about single-photon loss rate to convert the stabilization scheme into a probabilistic decoding of photon loss? We leave this question as a future direction for this analysis.

\section{Open Problem: Protected Qubits and Oscillators}\label{open:GKP-qec}
\paragraph{Protected qudits in circuit-QED:}
The error correction of GKP qubits is limited due to ancilla errors~\cite{royer2020stabilization}. See App.~\ref{ancilla_errors} for details. Thus, a better strategy would be to engineer a protection that does not require ancillary stabilization. Multiple 2D superconducting circuits can be engineered to protect the quantum information without ancillae. Examples of such circuits are Kerr-cat qubits~\cite{puri2017engineering}, and the $0-\pi$ qubit \cite{kitaev2006protected}. The Kerr-cat qubits are nonlinear oscillators with a double well potential that can be engineered using an actively driven SNAIL circuit~\cite{grimm2020stabilization}. This circuit offers biased noise protection against one type of Pauli noise on the Kerr-cat qubit Bloch sphere. That is, $p_z\gg p_x,p_y$ is possible in this qubit technology. The $0-\pi$ qubit is the first proposal for a protected qubit that passively protects quantum information against any type of noise. These circuits have not yet demonstrated complete protection from quantum errors. 

There is an inverse relationship between inherent protection against errors and the ease of controlling the qubits. Inherent protection against any type of noise means that it is hard for the environment to manipulate the quantum state on a protected basis. For example, in a Kerr-cat qubit, if $p_z\gg p_x,p_y$ then the $\mathrm{X}$-basis gates like $\mathrm{CCX},\mathrm{CX}, \mathrm{X}(\theta)$ can be more erroneous or complicated in terms of hardware efficiency compared to the $\mathrm{Z}$-basis gates like $\mathrm{Z}(\theta),\mathrm{CZ},\mathrm{CCZ}$. Thus, for a completely protected qubit, like the $0-\pi$, control on any basis would be equally hard and much less efficient compared to a transmon or fluxonium qubit. This feature has contributed to the lack of experimental evidence of such a qubit realization in the circuit parameter regime where complete portection from errors can be claimed~\cite{gyenis_experimental_2021}. 

\begin{myframe}
\singlespacing
\begin{quote}
Recent works~\cite{sellem2025dissipative,nathan2024self,geier2024self} have tried to tackle the problem of protected control of qubits using superconducting circuits inspired by the GKP codes. Can the understanding of probabilistic correction of photon loss discussed in this chapter help in a more efficient design of such qubits? In addition, can the error-corrected gates discussed in the next section inspire protected gates on such qubits?
\end{quote}
\end{myframe}

\doublespacing

\paragraph{Oscillator error correction:}
 %We have formally described the multi-mode GKP codes~\cite{harrington2001achievable,royer2022encoding,conrad2022gottesman,conrad2024lattices} in Apps.~\ref{app:lattices}-\ref{app:mutli-mode_QEC}. 
 In this chapter, we have discussed GKP codes from the perspective of encoding a qubit in a single oscillator. However, the real advantage of the CV-DV architecture lies in using oscillators as oscillators~\cite{jordan2012quantum,crane2409hybrid}. Currently, an oscillator error correction strategy that could reduce errors in a single logical oscillator unit, with an increase in the number of physical oscillators, is absent~\cite{noh2020encoding,hanggli2021oscillator}. Developments in this area are crucial for oscillator-based useful quantum computing. In this context, we ask the following questions about what we have learned in this chapter.
 
 \begin{myframe}
\singlespacing
 \begin{quote}
 It has been shown that oscillator codes have no threshold against Gaussian displacement noise channels~\cite{hanggli2021oscillator}. Is this true for the case of photon loss? Could the understanding of probabilistic correction of photon loss be developed further with multi-mode codes to encode oscillators into many oscillators, enabling increased protection for the logical oscillator?     
 \end{quote}
\end{myframe}
  
\doublespacing

    \chapter{Control of an Error-Corrected Qudit in an Oscillator} \label{chapter:qec-control}
\begin{myframe}
\singlespacing
\begin{quote}
\textit{Is it possible to also control the error-corrected GKP states with protection against some faults?} To answer this question, we propose a high-fidelity measurement-free gate teleportation technique for logical operations on GKP bosonic codewords, advancing universal control of GKP qudits. The versatility of our non-abelian QSP framework discussed in Chapter~\ref{chapter:na-qsp} bridges the gap between the theoretically ideal and the experimentally realistic GKP codespace, significantly enhancing the fidelity of practical gate operations. 
\end{quote}
\end{myframe}

\doublespacing
 Further development of the GKP codes beyond an error-corrected memory requires universal control via state preparation, measurement, and gate operations. The phase-space instruction set is most suited for the translationally invariant grid codes, or as acronymed, the GKP codes. All our schemes are based on non-abelian sequence ($\mathrm{GCR}$) using the phase-space instruction set~\cite{ISA}. Fast gates in this instruction set can be realized in the low dispersive-coupling regime~\cite{eickbusch2022fast} which has the advantage of reducing errors associated with higher-order Kerr effects~\cite{eickbusch2022fast} that cannot be efficiently corrected using GKP states. In this section we derive analytical schemes for (i) error-detected qudit state preparation, (ii) end-of-the-line logical qubit readout, and (iii) high-fidelity logical single-qubit and two-qubit universal gate set. All schemes presented here improve upon the performance of the state-of-the-art theoretical schemes and experimental demonstrations. Our schemes are generalizable for arbitrary lattice spacing, thus yielding universal control for square, hexagonal, and rectangular GKP qubits, GKP qudits, and multi-mode GKP codes. Below we summarize our GKP control results before diving into the details. 

%Why is NA-QSP good for GKP
The ancilla-assisted finite-energy GKP readout in Ref.~\cite{hastrup2021improved} and stabilization scheme in Refs.~\cite{royer2020stabilization,de2022error} are also tied together using our non-abelian QSP-based composite sequence $\mathrm{GCR}$ in Sec.~\ref{ssec:GKLP-framework}. The stabilization scheme, used to achieve the record gain for beyond break-even logical lifetime in superconducting circuits \cite{sivak2023real}, was derived using dissipation engineering techniques. We not only give the first analytical explanation for the numerically optimized readout scheme but also tie it together with this independently derived stabilization scheme. Our framework is helpful in understanding error correction to the right logical state in the GKP manifold using this qubit-based dissipation scheme, which was only engineered to avoid leakage and not logical errors. In addition, in Sec.~\ref{ssec:logical-readout}  we give high-fidelity readout circuits when the GKP state has some residual (correctable) errors. These circuits are based on the QSP sequences $\mathrm{BB1}$ and $\mathrm{BB1}(\mathrm{GCR})$ introduced in Chapter~\ref{chapter:na-qsp}, and could yield better readout fidelity for GKP qubits with correctable displacement errors.

A key result of this Chapter is a pieceable error-corrected gate teleportation scheme, the first to correct errors during gate teleportation within a single system. This pieceable design not only corrects in-flight errors but also mitigates ancilla dephasing, suppressing otherwise uncorrectable ancilla-induced faults. Correctable errors are handled by an underlying error-corrected circuit we construct. As a result, our scheme enables high-fidelity, universal single-qubit logical rotations even with biased-noise ancillae like cat qubits\cite{mirrahimi2014dynamically,puri2019stabilized,grimm2020stabilization,ding2024quantum}. To our knowledge, it is the most effective approach for non-Clifford operations on GKP qubits to date~\cite{hastrup2021unsuitability}.

We restrict the derivation and discussion to the case of the square lattice GKP, where $x$ and $p$ are treated symmetrically, in Secs.~\ref{ssec:GKP-errors}-\ref{ssec:piecewise-teleportation}, while discussing generalizations to other qubit and qudit lattices in Sec.~\ref{ssec:GKP-extension}. As discussed in Chapter~\ref{chapter:GKP-qec}, for the code deformation from ideal GKP to finite-energy GKP states using the Gaussian envelope operator $\hat E=e^{-\Delta^2\hat n}$, we have,
\begin{align}
    \ket{\psi_\Delta}_\mathrm{GKP}=\hat E\ket{\psi}_\mathrm{GKP}, \hat A_\Delta=\hat E \hat A\hat E^{-1}.
\end{align}
Note that $\hat E$ is a non-unitary operation and hence, in experiments, we achieve this approximately using dissipation-based methods, discussed in App.~\ref{dissipation-engineering}. The practical envelopes resemble a cosine form, aligning more closely with Eq.(\ref{eq:GKP-bin-logical}) in Sec.\ref{ssec:GKP-States}. This envelope broadens delta functions into Gaussian wave packets with uncertainty $\Delta$, while the overall envelope itself carries an uncertainty of $1/\Delta$.

%summarizing the ideas
Our GCR-based schemes schemes reduce sensitivity to uncertainties in the position and momentum of finite-energy oscillator states. Thus, these schemes can be viewed as mappings from superpositions of idealized, infinitely squeezed position and momentum eigenstates to superpositions of realistic Gaussian states with finite uncertainty. This explains our findings in connection with GKP states. All GKP operations are well-defined for ideal GKP codes which are superpositions of infinitely squeezed states; our framework maps these operations to yield circuits that can come extremely close to the exact finite-energy GKP operations. In a previous work, an approach towards such finite-energy operations, in particular the logical entangling gates ($\mathrm{CX}_\mathrm{GKP}/\mathrm{CZ}_\mathrm{GKP}$), was suggested in~\cite{rojkov2023two} with a qutrit ancilla. In contrast, our construction for entangling gates yields comparable fidelity using two ancilla qubits. We also give decompositions to execute fast two-mode echoed Gaussian operations in the context of finite-energy GKP entangling operations in App.~\ref{app:finite-SUM}. We owe this improvement to the simplicity provided by the description of our framework. Both these approaches are better than implementing the gate designed for the ideal (infinite-energy) GKP code followed by several rounds of stabilization. Most importantly, all our single-mode circuits use the phase-space instruction set and the two-mode schemes use squeezing, and beam-splitters, in addition to the gates mentioned in Eq.~(\ref{eq:phasespaceIS}) for the phase-space instruction set. Proposals to realize these operations can be found in~\cite{chapman2023high,tsunoda2023error}. This indicates that our circuit construction is a closer approximation to the correct finite-energy GKP gates.

%gate-teleportation
We compare all our results against state-of-the-art theoretical schemes in terms of fidelity and feasibility, in the absence of errors. We also show that our analytical schemes achieve universal control of the GKP code robust to ancilla and cavity errors, surpassing the best-known fidelities in the presence of such faults. Note that, in this chapter, we will be using Wigner units $\hat x=\frac{\hat a+\hat a^\dagger}{2}$ unlike Chapter~\ref{chapter:GKP-qec}.

\section{High-Fidelity Error-Detected State Preparation}\label{ssec:GKP-errors}
Fault-tolerant preparation of the logical GKP states is an important resource for bosonic quantum error correction. Previously in Chapter~\ref{chapter:state-prep}, we described a protocol that prepares GKP states in small steps $\mathcal{S}_k,\mathcal{C}_k$, at the end of which the ancillary qubit was left in a deterministic state. In this section, we will investigate the performance of this scheme in the presence of cavity and qubit noise. We will then compare this fidelity with the scheme in Ref.~\cite{hastrup2021measurement} which can also benefit from mid-circuit error detection.

For the protocol presented in Sec.~\ref{ssec:GKP-States} we introduce mid-circuit detection on ancilla errors via post-selection upon outcome $\ket{g}$ after every round; resetting the qubit to $\ket{g}$ if the step has succeeded. During each gate in the circuit, we add photon loss at the rate $\kappa/2\pi=1/1000 \ \mu\textrm{s}^{-1}$, ancilla decay at the rate $\gamma/2\pi=1/200 \ \mu\textrm{s}^{-1}$, and ancilla dephasing at the rate $\gamma_\phi/2\pi=1/200 \ \mu\textrm{s}^{-1}$. With this, we run the protocol for $\Delta=0.34$~\cite{sivak2023real} for $10^5$ rounds. For each preparation round, we execute the GKP state preparation circuit $\mathcal{C}_1-\mathcal{C}_2$ shown in Fig.~\ref{fig:GKP-prep} with measurements after every $\mathcal{C}_k$. We throw away any round where we encounter a $-1$ outcome for a $Z$ measurement on the ancilla qubit and start over with the oscillator in the squeezed state. The success probability (fraction of rounds that will not be thrown away) for this simulation is $0.94$. Now, the average fidelity for this case was $0.96$ in comparison to $0.9969$, in comparison to $0.9969$ when no errors were introduced. On the other hand, the numerical scheme in Ref.~\cite{eickbusch2022fast} reports a simulated (and experimental) fidelity of $0.85$ for a numerically optimized circuit with the same $\Delta$ without post-selection. Post-selection with their circuit is possible but only at the end of the entire circuit whose length is comparable to our entire preparation circuit. Thus, the efficiency with which such a circuit can detect errors with post-selection would be low. Relative to Ref.~\cite{hastrup2021improved}, our scheme performs better due to a similar argument. In comparison to the first step of that protocol, the length of our snippet $\mathcal{C}_1$ is much smaller, reducing the probability of errors and thus improving the success probability and fidelity. Thus, we have shown numerically that our scheme performs better than state-of-the-art schemes for GKP preparation using a DV ancilla, even in the presence of errors.

\section{GKP Protocols in the Non-Abelian QSP Framework}\label{ssec:GKLP-framework}
\begin{figure}[tbh]
    \centering
    \includegraphics[width=\textwidth]{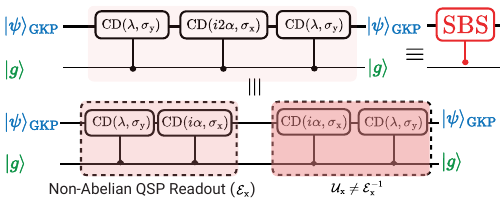}
    \caption[Finite-energy GKP readout and stabilization protocols in non-abelian QSP framework.]{\textbf{Finite-energy GKP readout~\cite{hastrup2021improved} and stabilization~\cite{royer2020stabilization} protocols in non-abelian QSP framework.} Interpretation of $\mathrm{SBS}$ circuit along the position quadrature as logical identity on the GKP codewords. The circuit is divided into the entangling and unentangling gadgets. The first half of this circuit $\mathcal{E}$ is the GKP readout circuit. Here, $\alpha=\frac{\sqrt{\pi}}{2\sqrt{2}}$ and $\lambda=-\alpha\Delta^2$.}
    \label{fig:GKp-framework}
\end{figure}

\paragraph{Analytical explanation for numerically-optimized readout scheme in Ref.~\cite{hastrup2021improved}:} Ref.~\cite{hastrup2021improved} shows a numerically optimized circuit for improved GKP readout over the ideal readout scheme. The ideal GKP readout scheme is explained in Sec.~\ref{ssec:logical-readout}. The ideal GKP readout circuit is a rotation gadget which intends to rotate the state by $2\pi$ ($\pi$) if the logical codeword is $\ket{0}_\mathrm{GKP}$ ($\ket{1}_\mathrm{GKP}$) using $e^{i\sqrt{\frac{\pi}{2}}\hat x\otimes\sigma_\mathrm{x}}$ with the qubit starting in $\ket{g}$. Thus, a readout circuit is nothing but an entanglement gadget. For finite-energy GKP, we can employ our entanglement gadget with the small GCR pre-correction $e^{i\sqrt{\frac{\pi}{2}}\Delta^2\hat p\otimes\sigma_\mathrm{y}}$. Thus, we give the logical readout circuits for logical bases X ($\hat v=\hat p$), Y ($\hat v=\hat x+\hat p$), and Z ($\hat v=\hat x$).
\begin{align}
    \mathcal{E}_{\hat v}\Big[\frac{\pi}{2},\sqrt{\frac{\pi}{2}},\Delta\Big]=e^{i\sqrt{\frac{\pi}{2}}\hat v_\perp\Delta^2\sigma_\mathrm{y}}e^{i\sqrt{\frac{\pi}{2}}\hat v\sigma_\mathrm{x}}.\label{eq:GKP-readout}
\end{align}
See Sec.~\ref{sec:preliminaries} for the definition of $\hat v_\perp$. Notice that for a square GKP code defined in Eqs.~\ref{eq:GKP-logical0},\ref{eq:GKP-logical1} and Eq.~(\ref{eq:GKP-bin-logical}), the displacement required to implement a logical $Y$ operation is longer than the corresponding displacements required for logical $X$ or $Z$ operations by a factor of $\sqrt{2}$. 

The GCR correction given here, which we obtained analytically from $\mathcal{E}=\mathrm{GCR}$ following the discussion in Sec.~\ref{ssec:ent-unent},  exactly matches the numerically-optimized correction in Ref.~\cite{hastrup2021improved}.
\paragraph{Dissipation-engineering based stabilization and correction of displacement errors:} 
Dividing the stabilization circuit $\mathrm{SBS}$ from Ref.~\cite{royer2020stabilization} into two halves (see Fig.~\ref{fig:GKp-framework}), the first half can be identified exactly as the entangling gadget used for readout as described in Eq.~(\ref{eq:GKP-readout}). The second half satisfies the constraints identified for the un-entangling gadget $\mathcal{U}(\pi/2,\sqrt{\pi/2},\Delta)$ in Sec.~\ref{ssec:Cat_States} since,
\begin{align}
    \pi/2\textrm{ (from }\mathcal{E})+\pi/2\textrm{ (from }\mathcal{U})=\pi,
\end{align}
as required by Eq.~(\ref{eq:ent_unent}). And thus, 
\begin{align}
\mathrm{SBS}_{\hat v}=\mathcal{E}_{\hat v}\Big[\frac{\pi}{2},\sqrt{\frac{\pi}{2}},\Delta\Big]\mathcal{U}_{\hat v}\Big[\frac{\pi}{2},\sqrt{\frac{\pi}{2}},\Delta\Big].    
\end{align}
 Note that, here $\mathcal{U}\neq\mathcal{E}^{-1}$, and so, this circuit is a logical GKP identity and not a universal identity. The circuit has a non-trivial back action if the oscillator is not in the GKP state which is the key to its success for error correction.

The error correction properties of this circuit were experimentally verified and qualitatively discussed in~\cite{sivak2023real}. Here, we provide a quantitative argument using our framework to compute the back-action on the oscillator and its effects on the GKP logical state with a displacement error. As discussed before, the $\mathrm{SBS}$ circuit applies a deterministic logical Pauli when the oscillator is in the GKP codespace. However, when the state is not in this logical codespace, the unentangling gadget needs to apply the correction along a different axis, depending on the displacement error $\epsilon$. Since this knowledge is not available to the stabilization circuit, it applies an autonomous back action on the oscillator depending on the qubit outcome. 

This back-action of the SBS circuit can be explained using non-abelian QSP as follows. In the event of a displacement error, say $\epsilon$, the GKP states are positioned at $m\sqrt{2\pi}+\epsilon$, where $m$ is any odd (even) integer for the peaks of $\ket{0_\Delta}_\mathrm{GKP}$ ( $\ket{1_\Delta}_\mathrm{GKP}$). Now, the effect of the entangling part of SBS is to rotate the qubit by $e^{i\sqrt{\frac{\pi}{2}}(m\sqrt{2\pi}+\epsilon)\sigma_\mathrm{x}}$. Thus, for an erroneous state $\ket{\psi}_\epsilon=\alpha\ket{0}_\epsilon+\beta\ket{1}_\epsilon$, where $\ket{0}_\epsilon,\ket{1}_\epsilon$ denote erroneous GKP states $\ket{0_\Delta},\ket{1_\Delta}$ with displacement error $\epsilon$, we have,
\begin{align}
   \mathcal{E}_{\hat x} \ket{\psi}_\epsilon\ket{g}&\approx e^{i\epsilon\sqrt{\frac{\pi}{2}}\sigma_\mathrm{x}}(\alpha\ket{0}_\epsilon\ket{g}+\beta\ket{1}_\epsilon\ket{e})\\&=-\alpha\ket{0}_\epsilon(\cos{\epsilon\sqrt{\pi/2}}\ket{g}+i\sin{\epsilon\sqrt{\pi/2}}\ket{e})\nonumber\\&\quad+\beta\ket{1}_e(\cos{\epsilon\sqrt{\pi/2}}\ket{e}+i\sin{\epsilon\sqrt{\pi/2}}\ket{g}).
\end{align}
We would like to remind the reader at this point that $\mathcal{U}_{\hat x}$, the second half of the circuit has a correction $e^{i2\lambda\hat p\sigma_\mathrm{y}}$ (with $\lambda=\alpha\Delta^2$) which depends on the final qubit state. The un-entanglement routine $\mathcal{U}_{\hat x}$ will correctly rotate the qubit state back to $\ket{g}$ for the first terms in each row. However, for the second term, where the qubit will be rotated to $\ket{e}$, the sign of the finite-energy correction is wrong. And hence, here the finite energy correction of $\mathcal{U}_{\hat x}$ will apply a back action of $e^{i4\lambda\hat p\sigma_\mathrm{y}}$. Thus, after $\mathcal{U}_{\hat x}$, the hybrid qubit-oscillator state takes the form,
\begin{align}       
&(\cos{\epsilon\sqrt{\pi/2}}\ket{g}+ie^{i4\lambda\hat p\sigma_\mathrm{y}}\sin{\epsilon\sqrt{\pi/2}}\ket{e})\alpha\ket{0}_\epsilon\nonumber\\&\quad-(\cos{\epsilon\sqrt{\pi/2}}\ket{g}+ie^{i4\lambda\hat p\sigma_\mathrm{y}}\sin{\epsilon\sqrt{\pi/2}}\ket{e})\beta\ket{1}_\epsilon\nonumber\\
&=(\cos{\epsilon\sqrt{\pi/2}}\ket{g}+ie^{i4\lambda\hat p\sigma_\mathrm{y}}\sin{\epsilon\sqrt{\pi/2}}\ket{e})(\alpha\ket{0}_\epsilon-\beta\ket{1}_\epsilon).
\end{align}
If $\epsilon=0$, this hybrid state is equal $(\alpha\ket{0}_\epsilon-\beta\ket{1}_\epsilon)\ket{g}=\ket{\bar\psi}_\epsilon\ket{g}$, where a deterministic logical Pauli operation has been applied to $\ket{\psi}$. Note that this Pauli operation can be tracked and hence does not play any role in stabilization. In the presence of error, as we can see, the probability for outcome $\ket{g}$ is not 1. At this point, if the qubit is measured, the probability of getting each possible outcome is, 
\begin{align}
   P_g&=\int_{-\infty}^\infty \ dp \ (\cos^2{(\epsilon\sqrt{\pi/2})}+\sin^2{(\epsilon\sqrt{\pi/2})}\sin^2{(4\lambda p)})\nonumber\\&\quad\times|\psi(p)|^2,\\
    P_e&=\int_{-\infty}^\infty \ dp \ \sin^2{(\epsilon\sqrt{\pi/2})}\cos^2{(4\lambda p)}|\psi(p)|^2.
\end{align}
Now, the back action in the event that the qubit is projected to state $\ket{g}$ is given by $W_{gg}$ (in the notation introduced in Sec.~\ref{sec:preliminaries}) for the SBS protocol in this case,
\begin{align}
&(\cos{(\epsilon\sqrt{\pi/2})}\mathrm{I}+i\sin{(4\lambda\hat p)}\sin{(\epsilon\sqrt{\pi/2})})\ket{\bar\psi}_\epsilon\\
&=(\mathrm{I}+i4\lambda\epsilon\sqrt{\pi/2}\hat p)\ket{\bar\psi}_\epsilon\nonumber\\&\quad +(\mathcal{O}(\epsilon^2)+\mathcal{O}(\lambda\epsilon^2)\hat p+\mathcal{O}(\lambda^2\epsilon)\hat p)\ket{\bar\psi}_\epsilon\\
&\approx e^{i\pi\epsilon\Delta^2\hat p}\ket{\bar\psi}_\epsilon= \mathrm{D}(-\pi\epsilon\Delta^2/2)\ket{\bar\psi}_\epsilon.
\end{align}
In the event of an error, this is a corrective displacement in the direction opposite to the error, as desired for the stabilization scheme. Thus, in the event of a $+1$ outcome, a single round of SBS partially corrects the error by applying a displacement of $-\pi\epsilon\Delta^2/2\approx-0.18\epsilon$ for $\Delta=0.34$. 

Similarly, the back action in the event when the qubit is projected to state $\ket{e}$ is given by $W_{eg}$ for the SBS unitary in this case,
\begin{align}
\cos{(4\lambda\hat p)}\ket{\bar\psi}_\epsilon
\end{align}
This back action is independent of the error parameter $\epsilon$, however, the probability of outcome increases with $\epsilon$. This operator applies a symmetrized displacement along the position axis. In the momentum basis, this is equivalent to a cosine envelope on the state. To second order in $\hat p$, the cosine envelope can be approximated as a Gaussian,
\begin{align}
    \cos{4\lambda\hat p}\approx e^{-8\lambda^2\hat p^2},
\end{align}
where $8\lambda^2=8\alpha^2\Delta^4=\pi\Delta^4$. We compare this envelope correction with the momentum part of the target Gaussian envelope $e^{-\Delta^2\hat n}=\exp{[-\Delta^2(\hat x^2+\hat p^2)]}$ and note that an $e$ outcome reduces the momentum uncertainty by $15\%$. Currently, we do not have an intuitive explanation for how this back action supports error correction or stabilization. The probability of both the back actions increases with an increase in the error $\epsilon$, as should be the case for any autonomous error correction scheme. Importantly, the probability of correction is maximum at $\epsilon=\sqrt{\pi/2}$, at half the distance of the GKP code. At $\epsilon\ge \sqrt{2\pi}$ the displacement error causes a logical error in the GKP subspace and hence the stabilization scheme seems to slow down the correction as the error approaches this value. Similarly, the stabilization of the logical $X$ basis corrects displacement errors and envelope errors along the momentum axis.

In summary, we have shown that $\mathrm{SBS}$ is an example of probabilistic (or autonomous) error correction. 

\section{Logical Readout with Correctable Errors}\label{ssec:logical-readout}
\begin{figure*}
    \centering
    \includegraphics[width=\linewidth]{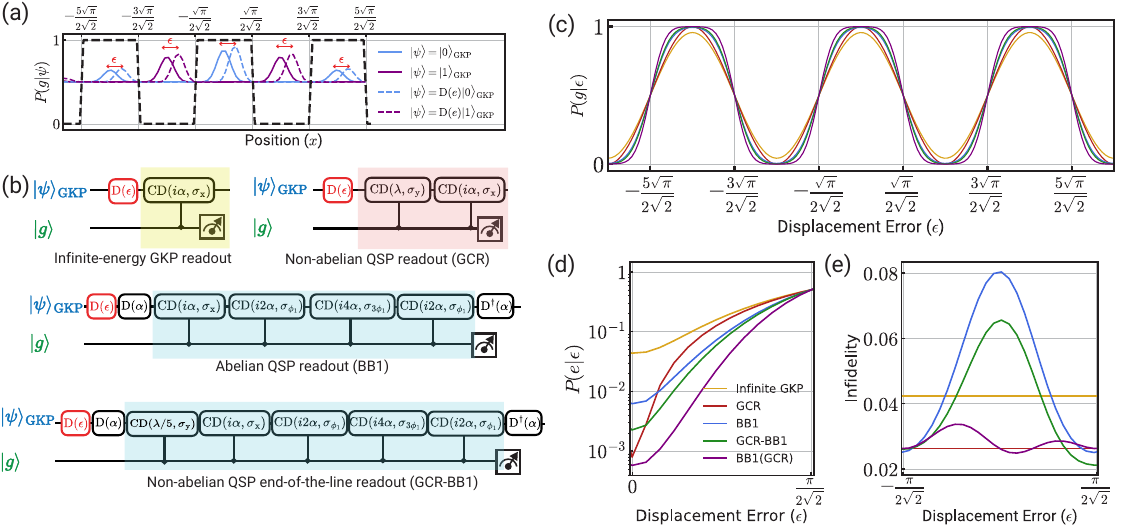}
    \caption[Logical Readout of GKP qubits with correctable errors.]{\textbf{Logical readout of GKP states with correctable errors.} The GKP readout procedure maps the logical state onto the ancilla qubit states $g$ and $e$, which are then measured.
\textbf{(a)} Solid curves show finite-energy GKP codewords; dotted curves show the corresponding displaced codewords. An ideal end-of-sequence pulse for reading out a state with a correctable displacement error $\epsilon$ is a square wave (black dashed line). It yields a correct measurement probability $P(g|0,\epsilon)$ (for $\sigma_\textrm{z}$ measurement) across all logical $\ket{0,\epsilon}$ states with $|\epsilon|<\sqrt{\pi/2}$, and similarly for $P(e|1,\epsilon)$ for $\ket{1,\epsilon}$.
\textbf{(b)} GKP readout schemes: the infinite-energy scheme corrects only peak locations; the non-abelian QSP readout (also found numerically in Ref.\cite{hastrup2021improved}) corrects Gaussian broadening; and the abelian BB1(90) QSP sequence corrects peak location ambiguity by flattening the cosine response.
\textbf{(c)} Readout probability of getting the right outcome $P(g|0,\epsilon)$ as a function of displacement error $\epsilon$ for different schemes: infinite-energy (no correction), GCR (Gaussian correction), and BB1 (peak location correction). Color legend is in (d).
\textbf{(d)} Zoom-in of (c) shows $1-P(g|0,\epsilon)$ within the Voronoi cell of $\ket{0}_\mathrm{GKP}$. Here, GCR-BB1 denotes enhancing BB1(90) with a conditional displacement, while BB1(GCR) corresponds to the composed sequence derived in Eq.(\ref{eq:GCR-BB1}). In BB1(GCR), the blue region circuit is replaced according to this equation. These sequences correct Gaussian uncertainty while producing a square response. Parameters $\alpha$ and $\lambda$ follow Fig.~\ref{fig:GKp-framework}.
\textbf{(e)} Readout infidelities: color coding matches (b,d). BB1 and GCR-BB1 show the highest infidelities, while finite-energy readout and BB1(GCR) achieve the lowest, mainly limited by envelope errors that could be further stabilized. All curves correspond to a GKP state width set by $\Delta=0.34$.}
    \label{fig:bb1-readout}
\end{figure*}
At the end of any quantum circuit, the logical qubits must be measured to determine their state.  In so-called dynamics circuits, such measurements may occur in the middle of the circuit so that a program branching decision (`measurement and feed forward') can be made.  Inevitably, there can be residual correctable errors in the logical code states being measured. For the GKP encoding, such errors might be correctable random displacements, that is, $\sqrt{\pi}/2\sqrt{2}$. See the dashed curves in Fig.~\ref{fig:bb1-readout}(a). These errors can lower the measurement fidelity since the ideal and finite-energy readout sequences are optimal at $\epsilon=0$ only. For instance, Fig.\ref{fig:bb1-readout}(c) shows that the probability of correct measurement decays as a cosine for infinite-energy readout, and similarly for finite-energy readout, following the circuits in Fig.\ref{fig:bb1-readout}(b). Fig.~\ref{fig:bb1-readout}(a) illustrates the ideal square-wave pulse sequence (black dashed line) achievable with correctable errors. In this section, we use QSP sequences to approximate this ideal square response.

The problem at hand is described by Fig.~\ref{fig:bb1-readout}(a) which suggests that we need to extract the one bit of information $\{0_\mathrm{GKP},1_\mathrm{GKP}\}$ where $0_\mathrm{GKP}$ ($1_\mathrm{GKP}$) corresponds to all states in the Hilbert space which are closer to logical $\ket{0}_\mathrm{GKP}$ ($\ket{1}_\mathrm{GKP}$) than to the opposite logical state. This can be understood from the readout sequence for infinite-energy states, and the argument carries over to all schemes. The infinite-energy sequence for logical $Z$ measurement is described as follows
\begin{align}
    \mathrm{CD}(i\alpha,\sigma_\mathrm{z})\ket{\psi}_\mathrm{GKP}\otimes\ket{+}&=\frac{1}{2}(
    \mathrm{D}(i\alpha)\pm
    \mathrm{D}(-i\alpha))\ket{\psi}_\mathrm{GKP}\otimes\ket{\pm})\\
    &=\frac{1}{2}\mathrm{D}(i\alpha)(
    \mathrm{I}\pm
    \mathrm{D}(-i2\alpha))\ket{\psi}_\mathrm{GKP}\otimes\ket{\pm})\\
    &=\frac{1}{2}\mathrm{D}(i\alpha)(
    \mathrm{I}\pm
    e^{-i4\alpha\hat x})\ket{\psi}_\mathrm{GKP}\otimes\ket{\pm})\\
    &=\mathrm{D}(i\alpha)\frac{
    \mathrm{I}\pm
    \mathrm{Z}_\mathrm{GKP}}{2}\ket{\psi}_\mathrm{GKP}\otimes\ket{\pm}).
\end{align}
Thus, this sequence applies a projective measurement modulo a displacement by $|\alpha|$. However, for this infinite-energy GKP case, the readout fidelity follows a cosine curve (shown in panel (c) of Fig.~\ref{fig:bb1-readout}) as a function of displacement $\epsilon$. The fidelity value for the no-error case is less than unity when the infinite-energy readout scheme is applied to the finite-energy GKP state (see Fig.~\ref{fig:bb1-readout}). The finite-energy readout, described in Sec.~\ref{ssec:GKLP-framework}, yields a readout fidelity that follows a similar curve with a peak value (case of no error) closer to unity. In the presence of displacement errors $\epsilon$, the qubit is rotated to a different basis other than the $\sigma_\mathrm{z}$ basis. Thus, for varying $\epsilon$, the curves resemble a cosine. However, in the case of such residual (yet correctable) errors, we would like a readout sequence that yields a square wave response shown by the dotted black lines in Fig.~\ref{fig:bb1-readout}(a).

\paragraph{Abelian QSP readout:} We can achieve something close to the required square wave using our adaptation of BB1(90) in Eq.~(\ref{eq:BB1}) for hybrid oscillator-qubit control. See circuit in Fig.~\ref{fig:bb1-readout}(b). This circuit yields improved readout fidelity for all correctable error states compared to the finite-energy readout circuit, as shown by Fig.~\ref{fig:bb1-readout}(c). A zoomed-in version is shown in Fig.~\ref{fig:bb1-readout}(d) to assess the situation in the no error case of $\epsilon=0$. Note that the Helstrom bound for the given value of $\Delta=0.34$ is $\frac{1}{2}\big(1-\sqrt{1-|\braket{0|1}|^2}\big)\approx 10^{-4}$, far below the BB1 protocol readout error at $\epsilon=0$. Hence, we do not need to account for the non-orthogonality of the GKP logical Pauli states. The abelian end-of-the-line sequence works desirably well and is relatively robust for $\epsilon\neq 0$ case. However, its performance for the no error case $\epsilon=0$ is worse than the finite-energy readout sequence. 

\paragraph{Non-abelian end-of-the-line readout:} 
We now design an end-of-sequence correction that also accounts for finite-energy effects. However, its performance for the no error case $\epsilon=0$ is worse than the finite-energy readout sequence. Thus, it might be a good idea to think of a readout sequence for correctable errors where these finite-energy corrections are also taken care of. This is the exact problem we solved in Chapter~\ref{chapter:na-qsp} while composing GCR into BB1. As shown above, this routine works on par with the finite-energy correction. We use the $\mathrm{BB1}(\mathrm{GCR}(90))$ composite pulse with $|\alpha|=\sqrt{\pi}/2$ to achieve this optimal readout sequence. For low enough envelope size of $\Delta=0.34$, used for GKP experiments, this sequence is not much longer compared to the $\mathrm{BB1}(90)$ pulse. Thus, our non-abelian QSP sequence also gives a better GKP readout scheme for states with correctable errors while GCR-BB1 performs better than BB1 but worse than GCR(BB1) at the no error case of $\epsilon=0$. 

However, the sequence still requires four additional gates with amplitude proportional to $\Delta^2$. Therefore we have also studied another sequence in which a single conditional displacement is prepended to BB1. The amplitude of this GCR-type correction ($\lambda/4$) is optimized numerically to take into account the finite-energy corrections of the four rotations in BB1 collectively. This sequence is termed GCR-BB1 in the figure above. Fig.~\ref{fig:bb1-readout} shows that BB1(GCR) is the best sequence among all readout sequences given correctable errors and finite-energy code words.

\paragraph{Back-action:} Note that during each readout sequence, the GKP state is displaced by an amount $|\alpha|$ along the quadrature orthogonal to the one being measured. However, this displacement is deterministic and can be accounted for. Thus, accommodating for this displacement, we compute the fidelity of the resulting state with the initial erroneous state. A large infidelity would indicate worse back action from the respective circuit. See Fig.~\ref{fig:bb1-readout}(d) for the back action of all readout schemes discussed in this section. Note that ideally for an end-of-the-line readout sequence, whereafter the state will not be used again, we do not care about the back action on the state. However, if the back action is not very bad, the readout scheme can be repeated to further increase the measurement fidelity.

Given that the non-abelian end-of-the-line readout requires a larger circuit depth than the abelian readout, the theoretical improvement in fidelity may be difficult to realize in practice.  We conjecture that the circuit depth could be reduced if we squeeze the oscillator quadrature that is being read out. If the squeezing parameter is $r$, squeezing will make all required conditional displacements smaller by a factor of $e^{-r}$. However, squeezing itself is a time-consuming process and could induce more errors. The question is if the reduction in ancilla errors during the shorter readout circuit overcomes the increase in oscillator errors during squeezing. This process requires a larger truncated Hilbert space to be simulated, and hence we have not numerically tested this idea.

\section{Universal Qubit Rotations: Pieceable Gate Teleportation}\label{ssec:piecewise-teleportation}
\begin{figure*}
    \centering
    \includegraphics[width=\linewidth]{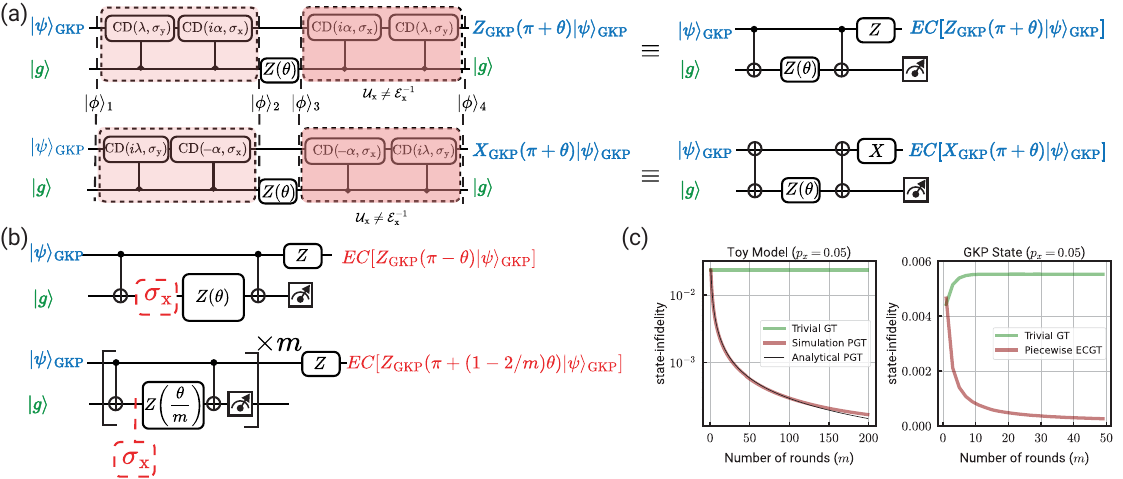}   \caption[Error-corrected GKP gate teleportation.]{\textbf{Error-suppressed GKP gate teleportation.} \textbf{(a)} Error-corrected gate teleportation of logical $Z(\theta)$ ($X(\theta)$) gate by an entangling-un-entangling sequence obtained from stabilizer of the logical $\{0,1\}$ ($\{+,-\}$) basis. \textbf{(b)} Toy model of a pieceable circuit to mitigate effects of biased-noise ancilla errors. \textbf{(c)} Comparison of trivial gate teleportation and pieceable gate teleportation for $\theta=\pi/4$ in the presence of ancilla errors for toy model (left) and GKP states (right). Here, the dotted line in the left panel presents the analytical curve for the state infidelity in the case of pieceable teleportation obtained from Eq.~(\ref{eq:PCGT}). The GKP state used for simulation results shown in the right panel simulation has an envelope size of $\Delta=0.34$. Note that, for the trivial gate teleportation, in the case of GKP, we teleport the gate at $m=1$ while applying stabilization for $m-1$ rounds.}
    \label{fig:GKP_Teleportation}
\end{figure*}

Arbitrary operations on finite-energy GKP states are generally not easily available. Recall that Pauli operations are simple phase space translations. However logical rotations are exponentials of Pauli's which (by the Pauli-Euler identity) can be written as a coherent superposition of identity and a phase space translation. Unfortunately, it is not possible to apply a classical control pulse that is in a superposition of zero amplitude (to achieve identity) and non-zero amplitude (to achieve the phase space displacement.  One method to circumvent this problem is to use ancillary systems to teleport gates into the GKP codespace by use of conditional displacements. However, this method limits the logical error due to the physical errors of the ancilla. Here, we devise a technique to teleport gates while simultaneously correcting errors in the oscillator. We devise a pieceable gate teleportation circuit that is protected against ancilla decay errors, yielding room for high-fidelity gate operations using a biased-noise ancilla. This protection against errors is different from the general path-independent mechanisms engineered for circuits using SNAP gates~\cite{ma2020path,ma2022algebraic,ReinholdErrorCorrectedGates}. Our construction does not require any hardware engineering feats such as chi matching~\cite{ReinholdErrorCorrectedGates,xu2024fault}(though these might yield further improvements). 
\paragraph{Error-corrected gate teleportation.}
The error-corrected gate teleportation sequence is constructed by realizing that the stabilization circuit is composed of entangling and unentangling gadgets, $\mathrm{SBS}=\mathcal{E}\mathcal{U}$. See Fig.~\ref{fig:GKp-framework}. By introducing a qubit gate in between the two gadgets, we can construct a phase-transfer circuit, as illustrated in Fig.~\ref{fig:GKP_Teleportation}(a). For logical $Z(\theta)$ gate, the circuit obeys the following equations, up to a global phase,
\begin{align}
\ket{\phi}_2&=\mathcal{E}_\mathrm{x}\ket{\phi}_1=\mathcal{E}_\mathrm{x}(a\ket{0}_\mathrm{GKP}+b\ket{1}_\mathrm{GKP})\otimes\ket{g}\\
    &~~~~~~~~~~~~~~~~=a\ket{0}_\mathrm{GKP}\otimes\ket{g}-b\ket{1}_\mathrm{GKP}\otimes\ket{e}\\
    \ket{\phi}_3&=Z(\theta)\ket{\phi}_1=e^{-i\frac{\theta}{2} \sigma_\mathrm{z}}\ket{\phi}_2=a\ket{0}_\mathrm{GKP}\otimes\ket{g}-e^{i\theta}b\ket{1}_\mathrm{GKP}\otimes\ket{e}\\
    \ket{\phi}_4&=\mathcal{U}_\mathrm{x}\ket{\phi}_2=a\ket{0}_\mathrm{GKP}\otimes\ket{g}-e^{i\theta}b\ket{1}_\mathrm{GKP}\otimes\ket{g}\\
    &~~~~~~~~~~~~~~~~=[Z(\pi+\theta)_\mathrm{GKP}(a\ket{0}_\mathrm{GKP}+b\ket{1}_\mathrm{GKP})]\otimes\ket{g}.
\end{align}
To perform logical $X(\theta)$ ($Y(\theta)$) rotations, one only needs to entangle the qubit with the logical $X$ ($Y$) eigenstates of the GKP code. This circuit corresponds to the stabilization of a different stabilizer operator. For example, the same circuit becomes a logical $X(\theta)$ gate if $\mathcal{E}_\mathrm{x}\rightarrow \mathcal{E}_\mathrm{p}, \ \mathcal{U}_\mathrm{x}\rightarrow \mathcal{U}_\mathrm{p}$. Similarly, for logical $Y(\theta)$ gate we have, $\mathcal{E}_\mathrm{x}\rightarrow \mathcal{E}_\mathrm{x+p}, \ \mathcal{U}_\mathrm{x}\rightarrow \mathcal{U}_\mathrm{x+p}$. The circuit can be easily changed to use ancilla $X(\theta)$ gates, \textit{mutatis mutandis}, if this is an easier gate for the DV ancillary system. This is the so-called phase transfer circuit.

If we ignore qubit errors, these circuits will yield $\ket{g}$ corresponding to the desired gate operation. The fidelity of these gates in the absence of any physical errors is $99.88\%$ while the probability of a successful logical gate operation is $0.9994$. These numbers are not unity due to SBS being correct only up to first order in $\Delta^2$. This is related to the fact that while devising GCR we ignored $\mathcal{O}(p^2)$ terms (see Sec.~\ref{sec:GCR}). \par 
\paragraph{Trivial gate teleportation:} Note that the ``error-corrected" feature of our sequence comes from the fact that it is constructed from the logical identity,
$\mathcal{E}_\mathrm{x/p/x+p}\mathcal{U}_\mathrm{x/p/x+p}$,
that is the SBS stabilization circuit, unlike trivial gate teleportation (where $\mathcal{U}_\mathrm{x/p/x+p}= \mathcal{E}_\mathrm{x/p/x+p}^{-1}$) where the two gadgets would have formed a universal identity. That is, the error-corrected gate teleportation circuit, in addition to applying the logical gate on logical GKP codewords, also applies a corrective back action on erroneous GKP states similar to the small-big-small stabilization circuits discussed in Sec.~\ref{ssec:GKLP-framework}.
\paragraph{Protection from biased-ancilla errors:} Our teleportation circuit, however, is severely affected in the presence of ancillary errors (the same as any teleportation circuit). For our circuit, ancilla errors tend to occur during the (relatively long duration) controlled displacement gates acting on the cavity. In the case of a biased-noise ancilla, the circuit will only be affected by one type of ancillary error since the other errors are largely suppressed~\cite{grimm2020stabilization,ding2024quantum}. For protection against a single type of ancillary error, we propose the pieceable circuit shown in Fig.~\ref{fig:GKP_Teleportation}(b). For example, without loss of generality, let us imagine a biased-noise ancilla where the dominant error is $\sigma_\mathrm{x}$. The effect of this gate is shown in the top circuit of Fig.~\ref{fig:GKP_Teleportation}(b). The $\sigma_\mathrm{x}$ error on the ancilla propagates to the oscillator as a logical rotation angle error of $2\theta$ on the GKP state. We propose to solve this problem by dividing the circuit into $m$ pieces where each piece applies a rotation by $\mathrm{Z}(\theta/m)$ as shown in the lower panel of Fig.~\ref{fig:GKP_Teleportation}(b). In this case, a single $\sigma_\mathrm{x}$ error will reduce the effect on the logical fidelity with increasing $m$. This circuit performs a random walk such that the average rotation of the gate is $\theta(1-2p_\mathrm{x})$ where $p_\mathrm{x}$ is the probability of $\sigma_\textrm{x}$ errors. The standard deviation of the rotation angle of the gate produced by this random walk after $m$ steps is $\sigma_m=\frac{|\theta|}{\sqrt{m}}2\sqrt{p_\mathrm{x}(1-p_\mathrm{x})}$ which becomes small for large $m$. If the standard deviation increases at a slower speed compared to the decrease in fidelity, we get an overall increase in the fidelity of the output state. After $m$ pieces of the circuits with rotations $\theta/m$, the state is rotated by the mean angle $\theta^\prime=\theta(1-2p_\mathrm{x})$. The fidelity of the resulting state with a state rotated by $\theta^\prime$ is given by,
\begin{align}
    \mathcal{F}=\sum_{k=0}^m {m\choose k}  (1-p_\mathrm{x})^{m-k}p_\mathrm{x}^k \cos^2{\Big[\theta(p_\mathrm{x}-k/m)\Big]}.\label{eq:PCGT}
\end{align}
We assume the fidelity for pure states is $\cos^2(\theta(1-2p_x)-\theta_k)$ where $\theta_k$ is the achieved rotation angle when $k$ bit flip errors occur. The systematic error in the mean rotation angle can be compensated by choosing to use angle $\theta^\prime = \theta/(1-2p_\mathrm{x})$. 
%\begin{align}
%    \mathcal{F}=\sum_{k=0}^m {m\choose k}  (1-p_\mathrm{x})^{m-k}p_\mathrm{x}^k \cos^2{\Big[2\theta^\prime(p_\mathrm{x}-k/m)\Big]},\label{eq:PCGT}
%\end{align}
 This calculation for the toy model assumes that the states are pure for analytical understanding. The metric used in the simulation of GKP states is the state fidelity computed using QuTiP~\cite{Johansson2013}\footnote{$\mathrm{Tr}(\sqrt{\rho_A}\rho_B\sqrt{\rho_A})$ for the density matrices $\rho_A,\rho_B$ of mixed states.}. The curves in Fig.~\ref{fig:GKP_Teleportation}(c) show that the decrease in infidelity is proportional to $1/m$. In the toy model, we apply errors only just before the $\mathrm{CZ}_\mathrm{GKP}$ gates with probability $p_x=0.05$ to emulate the case of GKP states where the $\mathrm{CD}$ gates are longer and more erroneous compared to the qubit rotation $Z(\theta)$. In the GKP simulation, we apply a $\sigma_\mathrm{x}$ error at a rather large rate of $\gamma\sim 1/22\mu\textrm{s}^{-1}$ during all conditional displacements (to emulate the probability $p_x=0.05$ during the large conditional displacements). We use the metric that a conditional displacement by a magnitude of $1$ takes time $\tau=1\mu\textrm{s}$ as outlined in App.~\ref{app:squeezing} and use $\tau$ to quote error probabilities.

We compare the method just described above against single-shot trivial gate teleportation followed by $m-1$ stabilization rounds in Fig.~\ref{fig:GKP_Teleportation}(c). The initial bump in the infidelity is due to the uncorrected gate teleportation step. The error introduced in this step is not corrected with further stabilization steps since it is a logical error. Note that, in the case of trivial gate teleportation, one could use the measurement outcome to check for ancilla errors more efficiently\footnote{the corrective back action renders the measurement outcomes less useful in terms of detecting errors.}, but that would make the protocol reliant on measurements which can be the slowest (or, most erroneous) part of the circuit. The pieceable circuit is not applied to the trivial gate teleportation since this circuit does not stabilize the GKP states. Thus, prolonged exposure to the trivial gate teleportation will decrease fidelity due to ancilla errors. However, as can be seen, this is not the case for our error-corrected gate teleportation (ECGT). Despite applying ECGT for multiple pieces ($m$), the logical error does not just stay constant but decreases. This indicates that ECGT has an error-correcting property. Thus, we have proven here that pieceable gate teleportation is a more efficient method to apply autonomously error-resilient single-qubit gate rotations in the presence of errors with a biased-noise ancilla. 

Just as in the stabilization circuit, the qubit is reset to $\ket{g}$ at the end of every piece in the circuit. This reset could be erroneous and this error has not been accounted for explicitly in our simulations. If the reset leaves the qubit in state $\ket{e}$, it has the same effect as a $\sigma_\mathrm{x}$ error on the ancilla at the beginning of the first conditional displacement. Since such an error has been accounted for, it indicates that if the total effect of ancilla errors during reset and  conditional displacements is low enough, we will see an improved fidelity for the pieceable error-corrected gate teleportation circuit.

\section{Entangling GKP Qubits: Extension of $\mathrm{GCR}$ to Multi-Modal Operations}
The two-qubit gates suggested in~\cite{gottesman2001encoding} for an infinite-energy GKP code have poor fidelity for the finite-energy code and require a few stabilization rounds to improve the error rate~\cite{rojkov2023two}. 
\paragraph{Single-qubit-ancilla:} For the finite-energy states, Ref.~\cite{rojkov2023two} derived the finite-energy version of the two-qubit entangling gate. This circuit, the same as the stabilization circuit, can also be derived using the non-abelian QSP extension for two modes. We discuss this extension here. For ideal GKP codes with support at positions $m\sqrt{\pi/2}, m\in \mathbb{Z}$, the conditional SUM gate displaces the second mode by the position $\pm \hat x$ of the first mode with the sign of the displacement determined by the state of the ancilla.  Equivalently, for each pair of peaks of the two GKP states, in the position (first mode) and momentum basis (second mode), respectively, the conditional SUM gate ($e^{i2\hat x\otimes\hat p}$) rotates the qubit via {$e^{il\pi \sigma_\mathrm{x}}$} by angle $2l\pi$, where $l$ is the product of the two integers defining the positions of the two peaks. This operation is equivalent to $(-1)^l \mathrm{I}$ on the joint oscillator-qubit state, and it applies a $\mathrm{CX}_\mathrm{GKP}$ gate on the two logical GKP codewords with the qubit going back to the original state ($\ket{g}$, in this case). The non-abelian correction for the entangling half of this operation $e^{i(\hat x\otimes\hat p)\otimes\sigma_\mathrm{x}}$  due to the envelope size $\Delta\neq 0$ with respect to the first (second) GKP qubit is given by, $e^{i\Delta^2\hat p\otimes\hat p\otimes \sigma_\mathrm{y}}(e^{-i\Delta^2\hat x\otimes\hat x\otimes \sigma_\mathrm{y}})$, assuming the ancilla starts in state $\ket{g}$. Thus, the GCR-type pre-correction due to both modes will be equal to,
\begin{align}
    S\equiv e^{-i\frac{\Delta^2}{2}(\hat x\otimes\hat x-\hat p\otimes\hat p)\otimes \sigma_\mathrm{y}}\label{eq:corr_sum}
\end{align}

The corresponding SBS-type circuit, where $B\equiv e^{i(2\hat x\otimes\hat p)\otimes\sigma_\mathrm{x}}$ and $S$ is given by Eq.~(\ref{eq:corr_sum}), will have rotated the qubit by an angle of $2\pi$ about $\sigma_\mathrm{x}$. In doing so, however, the SUM gate applies a logical controlled Pauli operation, just as the SUM gate applies a logical Pauli operation. This operation, the same as the stabilization circuit $\mathrm{SBS}$, is protected against biased-noise ancilla errors.  However, the fidelity of the Bell states prepared using this circuit is $\sim 0.90$ for $\Delta=0.34$ which indicates that this gate requires more terms in the non-abelian correction to reach higher fidelities. 

The fast conditional two-mode operations required for this operation can be achieved using Gaussian operations and weak dispersive coupling between the oscillator and qubit. Ref.~\cite{rojkov2023two} suggests that this sequence takes more gates (5 conditional two-mode operations equivalent to a SUM gate). However, we claim that this circuit can be achieved in three gate sequences given we can obtain fast $e^{i\frac{\Delta^2}{2}(\hat x\otimes \hat x-\hat p\otimes\hat p)\otimes\sigma_\mathrm{y}}$ gates using a two-mode extension of the echoed conditional displacement~\cite{campagne2020quantum,eickbusch2022fast,sivak2023real}. This decomposition is given in Ref.~\cite{ISA} for entangling oscillator gates in the weak dispersive regime, and we present it in App.~\ref{app:finite-SUM} in the context of two-qubit GKP operations. 

Fast echoed conditional displacement was obtained using (weak) dispersive coupling in the displaced frame. Similarly, we can obtain a fast echoed conditional SUM gate using (weak) dispersive coupling in a two-mode squeezed frame. The two-mode squeezing required for this operation can be obtained from single-mode squeezing and beam-splitters using Bloch-Messiah decomposition~\cite{ISA}. Thus, this circuit involves two single-mode squeezing operations and two beam-splitters to go to the two-mode squeezing frame using TMS$(\alpha,\pi)$ (see definition in Ref.~\cite{ISA}). In this new frame, the circuit uses evolution under the dispersive interaction in this frame for the duration $t_\mathrm{CX}\ge \frac{\Delta^2}{\chi\sinh{2\alpha}}+\frac{2\Delta^2}{\chi}$ where $\chi$ is the strength of the weak dispersive coupling. Thus, by increasing $\alpha$ one can make this process much faster. Hypothetically, $\lim_{\alpha\rightarrow \infty}t_\mathrm{CX}\ge\frac{2\Delta^2}{\chi}$ is allowed, however, in reality, we are limited to finite and much lower values of $\alpha$ due to unwanted state transitions in a nonlinear ancilla dispersively coupled to a resonator (oscillator)~\cite{ding2024quantum,eickbusch2022fast,sivak2023real}. 
\begin{figure}[t]
    \centering
    \includegraphics[width=\linewidth]{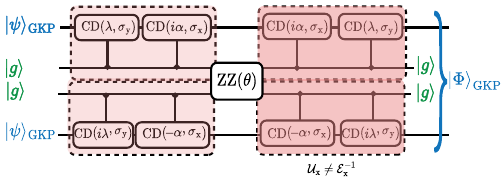}
    \caption[Pieceable GKP entangling operations.]{\textbf{Pieceable GKP entangling operations.} Circuit for logical Pauli operations using two ancillae. Here, $\ket{\Phi}=EC[\mathrm{ZZ}_\mathrm{GKP}(\pi+\theta)\ket{\psi,\psi}_\mathrm{GKP}]$. For $\theta=\pi/4$, we can prepare a Bell state if $\ket{\psi,\psi}_\mathrm{GKP}=\ket{++}_\mathrm{GKP}$. The fidelity of this Bell state preparation is $0.9997$ (as per Sec.~\ref{ssec:piecewise-teleportation}) and the success probability of the qubit outcome $\ket{g,g}$ is $P_g=(0.9993)^2=0.998$.}
    \label{fig:two-qubit-GKP}
\end{figure}

\paragraph{Two qubit ancillae:} Our alternative circuit using two-qubit ancillae does not have the issue of engineering conditional Gaussian operations. This circuit only uses conditional displacements and qubit gates (again, protected from biased-noise ancilla error piecewise gate teleportation). Essentially it is the error-corrected gate teleportation with two pairs of GKP oscillator state-DV qubit combinations. This circuit yields better fidelity compared to the single qubit ancillae scheme, in the absence of errors. It is similar to Fig.~\ref{fig:GKP_Teleportation}(a), except the middle gate can be a controlled Pauli operation between the two DV qubits given the initial states and final operation to be implemented. See App.~\ref{app:c-pauli}. However, this circuit cannot be converted into a pieceable circuit to yield high-fidelity gates in the presence of ancilla errors.

Nevertheless, we can achieve pieceable circuits for two-qubit logical Pauli rotations $\mathrm{P}_i\mathrm{P}_j(\theta)$ ($\mathrm{P}_i,\mathrm{P}_j\in\{\sigma_\mathrm{x},\sigma_\mathrm{y},\sigma_\mathrm{z}\}$). Fig.~\ref{fig:two-qubit-GKP} gives the circuit construction for $P_iP_j(\theta)=\mathrm{ZZ}_\mathrm{GKP}(\theta)$. These gates are inspired by the single-qubit case discussed in Sec.~\ref{ssec:piecewise-teleportation} and shown in Fig.~\ref{fig:GKP_Teleportation}(a), and thus, we can use the same arguments to perform two-qubit rotations in any basis, using the entangling and unentangling gadgets which apply SBS on both GKP qubits. Thus, we have achieved two-qubit universality with error-corrected gate teleportation. 

The pieceable version of this circuit is an error-corrected two-qubit entangling operation that is autonomously protected from ancilla errors and photon loss. In the absence of errors, this sequence yields a success probability of $P_g=0.9987$. We believe that the state fidelity of this process will be the same as the single-qubit gate teleportation scheme. For example, computing the fidelity of states prepared by SBS, for a system with two truncated oscillators (each with Hilbert space dimension of $50$) and two qubits, is $0.98$ (upper bounded by truncation issues and not the protocol). We achieve the same fidelity for the Bell pair constructed using our teleportation protocol. This fidelity is much lower than what is expected ($0.998$ from Sec.~\ref{ssec:piecewise-teleportation}), and we believe this is due to the Hilbert space constraints. The logical error probability increases with the use of two-qubit DV gates while it remains the same as the error in conditional displacement increases, same as the case of single-qubit gate teleportation analyzed in Sec.~\ref{ssec:piecewise-teleportation}.

In addition, this operation only requires us to perform fast conditional displacements, a combination of weak dispersive coupling, unconditional displacements, and two-qubit ancilla rotations. Thus, the circuit can be more feasible compared to the combination of weak dispersive coupling and unconditional two-mode Gaussian operations. Even though Gaussian operations are interpreted to be an easy resource for oscillators, not much work has been done to improve the fidelity of non-number-preserving operations like two-mode and single-mode squeezing. Thus, in this manuscript, we have only considered unconditional displacements, phase space rotations, beam-splitters, and ancilla qubit rotations as the set of instructions for two-mode phase space ISA. Note that these operations have been demonstrated with high fidelity in hybrid oscillator-qubit systems~\cite{campagne2020quantum,eickbusch2022fast,chapman2023high,lu_high-fidelity_2023}. 
\section{Generalization to Multi-Mode GKP Codes and GKP Qudits}\label{ssec:GKP-extension}
Our high-fidelity control operations will not only be instrumental in realizing multi-mode operations between various GKP qubits but also for the stabilization of multi-mode GKP codes. GKP qudits are encoded in an oscillator by changing the lattice spacing of the support (increasing the unit cell size in phase space to accommodate more than 2 code states) \cite{brock2024quantum}.
\paragraph{Error-detected state preparation:} The state preparation works by changing $\alpha$ to the required lattice spacing for qudits or an alternative qubit lattice.
\paragraph{Error-corrected gate teleportation:} The pieceable gate teleportation is also extendable since these gates are derived from the $\mathrm{GCR}$ interpretation of the stabilization routine. The stabilization circuits are extendable to any qubit or qudit lattice by choosing appropriate $\alpha$ in the same circuit. Thus, the same argument extends all our gate teleportation circuits to arbitrary qubit and qudit lattices.
\paragraph{High-fidelity logical readout:} The logical readout schemes for arbitrary lattices again follow from a change in $\alpha$. However, for efficient qudit readout circuits that take the least amount of time, we need access to DV qudits. For example, each circuit used in Fig.~\ref{fig:bb1-readout} can be extended to qudits using an ancilla qudit of the same dimension for readout. This extension may not have the same readout fidelity and is left for future work to analyze.

\section{Open Problem: Ancilla-Error-Transparent Protocols}
A promising direction towards fault-tolerant error-corrected control using the phase space instruction set is to engineer a conditional displacement gate that is transparent to ancilla errors and/or raises a flag in the presence of one. One popular strategy involves using a qutrit ancilla such that an ancilla decay raises a flag with qutrit in the $e$ state. This idea is inspired by the work that first appeared in Ref.~\cite{ReinholdErrorCorrectedGates} in the context of SNAP gates. In this technique, an ancilla error not only raises a flag but also leaves the oscillator unchanged in the event of a single error. This theory was more rigorously formulated in Refs.~\cite{ma2020path,ma2022algebraic}. In this open problem section, we lay out the problem in directly extending these methods to the case of logical gates for GKP states.

Natively, in superconducting circuits, conditional displacements are implemented using an echoed-conditional displacement Hamiltonian $H_\mathrm{CD}$~\cite{campagne2020quantum}, in units of $\hbar=1$,
\begin{align}
    H_{\mathrm{CD}}=-\frac{\chi}{2} \hat a^{\dagger} \hat a\sigma_{\mathrm{z}}-(\alpha(t) \hat a^{\dagger}+\alpha(t)^{*} \hat a)\sigma_\mathrm{z}-|\alpha(t)|^2\sigma_\mathrm{z}.
\end{align}
 To enable flags, we could modify this Hamiltonian to consider transitions directly between $\ket{g}$ and $\ket{f}$ levels of the transmon, such that $\sigma_{\mathrm{z}}=\mathrm{diag}[\chi_g,\chi_e,\chi_f]$. Thus, the choice of $\chi_g=-\chi_f=\chi_e=\chi$ (known as chi-matching) could yield an error-detectable gate which raises flag in the event of an ancillary error. In addition to the flag, the oscillator experiences a deterministic unconditional displacement depending on $\alpha,\chi$. In the context of this thesis, we ask,
\begin{myframe}
\singlespacing
\begin{quote}
How would such a protection from single ancilla decay perform if we replace a biased-noise ancilla for the pieceable protocols established in this chapter with a qutrit ancilla in the presence of chi matching?
\end{quote}
\end{myframe}

\doublespacing

    \chapter{Applications and Future Directions} \label{chapter:conc}

%oscillator-based applications
In this chapter, we discuss an application of CV codes for reduction in the resource overhead of fault-tolerance quantum computing. In particular, we focus on the reduction of resource overhead using oscillator codes. We describe the protocol in our work~\cite{singh2022high}, designed to prepare high-fidelity magic states, as an important resource for fault-tolerant logical non-Clifford operations. The improvement in this section is based on biased-noise qubits which can realize bias-preserving $\mathrm{CX}$ gates, a unique feature of CV systems~\cite{puri2020bias}. Such bias-preserving gates are not possible in a DV encoding~\cite{aliferis2008fault}, and thus, our protocol makes explicit use of the continuous variable nature of oscillators. This protocol can be used with CV codes like Kerr-cat codes~\cite{puri2017engineering,grimm2020stabilization} and dissipative-cat codes~\cite{mirrahimi2014dynamically}. Due to a lack of bias-preserving $\mathrm{CX}_\mathrm{GKP}$ gates for rectangular GKP codes, which could also be used to engineer a biased-noise architecture~\cite{zhang2023concatenation}, it is not possible to achieve as significant a reduction in resource overhead as for the cat codes. This discussion highlights some new concepts like concatenated CV-DV scalable error-correcting codes. In addition, we give some future applications for CV-DV architectures. As an open problem towards fault-tolerance, we pose the problem of local decoding of surface codes~\cite{delfosse2020hierarchical} via the probabilistic decoding described in Chapter~\ref{chapter:GKP-qec}. We also discuss the prospects of using oscillators as ancillary systems for intermediary tasks in an algorithm, like phase estimation.

\section{High-Fidelity Magic State Injection}\label{AQM}\label{sec:magic-state}
This section presents our article~\cite{singh2022high} which yields a quadratic reduction in the resource overhead of Fault-tolerant quantum computation. We use this opportunity to introduce some details related to the DV encoding known as surface codes, which will come in handy when discussing open problems in the next section. Here we use $\{X,Y,Z\}$ instead of $\{\sigma_\mathrm{x},\sigma_\mathrm{y},\sigma_\mathrm{z}\}$ to indicate Pauli operations on the logical cat codes abstracted as qubits. We will also use these terms interchangeably; $\mathrm{T}$ gates and $\mathrm{Z(\pi/8)}$, $\mathrm{S}$ gates and $\mathrm{Z(\pi/4)}$, $\ket{\mathrm{T}}$ states and magic states.

The resource cost of implementing fault-tolerant logical quantum computation is a major challenge in implementing useful quantum algorithms~\cite{fowler2012surface,reiher2017elucidating,o2017quantum,campbell2019applying,sanders2020compilation,babbush2021focus,gidney2021factor}. Several recent studies have shown that the structure of noise in the underlying qubit architecture can be leveraged to improve the performance of quantum error correction~\cite{tuckett2018ultrahigh,tuckett2019tailoring,ataides2021xzzx,darmawan2021practical,chamberland2020building,higgott2020subsystem,huang2020fault,guillaud2021error}. These studies motivate the design of new noise-aware protocols for resource-efficient logical operations for fault-tolerant quantum computation (FTQC). 

A significant resource overhead of practical quantum computing architectures is consumed performing non-Clifford gates. These are essential logical operations needed for universal quantum computing (see Sec.~\ref{DV_gates}). A versatile way of realizing non-Clifford gates is by teleportation where a high-fidelity resource state, called a magic state, is used by a Clifford gate teleportation (see Sec.~\ref{open-hierarchy}) circuit~\cite{bravyi2005universal}. High-quality resource states can be prepared with magic state distillation (MSD)~\cite{bravyi2005universal,reichardt2005quantum,bravyi2012magic,fowler2013surface,meier2012magic,jones2013multilevel,duclos2013distillation,duclos2015reducing,campbell2017unified,o2017quantum,haah2018codes,campbell2018magic,gidney2019efficient,litinski2019magic} where several copies of noisy magic states are consumed to produce a smaller number of copies with lower logical error rates.

The planar layout of the surface-code (SC) quantum computing architecture ~\cite{kitaev2003fault,dennis2002topological,bravyi1998quantum,fowler2012surface} makes it particularly appealing for experimental implementation and as such, significant effort has been dedicated to minimizing the resource cost of preparing magic states with the surface code. Additionally, magic state distillation protocols based on the surface code have been adopted in low-overhead schemes for fault-tolerant quantum computing based on finite rate quantum low-density parity-check codes~\cite{cohen2021lowoverhead}. Even with these considerable efforts, it remains that MSD is expected to occupy a large fraction of the resources of an SC architecture and it therefore presents a bottleneck in realizing quantum algorithms~\cite{fowler2013surface}.

In this section, we present a new protocol for preparing higher-fidelity input states for MSD protocols that is tailored for qubit architectures that experience biased noise such that bit-flips are far less likely than phase-flips. In our protocol, we use a physical two-qubit diagonal non-Clifford gate to prepare a magic state encoded in a two-qubit code capable of detecting a single dominant error. Therefore, the infidelity of the post-selected states that herald no error scales quadratically with the physical error probability when the bias is strong and physical error rates are modest. This is a quadratic reduction in the infidelity compared with more conventional approaches for state preparation~\cite{fowler2012surface,horsman2012surface,landahl2014quantum,li2015magic,Luoe2026250118}. Detecting more high-probability errors results in more states being discarded, but importantly this only results in a minute decrease in the success probability compared to other approaches based on post-selection~\cite{li2015magic}.

Our protocol follows a bottom-up approach for the design of fault-tolerant protocols. For example, our scheme utilizes a recently discovered, bias-preserving controlled-not ($\mathrm{CX}$) gate~\cite{puri2020bias} for detecting errors without affecting the noise bias of the system. This bias-preserving gate also enables us to encode the post-selected state into a high-distance error correcting code required for robust quantum computing while maintaining the quadratic improvement. Unlike the $\mathrm{CX}$, single- and two-qubit diagonal gates are trivially biased~\cite{aliferis2008fault}. Moreover, in the biased-noise superconducting Kerr-cat architecture, the two-qubit diagonal gates can be implemented with simple interactions and can in principle be much faster and higher fidelity than single qubit diagonal gates~\cite{puri2017engineering,puri2020bias,darmawan2021practical}. Consequently, we leverage two-qubit diagonal non-Clifford gates in this proposal. While, in practice, the dominant source of noise is independent perturbations on physical qubits, these independent errors can get correlated due to the action of the gate. For example, in the bias-preserving $\mathrm{CX}$ gate, a phase-flip error in the target qubit during the gate propagates to the control qubit, giving rise to correlated phase noise~\cite{puri2020bias,darmawan2021practical}. In contrast, the diagonal gates are transparent to phase errors in the qubits. Thus, the high-rate independent phase-flip events do not get correlated. Highly precise microwave control in the superconducting qubit platform also ensures that correlated errors due to control noise are rare events. The naturally low probability of correlated errors on diagonal gates ensures that high-fidelity preparation of magic states in our protocol is possible.

We incorporate our initialization protocol into a quantum-computing architecture based on the XZZX code~\cite{ataides2021xzzx, darmawan2021practical}; a surface code that is tailored to correct biased noise. With this setup, we find improvements in the fidelity of the injected magic state, leading to more effective MSD. For example, even with a modest $\mathrm{CX}$ gate infidelity of $\sim 0.7\%$, and average bias $O(10^3)$, we find that a raw XZZX magic state of size $5\times 25$ (equivalent to 441 data and ancilla qubits) can be prepared, with $\sim 94\%$ success rate, at an error rate of $\sim 0.1\%$. The average bias is defined as the total probability of phase-flip errors relative to that of other errors in the gate. After consuming these raw states in one round of the 15-to-1 distillation protocol~\cite{bravyi2005universal}, a single copy of a magic state can be produced at an error rate of $O(10^{-8})$. This error rate is, for example, sufficient for realizing quantum simulations with quantum advantage without further rounds of distillation~\cite{babbush2018encoding,childs2018toward,nam2019low}. 
On the other hand, the error rate after one round of distillation with raw magic states prepared using the standard scheme is two orders of magnitude larger. These numerical results correspond to the case when noise in the $\mathrm{CX}$ gates is an order of magnitude larger than other operations in the syndrome extraction circuit, as is typically the case with biased-noise cat qubits~\cite{darmawan2021practical}. When the $\mathrm{CX}$ gates are as noisy as other components in the circuit, the protocol proposed here gives a greater advantage over the standard approach. 
Other approaches have been studied for implementing non-Clifford gates with codes tailored to biased noise. In~\cite{webster2015reducing} for example, a magic state is initialized in the repetition code with success rate that decreases exponentially with the code size even in the absence of errors. This is in contrast to our proposal which prepares the magic state deterministically in the absence of errors and heralding errors only costs a small decrease in the success rate. Moreover, our scheme only requires two-qubit gates which are experimentally easy to realize and is effective even with modest amounts of bias achievable in near-term experiments. Proposals in Refs.~ \cite{chamberland2020building,guillaud2021error} on the other hand, use three-qubit entangling gates.

\begin{figure}
\centering
    \includegraphics[width=\textwidth]{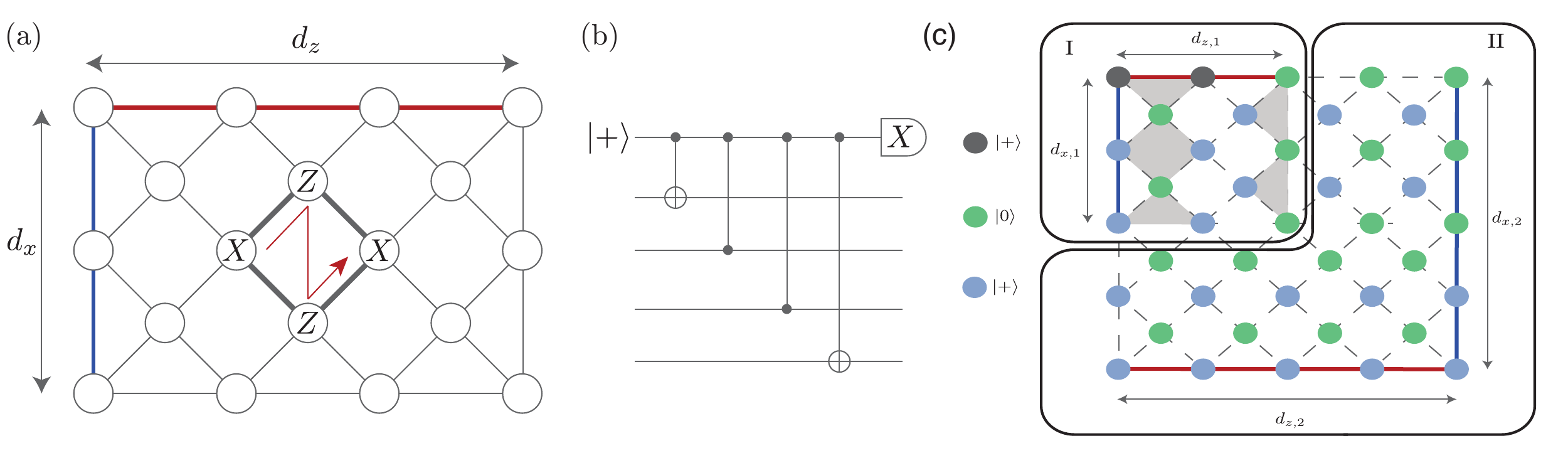}
\caption[Illustration of the protocol for magic state preparation in an XZZX code.]{\textbf{Illustration of the protocol for magic state preparation in an XZZX code.} (a) Rectangular XZZX code with data qubits on the vertices of a rotated grid. The stabilizers are the product of two Pauli $X$ and two Pauli $Z$ operators on qubits arranged on the vertices around each face. The distance to $X$ and $Z$ errors is $d_x$ and $d_z$ respectively. The logical qubit Pauli $X_\mathrm{L}(Z_\mathrm{L})$ are the product of Pauli $X (Z)$ on the qubits along the blue and red edges respectively. The order in which qubits are coupled to the ancilla at the center of each face (not shown) is indicated by the red arrow. (b) Circuit for stabilizer measurements. The ancilla is prepared in state $\ket{+}$, then coupled to data qubits with $\mathrm{CX}$ and $\mathrm{CZ}$ gates and finally read out in the $X$ basis. (c) In stage I the qubits in region I are initialized as shown, a $ZZ(\theta)$ gate is applied to the two grey qubits, and the stabilizers are measured twice. The faces shaded in grey mark the {\it{fixed stabilizers}} for stage I. After stage I is successful and a $d_{x,1}\times d_{z,1}$ magic state is prepared, qubits in region II are initialized as shown. Stage II is then implemented and the $d_{x,1}\times d_{z,1}$ state is grown to a $d_{x,2}\times d_{z,2}$ state, where stabilizers are measured for $d_m=d_{z,2}$ rounds. In stage I the qubits in region I are initialized as shown, a $ZZ(\theta)$ gate is applied to the two grey qubits, and the stabilizers are measured twice. The faces shaded in grey mark the {\it{fixed stabilizers}} for stage I. After stage I is successful and a $d_{x,1}\times d_{z,1}$ magic state is prepared, qubits in region II are initialized as shown. Stage II is then implemented and the $d_{x,1}\times d_{z,1}$ state is grown to a $d_{x,2}\times d_{z,2}$ state, where stabilizers are measured for $d_m=d_{z,2}$ rounds.}
\label{fig:xzzx}
\end{figure}

\subsection{The Protocol}
\label{Main}
We demonstrate our protocol with the XZZX code~\cite{ataides2021xzzx} defined on a rectangular lattice of size $d_x\times d_z$ shown in Fig~\ref{fig:xzzx}(a). Data qubits are placed on the vertices of the lattice, and $d_x$ and $d_z$ respectively denote the code distance with respect to pure $X$ and $Z$ errors. The stabilizers of the code are %the product of Pauli $X$, $Z$, $Z$, $X$ operators of 
of the form $X\otimes Z\otimes Z\otimes X$ on 
the qubits around each face, as shown in Fig~\ref{fig:xzzx}(a). The logical operator $X_\mathrm{L}$ is the product of Pauli $X$ operators of the qubits along a vertical edge and $Z_\mathrm{L}$ is the product of Pauli $Z$ operators of the qubits along a horizontal edge. The stabilizer measurement circuit is illustrated in Figure~~\ref{fig:xzzx}(b). An ancilla qubit, placed at the center of each face, is initialized in $\ket{+}$. Next, a sequence of $\mathrm{CX}$ and $\mathrm{CZ}$ gates is applied in the order shown in Fig.~\ref{fig:xzzx}(a), and finally the ancilla is measured in the $X$ basis. 

The injection protocol proceeds in two stages similar to that presented in~\cite{li2015magic}. In stage~I, a small XZZX code of size $d_{x,1}\times d_{z,1}$ is prepared in the magic state. Some errors are detected, but not corrected, at this stage. States where no errors are detected proceed to stage~II where the code is grown to a larger distance; $d_{x,2}\times d_{z,2}$. Our protocol goes beyond the preparation protocol in \cite{li2015magic} in that, as an intermediate step in stage~I, we prepare a two-qubit error detecting code that detects a single dominant error acting on the raw magic state before it is injected into the stage~I code. This gives a quadratic improvement to the fidelity of the input state.
The detailed steps in our protocol are given below. 

\paragraph{Stage I:}

Stage I proceeds over three separate steps.

\begin{itemize}
\item Step 1: Physical qubits in region I are initialized as shown in Fig~\ref{fig:xzzx}(c). The qubits marked in green and blue are initialized in state $\ket{0}$ and $\ket{+}$ respectively. The two qubits on the top left corner, marked in grey, are initialized in $\ket{+}$. In the following, the stabilizers on the faces shaded in grey will be referred to as {\it{fixed stabilizers}}.

\item Step 2: A two-qubit $ZZ(\theta)=e^{-i\theta Z\otimes Z}$ gate is applied on the two qubits at the top left which are highlighted in grey in Fig~\ref{fig:xzzx}(c).  

\item Step 3: All the stabilizers are measured twice and stabilizer measurement outcomes or syndromes are recorded. If the outcome of measuring any fixed stabilizer is ${-}1$ or if the measurement outcomes from the two rounds are not identical, then an error has been detected. In this case the state is discarded and stage I is restarted. Otherwise, the code is sent to stage II. 

\end{itemize}
Let us give some motivation for these steps. In the absence of errors, the initial product state in step 1 is the ${+}1$ eigenstate of the fixed stabilizers.

In step 2, the $ZZ(\theta)$ gate entangles the two grey qubits, while the rest of the qubits remain un-entangled. For a general angle $\theta$, which is not an integral multiple of $\pi/4$, this is a non-Clifford gate. We can think of the grey qubits as forming a two-qubit repetition code with $Z^\prime_\mathrm{L}=Z\otimes Z$ and $X^\prime_\mathrm{L}=X\otimes I$. In this picture, the effect of the physical $ZZ(\theta)$ gate is to non-transversally apply a logical $e^{-i\theta Z^\prime_\mathrm{L}}$ gate to the two-qubit repetition code. After this step, the state of the physical qubits on the $X_\mathrm{L}$ and $Z_\mathrm{L}$ edge is the ${+}1$ eigenstate of $\cos(2\theta)X_\mathrm{L}+\sin(2\theta)Y_\mathrm{L}$. Observe that in the absence of errors, the physical qubits remain in the ${+}1$ eigenstate of the fixed stabilizers.

The first measurement round of step 3 projects the system into an eigenspace of the stabilizers and the logical qubit is realized. In the absence of errors, the syndromes corresponding to the fixed stabilizers will be ${+}1$, while those corresponding to the unmarked stabilizers can be either ${+}1$ or ${-}1$. Moreover, in the absence of errors, measurement outcomes from the two measurement rounds in step 3 will be identical. Because the stabilizers commute with the logical operators, the  resulting logical qubit state is the ${+}1$ eigenstate of $\cos(2\theta)X_\mathrm{L}+\sin(2\theta)Y_\mathrm{L}$. Thus, when $\theta=\pi/8$, the $d_\mathrm{x,1}\times d_\mathrm{z,1}$ code is initialized in the logical magic state $|m\rangle_\mathrm{L}=\ket{0}_\mathrm{L}+e^{i\pi/4}\ket{1}_\mathrm{L}$. If the target state is $|{+}Y\rangle_\mathrm{L}$, then $\theta=\pi/4$ is used. Thus, by tuning $\theta$, arbitrary states in the $X-Y$ plane of the Bloch sphere can be prepared.

\paragraph{Stage II:}

Stage~II proceeds to encode the magic state into a larger surface code, pending an appropriate heralded outcome at stage~I~\cite{li2015magic}. Physical qubits in region II are initialized as shown in Fig.~\ref{fig:xzzx}(c). All the stabilizers of the $d_{x,2}\times d_{z,2}$ code are measured for $d_m$ rounds, and error correction is performed using standard decoding algorithms like minimum weight perfect matching~\cite{dennis2002topological,edmonds1965paths,kolmogorov2009blossom,ataides2021xzzx}. Subsequently, the state may be sent for MSD.

Let us remark that there is some freedom in choosing the initial state of qubits in regions I and II. The initial state pattern shown in Fig~\ref{fig:xzzx}(c) works well for the range of parameters used in section~\ref{results}. Appendix~\ref{alt_app} gives an example of an alternative pattern.

\subsection{Noise}
\label{noise}
Here we argue that our scheme is tolerant to a single dephasing error on a data qubit or an ancilla qubit during preparation, idling, or any of the gates, to a single measurement error, or to a single correlated dephasing error that occurs during $\mathrm{CX}$ and $\mathrm{CZ}$ gates. As a consequence, when bit-flip errors are absent, the preparation error rate is $O(p^2)$, with $p$ the probability of a dominant error. This improvement remains significant for realistic noise models with high but finite bias $\eta$, where $1/\eta$ ($\eta \gg 1$ ) is the factor by which the probability of a non-Z error is suppressed compared to that of the dominant $Z$ error. In this case, undetectable preparation errors can occur at rate $O(p/\eta)$. It follows that if $\eta$ is large relative to $p^{-1}$, we obtain a quadratic improvement in the fidelity of injected magic states at finite bias compared to standard injection protocols.
At very small $p$ we obtain an improvement by a factor of  $1 /\eta$ in preparation fidelity; $O(p/\eta)$. The competition between the contribution of infidelity due to high rate and low rate errors can be determined by numerical experiments such as those we describe in Section~\ref{results}.
For the following qualitative discussion, we concentrate on errors at stage~I because this will be the dominant source of infidelity given sufficiently large $d_{x,2}$ and $d_{z,2}$ at stage~II.

We assume a Pauli approximation to a biased circuit noise model. Each single-qubit operation, including preparation and idling, is followed by a Pauli error $ Q = \{I,X,Y,Z\}$ that occurs with probability $p_Q$. Faulty measurements are modelled by flipping a given
measurement outcome with probability $p_M$. Errors in two-qubit gates are modelled by applying a Pauli error $Q = Q_C \otimes Q_T$ with $Q_C,\, Q_T \in \{I, X,Y,Z\}$ with probability $P_Q$ before the gate where $Q_C$($Q_T$) denotes the error acting on the control(target) qubit of the gate. Our protocol is designed to be highly effective against $Z$-biased noise where $p_Z,\, p_{ZI},\, p_{IZ}, \,p_{ZZ}$, and $p_M$ are significantly larger than the probabilities of other non-trivial, i.e., non-identity, error events and we take $p_{ZZ}$ to be small in the $ZZ(\theta)$ gate following experimentally well-motivated arguments given below.% and $1-p_I$ and $1-p_{II}$ are small.

We now demonstrate that our protocol is robust against a single high-rate error event in a biased-noise architecture. Over steps 1-3, a $Z$ error on any of the qubits highlighted in grey and blue will cause the syndromes corresponding to the fixed stabilizers to change to ${-}1$. Thus, these errors are detected in step 3. A $Z$ error on the qubits marked in green before the first measurement round of step 3 will not cause a logical error. A $Z$ error on these qubits in the second measurement round of step 3 will result in a mismatch of the syndromes, corresponding to the unshaded stabilizers in region I, in the two measurement rounds.  Hence, this error is also detected in step 3. A $Z$ error on an ancilla or a measurement error will also be detected as it will either cause the outcome of measuring a fixed stabilizer to be ${-}1$ or cause a mismatch of stabilizer measurement outcomes from the first and second rounds.

So far we have ignored correlated errors introduced by the two-qubit gates. During a correlated error, two qubits simultaneously suffer from phase-flips with a probability that can be greater than the probability of independent phase-flips on the two qubits. In case of pure-dephasing noise, the $\mathrm{CX}$ or $\mathrm{CZ}$ gates acting between data and ancilla qubits do not lead to correlated errors on the data qubits. A correlated $Z\otimes Z$ error in any one of these gates in the first round of step 3 will either cause the outcome of measuring a fixed stabilizer to be ${-}1$ or cause a mismatch of stabilizer measurement outcomes and hence will be detected. Moreover, a $Z\otimes Z$ error in the second round will be corrected by subsequent rounds of error correction in stage II. A correlated $Z\otimes Z$ error in the $ZZ(\theta)$ gate will cause a logical error which will not be detected in either stage I or II. However, these are expected to be low-rate errors in superconducting biased-noise architecture since independent phase-noise in the qubits don't get correlated and control and crosstalk errors can be easily mitigated (see further discussion in section~\ref{discuss}). Thus, a $Z\otimes Z$ error in the $ZZ(\theta)$ gate will not limit the performance of the scheme in practice. There are several instances of independent errors occurring simultaneously on two or more qubits which will also not be detected. For example, simultaneous phase-flip errors during initialization of the two grey qubits will go undetected. 

In summary, we find that the proposed scheme is robust against a single $Z$ error during preparation, idling, or any of the gates, or a correlated $Z\otimes Z$ error in the $\mathrm{CX}$ and $\mathrm{CZ}$ gates, or a single measurement error. These errors are detected and discarded in stage I or corrected in stage II. Thus, our protocol has a finite success rate which decreases with an increase in the number of locations at which a fault can occur. Hence, for a high enough success rate, the distance of the code in stage I should not be too large.

In order to determine the scaling of the logical error rate as a function of the probability of high-rate errors, we consider a physically realistic noise model where each qubit is subject to independent phase-flip errors with identical probability $p$. In this case, $p_Z=p$ for the single-qubit operations, 
$p_{ZI}=p$, $p_{IZ}=p_{ZZ}=p/2$
%$p_{ZI}=p$ and $p_{IZ}=p_{ZZ}=p/2$
for the $\mathrm{CX}$ gates, and $p_{ZI}=p$, $p_{IZ}=p$, $p_{ZZ}=p^2$ for the diagonal gates. Errors in the measurement can also be assumed to be $p_M=O(p)$. Thus in the absence of non-$Z$ noise, the logical error rate of the injected magic state is $p_\mathrm{L}=O(p^2)$. The error-channel used to obtain this scaling is justified because in the bias-preserving $\mathrm{CX}$ gates a $Z$ error on the target qubit propagates as a combination of a $Z$ error on the target and a $Z\otimes Z$ error on the target and control qubits, giving $p_{IZ}$, $p_{ZZ}=p/2$~\cite{puri2020bias,darmawan2021practical}.
Such error-correlations cannot be trivially introduced in the diagonal gates since they can be implemented in an error-transparent manner using interactions that commute with physical $Z$ errors in qubits~\cite{puri2020bias}. Hence, the probability of two qubit $Z\otimes Z$ errors is the same as the probability of two independent $Z$ errors for the diagonal gates, $p_{ZZ}=p_{IZ}\cdot p_{ZI}=p^2$.

\paragraph{Noise modeling in simulations:}
We now describe the circuit noise model used to obtain the numerical results presented in the next section. In biased-noise qubits the $\mathrm{CX}$ gate is the slowest operation and total noise in the $\mathrm{CX}$ gate can be much greater than that in the diagonal two-qubit gates. In particular, in the Kerr-cat qubit architecture, the probability of phase-flip errors during the $\mathrm{CX}$ gate can be an order of magnitude greater than that of the $\mathrm{CZ}$ gate~\cite{darmawan2021practical} unless sophisticated control techniques are applied~\cite{xu2021engineering}. So we show numerical results for two noise models: (A) $\mathrm{CX}$ slower than $\mathrm{CZ}$, and (B) $\mathrm{CX}$ as fast as $\mathrm{CZ}$. In both these cases, for the diagonal $\mathrm{CZ}, ZZ(\theta)$ gates we use $p_{IZ},p_{ZI}$ and $ p_{ZZ}$ as described before, and the probability of other non-trivial two-qubit errors $=p/\eta$. For the single-qubit preparation errors, idling errors on data qubits while the ancillae are being measured, and errors on some of the qubits which idle during $\mathrm{CZ}$ gates, we use $p_Z=p $ and $p_X=p_Y=p/\eta$. Measurement errors are applied with probability $p+p/\eta$. To model the fast $\mathrm{CX}$ gate in (B) we use $p_{ZI},p_{IZ},p_{ZZ}$ as described before and the probability of other non-trivial two-qubit errors $=p/\eta$. In this case, the error channel applied to qubits which idle during the $\mathrm{CX}$ gate is identical to that applied to qubits which idle during the $\mathrm{CZ}$ gate. In (A), for the $\mathrm{CX}$ and single-qubit idling errors during this gate, we use the same channel as (B) but with $p$ replaced by $10p$. 

For numerical results, we use two biases $\eta=10^4$ and $\eta=10^3$, for which the average gate bias in the $\mathrm{CX}$ gate is $\sim 1667$ and $\sim 167$ respectively. The average gate bias is defined as the ratio of the sum of the probabilities of $I\otimes Z,Z\otimes I$ and $Z\otimes Z$ error to the sum of the probabilities of all other non-trivial errors. We start with a $d_{x,1}\times d_{z,1}=1\times 3$ code in stage I and grow it to a larger $d_{x,2}\times d_{z,2}$ code with $d_m=d_{z,2}$. 

For comparison, we also present the logical error rate and success rate obtained when the standard scheme based on using a single-qubit $Z(\theta)=e^{-i\theta Z}$ gate, as described in Appendix~\ref{standard}, is used. For the error model of this gate, we use $p_{Z}=p $ and the probability of other non-trivial single-qubit errors $=p/\eta$. We keep the probability of phase-flip error per qubit in the $ZZ(\theta)$ and $Z(\theta)$ gate to be the same, even though in practice the former can be smaller.

\subsection{Results}
\label{results}
\begin{figure}
\centering
    \includegraphics[width=\textwidth]{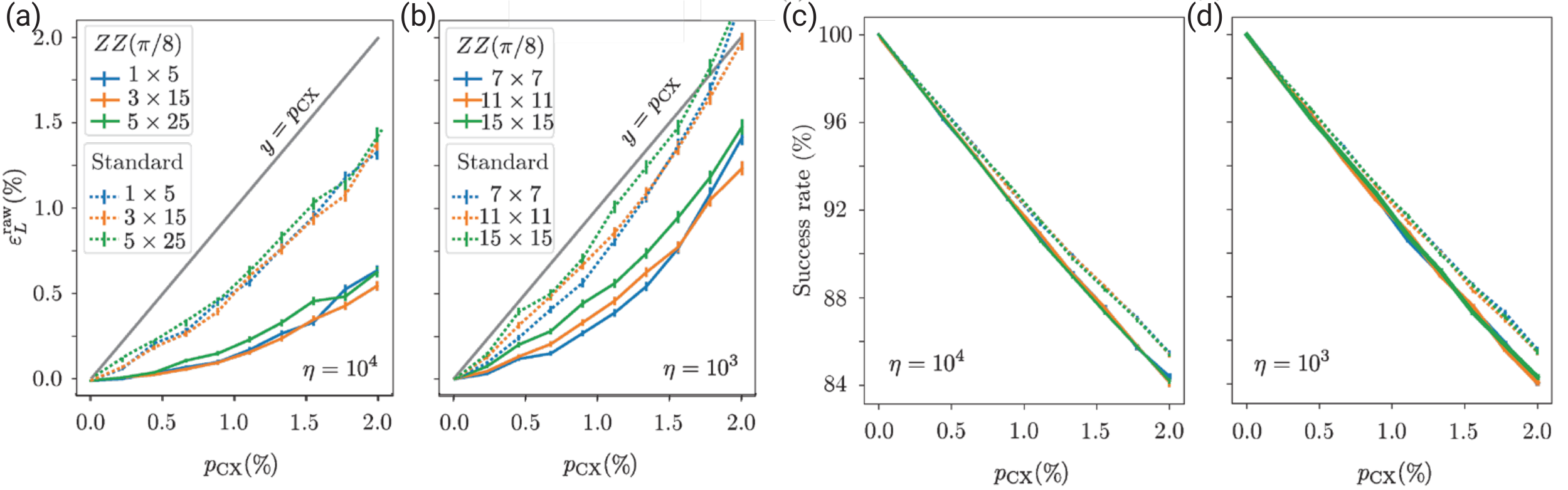}
\caption[Logical error rate and success probability for magic state injection using cat codes]{\textbf{Logical error rate and success probability for magic state injection using cat codes.} Logical error rate ($\varepsilon^\mathrm{raw}_L$) and success rate after $d_{m}$ rounds of error correction in stage II with noise model A ($\mathrm{CX}$ slower than $\mathrm{CZ}$) so that $p_{\mathrm{CX}}=20p+120p/\eta$. The bias is $\eta=10^4$ in (a,c) and $\eta=10^3$ in (b,d). The code size in stage I is $d_{x,1}\times d_{z,1}=1\times 3$. Stage II code sizes $d_{x,2}\times d_{z,2}$ are shown in the legend, with $d_{m}=d_{z,2}$. The results for our scheme are shown using solid lines and that for the standard approach are shown using dotted lines. Error bars indicate standard error of the mean. Each data point is generated with $10^5$ Monte-Carlo samples.}
\label{fig:plot1}
\end{figure}
Finally, we present numerical results that demonstrate the advantage of our scheme for logical magic state preparation, and subsequently for distillation with practical system parameters. 
Figure~\ref{fig:plot1} shows the total logical error rate $\varepsilon^\mathrm{raw}_L$ of the output XZZX magic state and success rate as a function of the total error rate of the physical $\mathrm{CX}$ gate $(p_{\mathrm{CX}})$ for the noise model (A) and for three different $d_{x,2}\times d_{z,2}$.  

Using our scheme, we find that when bias is large $\eta=10^4$, $\varepsilon^\mathrm{raw}_L$ is approximately independent of the code size and the curvature of $\varepsilon^\mathrm{raw}_L(p_{\mathrm{CX}})$ indicates a non-linear dependence of $\varepsilon^\mathrm{raw}_L$ on 
the physical error rate. This follows from the discussion in section~\ref{noise}, according to which the dominant source of uncorrectable errors is two phase-flip events, or two faulty-measurement outcomes, or a combination of these in the initial $1\times 3$ code. The deviations between $\varepsilon^\mathrm{raw}_L$ for different code sizes in Fig.~\ref{fig:plot1} are mainly due to small but non-zero bit-flip noise. By numerical fitting of the component of $Z_L$ error in $\varepsilon^\mathrm{raw}_L$ for $\eta=10^4$, we find that this component scales as $((4.48\pm 0.07) \times 10^3) p^2$
or $(11.2\pm 0.2) p_{\mathrm{CX}}^2$. In contrast, with the standard scheme, the curvature for $\varepsilon^\mathrm{raw}_L(p_{\mathrm{CX}})$ indicates a linear dependence on the physical error rate even if the bias is large. In this case, with numerical fitting, we find that the $Z_L$ component of error in $\varepsilon^\mathrm{raw}_L$ scales as $(11.6\pm0.5)p$ or $(0.58\pm 0.02) p_{\mathrm{CX}}$. Details for the fitting and different components of the total logical error rate are given in Appendix~\ref{decomp}.

Results in Fig.~\ref{fig:plot1}(a) show that $\varepsilon^\mathrm{raw}_L$ can be about an order of magnitude lower than the physical error rate of the noisiest gate in the system. For example, when $p_{\mathrm{CX}}=0.67\%$ and $\eta=10^4$, the infidelity of the injected magic state in the $3\times 15$ code is $= 0.07\%$. The probability of success is high $= 94.4\%$. For an order of magnitude lower bias $\eta=10^3$, $\varepsilon^\mathrm{raw}_L$ increases and is still somewhat independent of the code size in the given range of $p_{\mathrm{CX}}$. Moreover, due to greater contributions from the non-Z errors, the curve $\varepsilon^\mathrm{raw}_L(p_{\mathrm{CX}})$ starts to flatten out. 
Nonetheless, the scheme introduced here prepares a XZZX magic state with a significantly lower error rate than the standard approach for both $\eta=10^4$ and $\eta=10^3$. The ability to detect more errors with our scheme leads to a small decrease in the success rate compared to the standard approach.

\begin{figure}
\centering
    \includegraphics[width=\textwidth]{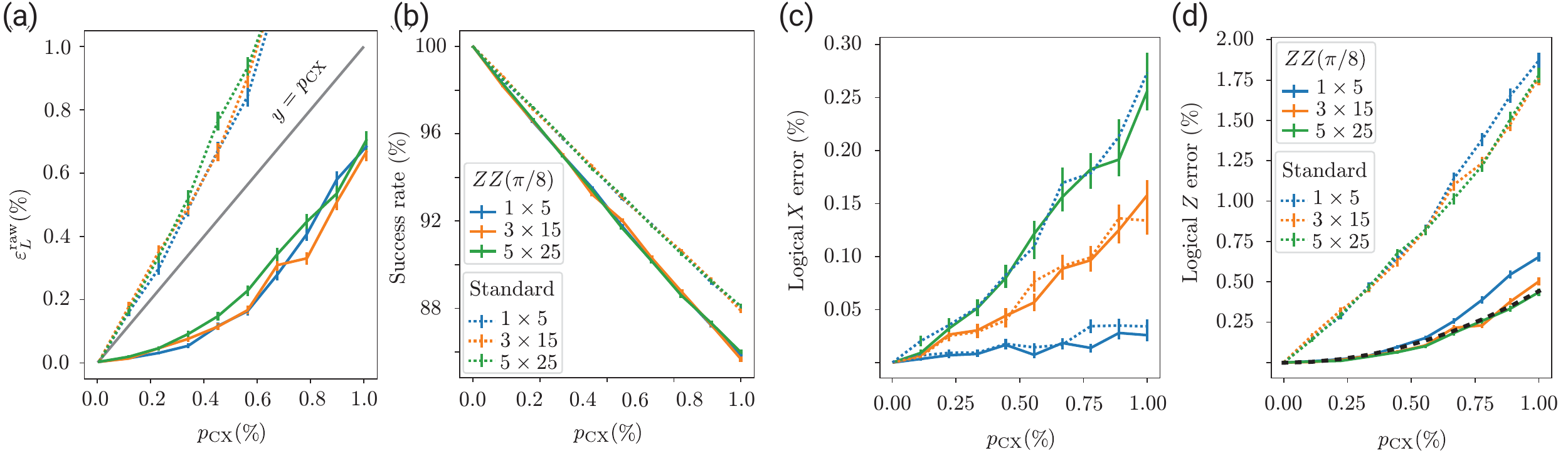}
\caption[Logical error rate and succcess probability using fast bias-preserving gates for biased-noise qubits.]{(a,b) Logical error rate ($\varepsilon^\mathrm{raw}_L$) and success rate after $d_{m}$ rounds of error correction in stage II with noise model B ($\mathrm{CX}$ as fast as $\mathrm{CZ}$) so that $p_{\mathrm{CX}}=2p+12p/\eta$. The bias is $\eta=10^4$ and the code size in stage I is $d_{x,1}\times d_{z,1}=1\times 3$. Stage II code sizes $d_{x,2}\times d_{z,2}$ are shown in the legend, with $d_{m}=d_{z,2}$. The results for our scheme are shown using solid lines and that for the standard approach are shown using dotted lines. Error bars indicate standard error of the mean. Each data point is generated with $10^5$ Monte-Carlo samples. (c,d)$X_L$ and $Z_L$ error rate in the magic state for $\eta=10^4$ for noise model (B). The black dashed lines in (d) is found by fitting $Z_L$ error rate in the magic state prepared using our scheme to $Ap^2$. We use the solid lines corresponding to $d_{x,2}\times d_{z,2}=3\times 15$ and $d_{x,2}\times d_{z,2}=5\times 25$ for the fit and find $A=(1.78\pm 0.06)\times 10^2$.}
\label{fig:plot4}
\end{figure}

In Fig.~\ref{fig:plot4}(a,b) we present $\varepsilon^\mathrm{raw}_L$ and success rate as a function of $p_{\mathrm{CX}}$ for the noise model (B). We use $\eta=10^4$ and again we find that the scheme based on $ZZ(\pi/8)$ gate outperforms the standard approach. For example, even when the physical error rate in the two-qubit gates is as high as 0.45$\%$, the infidelity of the injected $3\times 15$ magic state is five-fold lower $\sim 0.11\%$, while that with the standard scheme is higher $\sim 0.66\%$. 

The impact of our protocol becomes evident from the subsequent reduction in cost for MSD. If the infidelity of the raw injected state is $\varepsilon^\mathrm{raw}_L$, then after a round of 15-to-1 distillation protocol, the logical error rate can be made arbitrarily close to $35{(\varepsilon^\mathrm{raw}_L)}^3$, if sufficiently large code $d_{x,2}\times d_{z,2}$ is used so that errors in the distillation circuit are negligible~\cite{bravyi2005universal}. 
Consider Fig.~\ref{fig:plot1} and note that $\varepsilon^\mathrm{raw}_L=0.11\%$ or $35{(\varepsilon^\mathrm{raw}_L)}^3\sim 4.7\times 10^{-8}$ when $p_\mathrm{CX}= 0.67\%$, $\eta=10^4$, and $d_{x,2}\times d_{z,2}\times d_m=5\times 25\times 25$. From numerical simulations, we have confirmed that for the same noise channel the logical error rate for $d_m=25$ rounds of error correction with $5\times 25$ code is $\ll 10^{-8}$. Thus, we find that after one round of distillation, a magic state with error rate $O(10^{-8})$ can be realized with a $5\times 25$ XZZX code. In contrast, with the standard approach, for the same sized code and physical gate errors, $\varepsilon^\mathrm{raw}_L=0.33\%$, so that only an error rate of $O(10^{-6})$ will be possible after one round of distillation.

\subsection{Logical Error Decomposition}
\label{decomp}

\begin{figure}
\centering
    \includegraphics[width=\textwidth]{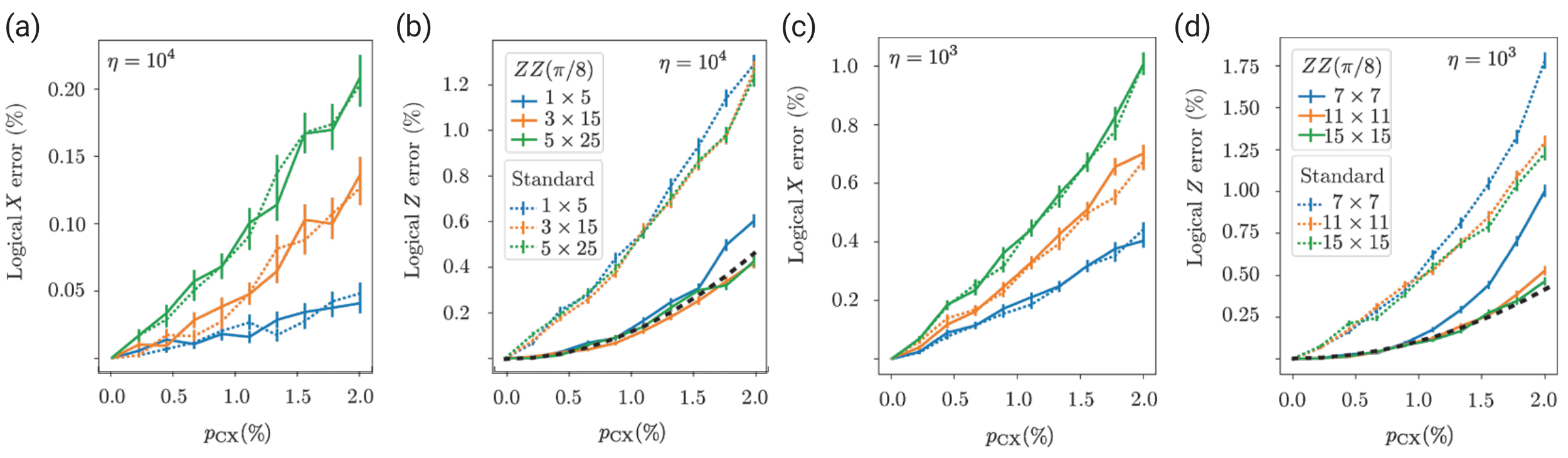}
\caption[Logical error rate and success probability for slow bias-preserving CX gates with biased-noise qubits.]{$X_L$ and $Z_L$ error rate in the magic state for $\eta=10^4$ (a,b) and $\eta=10^3$ (c,d) for noise model (A).  The black dashed lines in (b,d) is found by fitting $Z_L$ error rate in the magic state prepared using our scheme, at low $p$ and large distances, to $Ap^2$. In (b) we use the solid lines corresponding to $d_{x,2}\times d_{z,2}=3\times 15$ and $d_{x,2}\times d_{z,2}=5\times 25$ for the fit and find $A=(4.48\pm 0.07)\times 10^3$. In (d) we use the solid lines corresponding to $d_{x,2}\times d_{z,2}=11\times 11$ and $d_{x,2}\times d_{z,2}=15\times 15$ for the fit and find  $A=(4.34\pm 0.09)\times 10^3$.}
\label{fig:plot3}
\end{figure}

Figure~\ref{fig:plot3} shows the component of $X_L$ and $Z_L$ errors in the total error rate presented in  Fig.~\ref{fig:plot1} of the main text. For small $p$, we find a quadratic dependence of $Z_L$ errors on $p$ ($Ap^2$) when the scheme introduced in this section is used.
On the other hand, the dependence of $Z_L$ errors on $p$ is linear when the standard protocol is used. In Fig.~\ref{fig:plot3}(b) we fit $Z_L$ for $d_{x,2}\times d_{z,2}=3\times 15$ and $d_{x,2}\times d_{z,2}=5\times 25$ to $Ap^2$ and find $A=(4.48\pm 0.07)\times 10^3$. In Fig.~\ref{fig:plot3}(d) we fit $Z_L$ for $d_{x,2}\times d_{z,2}=11\times 11$ and $d_{x,2}\times d_{z,2}=15\times 15$ to $Ap^2$ and find $A=(4.34\pm 0.09)\times 10^3$. This confirms the analysis in section~\ref{noise}, according to which, $Z_L$ error rate, or equivalently $A$, should be independent of the code size in stage II if $d_{z,2}$ is large enough. Because of the initialization pattern chosen in stage II, the $X_L$ error rate is expected to grow with the distance $d_{z,2}$. This can be understood from the fact that bit-flip errors on any one of the $d_{z,2}$ qubits in the top row of block II will be un-correctable. However, since the bias is large, failure due to such error events is not too large. It is possible to prevent such errors from accumulating, especially when the bias is small, by using a larger $d_{x,1}$ in stage I or by using an alternative initialization strategy in stage II, as discussed in the Appendix~\ref{alt_app}.

Figures~\ref{fig:plot4}(c,d) show the component of $X_L$ and $Z_L$ errors in the total error rate presented in  Fig.~\ref{fig:plot4}(a,b) of the main text. We fit $Z_L$ for $d_{x,2}\times d_{z,2}=3\times 15$ and $d_{x,2}\times d_{z,2}=5\times 25$ to $Ap^2$ and find $A=(1.78\pm 0.06)\times 10^2$.

\subsection{Standard Protocol Based on the Single-Qubit $Z(\theta)$ Gate}
\label{standard}
The numerical results corresponding to the standard scheme used in Figs.~\ref{fig:plot1},\ref{fig:plot4} were produced by modifying the steps in Stage I of the protocol described in the main text as follows: 

\begin{itemize}
\item Step 1: Physical qubits in region I are initialized as shown in Fig~\ref{fig:alt}(a).  
\item Step 2: A $Z(\theta)=e^{-i\theta Z}$ gate is applied on the qubit on the top left, highlighted in grey in Fig~\ref{fig:alt}(a). The fixed stabilizers are shown in grey.
\item Step 3: All the stabilizers are measured twice and stabilizer measurement outcomes or syndromes are recorded. If the outcome of measuring any fixed stabilizers is ${-}1$ or if the measurement outcomes from the two rounds are not identical, then an error has been detected. In this case the state is discarded and stage I is started afresh. Otherwise, the code is sent to stage II. 
\end{itemize}

\subsection{Possibilities for Further Optimization in the XZZX Code and Other Surface Codes}
\label{alt_app}
\begin{figure}
\centering
    \includegraphics[width=\textwidth]{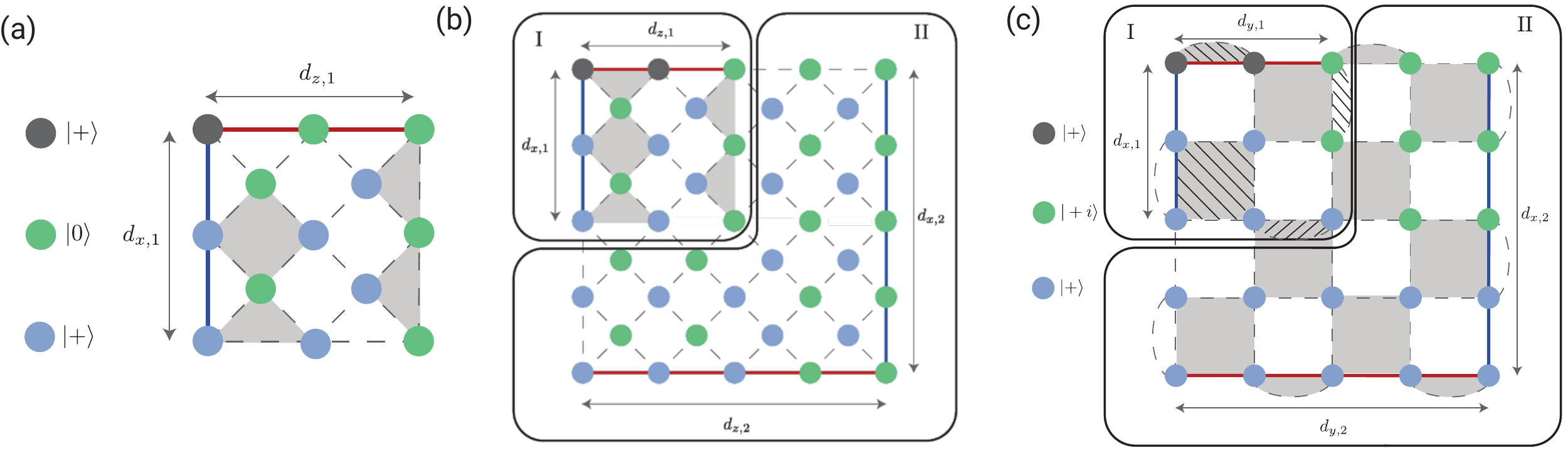}
\caption[Alternative initialization schemes for magic state injection.]{(a) Qubit arrangement in stage I of the standard scheme used for comparison in this section. The faces shaded in grey mark the fixed stabilizers for stage I. Stage II is identical to Fig.~\ref{fig:xzzx}(c). (b) Illustration of the protocol for preparing the magic state in the XZZX code with alternate stage II initialization pattern. The faces shaded in grey mark the fixed stabilizers for stage I. (c) Arrangement of qubits for preparing the magic state $\cos(\pi/8)\ket{+i}_L-i\sin(\pi/8)\ket{-i}_L$ in the tailored surface code. This code has two types of stabilizers: product of Pauli $Y, Y, Y, Y$ on the qubits around the white squares and product of Pauli $X, X, X, X$ on the qubits around the grey squares. At the boundaries the stabilizers are product of $X,X$ and $Y,Y$ on two qubits.  The fixed stabilizers for stage I are marked using black lines. The $ZZ(\theta)$ gate is applied to the two grey qubits on the top left.}
\label{fig:alt}
\end{figure}

Our protocol can be understood as preparing a $1\times 2$ surface code magic state directly by using a physical two-qubit operation $ZZ(\theta)$. Next, the $1\times 2$ code is grown into a $d_{x,1}\times d_{z,1}$ code in stage I in a standard way and all the stabilizers are measured twice. Only when no errors are detected, the $d_{x,1}\times d_{z,1}$ code is grown into $d_{x,2}\times d_{z,2}$ code and subsequent rounds of error correction are performed. In both the growing steps, the initial state of the qubits (apart from the qubits forming the original $1\times 2$ code) is chosen so that the logical operators grow correctly and to maximize the number of errors that can be detected or corrected. For example, an alternate initialization pattern is shown in Fig~\ref{fig:alt}(b) which would be more beneficial when noise is not too strongly biased. While we mainly focused on the XZZX code, this basic procedure outlined above can also be applied to other surface code families, like the tailored surface code. The main common component is to start with two qubits in $\ket{+}\otimes \ket{+}$ state and place them in the magic state of a $1\times 2$ SC using the two-qubit $ZZ(\theta)$ gate. To illustrate, a possible arrangement of qubit states for the tailored surface code is shown in Fig.~\ref{fig:alt}(c).

\subsection{Protocol with $ZZZ(\theta)$ Gate}\label{zzz}
In biased-noise cat qubits it is possible to realize a three-qubit $ZZZ(\theta)=e^{-i\theta Z\otimes Z\otimes Z}$ gate. It can be activated parametrically via four-wave mixing and can be easily implemented with the current circuit-QED toolbox~\cite{puri2020bias}. In fact, operations requiring similar interactions have already been realized in several experiments~\cite{leghtas2015confining,touzard2018coherent,grimm2020stabilization,lescanne2020exponential}.
With such a gate, it is possible to directly prepare a $1\times 3$ code in the magic state. Following the procedure in section~\ref{Main}, the $1\times 3$ code can be first grown to a $d_{x,1}\times d_{z,1}$ code by measuring the stabilizers thrice in stage I, and the state post-selected on no error detection can be grown to a $d_{x,2}\times d_{z,2}$ code in stage II. When the bias is large and the probability of three-qubit phase-flip error in the $ZZZ(\theta)$ gate is small, the probability of a logical error scales as $O(p_\mathrm{phy}^3)$. Alternatively, error detection in stage I can be skipped, and the $1\times 3$ code can be directly grown into a $d_{x,2}\times d_{z,2}$ code. In this case, the logical error probability is dominated by the failure rate of the $1\times 3$ code and scales as $O(p_\mathrm{phy}^2)$. In general, the protocol can be adapted to use a $k$-qubit $Z^k(\theta)$ gate. 

\subsection{Summary and Discussion}
\label{discuss}
To summarize this section, we have introduced a protocol to prepare raw encoded states with a low error rate by exploiting features of biased-noise hardware. This, in turn, reduces the overhead cost of MSD for such systems.

The protocol is robust against the typical errors of a biased circuit noise model. To gain an advantage over the standard protocol, the probability of two-qubit correlated phase-flip errors in the $ZZ(\theta)$ gate must be low relative to the probability of two independent single-qubit phase-flip errors. 
  We expect this to be the case with Kerr-cat qubits. 
  
 While correlated phase-flip errors may be induced due to virtual transitions to the excited states caused by the microwave drive that realizes the $ZZ(\theta)$ gate, such noise can be mitigated by pulse shaping or by adding counter-diabatic drives~\cite{xu2021engineering}. Another source of correlated errors is crosstalk, which can be mitigated by appropriate frequency arrangement of qubits~\cite{gambetta2017building}. Thus, while we do not believe correlated errors will be a significant issue, further investigation in mitigating such errors is called for, which will be made possible by rapid advances in biased-noise qubit technology.

We expect that the simple protocol we have proposed can be widely generalized and adapted to other magic state preparation schemes. For example, it might be interesting to determine if further improvements can be achieved by combining our ideas with recent developments using flag qubits~\cite{chamberland2019fault,chamberland2020very}. We could also consider using the state-preparation protocol with other codes, and we expect that there may be some room for optimization of the initialization strategy we have presented. We discuss these suggestions in Appendix~\ref{alt_app}.

Our work shows the value of carefully analyzing the circuit operations that are available with the underlying platform to ease the requirements of fault-tolerant quantum logical operations. To begin with, with the architecture we have considered here, we might expect to obtain an additional order of magnitude reduction in the preparation error by using a three-qubit $ZZZ(\theta)$ entangling gate. We discuss this gate in Appendix~\ref{zzz}. Moving forward, the discovery of better multi-qubit entangling gates that can be built using near-term technology could give us better error-corrected devices that are essential for practical quantum computing. 

\section{Open Problems}\label{AOM}
We now give some prospects related to CV-DV concatenation similar to the one considered in Sec.~\ref{sec:magic-state}, where a CV qubit encoding is abstracted as data qubits and ancilla qubits for a DV code. In particular, we will focus on open problems related to the case where the CV encoding is a GKP qubit. While the discussion on applications has been limited to oscillators used as a logical qubit, the true potential of oscillators lies in them being used as oscillators in quantum computation. To demonstrate useful quantum computation with oscillators, one requires error-corrected oscillators. A strategy to encode oscillators such that errors reduce with an increase in the number of physical oscillators is missing from the literature currently. We have briefly stated this issue as an open problem in Chapter~\ref{open:GKP-qec}. Thus, we do not dive into oscillator-based algorithms or simulations in this thesis. However, as a future insight, we outline a protocol to use them for phase estimation in this section. While we do not give a complete circuit-depth analysis, this result just hints at how techniques developed in this thesis could be used to develop quantum algorithms using oscillators.

\subsection{Hierarchical or `Lazy' Decoding via Probabilistic Decoding in CV-DV Concatenation}
Given the introduction to surface codes in Sec.~\ref{sec:magic-state} and probabilistic decoding in Chapter~\ref{chapter:GKP-qec}, we ask if it is possible to use the two in harmony to reduce the classical decoding time. Surface codes are large lattices that use matching algorithms to decode stabilizer measurement outcomes into the most likely error chain. These decoding strategies take a considerable amount of classical post-processing time and also present one of the bottlenecks in the computation speed of quantum computers. Multiple strategies have been used to make the decoding of DV codes like surface codes faster in recent years~\cite{wu2023fusion,higgott2025sparse}. One such attempt was at a hierarchical decoding strategy where a \emph{lazy decoder} acts as a pre-decoder to correct for easy error configurations. On top of this lazy decoder lies a more sophisticated decoding unit which is used when the lazy decoder cannot reach a verdict. This method was shown to achieve reductions in decoding hardware requirements.

The probabilistic decoding using autonomous dissipation described in Chapter~\ref{chapter:GKP-qec}, could be used to replace the lazy decoder if we use CV codes at the base layer of the DV encoding. Such CV-DV concatenation helps one to reduce the logical error probability achieved by only CV encoding (since CV codes are not proven to be scalable). The idea is to use probabilistic decoding (described in Chapter~\ref{chapter:GKP-qec}) to lower the strain on classical post-processing. In some cases, such concatenation has been shown to also lower qubit overhead~\cite{darmawan2021practical}.

With this hope, we propose the following different architectures to be studied for a GKP-based CV-DV concatenation,
\begin{itemize}
    \item In the past, the CV-DV concatenation schemes have been studied using a GKP data qubit and a GKP ancilla qubit. An autonomous-dissipation-based lazy decoder recommended here requires a biased-noise ancilla like the cat code for stabilizer measurements. Could a cat qubit ancilla be used to apply an autonomous dissipation into a four-mode GKP encoding in various patches of the surface code, on top of which a sophisticated surface code decoding is performed while maintaining the distance of the surface code?
    \item Such GKP-surface-code concatenation could prove helpful in dealing with ancilla errors that yield an uncorrectable displacement error on the GKP code; the question with an unintuitive answer is as follows. How does a stabilizer measurement of the DV code on the CV data qubits give any intuition about photon losses not yet corrected by the trickle-down approach shown in our work~\cite{sivak2023real}? 
    \item Finally, since the current schemes for GKP error correction require a biased-noise ancilla, the real question is: How does a CV-DV concatenation using GKP qubits as data qubits and GKP or cat qubits as ancilla qubits compare against the one using, say cat data qubits and cat ancilla qubits? If the latter works better, we would be better off using a cat-only CV-DV concatenation. However, a GKP qubit yields optimal protection against photon loss and has shown very promising results in experiments~\cite{campagne2020quantum,sivak2023real,brock2024quantum} over the last few years. If optimized carefully, CV-DV concatenation with the GKP codes could yield a significant reduction in resource overhead of fault-tolerant quantum computing.
\end{itemize}
\subsection{Quantum Phase Estimation}\label{sssec:quantum-phase-estimation}
\begin{figure}[htb]
    \centering
    \includegraphics[width=\linewidth]{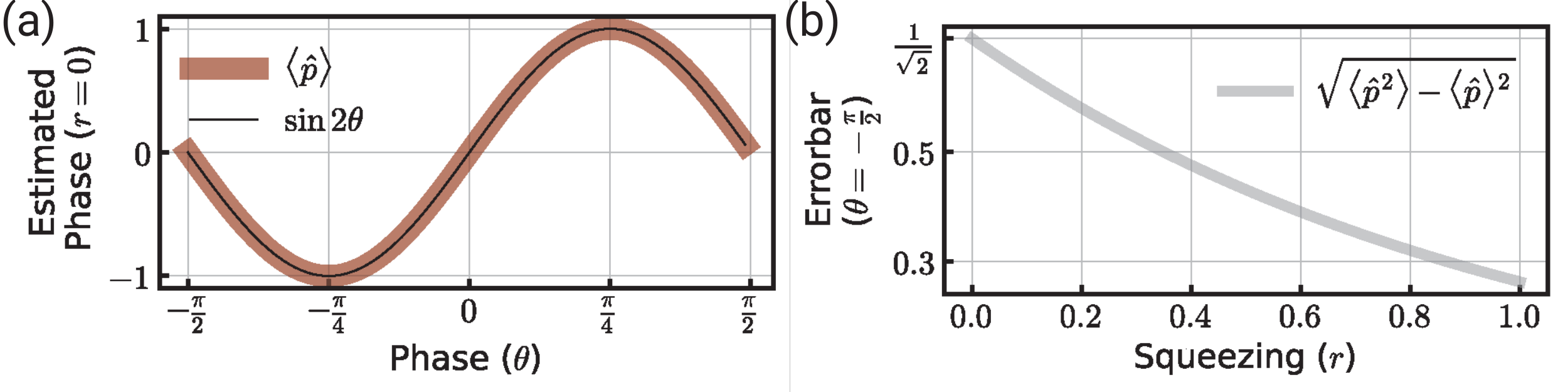}
    \caption[Quantum phase estimation in hybrid oscillator-qubit architecture.]{\textbf{Quantum phase estimation in hybrid oscillator-qubit architecture.} \textbf{(a)} Phase estimation can be performed using the control sequence described in Sec.~\ref{sssec:quantum-phase-estimation} as confirmed by this figure, for $\alpha=1$ in Eqs.~(\ref{eq:start_phase_est}-\ref{eq:end_phase_est}). \textbf{(b)} The error bar of this operation relies on the squeezing parameter $r$, upper bounded by $1/\sqrt{2}$ at $r=0$. This upper bound can only be improved by repeated application of the protocol.}
    \label{fig:phase_est}
\end{figure}
The eigenvalue of a unitary $U=e^{i\theta \hat{n}\cdot\vec\sigma}$ on an eigenstate $\ket{\psi}$ can be estimated in the hybrid oscillator-qubit architecture. In this section, we present an algorithm for this phase estimation with the help of an ancillary oscillator using the phase space instruction. For this purpose, we need to construct the following controlled-unitary $CU$ operation $C_xU=e^{if(\theta)\hat{x}\otimes\hat{n}\cdot\vec\sigma}$, where $f(\theta)$ is a known polynomial in $\theta$, using conditional displacements and qubit rotations (including $U$). Such an operation boosts the momentum of the oscillator state conditioned on the qubit and $f(\theta)$. The measurement of momentum boost using techniques like homodyne measurement can then be used to determine the eigenvalue $\theta$. This technique was  outlined in~\cite{Liu2016e} to find the eigenvalue of a Hamiltonian $H$ ($=\theta\hat n\cdot\vec{\sigma}$). The authors construct the conditional gate $e^{i\hat x\otimes \hat H}$ by assuming the availability of elementary gates like $e^{i\hat x\otimes h_k}$ (i.e., $f(\theta)=\theta$) such that $\prod_{k}e^{ih_k}=e^{iH}$ and $[h_k,h_k^\prime]=0$. Building on Ref.~\cite{Liu2016e}, we construct the $C_xU$ using \emph{non-commuting gates} from the phase-space instruction architecture, assuming only the availability of $U$ and leveraging \emph{arbitrary conditional displacements and rotations} instead of specific hybrid gates. The precision of this phase estimation technique will thus not only depend on the squeezing of the bosonic mode (illustrated in Ref.~\cite{Liu2016e}) but also on the approximation to which $C_xU$ can be constructed using the instruction set. 

For simplicity, we take $U=e^{i\theta\sigma_\mathrm{y}}$, and the construction below, \textit{mutatis mutandis}, follows for the case of arbitrary $\hat{n}\cdot\vec\sigma$ (see derivation in App.~\ref{app:phase-est}). 

\begin{equation}
e^{i\alpha\hat{x}\hat{n}\cdot\sigma_\mathrm{z}}e^{i\theta\sigma_\mathrm{x}}e^{-i\alpha\hat{x}\hat{n}\cdot\sigma_\mathrm{z}}e^{-i\theta\sigma_\mathrm{x}} =e^{i\hat{g}\hat{n'}\cdot\sigma}=C_xU
\end{equation}
where,
\begin{align}
\cos{\hat{g}}&=1-2\sin^2(\alpha\hat{x})\sin^2(\theta)\\
\cos{\hat{c}}&=\cos{\alpha\hat{x}}\cos{\theta}\label{eq:start_phase_est}\\
n'_\mathrm{z} &=\frac{\sin^2{\theta}\sin{2\alpha \hat{x}}}{\sin{\hat{g}}\sin^2{c}}\\
n'_\mathrm{y} &=\frac{\sin{2\theta}\sin{2\alpha\hat{x}}}{2\sin{\hat{g}}}\\
n'_\mathrm{x} &=\frac{\sin^2{\alpha\hat{x}}\sin{2\theta}}{\sin{\hat{g}}\sin^2{c}}.\label{eq:end_phase_est}
\end{align}
In Fig.~\ref{fig:phase_est} we show a proof of this calculation using a simulation that satisfies $\braket{p}=\alpha\sin{2\theta}$. The figure also shows that the standard deviation for the protocol is upper bounded by $1/\sqrt{2}$ at no squeezing when $r=0$, where $r$ is the momentum-squeezing parameter. The precision of the homodyne measurement could be made better via repeated measurements, and the cost of this repetition should be compared against qubit and time overhead in DV phase-estimation circuits. Here, we could employ the GCR-composed BB1 schemes developed in Sec.~\ref{sec:GCR} to perform this task using qubits for bit-wise measurement of the mean position value of the squeezed oscillator state. We will not discuss the details of a circuit-depth analysis for this strategy in this section.

\subsection{Random Walks in Quantum Phase Space and Deterministic State Preparation}
A quantum random walk in hybrid oscillator-qubit systems is a phenomenon where using a small displacement we can get an effectively large displacement but with extremely small probability. For example, a small displacement conditioned on $\sigma_z$, followed by measurement of $\sigma_x$, makes an even cat with a very high probability but with an extremely small probability, it makes an odd cat that has a very high overlap with Fock $\ket{1}$ state. In the latter case, this means the probability density will have a small amplitude at $x,p=1$ despite the displacement being $dx,dp\lll 1$. This effect was first illustrated and analyzed in Ref.~\cite{aharonov1993quantum}, and it truly captures the essence of quantum interference in phase space. 

Here, we will analyze this idea from the perspective of deterministic state preparation along the lines of the discussion in Sec.~\ref{ssec:universal}. Instead of relying on increasing the probability of a certain measurement, we would like to increase the probability of that outcome only using conditional displacements, i.e., a unitary channel (see Chapter~\ref{chapter: paper0}). A non-trivial back action on the oscillator can occur when the qubit completes a non-trivial loop/path in the oscillator-qubit Hilbert space, through only conditional displacements. More importantly, we consider the series of alternate displacements conditioned on $\sigma_\mathrm{x}$ and momentum boost conditioned on $\sigma_\mathrm{y}$, similar to Eq.~(\ref{eq:CD_circuit}).

After a conditional displacement, we have the following state $\ket{\psi_1}$,
\begin{align}
    e^{-i2\alpha\hat p\sigma_\mathrm{x}}\ket{0}_\mathrm{vac}\ket{g}&=\mathcal{N}\frac{(\ket{\alpha}\ket{+}+\ket{-\alpha}\ket{-})}{\sqrt{2}}\\
\end{align}
Now, we have a choice to make using a multi-sided coin with a $R_\mathrm{y}$ or an $R_\mathrm{z}$ rotation. Below, we will analyze both cases.
\begin{align}
R_\mathrm{y}(\theta)\ket{\psi_1}&=\frac{1}{2}\Big[\cos{\frac{\theta}{2}}\ket{\alpha}+\sin{\frac{\theta}{2}}\ket{-\alpha}\Big]\ket{+}\nonumber\\&-\frac{1}{2}\Big[\sin{\frac{\theta}{2}}\ket{\alpha}-\cos{\frac{\theta}{2}}\ket{-\alpha}\Big]\ket{-}\\
R_\mathrm{z}(\theta)\ket{\psi_1}&=\frac{1}{2}\Big[\cos{\frac{\theta}{2}}\ket{\alpha}-i\sin{\frac{\theta}{2}}\ket{-\alpha}\Big]\ket{+}\nonumber\\&+\frac{1}{2}\Big[\cos{\frac{\theta}{2}}\ket{-\alpha}-i\sin{\frac{\theta}{2}}\ket{\alpha}\Big]\ket{-}\\
&=\Bigg[\frac{\ket{\alpha}+\ket{-\alpha}}{2}\Bigg]\ket{g}+e^{i\theta}\Bigg[\frac{\ket{\alpha}-\ket{-\alpha}}{2}\Bigg]\ket{e}
\end{align}
If we do not measure the qubit after this step, then we track the random walk using unitary evolution. The top row allows interference between $\ket{\pm\alpha}$ which in turn yields $\braket{\sigma_\mathrm{x}}\neq 0,\braket{\sigma_\mathrm{y}}\neq 0$. On the other hand, the bottom row does not allow such interference which in turn leaves $\braket{\sigma_\mathrm{x}}\rightarrow 0$ and $\braket{\sigma_\mathrm{y}}\rightarrow 0$ as required for a pure rotation on the Bloch sphere. If we only want to rotate the qubit conditioned on the oscillator state, we should use the latter. 

Note that if the oscillator starts in $\ket{0}_\mathrm{vac}\ket{e}$, and the unit of alternate $\mathrm{CD}s$ used can be combined as 
\begin{align}
   x\sigma_\mathrm{x}+p\sigma_\mathrm{y}=\hat a\sigma_{+}+\hat a^\dagger \sigma_{-}, 
\end{align}
then the evolution under such a circuit should conserve parity conservation of the joint oscillator-qubit system as expected for the Jaynes-Cummings Hamiltonian. Note that $\ket{0}_\mathrm{vac}$ is an even parity state while $\ket{e}$ is an odd parity state. The joint parity of the hybrid system is odd in this case. Thus, at the end of the circuit, the oscillator will remain in an even parity state if the qubit was measured to be in state $\ket{e}$, but will switch to an odd-parity state if the qubit was measured to be in state $\ket{g}$. 

Similarly, let the oscillator start in $\ket{0}_\mathrm{vac}\ket{g}$, and the unit of alternate $\mathrm{CD}s$ used to be combined as 
\begin{align}
   x\sigma_\mathrm{x}-p\sigma_\mathrm{y}=\hat a\sigma_{-}+\hat a^\dagger \sigma_{+}.
\end{align}
In this case, the oscillator will remain in an even parity state if the qubit was measured to be in state $\ket{g}$, but will switch to an odd-parity state if the qubit was measured to be in state $\ket{e}$, as expected for the anti-Jaynes-Cummings Hamiltonian.

\begin{myframe}
\singlespacing
\begin{quote}
We conjecture that a systematic rotation of the qubit by an angle of $m\pi$ on the Bloch sphere via only CDs in the alternating sequence similar to Eq.~(\ref{eq:CD_circuit}) exists such that the random path followed by the hybrid state transfers the oscillator state from $\ket{0}$ to $\ket{m}$. For example, to create an even (odd) Fock state $\ket{2m}$ ($\ket{2m+1}$) for $m\in\mathbb{Z}$, we need to rotate the qubit by $2m\pi$ ($(2m+1)\pi$) about $\sigma_\mathrm{z}$ or $\sigma_\mathrm{x}$. That is, the oscillator-qubit state transitions from $\ket{0}\ket{e}$ to $\ket{2m}\ket{e}$ ($\ket{2m+1}\ket{g}$) or $\ket{0}\ket{g}$ to $\ket{2m}\ket{g}$ ($\ket{2m+1}\ket{e}$). As discussed in Sec.~\ref{ssec:universal}, the most straightforward circuit for doing so is a trotterization of the JC or AJC Hamiltonian. However, in this section, we also show that trotterization is not the most efficient way to approach this problem with the example of circuits with circuit depth $N=\{1,2,3\}$. We would like to know if combining the methods used in a random walk to increase the probability of observing the rarer event, we can find a constructive algorithm for Fock state preparation using CDs without trotterization of JC or AJC. More specifically, the question is, what happens if $\phi\in [0,2\pi]$ and not just $\phi\in\{0,\pi/2,\pi,3\pi/2\}$ in a circuit composed of CDs where $CD=e^{i\vec{\beta}\sigma_\mathrm{x}}$ with $\vec{\beta}=\beta(\cos{\phi}\hat x-\sin{\phi}\hat p)$?
\end{quote}
\end{myframe}

\doublespacing

The interface of quantum random walks and deterministic state preparation can be, not only helpful in understanding universal state preparation analytically but will also give insights into analyzing the joint oscillator-qubit Hilbert space. We present some arguments along these lines in App.~\ref{app:conc} and leave a systematic study of such random walk-based preparation schemes as a future prospect. 

% etc
    \chapter{Open Question Summary and Conclusion}\label{sec:summary}
This thesis explores key aspects of hybrid continuous-variable (CV) and discrete-variable (DV) systems, with a focus on using the non-trivial formalism of CV phase space and improving CV-DV architectures. The contributions presented here provide new insights into the classification of CV operations, oscillator state control, error correction, and fault tolerance, highlighting open problems that could inform future research.  

Chapter~\ref{chapter: paper0} introduces CV, DV, and hybrid quantum systems, noting the absence of a classification of CV operations analogous to the Clifford hierarchy for DV (qubit) operations. Given the role of the Clifford hierarchy in fault tolerance, we propose a similar hierarchy for oscillator-based codes in Sec.~\ref{open-hierarchy} which we call the Gaussian hierarchy. Understanding the relationship between the Gaussian hierarchy and the Clifford hierarchy could clarify how non-Gaussian and non-Clifford operations differ fundamentally. Specifically, we show that $\mathcal{G}_n \not\subset \mathcal{C}_n$, and thus, ask: Is $\mathcal{C}_n \subset \mathcal{G}_n$ for $n \geq 4$? Is this true for $n = 3$? Establishing this relationship could deepen our understanding of non-classical operations in phase space and their role in hybrid quantum systems. A broader question is how such a hierarchy might impact CV analogs of the Solovay-Kitaev theorem~\cite{SK,arzani2025can} and transversality~\cite{bravyi2013classification} in multi-mode CV codes.  

In Chapter~\ref{chapter:na-qsp}, we examine non-abelian quantum signal processing (QSP) from the perspective of oscillator control. We introduce the first composite pulse sequence, GCR, which outperforms the best-known abelian QSP sequence (BB1) by a factor of 4.5 in duration while maintaining comparable performance. This result demonstrates that non-abelian QSP can provide more efficient control of oscillator-qubit systems, improving the manipulation of quantum states in phase space. Extending the framework of quantum singular value transformation (QSVT) to non-abelian QSP could unify control protocols for hybrid systems. We suggest that this formalism might be generalizable to two-qubit systems and multi-mode oscillator states, offering a path toward more flexible control schemes for multiple quantum systems.  

Chapter~\ref{chapter:state-prep} explores the application of the GCR sequence for deterministic state preparation of oscillator states, including squeezed states, cat states, GKP states, four-legged cat states, and Fock states. We propose a quantum random walk approach to Fock state preparation, hypothesizing that a systematic qubit rotation by $m\pi$ using conditional displacements (CDs) could transfer the oscillator state from $\ket{0}$ to $\ket{m}$. This method could enable efficient generation of even and odd Fock states without relying on trotterization of the JC or AJC Hamiltonian. An open question is whether adjusting the displacement phase $\phi \in [0, 2\pi]$ (rather than restricting it to $\{0, \pi/2, \pi, 3\pi/2\}$) could improve the fidelity of state preparation, leveraging the rare-event enhancement seen in quantum random walks~\cite{aharonov1993quantum}. These results suggest a deeper connection between phase space dynamics and efficient oscillator state generation.  

Chapter~\ref{chapter:GKP-qec} provides the first analytical insights into probabilistic error correction of photon loss using GKP stabilization schemes, building on recent experimental work~\cite{sivak2023real, brock2024quantum}. Our analysis raises two key questions: While oscillator codes have no threshold against a random displacement channel~\cite{hanggli2021oscillator}, is this also true for photon loss? Could encoding oscillators into oscillators enable photon-loss-protected states? Can insights from probabilistic correction of photon loss correction developed in this thesis improve the design of error-protected qubits, such as those based on GKP codes, in superconducting circuits~\cite{sellem2025dissipative, nathan2024self, geier2024self}? These questions are central to developing more robust hybrid architectures that leverage the phase space structure of oscillators for improved error resilience.  

In Chapter~\ref{chapter:qec-control}, we present high-fidelity circuits for error-detected state preparation, logical readout with residual errors, and pieceably fault-tolerant gate teleportation in GKP codes. Our gate teleportation circuit tolerates biased noise in the ancilla, without the need for conditional displacement gates that are transparent to ancilla errors. That said, the availability of an error-transparent gate could improve our protocol by removing the requirement of a biased noise ancilla, simplifying the hardware requirements. Inspired by Ref.~\cite{ReinholdErrorCorrectedGates}, we ask whether such error-transparent gates can yield sufficient protection with our pieceable protocols using a qutrit ancilla in place of a biased-noise ancilla. 

Chapter~\ref{chapter:conc} discusses a protocol using a concatenation of CV cat codes with scalable DV surface codes that could reduce the resource overhead of universal fault-tolerant quantum computing. Cat codes are natural candidates for this protocol due to their bias-preserving $\mathrm{CX}$ gates; rectangular GKP codes could also be suitable if bias-preserving $\mathrm{CX}$ gates can be engineered—an open problem. For a CV-DV concatenation scheme with respect to GKP codes, we ponder over the advantages in terms of decoding requirements. For example, replacing the lazy decoder in hierarchical decoding~\cite{delfosse2020hierarchical} with autonomous dissipation could enhance performance. Key questions in this direction include: How does stabilizer measurement of the DV code provide insight into photon loss correction in a CV-DV concatenation (with GKP codes as data qubits)? Does concatenation improve photon loss correction? How does a scheme using GKP data and cat ancilla qubits compare to one using only cat qubits? We note that pending practical oscillator error correction, oscillator-based quantum computation remains limited in its ability to extract useful information from quantum systems. 

To conclude, this thesis develops key steps that could drive progress in CV-DV architectures. Specifically, we have
\begin{enumerate}
\item proposed a Clifford-like hierarchy, which we call the Gaussian hierarchy, for CV operations to clarify the structure of CV operations,
\item introduced composite pulse sequences in the class of non-abelian QSP to improve hybrid state control,
\item provided analytical insights into deterministic oscillator state preparation, consistent with numerical optimization,  
\item developed state-of-the-art high-fidelity GKP error correction and control schemes, and
\item explored the potential of CV-DV concatenation to reduce fault-tolerant resource overhead.
\end{enumerate}
In particular, contributions numbered $2,3,4$ cover the main results of this thesis captured in Chapters~\ref{chapter:na-qsp}-\ref{chapter:qec-control}. Our work deepens the understanding of phase space dynamics in hybrid systems and proposes efficient control, error correction, and fault-tolerant quantum computing in CV-DV architectures.
    
    \appendix
    
\chapter{Superconducting CV-DV Platform}\label{App:Phys_Imp}
We will primarily discuss the architectures of a superconducting system in addition to some other experimental aspects of CV-DV quantum computing. This restriction is made due to my collaborations in superconducting experiments and in the theory of device physics reported in Refs.~\cite{singh2024impact,sivak2023real,brock2024quantum,nguyen2022blueprint}.
\begin{figure}[htb]
    \centering
    \includegraphics[width=\linewidth]{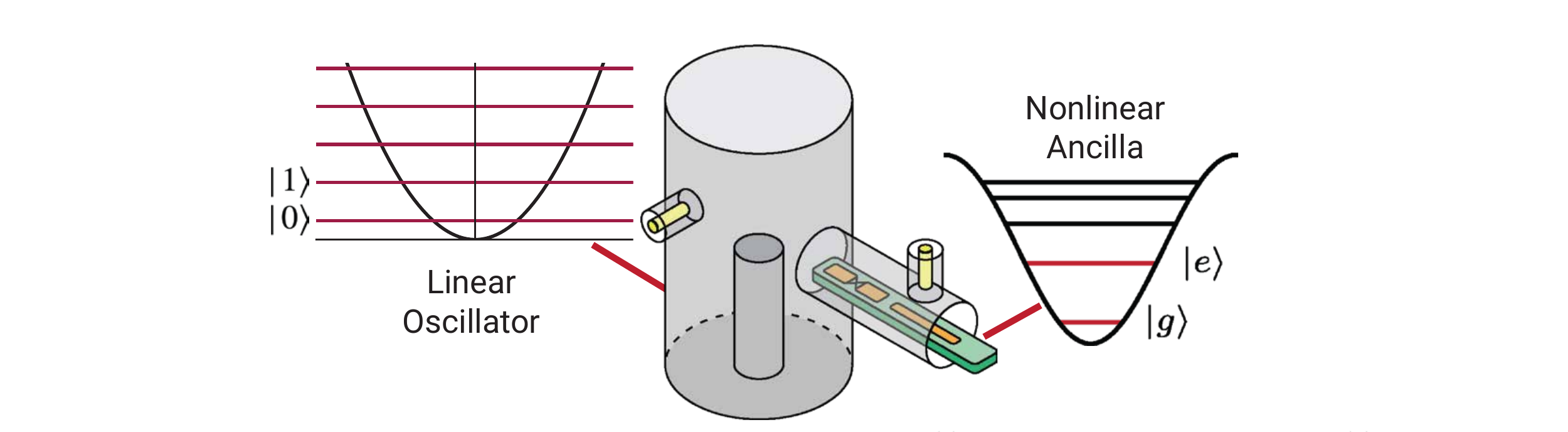}
    \caption[Illustration of a coupled CV-DV physical system]{Illustration of a 3D superconducting cavity (oscillator) coupled to a nonlinear ancilla. The lowest two ($d$) levels of the nonlinear ancilla form the qubit (qudit) subspace, while the electromagnetic modes in the 3D cavity form the harmonic oscillator. In this case the ancilla resembles the potential of a transmon, a popular qubit realization in superconducting circuits. The oscillator modes have high coherence but are harder to control. The DV systems are easier to control but have low lifetimes. In a hybrid architecture, we can use the ancilla as controllers of the oscillators. In this thesis, we focus on an efficient control techniques where the ancilla is in a deterministic state after  short snippets of the circuit. If the ancilla is reset at this point then the total circuit will suffer much less errors. The idea is that, an error in a small snippet of the circuit does not harm the fidelity of operations by a large amount, if the qubit is reset. This is because reset allows to protect the rest of the circuit from being affected by the errors occurring in the early stages of the circuit. Figure inspired from our work~\cite{sivak2023real}.}
    \label{fig:phys_sys}
\end{figure}
\section{DV systems}
\paragraph{Transmon~\cite{Koch2007}:} These are simplest qubit architectures realized using a single Josephson junction (JJ) and a capacitor connected in parallel. These qubits offer fast and efficient control of the quantum information but have high decay rates. 
\paragraph{Fluxonium~\cite{manucharyan2009fluxonium}:} These are nonlinear systems with an inductor in parallel with the transmon circuit. The inductor can be realized using Josephson junction (JJ) array with $100$ elements or a strip of granular aluminum (GrAl). The inductive shunt protects the qubit against charge noise. Such qubits can have very long coherence time but engineering gates for them becomes harder. Recently there has some advancements in the control (gates and readout) of these qubits~\cite{ding2023high,zhang2024tunable,nguyen2022blueprint}. These qubits have the flexibility to realize a biased-noise error model such that probability ($p_z$) of $\sigma_\mathrm{z}$ errors is much larger than probabilities ($p_x,p_y$) of a $\sigma_\mathrm{x}$ or $\sigma_\mathrm{y}$ error. In a separate work, not presented in this dissertation, we theoretically studied the efficiency of readout in the presence of parasitic modes from the JJ array used for the inductive shunt in these less simpler devices~\cite{singh2024impact}. Our work extends to any superconducting circuit with multiple Josephson junctions, or even a superconducting chip with spurious modes.

There are protected qubits, like the Kerr-cat qubits (partially protected) and $0-\pi$ qubits, as discussed at the end of Chapter~\ref{chapter:GKP-qec}.
\section{CV}
\begin{figure}[tbh]
    \centering
    \includegraphics[width=0.5\textwidth]{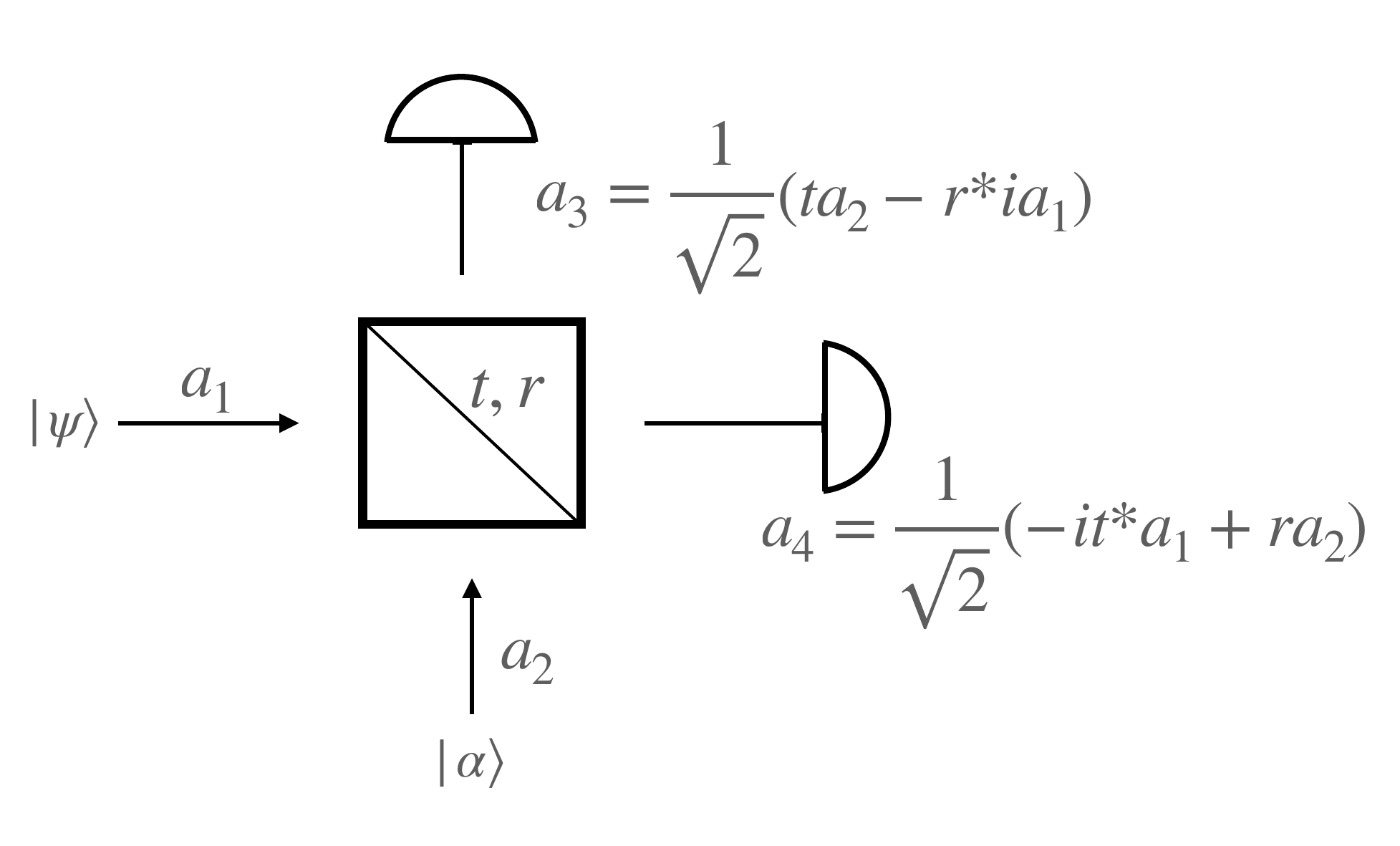}
    \caption[Homodyne detection]{Homodyne detection involves a beam splitter (BS) with transmission coefficient `t' and reflection coefficient `r,' and two detectors at the output ports of the BS. For balanced homodyne detection, $|t|^2=|r|^2=\frac{1}{2}$.}
    \label{fig:BS}
\end{figure}

The CV systems include different types of resonators with a millisecond long lifetime~\cite{ganjam2024surpassing}. Niobium cavities~\cite{milul2023superconducting} coupled to DV systems (qubits) have shown tens of millisecond long lifetimes but the size of these cavities is too big to be practical for quantum computing purposes. Some cavities have shown second long lifetimes~\cite{posen2023high} without any consideration to couple them to qubits.
\paragraph{Homodyne Detection:}
\label{sec:homodyne-detection}
Homodyne detection can be seen as a projective measurement of the phase-space quadratures. The setup for homodyne detection includes a beam splitter with transmission coefficient (t) and reflection coefficient (r), and two photodetectors (see figure~\ref{fig:BS}). The quantity of interest here is
\begin{equation}
    \langle \hat n_3\rangle -\langle \hat n_4\rangle
\end{equation}
the difference of mean photon numbers detected by the two detectors for any input state.
The state to be measured ($\ket{\psi}$) is sent through one port while a coherent state ($\ket{\alpha}$) is sent through another port. In the Heisenberg picture, the photon number operators at the output ports are, 
\begin{eqnarray}
    \hat n_3&=&\frac{1}{2}\left [|t|^2 \hat n_2+|r|^2 \hat n_1-i(t^*r\hat a_2^\dagger \hat a_1-r^*t\hat a_1^\dagger \hat a_2) \right]
    \,,
    \nonumber\\&&\\
    \hat n_4&=&\frac{1}{2} \left[|t|^2 \hat n_1+|r|^2 \hat n_2-i(t^*r\hat a_1^\dagger \hat a_2-r^*t\hat a_2^\dagger \hat a_1)\right ],\nonumber\\    
\end{eqnarray}
and the mean photon number is given by
\begin{eqnarray}
    \langle n_4\rangle &=&\frac{1}{2}\left [
    \rule{0pt}{2.4ex}
    |t|^2\bra{\psi}\hat n_1\ket{\psi}+|r|^2 |\alpha|^2
    \right. ~~~~~~~
    \nonumber\\
    && ~~~~~~ \left. -i \bra{\psi}(t^* r\alpha) a_1^\dagger-(tr^*\alpha^*) a_1\ket{\psi} 
    \rule{0pt}{2.4ex} \right]
    \nonumber.\\&&
\end{eqnarray}
Assuming a balanced homodyne detection with $t=r$ and $\alpha=|\alpha|e^{i\phi}$
\begin{equation}
     \langle n_3\rangle-\langle n_4\rangle=|\alpha|\bra{\psi} (a^\dagger e^{i\phi}-ae^{-i\phi})\ket{\psi}.
\end{equation}
Thus, the difference between mean photon numbers of each detector is the mean value of the phase-space quadrature along an axis dependent on the phase of the coherent state. Homodyne detection on identically prepared states reconstructs the quadrature probability distribution. 
\section{DV-CV}
\paragraph{Dissipative cat qubits:} These qubits encode a cat code in an oscillator using dissipation based stabilization. such stabilization emulates a dissipator $\mathcal{D}[a^n-\alpha^n]$ to stabilize an n-legged cat states. Dissipative cats are the first hybrid bosonic-qubit systems in which QEC was achieved beyond the break-even point~\cite{ofek2016extending}. These codes have recently been used for error correction in concatenation with repetition codes~\cite{putterman2024hardware}, an approach which could, in principle, reduce the space time overhead of error correction compared to using 2D surface codes with transmons~\cite{acharya2024quantum}.
\paragraph{Dissipative GKP qubits:} Two experiments~\cite{sivak2023real,campagne2020quantum} have achieved QEC gain of GKP memory beyond break-even. These experiments are based on an approach where the dissipator of the finite energy GKP code is engineered to drive the resonator into the ground state of the GKP codespace using an ancillary qubit. The experiments have shown that current limits on the QEC gain are due to errors in the DV ancillary systems. It has been proposed~\cite{royer2020stabilization} that with use of biased-noise ancilla, like Kerr cat qubits or dissipative cat qubits, one could protect GKP code from ancillary errors\footnote{Recently preliminary of such a system has been presented in Ref.~\cite{ding2024quantum}}. These circuits make use of conditional displacements. An ideal scenario would be to engineer a conditional displacement gate which is tolerant to ancilla errors upto some order, like the what was realized for SNAP gate~\cite{ReinholdErrorCorrectedGates}. This direction is presented as an open problem in Chapter~\ref{chapter:qec-control}.
\paragraph{Dissipative qudits} There have been several experiments recently realizing qudit systems~\cite{brock2024quantum,yu2025schrodinger,omanakuttan2024fault,wetherbee2024mathematical}. In particular, in Ref.~\cite{brock2024quantum}, we achieve beyond-break even error correction performance for GKP qutrits and ququarts using reinforcement learning based otimization for error correction of GKP codes, inspired by Ref.~\cite{sivak2023real}. The scheme used here is also based on dissipation engineering.

\chapter{Supplementary for Chapter~\ref{chapter:na-qsp}}\label{app:comp_err}
In this appendix, we compute the error analysis for the univariate QSP sequence $\mathrm{BB1}(2\theta)$ and the bivariate non-abelian QSP sequence $\mathrm{GCR}(2\theta)$. We begin by describing the metrics used in the Chapter~\ref{chapter:na-qsp} and the subsequent appendices. The equations extracted in the latter sections have been used in plots shown in Secs.~\ref{sec:composite_error}.
\section{Performance Metrics}\label{performance metrics}
Chapters~\ref{chapter:na-qsp}-\ref{chapter:state-prep} and~\ref{chapter:qec-control} focus on extracting a single bit of information from the oscillators (CV systems) via qubits (DV systems). Here, we define the metrics which will be helpful in assessing the performance of the various sequences used towards this goal. One of the examples of this task is to distinguish between a $\ket{\alpha_\Delta}$ and $\ket{-\alpha_\Delta}$, where $\ket{\pm\alpha_\Delta}$ is a squeezed coherent state defined in Eq.~(\ref{eq:Gaussian}). 
Let us suppose the QSP sequence used for this task is denoted by $U$ has been applied to the hybrid oscillator-qubit system $\ket{g}\otimes\ket{\pm\alpha_\Delta}$. The most general statement we can write about this operation is as follows,
\begin{align}
    U(\ket{g}\otimes\ket{\alpha_\Delta})&=\beta_{+g}\ket{\psi_{+g}}\ket{g}+\gamma_{+e}\ket{\phi_{+e}}\ket{e}\\
    U(\ket{g}\otimes\ket{-\alpha_\Delta})&=\gamma_{-g}\ket{\phi_{-g}}\ket{g}+\beta_{-e}\ket{\psi_{-e}}\ket{e}
\end{align}
We have two requirements here, (1) to determine whether the oscillator is in state $\ket{\alpha_\Delta}$ or $\ket{-\alpha_\Delta}$ with maximum fidelity, and (2) this measurement should be quantum non-demolition (QND) in that it has minimal back action on the oscillator state. We will assume that the two states are symmetric in that $\beta_{+g}=\beta_{-e}\equiv \beta, \braket{+\alpha_\Delta|\psi_{+g}}=\braket{-\alpha_\Delta|\psi_{-e}}\equiv \braket{\alpha|\psi}$ and $\gamma_{+e}=\gamma_{-g}\equiv \gamma,\braket{+\alpha_\Delta|\phi_{+e}}=\braket{-\alpha_\Delta|\phi_{-g}}\equiv \braket{\alpha|\phi}$. This indicates that $|\beta|^2$ decides the ability to successfully (unsuccessfully) deduce the sign of the mean position of the oscillator using the qubit outcome. Thus, we give the following two 
quantities. 
\begin{itemize}
    \item The \emph{fidelity} of this measurement strategy, 
\begin{align}
    &\frac{|\beta_{+g}|^2-|\gamma_{+e}|^2+|\beta_{-e}|^2-|\gamma_{-g}|^2}{2}\\
    =&|\beta|^2-|\gamma|^2=1-2|\gamma|^2=1-2P_e(U),
\end{align}
where $P_e(U)$ is the probability of failure to rotate the qubit to the state(s) predicted by the mean position of the oscillator (for example, $\ket{g}$ ($\ket{e}$) if oscillator is in $\ket{\alpha_\Delta}$ ($\ket{-\alpha_\Delta}$) state).
\item Another quantity is the \emph{QNDness} of this measurement strategy which quantifies the back action of this strategy on the oscillator state,

\begin{align}
&\frac{|\beta_{+g}|\braket{+\alpha_\Delta|\psi_{+g}}|^2+|\gamma_{+e}|\braket{+\alpha_\Delta|\phi_{+e}}|^2+|\beta_{-e}|\braket{-\alpha_\Delta|\psi_{-e}}|^2+|\gamma_{-g}|\braket{-\alpha_\Delta|\phi_{-g}}|^2}{2}\\
&=|\beta|^2|\braket{\alpha|\psi}|^2+|\gamma|^2|\braket{\alpha|\phi}|^2.
\end{align}

The first (second) term represents the fidelity of the oscillator with the original state in the event of a success (failure). Thus, this quantity presents the fidelity of the oscillator state in the event of a qubit reset, which we call the hybrid state fidelity $F_\mathrm{H}(U)$ for the specific QSP sequence $U$.
\end{itemize}
Thus, based on these metrics, for each QSP sequence used in this thesis, we will quote the $P_e(U)$ and $F_\mathrm{H}(U)$.
\subsection{No QSP correction}\label{app:no_corr}
Here, we start with a state $\ket{\psi_1}=\ket{g}\otimes\ket{\alpha}$ and analyze the effect of applying the conditional momentum boost $e^{-i\frac{\pi}{4\alpha}\hat x\sigma_\mathrm{x}}$. Ignoring normalization factors, we have,
\begin{align}
\braket{x|\psi_1}&=\Big(\frac{2}{\pi}\Big)^\frac{1}{4}e^{-(x-\alpha)^2}\ket{g}\label{psi_1x}\\
\braket{x|\psi_2}&=\Big(\frac{2}{\pi}\Big)^\frac{1}{4}e^{-i\frac{\pi}{4\alpha}\hat x\sigma_\mathrm{x}}e^{-(x-\alpha)^2}\ket{g}
\end{align}
Now, the task is to compute the overlap with the desired state $\ket{-i}\otimes\ket{\alpha}$ which is given by,
\begin{align}
&=\Big(\frac{2}{\pi}\Big)^\frac{1}{2}\int_{-\infty}^{\infty}dx\quad \braket{-i|e^{-i\frac{\pi}{4\alpha}x\sigma_\mathrm{x}}e^{-2(x-\alpha)^2}|g}\label{eqn_1a}\\
&=\Big(\frac{2}{\pi}\Big)^\frac{1}{2}\frac{1}{2}\int_{-\infty}^{\infty}dx\quad (e^{i\pi/4}e^{-i\frac{\pi}{4\alpha}x}+e^{-i\pi/4}e^{i\frac{\pi}{4\alpha}x})e^{-2(x-\alpha)^2}\\
&=\Big(\frac{2}{\pi}\Big)^\frac{1}{2}\frac{1}{2}\int_{-\infty}^{\infty}dx\quad (e^{-i\frac{\pi}{4\alpha}(x-\alpha)}+e^{i\frac{\pi}{4\alpha}(x-\alpha)})e^{-2(x-\alpha)^2}\\
&=\Big(\frac{2}{\pi}\Big)^\frac{1}{2}\int_{-\infty}^{\infty}dx\quad \cos{\Big(\frac{\pi}{4\alpha}(x-\alpha)\Big)}e^{-2(x-\alpha)^2}\\
&=\Big(\frac{2}{\pi}\Big)^\frac{1}{2}\int_{-\infty}^{\infty}dx\quad \Big(1-\frac{\pi^2}{32\alpha^2}y^2\Big)e^{-2y^2}\\
&=1-\Big(\frac{2}{\pi}\Big)^\frac{1}{2}\frac{\pi^2}{32\alpha^2}\int_{-\infty}^{\infty}dx\quad y^2e^{-2y^2}\\
&=1-\frac{\pi^2}{128\alpha^2}.
\end{align}
Therefore, the hybrid state fidelity $F_\mathrm{H}$ for large $|\alpha|$ is equal to,
\begin{align}
F_\mathrm{H}(\mathrm{no-QSP})\approx\Big|1-\frac{\pi^2}{128\alpha^2}\Big|^2\approx 1-\frac{\pi^2}{64\alpha^2}.\label{eqn:fidelity}
\end{align}
Since the operation $e^{i\frac{\pi}{4}x\sigma_\mathrm{x}}$ applies no back action on the position basis, the probability of failure in this case is the same as the reset fidelity computed above.
\subsection{Bivariate Sequence: $\textrm{GCR}(2\theta)$}\label{app:error-analysis}
Let us proceed with the calculation of $P_g=1-P_e$. For simplicity, we also define $y\Delta=(x-\alpha)$ and use $\lambda=\frac{\theta\Delta^2}{|\alpha|}$. Using the Taylor expansions from eqs.~\ref{eq:U}-\ref{eq:V} to write,
{\allowdisplaybreaks
\begin{align}
\braket{x|UV|\alpha_\Delta,g}=&\sum_{m=0}^{\infty}\frac{[i\theta(x-\alpha)\sigma_\mathrm{x}]^m}{|\alpha|^mm!}\sum_{n=0}^\infty \Big(-\frac{\lambda\sigma_\mathrm{y}}{2\Delta}\Big)^n \frac{1}{n!}H_n\Big(\frac{x-\alpha}{\Delta}\Big)\alpha_\Delta(x)\ket{g}\\
=&\sum_{m=4\mathbb{Z},n=2\mathbb{Z}}^\infty \frac{(\lambda/\Delta)^{m+n}}{2^n n!m!}y^m\Big[H_{n}(y)-\frac{(\lambda/\Delta)\sigma_\mathrm{y}}{2(n+1)}H_{n+1}(y)\Big]\alpha_\Delta(x)\ket{g}\\
    &+\sum_{m=4\mathbb{Z}+1,n=2\mathbb{Z}}^\infty \frac{i(\lambda/\Delta)^{m+n}}{2^n n!m!}y^m\Big[H_{n}(y)\sigma_\mathrm{x}-\frac{i(\lambda/\Delta)\sigma_\mathrm{z}}{2(n+1)}H_{n+1}(y)\Big]\alpha_\Delta(x)\ket{g}\\
&+\sum_{m=4\mathbb{Z}+2,n=2\mathbb{Z}}^\infty \frac{-(\lambda/\Delta)^{m+n}}{2^n n!m!}y^m\Big[H_{n}(y)-\frac{(\lambda/\Delta)\sigma_\mathrm{y}}{2(n+1)}H_{n+1}(y)\Big]\alpha_\Delta(x)\ket{g}\\
    &+\sum_{m=4\mathbb{Z}+3,n=2\mathbb{Z}}^\infty \frac{-i(\lambda/\Delta)^{m+n}}{2^n n!m!}y^m\Big[H_{n}(y)\sigma_\mathrm{x}-\frac{i(\lambda/\Delta)\sigma_\mathrm{z}}{2(n+1)}H_{n+1}(y)\Big]\alpha_\Delta(x)\ket{g}
\end{align}
}
We have broken down the product terms $\braket{x|UV|\alpha_\Delta,g}$ into eight groups corresponding to the combinations of $m\in\{4\mathbb{Z},4\mathbb{Z}+1,4\mathbb{Z}+2,4\mathbb{Z}+3\}$ and $n,n+1$ s.t. $n\in 2\mathbb{Z}$. Now satisfying the requirements from our framework we have, $\sigma_\mathrm{y}\ket{g}=i\sigma_\mathrm{x}\ket{g},\sigma_\mathrm{z}\ket{g}=\ket{g}$,
{\allowdisplaybreaks
\begin{align}    \braket{x|UV|\alpha_\Delta,g}
&=\sum_{m=4\mathbb{Z},n=2\mathbb{Z}}^\infty \frac{(\lambda/\Delta)^{m+n}}{2^n n!m!}y^m\Big[H_{n}(y)-\frac{i(\lambda/\Delta)\sigma_\mathrm{x}}{2(n+1)}H_{n+1}(y)\Big]\alpha_\Delta(x)\ket{g}\\
    &+\sum_{m=4\mathbb{Z}+1,n=2\mathbb{Z}}^\infty \frac{(\lambda/\Delta)^{m+n}}{2^n n!m!}y^m\Big[iH_{n}(y)\sigma_\mathrm{x}+\frac{(\lambda/\Delta)}{2(n+1)}H_{n+1}(y)\Big]\alpha_\Delta(x)\ket{g}\\
&+\sum_{m=4\mathbb{Z}+2,n=2\mathbb{Z}}^\infty \frac{(\lambda/\Delta)^{m+n}}{2^n n!m!}y^m\Big[-H_{n}(y)+\frac{i(\lambda/\Delta)\sigma_\mathrm{x}}{2(n+1)}H_{n+1}(y)\Big]\alpha_\Delta(x)\ket{g}\\
    &+\sum_{m=4\mathbb{Z}+3,n=2\mathbb{Z}}^\infty \frac{(\lambda/\Delta)^{m+n}}{2^n n!m!}y^m\Big[-iH_{n}(y)\sigma_\mathrm{x}-\frac{(\lambda/\Delta)}{2(n+1)}H_{n+1}(y)\Big]\alpha_\Delta(x)\ket{g}
\end{align}
}
It is clear from the above expression that the $m+n\in 2\mathbb{Z}+1$ terms reduce the probability of success, taking $\ket{g}\rightarrow \ket{e}$. Thus, we can rewrite the above expression as,

\begin{align}
    \braket{x|UV|\alpha_\Delta,g}
&=\sum_{m+n\in2Z}(-1)^{\nu_m}c_{n,m}y^mH_n(y)e^{-y^2}\ket{g}\nonumber\\&\quad+i\sum_{m+n\in 2Z+1}(-1)^{\mu_m}c_{n,m}y^mH_n(y)e^{-y^2}\ket{e},\\
\mathrm{where}\quad    c_{n,m}&=\frac{(\lambda/\Delta)^{m+n}}{2^n n!m!}, \ \nu_m:\mod(m,4)\ge 2, \  \mu_m:\mod(m,4)==(0 \  \mathrm{or}\ 3).
\end{align}
Here, $\mu_m,\nu_m$ are conditional variables that are equal to $1$ if the condition representing them is true else they are $0$. All terms yield a well-bounded Gaussian integral with decreasing contribution to the success probability for increasing $m+n$, assuming $\lambda/\Delta\ll 1$. Thus, the total error of the process is also bounded and we can focus on the leading order term. We will extract terms up to $\mathcal{O}(\lambda^6/\Delta^6)$ as these will contribute to the leading order terms in the failure probability, as shown below,

\allowdisplaybreaks{
\begin{align}   \braket{x|UV|\alpha_\Delta,g}=&e^{-y^2}[c_{00}H_0(y)\ket{g}-ic_{10}H_1(y)\ket{e}+c_{20}H_2(y)\ket{g}-ic_{30}H_3(y)\ket{e}+c_{40}H_4(y)\ket{g}\nonumber\\
&\quad\quad-ic_{50}H_5(y)\ket{e}+c_{60}H_6(y)\ket{g}-ic_{70}H_7(y)\ket{e}+c_{80}H_8(y)\ket{g}\\
&+ic_{01}yH_0(y)\ket{e}+c_{11}yH_1(y)\ket{g}+ic_{21}yH_2(y)\ket{e}+c_{31}yH_3(y)\ket{g}\nonumber\\
&\quad\quad+ic_{41}yH_4(y)\ket{e}+c_{51}yH_5(y)\ket{g}+ic_{61}yH_6(y)\ket{e}+c_{71}yH_7(y)\ket{g}\\
&-c_{02}y^2H_0(y)\ket{g}+ic_{12}y^2H_1(y)\ket{e}-c_{22}y^2H_2(y)\ket{g}+ic_{32}y^2H_3(y)\ket{e}\nonumber\\
&\quad\quad-c_{42}y^2H_4(y)\ket{g}+ic_{52}y^2H_5(y)\ket{e}-c_{62}y^2H_6(y)\ket{g}\\
&-ic_{03}y^3H_0(y)\ket{e}-c_{13}y^3H_1(y)\ket{g}-ic_{23}y^3H_2(y)\ket{e}-c_{33}y^3H_3(y)\ket{g}\nonumber\\
&\quad\quad-ic_{43}y^3H_4(y)\ket{e}-c_{53}y^3H_5(y)\ket{g}\\
&+c_{04}y^4H_0(y)\ket{g}-ic_{14}y^4H_1(y)\ket{e}+c_{24}y^4H_2(y)\ket{g}-ic_{34}y^4H_3(y)\ket{e}\nonumber\\
&\quad\quad+c_{44}y^4H_4(y)\ket{g}\\
&+ic_{05}y^5H_0(y)\ket{e}+c_{15}y^5H_1(y)\ket{g}+ic_{25}y^5H_2(y)\ket{e}+c_{35}y^5H_3(y)\ket{g}\\
&-c_{06}y^6H_0(y)\ket{g}+ic_{16}y^6H_1(y)\ket{e}-c_{26}y^6H_2(y)\ket{g}\\
&-ic_{07}y^7H_0(y)\ket{e}-c_{17}y^7H_1(y)\ket{g}\\
&+c_{08}y^8H_0(y)\ket{g}]+\mathcal{O}(\lambda^9/\Delta^9)
   \end{align}
As suggested earlier, the first order terms in $y$ exactly cancel since $ic_{10}H_1(y)=ic_{01}yH_0(y)$, and hence there is no effect on the final state from terms that are degree $1$ in $\lambda/\Delta$. Defining $\chi=\lambda/\Delta$, we have
\begin{align}
\braket{x|UV|\alpha_\Delta,g}=&\mathcal{N}\Bigg\{1+\chi^2\Bigg[y^2-\frac{1}{4}\Bigg]+\chi^4\Bigg[-\frac{y^4}{6}-\frac{y^2}{4}+\frac{1}{32}\Bigg]+\chi^6\Bigg[-\frac{y^6}{90}+\frac{y^4}{24}+\frac{y^2}{32}-\frac{1}{384}\Bigg]\nonumber\\&+\chi^8\Bigg[\frac{y^8}{2520}+\frac{y^6}{360}-\frac{y^4}{192}-\frac{y^2}{384}+\frac{1}{6144}\Bigg]+\mathcal{O}(\chi^{10})\Bigg\}e^{-y^2}\ket{g}\label{eq:g}\\
&+\mathcal{N}\Bigg\{i\chi^3\Bigg[\frac{2y^3}{3}\Bigg]-i\chi^5\Bigg[\frac{y^3}{6}\Bigg]+\mathcal{O}(\chi^7)\Bigg\}e^{-y^2}\ket{e}\label{eq:e}
\end{align}
}
Here, $\mathcal{N}$ is the normalization constant which will be computed using $\braket{\psi|\psi}=1$.

\paragraph{Figures of merit:}

\begin{itemize}
\item \textit{Success probability.} 

The success probability of rotating the qubit by $I$ is only affected by $\mathcal{O}(\chi^3)$ and $\mathcal{O}(\chi^5)$ terms in the expansion of $UV$. Using the variable transformation $dx=\Delta dy$ and $\int_{-\infty}^{\infty}dy \  y^{2n}e^{-2y^2}=\sqrt{\frac{\pi}{2}}\frac{(2n-1)!!}{(4)^n}$ effect of these terms can be approximated as ,
{\allowdisplaybreaks
\begin{align}
    P_g&=\int_{-\infty}^\infty  dx \  |\braket{x,g|UV|\alpha,g}_\Delta|^2\\&=\mathcal{N}^2\Delta\int_{-\infty}^\infty dy \ e^{-2y^2}\Big(1+\chi^2\Big[2y^2-\frac{1}{2}\Big]\nonumber\\&+\chi^4\Big[\frac{2y^4}{3}-y^2+\frac{1}{8}\Big]+\chi^6\Big[-\frac{16}{45}y^6-\frac{y^4}{3}+\frac{y^2}{4}-\frac{1}{48}\Big]\nonumber\\&+\chi^8\Big[-\frac{3}{140}y^8+\frac{17}{180}y^6+\frac{y^4}{32}-\frac{5}{192}y^2+\frac{5}{3072}\Big]\nonumber\\&+\mathcal{O}(\chi^{10})\Big)\\
    &=\mathcal{N}^2\Delta\sqrt{\frac{\pi}{2}}(1-5\chi^6/48+11\chi^8/768+\mathcal{O}(\chi^{10}))\label{eq:succ_norm}
\end{align}
where
\begin{align} 1/\mathcal{N}^2&=P_g+\Delta\int_{-\infty}^\infty dy \ e^{-2y^2} \Big(\frac{4}{9}\chi^6y^6-\frac{2}{9}\chi^8y^6+\mathcal{O}(\chi^{10})\Big)
\\&=\sqrt{\frac{\pi}{2}}\Delta(1-29\chi^8/768+\mathcal{O}(\chi^{10}))\label{eq:norm}
\end{align} 
Thus, we get the probability of making an incorrect rotation as the probability of ending in qubit state $\ket{e}$ at the end of $UV$,
\begin{align}
    P_e&=\frac{5\chi^6/48-5\chi^8/96}{1-29\chi^8/768}+\mathcal{O}(\chi^{10})
\end{align}
where
$\chi=\frac{\pi\Delta}{4|\alpha|}=\frac{\theta\Delta}{|\alpha|},$
and $2\theta$ is the angle by which the qubit is rotated on the Bloch sphere. Hence, the probability of making an erroneous rotation has been proved to scale as $\sim\chi^6/10$. In contrast to the traditional schemes for composite pulses with classical variables, the error terms for quantum control variables scale with $\Delta/\alpha$ instead of error $e=|x-\alpha|$ due to the Gaussian-weighted distribution of error over this range. For a given state, the variables $\Delta,\alpha$ are fixed, and hence the success probability is also fixed. As $\alpha\rightarrow 0$, the curve deviates since higher order terms start come into play. This is not an issue since neither the scheme nor the small $\chi$ approximation are well-suited for $\alpha\rightarrow 0$ limit.

\item \textit{Post-selected fidelity.} Next, we quantify the back action on the oscillator state conditioned on the qubit being in the desired state using $\mathcal{F}_{\mathrm{success}}$. The final oscillator state  conditioned upon the qubit being in $\ket{g}$ state $\psi_\textrm{final}(x)$ is given by the Eq.~(\ref{eq:g}). This yields the fidelity upon success that is the fidelity of the oscillator state when the ancilla is in $\ket{g}$.
\begin{align}
F_\mathrm{ps}&=| \braket{\alpha|\psi_\Delta}|^2:\psi_\Delta(x)=\mathcal{N}_g\braket{x,g|UV|\alpha,g}_\Delta,\end{align}
where $1/\mathcal{N}^2_g$ is obtained from dividing Eq.~(\ref{eq:succ_norm}) by $\mathcal{N}^2=\Delta\sqrt{\frac{\pi}{2}}(1-5\chi^6/48+11\chi^8/768)+\mathcal{O}(\chi^{10})$
\begin{align}
|\braket{\alpha_\Delta|\psi_{final}}|^2&=\sqrt{\frac{2}{\pi}}\mathcal{N}^2_g\Delta\Bigg|\int_{-\infty}^\infty dy\quad e^{-2y^2} \Bigg\{1+\chi^2\Bigg[y^2-\frac{1}{4}\Bigg]\nonumber\\&\quad+\chi^4\Bigg[-\frac{y^4}{6}-\frac{y^2}{4}+\frac{1}{32}\Bigg]+\chi^6\Bigg[-\frac{y^6}{90}+\frac{y^4}{24}+\frac{y^2}{32}-\frac{1}{384}\Bigg]\nonumber\\&\quad+\chi^8\Bigg[\frac{y^8}{2520}+\frac{y^6}{360}-\frac{y^4}{192}-\frac{y^2}{384}+\frac{1}{6144}\Bigg]+\mathcal{O}(\chi^{10})\Bigg\}\nonumber\\&=\frac{|1-\chi^4/16+\chi^6/48-\chi^8/1536|^2}{1-5\chi^6/48+11\chi^8/768}\nonumber\\&\quad+\mathcal{O}(\chi^{10})\\
   \therefore 1-F_\mathrm{ps}&=\frac{\chi^4/8-\chi^6/8+\chi^8/64}{1-5\chi^6/48+11\chi^8/768}+\mathcal{O}(\chi^{10})
\end{align}
\item \textit{Hybrid fidelity.} If the failure probability is low enough, we can afford to ignore the outcome of the qubit and let it reset. In this case, the fidelity of the oscillator state is bounded as follows,
\begin{align}
  1-F_\mathrm{H}&=| \braket{\alpha,g|UV|\alpha, g}_\Delta|^2=\chi^4/8-\chi^6/48+\mathcal{O}(\chi^{8})
\end{align}
 Thus we see that the post-selected infidelity and reset infidelity both scale as $\chi^4/8$ for $\chi\ll 1$.} 
\end{itemize}

\subsection{Univariate Sequence: $\textrm{BB1}
(2\theta)$}\label{app:err_BB1}
We perform error analysis for the composite pulse sequence using quantum variables adapted from the well-known BB1$(2\theta)$ sequence~\cite{wimperis1994broadband}. Here, the fidelity needs to be computed for the Gaussian-weighted error terms obtained by Taylor expanding Eq.~(\ref{eq:BB1}) when applied on the state $\ket{\psi}\otimes\ket{g}$ where $\braket{x|\psi}=e^{-\frac{(x-\alpha)^2}{\Delta^2}}$. We will continue to use the pre-defined shorthand notations $y\Delta=x-\alpha,\chi=\theta\Delta/|\alpha|$ from Chapter~\ref{chapter:na-qsp}. Note that, we have used the Taylor expansion $f(a+\varepsilon)=\sum_{n=0}^\infty \frac{\varepsilon^n}{n!}f^n(a)$, where $f^n(a)$ is the $n^\textrm{th}$ derivative of the rotation $f(a)=\cos{a}I+i\sin{a}\sigma_\phi$. We want Eq.~(\ref{eq:BB1_err}) to be equivalent to identity upto $\mathcal{O}(\chi^k)$ where $k\ge 2$ determines the order of error cancellation. 

\begin{align}
    \ket{\psi_\textrm{BB1}}=&\mathrm{R}_{\phi_1}\Bigg(-\frac{\pi}{|\alpha|}\hat x\Bigg)\mathrm{R}_{3\phi_1}\Bigg(-\frac{2\pi}{|\alpha|}\hat x\Bigg)\mathrm{R}_{\phi_1}\Bigg(-\frac{\pi}{|\alpha|}\hat x\Bigg) \mathrm{R}_0\Bigg(-\frac{2\theta}{|\alpha|}(\hat x-\alpha)\Bigg)\ket{\psi}\otimes\ket{g}\\
     \braket{x|\psi_\textrm{BB1}}=&\Bigg[\cos{\frac{\pi\alpha}{2|\alpha|}}I+i\sin{\frac{\pi\alpha}{2|\alpha|}}\sigma_{\phi_1}-\frac{\pi}{2\theta}\chi y\Big(\sin{\frac{\pi\alpha}{2|\alpha|}}I-i\cos{\frac{\pi\alpha}{2|\alpha|}}\sigma_{\phi_1}\Big)\Bigg]\nonumber\\&\times\Bigg[\cos{\frac{\pi\alpha}{|\alpha|}}I+i\sin{\frac{\pi\alpha}{|\alpha|}}\sigma_{3\phi_1}-\frac{\pi}{\theta}\chi y\Big(\sin{\frac{\pi\alpha}{|\alpha|}}I-i\cos{\frac{\pi\alpha}{|\alpha|}}\sigma_{3\phi_1}\Big)\Bigg]\nonumber\\&\times\Bigg[\cos{\frac{\pi\alpha}{2|\alpha|}}I+i\sin{\frac{\pi\alpha}{2|\alpha|}}\sigma_{\phi_1}-\frac{\pi}{2\theta}\chi y\Big(\sin{\frac{\pi\alpha}{2|\alpha|}}I-i\cos{\frac{\pi\alpha}{2|\alpha|}}\sigma_{\phi_1}\Big)\Bigg]\nonumber\\&\times\Bigg[\cos{\frac{\theta\alpha}{|\alpha|}}I+i\sin{\frac{\theta\alpha}{|\alpha|}}\sigma_\textrm{x}-\chi y\Big(\sin{\frac{\theta\alpha}{|\alpha|}}I-i\cos{\frac{\theta\alpha}{|\alpha|}}\sigma_\textrm{x}\Big)\Bigg]\mathrm{R}_0\Bigg(2\theta\frac{\alpha}{|\alpha|}\Bigg)e^{-y^2}\ket{g}\nonumber\\&\quad+\mathcal{O}(\chi^2)\label{eq:BB1_err}.
\end{align}
Simplifying this, using $\cos{(\pi\alpha/|\alpha|)}=-1,\cos{(\pi\alpha/2|\alpha|)}=0,\sin{(\pi\alpha/|\alpha|)}=0,\sin{(\pi\alpha/2|\alpha|)}=\alpha/|\alpha|$, we have $\braket{x|\psi_\textrm{BB1}} $

 \begin{align}
     &=\mathcal{N}\sum_{m=0}^\infty \frac{\sigma_{\phi_1}}{m!}\Big[i\frac{\pi}{2\theta}\chi y\sigma_{\phi_1}\Big]^m\sum_{n=0}^\infty\frac{1}{n!}\Big[i\frac{\pi}{\theta}\chi y\sigma_{3\phi_1}\Big]^n\sum_{o=0}^\infty \frac{\sigma_{\phi_1}}{o!}\Big[i\frac{\pi}{2\theta}\chi y\sigma_{\phi_1}\Big]^o\sum_{o=0}^\infty \frac{1}{p!}\Big[i\chi y\sigma_\textrm{x}\Big]^pe^{-y^2}\ket{g}.
\end{align}
Here $\mathcal{N}$ is the normalization constant. This expression simplifies as follows for terms up to third order, $\braket{x|\psi_\textrm{BB1}}$
\allowdisplaybreaks{
\begin{align}
=&\mathcal{N}\Big\{1+i\chi y\Big[\frac{\pi}{\theta}(\sigma_{\phi_1}+\sigma_{\phi_1}\sigma_{3\phi_1}\sigma_{\phi_1})+\sigma_\textrm{x}\Big]-\chi^2y^2\Big[\frac{1}{2}+\frac{\pi^2}{\theta^2}+\frac{\pi^2}{2\theta^2}(\sigma_{\phi_1}\sigma_{3\phi_1}+\sigma_{3\phi_1}\sigma_{\phi_1})\nonumber\\&+\frac{\pi}{\theta}(\sigma_{\phi_1}\sigma_\textrm{x}+\sigma_{\phi_1}\sigma_{3\phi_1}\sigma_{\phi_1}\sigma_\textrm{x})\Big]-i\chi^3y^3\Big[\frac{2\pi^3}{3\theta^3}+\frac{\pi}{2\theta}(\sigma_{\phi_1}+\sigma_{\phi_1}\sigma_{3\phi_1}\sigma_{\phi_1})-\frac{\pi^3}{4\theta^3}\sigma_{\phi_1}\sigma_{3\phi_1}\sigma_{\phi_1}\nonumber\\&+\Big(\frac{\pi^2}{\theta^2}+\frac{1}{6}\Big)\sigma_\textrm{x}+\frac{\pi^3}{4\theta^3}\sigma_{3\phi_1}+\frac{\pi^2}{2\theta^2}(\sigma_{2\phi_1}+\sigma_{-2\phi_1})\Big]+\mathcal{O}(\chi^4)\Big\}e^{-y^2}\ket{g}\\
 =&\mathcal{N}\Big\{1+i\chi y\Big[\frac{\pi}{\theta}(\sigma_{\phi_1}+\sigma_{-\phi_1})+\sigma_\textrm{x}\Big]-\chi^2y^2\Big[\frac{1}{2}+\frac{\pi^2}{\theta^2}+\frac{\pi^2}{2\theta^2}(e^{i2\phi_1\sigma_\textrm{z}}+e^{-i2\phi_1\sigma_\textrm{z}})\nonumber\\&+\frac{\pi}{\theta}(e^{i\phi_1\sigma_\textrm{z}}+e^{-i\phi_1\sigma_\textrm{z}})\Big]-i\chi^3y^3\Big[\Big(\frac{2\pi^3}{3\theta^3}+\frac{\pi}{2\theta}\Big)(\sigma_{\phi_1}+\sigma_{-\phi_1})-\frac{\pi^3}{4\theta^3}\sigma_{-\phi_1}+\Big(\frac{\pi^2}{\theta^2}+\frac{1}{6}\Big)\sigma_\textrm{x}\nonumber\\&+\frac{\pi^3}{4\theta^3}\sigma_{3\phi_1}+\frac{\pi^2}{2\theta^2}(\sigma_{2\phi_1}+\sigma_{-2\phi_1})\Big]+\mathcal{O}(\chi^4)\Big\}e^{-y^2}\ket{g}\\
=&\mathcal{N}\Big\{1+i\chi y\Big[\frac{2\pi}{\theta}\cos{\phi_1}\sigma_\textrm{x}+\sigma_\textrm{x}\Big]-\chi^2y^2\Big[\frac{1}{2}+\frac{\pi^2}{\theta^2}+\frac{\pi^2}{\theta^2}\cos{2\phi_1}+\frac{2\pi}{\theta}\cos{\phi_1}\Big]\nonumber\\&-i\chi^3y^3\Big[\Big(\frac{4\pi^3}{3\theta^3}+\frac{\pi}{\theta}\Big)
\cos{\phi_1}\sigma_\textrm{x}-\frac{\pi^3}{4\theta^3}\sigma_{-\phi_1}+\Big(\frac{\pi^2}{\theta^2}+\frac{1}{6}\Big)\sigma_\textrm{x}+\frac{\pi^3}{4\theta^3}\sigma_{3\phi_1}+\frac{\pi^2}{\theta^2}\cos{2\phi_1}\Big]\nonumber\\&+\mathcal{O}(\chi^4)\Big\}e^{-y^2}\ket{g}.
\end{align}
}
 For simplification we have used $\sigma_{\phi}=e^{-i\phi\sigma_\textrm{z}}\sigma_\textrm{x}$ to deduce that $\sigma_{\phi_1}\sigma_{3\phi_1}\sigma_{\phi_1}=\sigma_{-\phi_1}$. We observe that both first- and second-order terms cancel out with the choice of $\phi=\pm\cos^{-1}(-\theta/2\pi)$ and using $\cos{2\phi}=2\cos^2{\phi}-1$. 

\begin{align}
\therefore\braket{x|\psi_\textrm{BB1}} &=\mathcal{N}\Big\{1-i\chi^3y^3\Big[\Big(-\frac{\pi^2}{6\theta^2}+
\frac{1}{24}\Big)\sigma_\textrm{x}+\frac{\pi^3}{4\theta^3}(\sin{3\phi_1}+\sin{\phi_1})\sigma_\textrm{y}\Big]+\mathcal{O}(\chi^4)\Big\}e^{-y^2}\ket{g}\\
&=\mathcal{N}\Big\{\Big[1+\mathcal{O}(\chi^4)\Big]e^{-y^2}\ket{g}+\Big[-i\chi^3y^3\Big(-\frac{\pi^2}{6\theta^2}+
\frac{1}{24}+i\frac{\pi^3}{2\theta^3}\cos{\phi_1}\sin{2\phi_1}\Big)\nonumber\\&\quad+\mathcal{O}(\chi^5)\Big]e^{-y^2}\ket{e}\Big\}\\
&=\mathcal{N}\Big\{\Big[1+\mathcal{O}(\chi^4)\Big]e^{-y^2}\ket{g}+\Big[-i\chi^3y^3\Big(-\frac{\pi^2}{6\theta^2}+
\frac{1}{24}+i\frac{\pi}{4\theta}\sqrt{1-\frac{\theta^2}{4\pi^2}}\nonumber\\&\quad+\mathcal{O}(\chi^5)\Big]e^{-y^2}\ket{e}\Big\}
\end{align}

For $\theta=\pi/2$, the failure probability of this error cancellation scheme is given by the probability of obtaining $\ket{e}$ in the outcome. 
\begin{align}
    P_e&=\int_{-\infty}^\infty \ dy \ \Bigg|-\frac{\pi^2}{6\theta^2}+
\frac{1}{24}+i\frac{\pi}{4\theta}\sqrt{1-\frac{\theta^2}{4\pi^2}}\Bigg|^2\chi^6y^6e^{-2y^2}\nonumber\\&\quad+\mathcal{O}(\chi^8)\\&\quad=\Bigg|-\frac{\pi^2}{6\theta^2}+
\frac{1}{24}+i\frac{\pi}{4\theta}\sqrt{1-\frac{\theta^2}{4\pi^2}}\Bigg|^2\Bigg(\frac{15}{64}\Bigg)\chi^6+\mathcal{O}(\chi^8)
\end{align}
For $\theta=\pi/2$ i.e. $\textrm{BB1}(180)$, we get,
\begin{align}
    P_e&=\Bigg|\frac{5}{8}-i\frac{\sqrt{15}}{8}\Bigg|^2\Bigg(\frac{15}{64}\Bigg)\chi^6+\mathcal{O}(\chi^8)=0.15\chi^6+\mathcal{O}(\chi^8)
\end{align}
For $\theta=\pi/4$ i.e. $\textrm{BB1}(90)$, we get,
\begin{align}
    P_e&=\Bigg|\frac{63}{24}-i\frac{\sqrt{63}}{8}\Bigg|^2\Bigg(\frac{15}{64}\Bigg)\chi^6+\mathcal{O}(\chi^8)=1.85\chi^6+\mathcal{O}(\chi^8)
\end{align}
We see that the failure probability is worse for $\textrm{BB1}(2\theta)$ compared to $\textrm{GCR}(2\theta)$ for $2\theta=90^\circ$ whereas it is comparable for both when $2\theta=180^\circ$. For our purpose, we will primarily be using $\textrm{BB1}(90)$. 

Finally, the reset fidelity in both cases is also important when using our formalism. Note that in both cases, $\textrm{BB1}$ and $\textrm{GCR}$, the second order term disappears in the final integral. The reset fidelity expressions in Eqs.~\ref{eq:bb1_180}-\ref{eq:bb1_90} have been computed using the coefficient of $\chi^4,\chi^6$ in the Taylor expansion using Mathematica. For $2\theta=90^\circ$, we have, the additional terms in the Taylor expansion are as follows,
\begin{align}
    \braket{x,g|\psi_\textrm{BB1}}&=1-i\frac{\sqrt{15}}{8}\chi^4y^4-\frac{7}{9}\chi^6y^6+i\frac{5\sqrt{5}}{8\sqrt{3}}\chi^6y^6\nonumber\\&\quad+\mathcal{O}(\chi^6)\label{eq:bb1_180}
\end{align}
This yields the hybrid state fidelity to be equal to for $2\theta=180^\circ$,
\begin{align}
    F_\textrm{H}&=\Bigg|\int_{-\infty}^\infty \ dy \ \Big[1-i\frac{\sqrt{15}}{8}\chi^4y^4-\frac{7}{9}\chi^6y^6+i\frac{5\sqrt{5}}{8\sqrt{3}}\chi^6y^6+\mathcal{O}(\chi^8)\Big]e^{-2y^2}\Bigg|^2\\
    &=\Bigg|1 -i\frac{\sqrt{15}}{8}\Bigg(\frac{3}{16}\Bigg)\chi^4-\frac{7}{9}\Bigg(\frac{15}{64}\Bigg)\chi^6+i\frac{5\sqrt{5}}{8\sqrt{3}}\Bigg(\frac{15}{64}\Bigg)\chi^6+\mathcal{O}(\chi^8)\Bigg|^2\\&\quad=1-\frac{105}{288}\chi^6+\mathcal{O}(\chi^8)\\
    1-F_\textrm{H}&=0.37\chi^6+\mathcal{O}(\chi^8),
\end{align}
Since the fourth-order term is purely imaginary there is no fourth-order term contributing to the infidelity of the state. Thus, in this case the infidelity scales as $\chi^6$.
We can repeat this exercise for $2\theta=90^\circ$, and find,
\begin{align}
    1-F_\textrm{H}=15.6\chi^6+\mathcal{O}(\chi^6).\label{eq:bb1_90}
\end{align}
\begin{figure}
    \centering
    \includegraphics[width=0.9\linewidth]{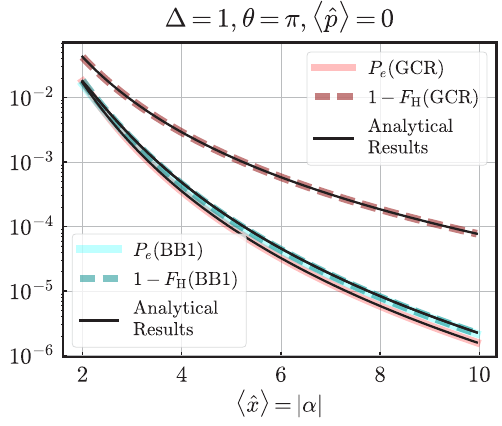}
    \caption[Comparison of $\mathrm{GCR}(\theta)$ and $\mathrm{BB1}(\theta)$ schemes for $\theta=\pi$.]{\textbf{Comparison of $\mathrm{GCR}(\theta)$ and $\mathrm{BB1}(\theta)$ schemes for $\theta=\pi$.} This scheme confirms our analytical understanding which shows that the failure probability for this case will be comparable for both the schemes. The figure can be contrasted against Fig.~\ref{fig:Correctness}(a) where the two failure probabilities were off by an order of magnitude.}
    \label{fig:BB1_180}
\end{figure}

Hence, in this appendix, we have confirmed that the performance of $\mathrm{GCR}(2\theta)$ is on par with $\mathrm{BB1}(2\theta)$ in terms of success probability while yielding a lower fidelity, however, at a much lower circuit-depth, as claimed in Sec.~\ref{sec:composite_error}. In Fig.~\ref{fig:BB1_180}, we plot the comparison of $\mathrm{GCR}(2\theta)$ for $\theta=\pi/2$ which can be contrasted with Fig.~\ref{fig:Correctness}(a) for $\theta=\pi/4$.

\section{Variations of GCR($\theta$): $\braket{\hat p}\neq 0$, $\braket{\hat x}\neq 0$ and $\Delta>1$.}\label{ssec:pnot0} 
In Sec.~\ref{sec:GCR}, we focused on the case of $\alpha\in\mathbb{R}$, however, our scheme is generalizable to arbitrary coherent states. For states in Eq.~(\ref{eq:Gaussian}) where $\alpha$ is not real, i.e., the state is not located on the position axis of the oscillator phase space, or where $\Delta>1$, i.e., a squeezed coherent state, $\mathrm{GCR}$ requires simple modifications as follows. For the latter, we simply choose $\hat v_1=\hat p$ in Eq.~(\ref{eq:GCR}). Let us discuss the former case of $\braket{\hat p}\neq 0$ and $\braket{\hat x}\neq 0$. Up to a normalization constant and a phase factor,
\begin{align}
\braket{x|\alpha_\Delta,i\beta_\Delta}=e^{i\beta\hat x}e^{-\frac{(x-\alpha)^2}{\Delta^2}}.
\end{align}
Let us see how we can modify $\mathrm{GCR}(\theta)$ to rotate the qubit using this state. We have, 
\begin{align}
    &\mathrm{GCR}(\theta)\ket{+\alpha_\Delta,i\beta_\Delta;g}=\mathrm{GCR}(\theta)e^{i\beta\hat x}\ket{g,+\alpha_\Delta}\\
    &\quad\quad=e^{i\beta\hat x}\mathrm{GCR}(\theta)e^{-i\frac{\lambda}{2}\beta\sigma_\mathrm{y}}\ket{g,+\alpha_\Delta} \\
    &\quad\quad=e^{i\beta\hat x}\mathrm{GCR}(\theta)e^{-i\frac{\theta\Delta^2}{2|\alpha|}\beta\sigma_\mathrm{y}}\ket{g,+\alpha_\Delta}. \label{eq:pnot0}
\end{align}
Therefore, for the correction to work in this case, we need to apply an initial rotation on the qubit equivalent to $e^{i\theta\Delta^2\frac{\beta}{2|\alpha|}\sigma_\mathrm{y}}$. In Fig.~\ref{fig:Correctness}(b) we show the numerical results for this protocol as proof. This variation gives the generalization of $\mathrm{GCR}$ to coherent states located along arbitrary vectors in the phase space of a quantum oscillator. We use these calculations in the preparation of four-legged cat states in Sec.~\ref{ssec:universal}.

\chapter{Supplementary for Chapter~\ref{chapter:state-prep}}\label{app:state-prep}
\section{Squeezing}\label{app:squeezing}
Here, we look at the question of modifying the position uncertainty $\delta x$ of an oscillator state $\psi(x)=e^{-\frac{(x-\alpha)^2}{4{\delta x}^2}}$. Without loss of generality, we will use the vacuum state with $\alpha=0, \ 4{\delta x}^2=1$. The action of squeezing this state along position is equivalent to,
\begin{align}
\textrm{S}(\Delta)\ket{\psi}=\int_{-\infty}^{\infty}dp e^{-\frac{\delta_x^2 p^2}{4}}  \ket{p}=\int_{-\infty}^{\infty}dx e^{-\frac{x^2}{4\delta x^2}}  \ket{x}, 
\end{align}
where $\ket{p},\ket{x}$ are eigenstates of $\hat x,\hat p$ and $4\delta x^2>1$. 
\subsection{Correctness Metrics}
\paragraph{Squeezing in $\mathrm{dB}$.} For comparison with~\cite{eickbusch2022fast} we use,
\begin{align}
    S(r)=10\log_{10}(e^{r/2}), \ S_p=S(\log{4{\delta x}^2})=S(\log{4{\Delta}^2}), \ 
    S_x=S(\log{4/{\Delta}^2}),
\end{align}
where $r$ is the parameter used for the bosonic squeezing operation~\cite{ISA}. Here, $\delta x, \delta p$ are the uncertainties in position and momentum, respectively, as defined in Sec.~\ref{CV}.
\paragraph{Fisher information:}
Fisher information of a Gaussian state, such as, squeezed states is given by~\cite{hastrup2021unconditional}, 
\begin{align}
    F=2/\big(\braket{\hat x^2}-\braket{x}^2\big)=2/{\delta x}^2.
\end{align}
\paragraph{Circuit-duration and circuit-depth:} We plot the squeezing curves with respect to circuit-duration, that is the time taken by the circuit instead of the gate count. It is because the error and speed of a conditional displacement gate depends on the length of the conditional displacement. This dependence is computed given access to the Hamiltonian $H_{\textrm{CD}}=\chi (\gamma_0 a^\dagger-\gamma_0^* a)$ where $\frac{\chi}{2\pi}=50$ kHz and $|\gamma_0|=20\implies T_{|\gamma_i|}= \frac{|\gamma_i|}{\chi |\gamma_0|}$. Note that, duration of the conditional displacements are lower bounded by $T_{|\gamma_i|<0.024}=48\text{ }$ns. This duration includes the necessary components for an echoed conditional displacement, an unconditional displacement $|\alpha_0|$ ($24$ ns) and a mid-circuit qubit rotation ($24$ ns). For details see Ref.~\cite{eickbusch2022fast}.
\paragraph{Sum of two Gaussian functions:} 
Consider the sum of two Gaussian functions,
\begin{align}
    \mathcal{N}[e^{-\frac{(x-\alpha)^2}{\Delta^2}}+e^{-\frac{(x+\alpha)^2}{\Delta^2}}],
\end{align}
where $\mathcal{N}=(2/(\pi\Delta^2))^{1/4}/\sqrt{2(1+e^{-2\alpha^2/\Delta^2})}$. After each application of squeeze operator $\mathcal{S}$, as described in Sec.~\ref{sec:squeezing} we create a superposition of Gaussian functions which resembles a wider Gaussian function in the position basis ($\mathcal{N^\prime}e^{-x^2/{\Delta^\prime}^2}$). We use the Python package scipy.optimize() to estimate the $\Delta$ corresponding to this output state. We can also directly use the function variance(), on the output state, in QuTip~\cite{Johansson2013}. Alternatively, there are other numerical methods such as Pade's approximation~\cite{baker2012pade} which can be used here. 

\subsection{Squeezing with GCR}
Our protocol outlined in Sec.~\ref{sec:squeezing} can be described as follows. Here, we assume that vacuum is $\psi(x)=e^{-\frac{(x-\beta)^2}{\Delta^2}}$ such that $\Delta^2=1,\beta=0$. After a $\mathrm{CD}(\alpha,\sigma_\textrm{x})$, we have, 
\begin{align}
  \braket{\sigma_\textrm{z}}&=\cos{\theta}=\frac{2e^{-(x-\alpha)^2/\Delta^2-(x+\alpha)^2/\Delta^2}}{e^{-2(x-\alpha)^2/\Delta^2}+e^{-2(x+\alpha)^2/\Delta^2}},\\
  &=\frac{2}{e^{-4\alpha x/\Delta}+e^{4\alpha x/\Delta}}=\mathrm{sech}{4\alpha x/\Delta^2},\\
  \braket{\sigma_\textrm{y}}&=\sin{\theta}\sin{\phi}=0,\\
  \braket{\sigma_\textrm{x}}&=\sin{\theta}\cos{\phi}=1-\braket{\sigma_\textrm{z}}^2=\tanh{4\alpha x/\Delta^2}.
\end{align}
When $4\alpha/\Delta^2$ is small, $\sigma_\mathrm{x}$ varies linearly with $x$ across the support of $\psi(x)$, i.e., for $|x| \leq 2\delta x \lesssim \Delta$. In this regime, applying the rotation $\mathrm{R}\textrm{y}(-4\alpha \hat{x}/\Delta^2) = e^{i(2\alpha \hat{x}/\Delta^2)\sigma\mathrm{y}}$ drives $\braket{\sigma_\textrm{x}} \to 0$, as $\tanh(4\alpha x/\Delta^2)$ remains approximately linear. More precisely, the various expectation values take the following form after this corrective rotation,
\begin{align}
  \braket{\sigma_\mathrm{z}}&=\tanh{\frac{4\alpha x}{\Delta^2}}\sin{\frac{4\alpha x}{\Delta^2}}+\mathrm{sech}{\frac{4\alpha x}{\Delta^2}}\cos{\frac{4\alpha x}{\Delta^2}},\\
  &=1-\mathcal{O}(x^6)\\
  \braket{\sigma_\mathrm{y}}&=0,\\
  \braket{\sigma_\mathrm{x}}&=\tanh{\frac{4\alpha x}{\Delta^2}}\cos{\frac{4\alpha x}{\Delta^2}}-\mathrm{sech}{\frac{4\alpha x}{\Delta^2}}\sin{\frac{4\alpha x}{\Delta^2}}\\
  &=\mathcal{O}(x^3).\label{eq:new_sx}
\end{align}
 The composite pulse sequence $\mathrm{GCR}(2\theta)$ described in Sec.~\ref{sec:GCR} is exactly based on this principle, if analyzed in the momentum basis, since the $\frac{\theta}{|\alpha|}$ is small for large $|\alpha|$.

\textbf{Choice of $\alpha_k$:} We must choose $\alpha$ to ensure a linear slope for $\sigma_\mathrm{x}$ across $|x|\le 2\delta x \lesssim \Delta$. Since $\alpha$ controls the rate of squeezing convergence in each $\mathcal{S}_k$ step (see Fig.~\ref{fig:squeezing}(a)), it should be as large as possible. At the same time, the slope $4\alpha/\Delta^2$ must decrease with increasing $\Delta$ to preserve linearity of $\braket{\sigma\textrm{x}}$ over the support of $\psi(x)$. To be exact, the slope should be atleast, 
\begin{align}
\frac{4|\alpha|}{\Delta^2}\ll \frac{1/\mathcal{N}}{\textrm{FWHM}}&=\frac{{(2/\pi\Delta^{2}})^{1/4}}{2\Delta\sqrt{\ln 2}}=\frac{0.53}{\Delta^{3/2}}, \\
\implies |\alpha|&\ll 0.13\Delta^{1/2}.
\end{align}
 Here, $\mathcal{N}$ is the normalization constant, representing inverse of the peak of $\psi(x)$, and FWHM is the full width at half maximum. To ensure efficient unentangling by $\mathcal{S}_k$, we require $|\alpha|_{k+1} \ll 0.13\Delta_k^{1/2}$. As $\Delta_k$ increases, this upper bound tightens, implying that displacements must shrink at each step to maintain fidelity—slowing the squeezing rate. Faster convergence may be possible by operating in the $S_x \neq -S_p$ regime, as in Ref.~\cite{hastrup2021unconditional}. We fit $\braket{\sigma_\textrm{y}}$ using $\mathcal{S}_k$ with $|\alpha|0 = 0.13$ and $|\alpha_k| = 0.06\Delta^2$ for $k \neq 0$, yielding optimal correction for the approximately linear $\braket{\sigma\textrm{y}}$ slope. This method has been used to obtain Figs.~\ref{fig:squeezing}(c,d). Although this fit is not completely analytical and requires simple numerical techniques, our prediction of the slope will yield a seed for optimization of the protocol that gives a much faster convergence compared to optimization techniques where this value is arbitrary, as is the case in Ref.~\cite{hastrup2021unconditional}.

\subsection{Comparison with Previous Work}
Ref.~\cite{hastrup2021unconditional} demonstrates that allowing large conditional displacements enables a faster protocol: first preparing a large odd cat state, then displacing it toward vacuum while managing the Gaussian envelope. This accelerates squeezing because large cat states disentangle the qubit easily, and the subsequent displacement toward the origin flattens $\psi(x)$ while squeezing $\psi(p)$. This operates in the non-commuting regime, as the conditional displacements do not commute—highlighting the power of the non-abelian QSP framework. However, due to the absence of an analytical non-abelian QSP scheme, the authors rely on numerical optimization to disentangle the qubit and oscillator. In the $L = 0.45$ regime, for a $4.7\ \mu\textrm{s}$ circuit duration, the protocol achieved a state-of-the-art infidelity of $\sim 0.009$ for $S_p = 8.5\ \mathrm{dB}$ squeezing and $S_x = -9.9\ \mathrm{dB}$ anti-squeezing—note that $S_x \neq -S_p$.

In contrast, our protocol achieves $8.5\ \mathrm{dB}$ squeezing and $-8.4\ \mathrm{dB}$ anti-squeezing with a total displacement amplitude $\sum_i |\alpha_i| \sim 5.7\ \mu\textrm{s}$ and infidelity $\sim 0.003$. For the faster variant, our circuit duration aligns with that of Ref.\cite{hastrup2021unconditional}, though the underlying approach is distinct. Our protocol, as explained, gradually widens the gaussian wave function always centered at vacuum. Slowing down the squeezing rate of our protocol—for instance, by setting $k = 1/4$—yields improved fidelity at $6\ \mathrm{dB}$ squeezing using the same number of steps and total displacement. This trade-off between fidelity and circuit depth is illustrated in Fig.\ref{fig:squeezing}(e). 

Unlike our versatile scheme, the protocol in Ref.~\cite{hastrup2021unconditional} is limited to the regime of shorter circuit depth at the cost of fidelity, due to poor approximation of large odd cat states by Gaussian wavefunctions. Their strategy begins in the large-cat regime and numerically displaces toward the origin—whereas we start from vacuum and build up broader vacuum-like states through small odd cats. That approach faces two major issues: (1) unentanglement becomes inefficient as the Gaussian lobes begin to overlap, and (2) the overall envelope, which peaks at the origin, must be handled numerically. This results in inefficiencies, clearly visible in the final state’s dip at the center (see Ref.~\cite{hastrup2021unconditional})—an unphysical feature not characteristic of true squeezed states. Their reliance on starting with large cats, which is the root cause of this efficiency, stems from the absence of an analytical unentangling scheme—such as our $\mathrm{GCR}$—that can unentangle qubits from oscillators with high precision in the small-cat regime.

\begin{figure}
  \centering
  \includegraphics[width=\linewidth]{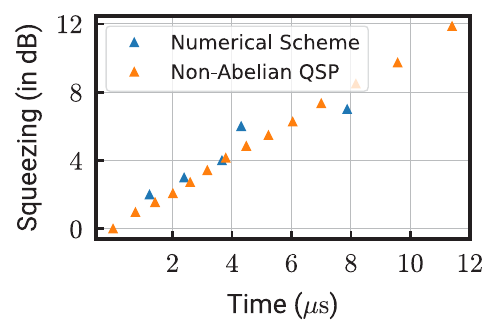}
  \caption{Squeezing of vacuum achieved with fidelity $\mathcal{F}>0.99$ using non-abelian QSP as discussed in Sec.~\ref{sec:squeezing} and the numerically optimal scheme in Ref.~\cite{eickbusch2022fast}.}
  \label{fig:numerical_squeezing}
\end{figure}

Finally, our scheme is on par with numerically optimized schemes shown in Ref.~\cite{eickbusch2022fast}. See Fig.~\ref{fig:numerical_squeezing} for comparison. 
\subsection{Comparison with Trotterization}
Let us look at a sequence of $\textrm{CD}$s similar to $\textrm{GCR}$. This sequence is given by the following sequence using the BCH formula, 
\begin{align}
(e^{i\epsilon/N x\sigma_\mathrm{x}}e^{i\epsilon/N p\sigma_\mathrm{y}}e^{-i\epsilon/N x\sigma_\mathrm{x}}e^{-i\epsilon/N p\sigma_\mathrm{y}})^N\ket{0,g}=e^{i\epsilon^2(xp+px)\sigma_\mathrm{z}}\ket{0,g}+\mathcal{O}(\epsilon^{4N})\\
&=e^{i\epsilon^2(xp+px)}\ket{0,g}+\mathcal{O}(\epsilon^{4N})
\end{align}
For a squeezed state, this sequence can be changed to the following for N=1 without loss of generality,
\begin{align}
e^{i(2\epsilon/\Delta^2) x\sigma_\mathrm{x}}e^{i(2\epsilon p)\sigma_\mathrm{y}}e^{-i(2\epsilon/(\Delta^2) x\sigma_\mathrm{x}}e^{-i\epsilon p\sigma_\mathrm{y}}\ket{0,g}&=e^{i\epsilon^2/(\Delta^4)(xp+px)\sigma_\mathrm{z}}\ket{0,g}\\
&=e^{i\epsilon^2/(\Delta^4)(xp+px)}\ket{0,g},
\end{align}
where $\Delta,\epsilon$ for $N^\mathrm{th}$ patch of the sequence relies on $\Delta,\epsilon$ from the $N^\mathrm{th}$ round. Here $\Delta>1$, that is, we squeeze along the $p$ quadrature for precision in $p$ measurements. %We can derive the above formula using the squeezing gadget to prepare a small even cat given in Sec.~\ref{sec:squeezing}. 

Now, the unentanglement of the qubit depends on (1) the order of error cancellation in the BCH sequence (2) the magnitude of $\epsilon$. Note that, increasing N and decreasing $\epsilon$, both decrease the acceleration of squeezing. The first because it effectively squeezes and anti-squeezes even if $N$ is increased. The second one is obvious. Since with each step $\Delta$ increases, it is legitimate to increase $\epsilon$ carefully such that the overall un-entanglement is not affected. For this purpose we use $\epsilon=\delta p/2$ as long as we maintain that $\delta x=\Delta/2$ decreases by the same factor as the increase in $\delta p$. We use the factor of $1/2$ since we want to maintain that the resulting Gaussian function (approximated from a small cat) is smaller than the width of the Gaussian function (of the squeezed state or vacuum) that we started with. Note that, $N$ should increase with $\epsilon$ in order to maintain a fixed fidelity. If we do not impose this condition then the fidelity decreases, however it does so gradually. The sequences using this idea prepare high-fidelity squeezing, however, this is achieved with large circuit duration. For example, for squeezing of $8.52dB$, trotterization requires a circuit duration of $40.67\mu\mathrm{s}$.

\section{Cat State Preparation}
Now, we compute fidelities of cat states against the output of the cat state preparation circuits in Fig.~\ref{fig:Cat_states}(a) for large cats and Fig.~\ref{fig:squeezing}(a) for small cats. Cat states are described as superposition of two diametrically opposite conditions, for example, $\ket{\text{Dead Cat}}+\ket{\text{Alive Cat}}$ or $\ket{0}^{\otimes n}+\ket{1}^{\otimes n}$. In CV architecture, cat states are defined as superposition of states located diametrically opposite in phase space with respect to the origin,
\begin{align}
\ket{C_{\pm\alpha}}&\propto(\mathrm{D}(\alpha)\pm \mathrm{D}(-\alpha))\ket{0}_\mathrm{vac}\approx\frac{\ket{\alpha}\pm\ket{-\alpha}}{\sqrt{2}},\\
\psi(x)&=\braket{x|C_{\pm\alpha}}=\Bigg(\frac{2}{\pi}\Bigg)^{1/4}(e^{-(x-\alpha)^2}\pm e^{-(x+\alpha)^2}){\sqrt{2}}.
\end{align}

\paragraph{Non-deterministic preparation:}
Preparation of cat states $\ket{C_{\pm\alpha}}$ requires one to entangle the cavity state in vacuum ($\ket{0}_\mathrm{vac}$) and the qubit in $\frac{\ket{g\pm i e}}{\sqrt{2}}$ state using $\mathrm{CD}(\alpha,\sigma_\mathrm{z})$. 
\begin{align}
\ket{\psi1}=\mathrm{CD}(\alpha,\sigma_\mathrm{x})\ket{0}\ket{\pm}\propto\ket{\alpha}\ket{+i}\pm\ket{-\alpha}\ket{-i}.\\
\label{eqn:entangled}
\end{align}
Now, rotating the qubit state along $\sigma_\mathrm{x}$ axis by $\pm\pi/2$ will give us even and odd cat states entangled with $\ket{g}$ and $\ket{e}$, respectively, if the qubit was initially in the state $\frac{\ket{g+e}}{\sqrt{2}}$.
\begin{align}
\ket{\psi2}&=\mathrm{R}_{0}(-\pi/2)\ket{\psi1}\propto\ket{\alpha}\ket{+i}\pm \ket{-\alpha}\ket{-i}\\
&\propto(\ket{\alpha}\pm\ket{-\alpha})\ket{g}-i(\ket{\alpha}\mp\ket{-\alpha})\ket{e}
\end{align}
Upon measurement of the qubit, the cavity will yield odd or even cats each with probability $\frac{1}{2}$.

\paragraph{Deterministic preparation:} The above protocol is probabilistic with a success probability of $P_g=0.5$. Ideally, we would like deterministic protocols which produce cat states with $100\%$ probability in the absence of any error. 
An insightful way to look at this problem is shown in Fig.~\ref{fig:Cat_states}(a,b). 

For a large cat state peaked with $|\alpha|^2>4$, the qubit is un-entangled from the cavity only when the spin of the qubit is polarized in a single direction globally, irrespective of the oscillator's position. This would require rotating the qubit entangled with cavity-state $\ket{\pm\alpha}$ by $\pm\frac{\pi}{2}$ about the $\sigma_\mathrm{x}$ axis or $\sigma_\mathrm{y}$ axis on the Bloch sphere. 
Note that, a momentum boost on this cavity-qubit state $e^{i\beta \hat x\sigma_\mathrm{z}}$ can be seen as a position dependent rotation by an angle $-2\beta \hat x$ about the $\sigma_\mathrm{z}$ axis. Using the identity $\mathrm{R}_{\phi}(\theta)=\mathrm{R}_{\pi/2-\phi}(-\pi/2)\mathrm{R}_\textrm{z}(\theta)\mathrm{R}_{\pi/2-\phi}(\pi/2)$, we have, 
\begin{align}
\ket{\psi4}&=\mathrm{R}_{\pi/2}(-2\beta \hat x)\ket{\psi1}\label{eqn:untangled}=\mathrm{R}_{0}(-\pi/2)\mathrm{R}_\textrm{z}(-2\beta \hat x)\mathrm{R}_{0}(\pi/2)\ket{\psi1}\\
&=\mathrm{R}_{0}(-\pi/2)\mathrm{R}_\textrm{z}(-2\beta \hat x)\mathrm{R}_{0}(\pi/2)\ket{\psi1}=\mathrm{R}_{0}(-\pi/2)\mathrm{R}_\textrm{z}(-2\beta \hat x)\ket{\psi2}\\
&=\mathrm{R}_{0}(-\pi/2)e^{i\beta \hat x \sigma_\mathrm{z}}\ket{\psi2}.\label{eqn:theta}\\
\end{align}
The final rotation performs a global rotation about the $\sigma_\mathrm{z}$ axis and can be skipped. Let us call the state $\ket{\psi4}$ minus this rotation as $\ket{\psi3}$. In order to align the qubit state entangled with the cavity-state at peaks of the Gaussian, the momentum boost $e^{i\beta \hat x\sigma_\mathrm{z}}$ should yield $\mathrm{R}_\mathrm{z}(\pi\alpha/(2|\alpha|))$ which implies $\beta=-\pi/4|\alpha|$. Cavity-qubit state $\ket{\psi4}$ is shown in Fig.~\ref{fig:Cat_states}(a) where the initial qubit state is $\ket{g}$ such that the protocol prepares an even cat with $|\alpha|^2$. Notice that, $\braket{\sigma_\mathrm{z}}=\cos\frac{\pi x}{2|\alpha|}$ and $\braket{\sigma_\mathrm{x}}=\sin\frac{\pi x}{2|\alpha|}$, such that the qubit spin polarization is in the xz-plane ($\ket{+}$) at $x=\pm 3$ as intended.

\subsection{Fidelity of Deterministic Preparation without QSP Correction}\label{app:cat_I}
Here, we rewrite the states $\ket{\psi1}-\ket{\psi4}$ in the position basis for the preparation of an even cat, ignoring normalization factors,
\begin{align}
\braket{x|\psi1}&=e^{-(x-\alpha)^2}\ket{g}+e^{-(x+\alpha)^2}\ket{e}\label{psi1x},\\
\braket{x|\psi2}&=e^{-(x-\alpha)^2}\ket{-i}-ie^{-(x+\alpha)^2}\ket{+i},\\
\braket{x|\psi3}&=e^{-i\frac{\theta(x)}{2}\sigma_\mathrm{z}}(e^{-(x-\alpha)^2}\ket{-i}-ie^{-(x+\alpha)^2}\ket{+i}),\\
\braket{x|\psi4}&=e^{i\frac{\pi}{4}\sigma_\mathrm{x}}e^{-i\frac{\theta(x)}{2}\sigma_\mathrm{z}}(e^{-(x-\alpha)^2}\ket{-i}-ie^{-(x+\alpha)^2}\ket{+i}),
\end{align}
where $\frac{\theta(x)}{2}=-\beta x$. We would have prepared a cat if $\theta(x)=\frac{\pi}{2}\frac{|x|}{x}$ (for large cats where the overlap between the two Gaussian curves is insignificant),
\begin{align}
\ket{\psi_\text{cat}}&\propto e^{i\frac{\pi}{4}\sigma_\mathrm{x}}(e^{-i\frac{\pi}{4}\sigma_\mathrm{z}}e^{-(x-\alpha)^2}\ket{-i}-ie^{i\frac{\pi}{4}\sigma_\mathrm{z}}e^{-(x+\alpha)^2}\ket{+i})\\
&=e^{i\frac{\pi}{4}\sigma_\mathrm{x}}(e^{-i\frac{\pi}{4}}e^{-(x-\alpha)^2}\ket{+}+e^{-i\frac{\pi}{4}}e^{-(x+\alpha)^2}\ket{+})\\
&=e^{-(x-\alpha)^2}\ket{+}+e^{-(x+\alpha)^2}\ket{+}\\
&=\ket{C_{+\alpha}}\ket{+}.
\end{align}
Therefore, the overlap between $\braket{x|\psi4}$ and $\braket{x|\psi_\text{cat}}$ can be computed approximately, neglecting the overlap between the two Gaussian curves, as
\begin{align}
&\approx\Big(\frac{2}{\pi}\Big)^\frac{1}{2}\frac{1}{2}\int_{-\infty}^{\infty}dx\quad e^{-2(x-\alpha)^2}\braket{+|e^{i\frac{\pi}{4}\sigma_\mathrm{x}}e^{-i\frac{\theta(x)}{2}\sigma_\mathrm{z}}|-i}\nonumber\\
&\quad\quad\quad-ie^{-2(x+\alpha)^2}\braket{+|e^{i\frac{\pi}{4}\sigma_\mathrm{x}}e^{-i\frac{\theta(x)}{2}\sigma_\mathrm{z}}|+i}\label{eqn1a}\\
&=\Big(\frac{2}{\pi}\Big)^\frac{1}{2}\frac{1}{2}\int_{-\infty}^{\infty}dx\quad e^{\frac{i\pi}{4}}e^{-2(x-\alpha)^2}\braket{+|e^{-i\frac{\theta(x)}{2}\sigma_\mathrm{z}}|-i}\nonumber\\
&\quad\quad\quad +e^{-\frac{i\pi}{4}}e^{-2(x+\alpha)^2}\braket{+|e^{-i\frac{\theta(x)}{2}\sigma_\mathrm{z}}|+i}\label{eqn1b}\\
&=\Big(\frac{2}{\pi}\Big)^\frac{1}{2}\frac{1}{2}\int_{-\infty}^{\infty}dx\quad e^{-2(x-\alpha)^2}\cos\Big(\frac{\pi}{4}-\frac{\theta(x)}{2}\Big)\nonumber\\
&\quad+e^{-2(x+\alpha)^2}\cos\Big(\frac{\pi}{4}+\frac{\theta(x)}{2}\Big)\label{eqn:add_term}\\
&=\Big(\frac{2}{\pi}\Big)^\frac{1}{2}\int_{-\infty}^{\infty}dx\quad e^{-2(x-\alpha)^2}\Big(1-\frac{(\beta x-\pi/4)^2}{2}\Big)\label{eqn3}\\
&=1-\frac{\pi^2}{32}\Big(\frac{2}{\pi}\Big)^\frac{1}{2}\int_{-\infty}^{\infty}dx\quad e^{-2(x-\alpha)^2}(\frac{x}{\alpha}-1)^2\\
&=1-\frac{\pi^2}{32\alpha^2}\Big(\frac{2}{\pi}\Big)^\frac{1}{2}\int_{-\infty}^{\infty}dx\quad e^{-2(x-\alpha)^2}(x-\alpha)^2\\
&=1-\frac{\pi^2}{128\alpha^2}.
\end{align}
 Eqs.~\ref{eqn3} uses $\int_{-\infty}^\infty dx\quad e^{-2(x-\alpha)^2}\cos(\beta(x-\alpha))=\int_{-\infty}^\infty dx\quad e^{-2(x+\alpha)^2}\cos(\beta(x+\alpha))$. Therefore, the fidelity for large cats is equal to,
\begin{align}
\mathcal{F}\approx\Big|1-\frac{\pi^2}{128\alpha^2}\Big|^2\approx1-\frac{\pi^2}{64\alpha^2}.\label{eqn:cat_fidelity}
\end{align}
The cavity state $\ket{\psi4}$ is not completely un-entangled from the qubit because the rotation angle varies continuously with x and has the correct values only at $x=\pm \alpha$. In an attempt to rotate the qubit in $\ket{\psi1}$ by $\pm\frac{\pi}{2}$ at $x=\pm\alpha$ we have over- and under-rotations at $|x|\ne\alpha$. The fidelity can be increased for large cats if the magnitude of the position-dependent rotation could be fixed to $\frac{\pi}{2}$. Note that this error is same as the case of no-QSP correction for rotation gadgets computed in App.~\ref{app:no_corr}. Thus, this calculation indicates that the correction from GCR and BB1 will be similar to rotation gadgets and hence we will not repeat this calculation for the preparation of cat states.
\subsection{The Problem with Small Cat States} \label{app:small_cats}
Cat states with a small number of photons do not obey the fidelity value given by Eq.~(\ref{eqn:cat_fidelity}) mainly because $\braket{\alpha|-\alpha}\rightarrow 0$ is not true in this case. Given that there is significant overlap for `small cats', the qubit state polarization in $\ket{\psi1}$ is no longer depicted by Fig.~\ref{fig:Cat_states}(b), that is, all $\ket{+}$ for $x>0$ and all $\ket{-}$ for $x<0$. Instead, the spin polarization is given by Fig.~\ref{fig:squeezing}(b) in this case. We have given the expressions for these spin polarizations in App.~\ref{app:squeezing} in reference to the squeezing gadgets. Here, we discuss the fidelity for the case of preparing odd and even cat states using this scheme. 

For small cats, the fidelity of an odd cat will always be lower than the fidelity of an even cat with the same number of photons for smaller $\alpha$. The reason for this difference in fidelity can be justified by analyzing the extra term that arises when $\braket{\alpha|-\alpha}\ne 0$. 
\paragraph{Small cat protocol with $\beta=-2\alpha$:}\label{Fidelity_small}
 The extra term in the overlap of $\braket{x|\psi4}$ and $\braket{x|\psi_{\text{cat}}}$ for the analogue of Eq.~(\ref{eqn1b}) when $\beta=-2\alpha$ is,
\begin{align}
&\sqrt{\frac{1}{2\pi}}\int_{-\infty}^\infty dx \ e^{-2(x^2+\alpha^2)}(\braket{+|e^{i\frac{\pi}{4}\sigma_\mathrm{x}}e^{-i\beta x\sigma_\mathrm{z}}|+i}-i\braket{+|e^{i\frac{\pi}{4}\sigma_\mathrm{x}}e^{-i\beta x\sigma_\mathrm{z}}|-i})\nonumber\\
&=\frac{e^{-2\alpha^2}}{\sqrt{2\pi}}\int_{-\infty}^\infty dx \ e^{-2x^2}(\cos\Big(\frac{\pi}{4}-\frac{\theta(x)}{2}\Big)\pm \cos\Big(\frac{\pi}{4}+\frac{\theta(x)}{2}\Big)\Big).
\end{align}
This correction is subtracted from the overlap of odd cats while it is added in the case of even cats. Now, computing the integral,
\begin{align}
&\frac{e^{-2\alpha^2}}{\sqrt{2\pi}}\int_{-\infty}^\infty dx \ e^{-2x^2}\cos\Big(\frac{\pi}{4}-\frac{\theta(x)}{2}\Big)=\frac{e^{-2\alpha^2}}{\sqrt{2\pi}}\int_{-\infty}^\infty dx\quad e^{-2x^2}\Big(1-\frac{(\beta x-\pi/4)^2}{2}\Big)\\
&=\frac{e^{-2\alpha^2}}{2}-\frac{e^{-2\alpha^2}}{\sqrt{2\pi}}\frac{\beta^2}{2}\int_{-\infty}^\infty dx\quad e^{-2x^2}(x^2+\frac{\pi^2}{16\beta^2})\\
&=\frac{e^{-2\alpha^2}}{2}\Big(1-\frac{\alpha^2}{2}-\frac{\pi^2}{32}\Big).\\
\end{align}
Since $\frac{e^{-2\alpha^2}}{\sqrt{2\pi}}\int_{-\infty}^\infty dx\quad e^{-2x^2}\cos\Big(\frac{\pi}{4}-\frac{\theta(x)}{2}\Big)=\frac{e^{-2\alpha^2}}{\sqrt{2\pi}}\int_{-\infty}^\infty dx\quad e^{-2x^2}\cos\Big(\frac{\pi}{4}+\frac{\theta(x)}{2}\Big)$, and the normalization constant $\mathcal{N}$ of the cat state including the overlap $\braket{\alpha|-\alpha}$ is given by,
\begin{align}
\mathcal{N}=\Big(\frac{1}{2\pi}\Big)^{\frac{1}{4}}\frac{1}{\sqrt{1+e^{-2\alpha^2}}},
\end{align}
instead of just $\Big(\frac{1}{2\pi}\Big)^{\frac{1}{4}}$. Therefore the fidelity for even cats is,
\begin{align}
\mathcal{F}_{\text{even}}&\approx\Big|\frac{1}{\sqrt{1+e^{-2\alpha^2}}}\Big(1-\frac{\alpha^2}{2}-2\alpha^4+\frac{\pi\alpha^2}{2}-\frac{\pi^2}{32}\Big)+\frac{e^{-2\alpha^2}}{\sqrt{1+e^{-2\alpha^2}}}\Big(1-\frac{\alpha^2}{2}-\frac{\pi^2}{32}\Big)\Big|^2,\label{Fidelity_small_even}
\end{align}
while for odd cats it is,
\begin{align}
\mathcal{F}_{\text{odd}}&\approx\Big|\frac{1}{\sqrt{1+e^{-2\alpha^2}}}\Big(1-\frac{\alpha^2}{2}-2\alpha^4+\frac{\pi\alpha^2}{2}+\frac{\pi^2}{32}\Big)-\frac{e^{-2\alpha^2}}{\sqrt{1+e^{-2\alpha^2}}}\Big(1-\frac{\alpha^2}{2}-\frac{\pi^2}{32}\Big)\Big|^2.\label{Fidelity_small_odd}
\end{align}
Here, the first addend represents the overlap when $\beta=-2\alpha x$ while the second addend is the correction due to the overlap. Thus, it is clear that the fidelity for odd cats is lower than even cats, and the difference becomes exponentially significant as $\alpha$ decreases. Consider a superposition of sum of $\ket{\pm\alpha}$ and difference of $\ket{\pm\alpha}$, each entangled with $\ket{\pm}$ qubit states (say), respectively. Due to this significant difference in normalization of the two states, the probability of projecting the oscillator onto the even small cat states will always be more than projecting onto the odd small cat, upon qubit measurement. 

This conclusion highlights the general problem with preparing an odd small cat state even with the QSP schemes engineered in this thesis (see Chapter~\ref{chapter:na-qsp}). We come across this problem when preparing Fock states. The connection between the preparation of cat state and Fock states is that, for small $\alpha$, the above problem corresponds to $\ket{0}\rightarrow \sim \ket{0} (\sim \ket{1})$ when we project the oscillator onto the even (odd) cat states. In Sec.~\ref{ssec:universal} we tackle this problem using what we call the amplification gadget, also engineered via a combination of non-abelian composite pulses.
%%%%%%%%%%%%%%%%%%%%%%%%%%%%%%%%%%%%%%%%%%%%%%%%%%%
\section{GKP Logical Pauli States}\label{app:GKP-prep}
\begin{figure}
  \centering \includegraphics[width=0.5\textwidth]{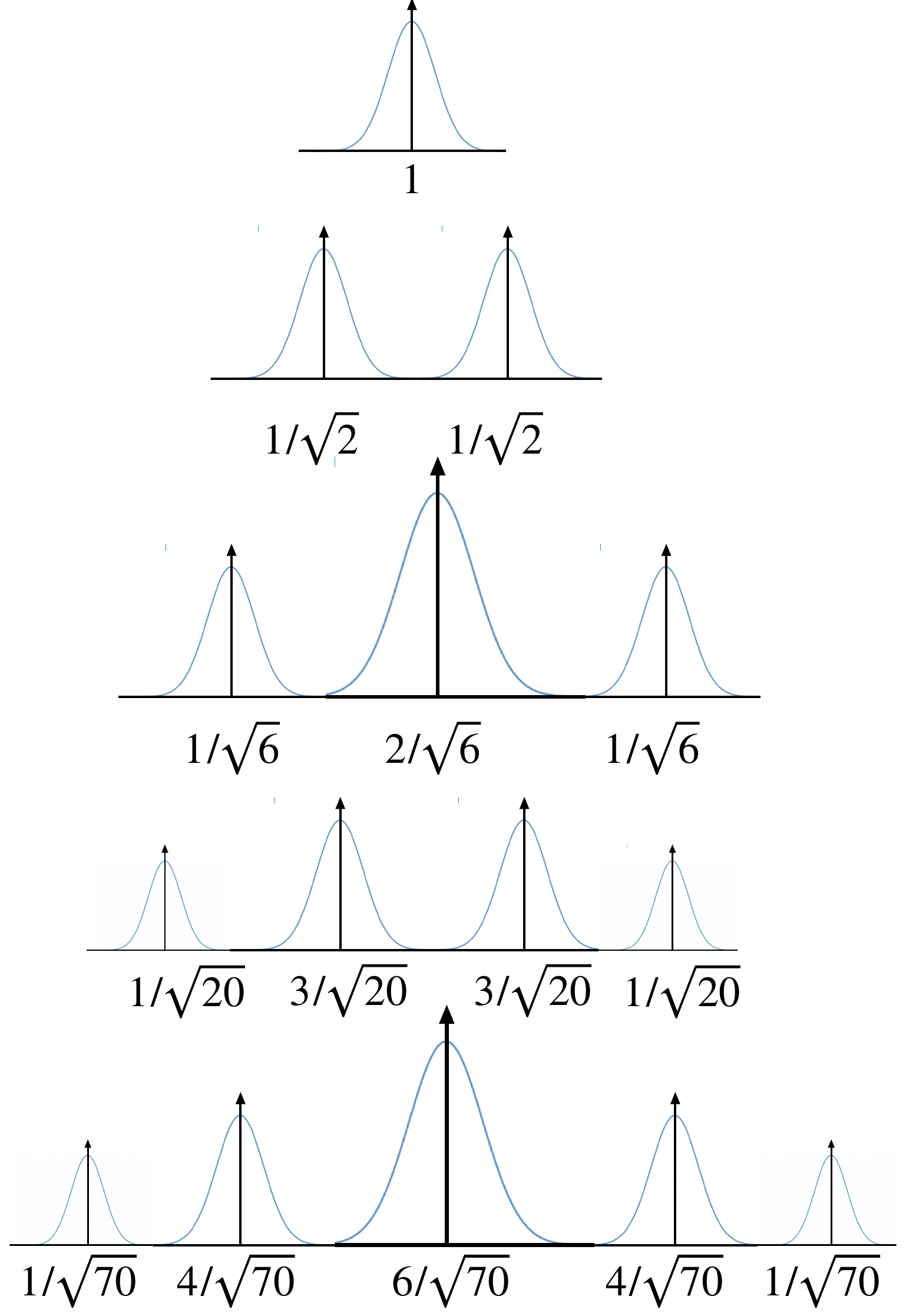}
  \caption{The probability corresponding to the various finite-energy basis states in superposition generated after repeated cat state transfer circuit are related to Pascal's triangle as shown here. This defines the state prepared by the circuit shown in Fig.~\ref{fig:GKP-prep}.}
  \label{fig:pascal}
\end{figure}

Here, we derive the numerical circuit presented in Sec.~\ref{ssec:GKP-States} for the preparation of GKP states. From Eqs.~\ref{eq:GKP-logical0} and~\ref{eq:GKP-logical1}, it is clear that we need superposition of the finite energy basis states $\ket{\alpha}_\Delta$ where $\alpha=m\sqrt{2\pi}, m\in 2\mathbb{Z}$ ($\ket{0}_\mathrm{GKP},\ket{+}_\mathrm{GKP},\ket{+i}_\mathrm{GKP}$) or $m\in 2\mathbb{Z}+1$ ($\ket{1}_\mathrm{GKP},\ket{-}_\mathrm{GKP},\ket{-i}_\mathrm{GKP}$). This is indeed doable with repeated use of the cat state preparation circuit $\mathcal{C}$. This circuit, however, prepares a state that is different from the GKP state defined in Eqs.~\ref{eq:GKP-logical0}-\ref{eq:GKP-logical1}. This definition has specific coefficients (that has a Gaussian dependence) for each finite-energy basis state $\ket{\alpha}_\Delta$. In this appendix, we compare the coefficients of the final state constructed by our scheme with the desired GKP state (with a Gaussian envelope) to show the relationship between fidelity ($F_\mathrm{H}$) and number of repetitions of $\mathcal{C}$ or circuit-depth ($N$) using only cat-state-transfer circuits. We also discuss how this motivates appending the stabilization scheme to the circuit in Fig.~\ref{fig:GKP-prep} to achieve the same fidelity ($F_\mathrm{H}$) with lower $N$.  
\paragraph{State fidelity v/s circuit depth:}
The superposition coefficients generated by $\mathcal{C}_k$ (see Fig.~\ref{fig:GKP-prep}(a)) arise from recursively splitting the vacuum state via conditional displacements. These follow a binomial distribution, matching Pascal’s triangle (see Fig.~\ref{fig:pascal}): after $N$ cat-state transfer steps, the $m^\text{th}$ peak has amplitude $\sqrt{{N \choose m}/2^N}$. In contrast, the target GKP state requires amplitudes proportional to $k e^{-\pi m^2 \Delta^2/2}$ at positions $m\sqrt{\pi}$, where $k$ is a normalization factor. We have ignored the common factor of $(2/\pi)^{1/4}$.

For a given Gaussian width $\Delta$ and $Z_\mathrm{GKP}$ codeword $\mu$, the optimal number of cat-splitting steps $N$ satisfies,
\begin{align}
  \sqrt{ {N\choose m+\mu}\Big/2^N}=\exp{-\pi\frac{(2m+\mu-N/2)^2\Delta^2}{4}},
\end{align}
for $m\in\mathbb{Z}$. Applying Stirling's approximation, this condition becomes,
\begin{align}
  \implies &\ln{\Big(\frac{N}{2}\Big)^N\frac{1}{m^m (N-m)^{N-m}}}=-2\pi(m-N/2)^2\Delta^2,\label{eq:stirling}\\
  \implies&\frac{1/2}{\Big(\frac{m}{N}\Big)^{m/N} \Big(1-\frac{m}{N}\Big)^{1-m/N}}=\exp{-\frac{2\pi\Delta^2(m-N/2)^2}{N}}\\
  \implies& \frac{x^{-x}(1-x)^{-(1-x)}}{2}=e^{-2N\pi\Delta^2(x-0.5)^2},
\end{align}
 $x=m/N$ where $x\le 1$. This is a transcendental equation that must be solved numerically to obtain $N = f(\Delta)$. Using Newton-Raphson iteration, we find that the overlap is maximized when $N\Delta^2 \approx 0.32$. Table~\ref{tab:GKP-prep-circuit-depth} lists the optimal values of $N$ for various $\Delta$ yielding fidelity $\mathcal{F} \ge 0.98$.
 
\paragraph{Un-entanglement:} For un-entanglement after $k=2$, the angle of rotation is not so straightforward, as shown in the Sec.~\ref{ssec:GKP-States} with the help of Fig.~\ref{fig:GKP-prep}(a). To determine the optimal angle for un-entanglement with $\mathcal{C}_k$, let us define the following abstract fidelity and normalization functions in terms of $a$, the magnitude of conditional displacement used for rotation, 
\begin{align}
  F&=\frac{\Big[\frac{\pi}{4}-a\sqrt{\pi}x\Big]^2+\sum_{i\in 2\mathbb{Z}}^{2\le i \le x} \Big[{k\choose i/2}a(x-i)\Big]^2}{1+\sum_{i\in 2\mathbb{Z}}^{2\le i \le x} {k\choose i/2}^2},\\
  \text{where,}\quad x&=\frac{k+1}{2}+1\quad\text{if}\quad k\in 2\mathbb{Z}+1,\\
  &=\frac{k}{2}+1\quad\text{if}\quad k\in 2\mathbb{Z}.
\end{align}
The fidelity function here only uses the left half since the effect from the right half will be the same. We do not take into account the central peak if $k$ is even because the central peaks will not rotate the qubit at all. Thus, we need to minimize the following expression which measures the infidelity of the qubit states entangled with the center of each peak to the target qubit state $\ket{+}$,
\begin{align}  
  \text{min}_a \Big[1-F\Big].\\
\end{align}
 The constraint is $a\le\sqrt{\pi}/4k$. One can verify that the minima are indeed located at $a=\frac{\sqrt{\pi}}{4k}$ if $k\le 3$. 

\paragraph{Appending SBS:} The state generated above do not have the right coefficients to yield peaks on the squeezed state in superposition farthest from the origin, (notice the absence of fringes in these peaks in the last Wigner graph before in Fig.~\ref{fig:GKP-prep}(a)). This can be resolved by appending a single round of $\mathrm{SBS}$ rounds. The advantage of using this scheme is two-fold. Firstly, it prepares the state with a higher fidelity compared to when only using $\mathrm{SBS}$ followed by a logical $Z_\mathrm{GKP}$ measurement, in the presence of errors. This is because, firstly, we are not relying on on ancilla measurement outcome at any step. In such a scenario, mid-circuit error detection can be used to detect errors on the ancilla. Secondly, we are not using the slow convergence of $\mathrm{SBS}$ (see Fig.~\ref{fig:GKP-prep}(b)). The convergence is faster here because we resort to $\mathrm{SBS}$ (single round) only after the overlap of the final state from our scheme with the target state in the GKP codespace is high enough using the faster circuit snippets $\mathcal{C}_k$. 
\paragraph{State preparation from vacuum:} If we repeat the scheme described above with $\Delta=1$, that is, vacuum in the oscillator, then we will prepare a momentum-squeezed state at the end of $\mathcal{C}_4$. Post this, we determine the squeezing of the state-prepared $\delta p$. We use this finite-energy parameter to repeat the protocol in the momentum quadrature. This process prepares a magic state with fidelity $0.85$ and success probability $0.90$. The decrease in fidelity is because the unentangling gadget yields a low success probability of $0.94$ when generating a squeezed cat from the squeezed vacuum prepared in this manner (rather than using the squeezing gadget). Note that, this circuit, however, would not need $\mathrm{SBS}$ to be appended at the end. This circuit is worse in terms of hybrid fidelity than the circuit presented in Fig.\ref{fig:GKP-prep}(b), and thus it is not discussed in the Sec.~\ref{ssec:GKP-States}.
\paragraph{Arbitrary GKP state:}
We have shown high-fidelity preparation of logical Pauli eigenstates of the GKP codespace. The magic state from the vacuum is a good resource for non-Clifford operations, however, to demonstrate universal state transfer we also need to show the preparation of arbitrary GKP code words. This can be done using a qubit-cavity state transfer technique restricted to GKP states (also used in Ref.~\cite{hastrup2021measurement}). Our method is more straightforward due to the analytical understanding we have developed using non-abelian QSP. We start with $\ket{0}_\mathrm{GKP}$ state in the cavity and the qubit in a desired state $a\ket{g}+b\ket{e}$. Next we can apply a finite-energy logical $Z_\mathrm{GKP}$ operation conditioned on the qubit state using $e^{i\sqrt{\pi/2}\hat{x}\sigma_\mathrm{z}}$. Thus, we have the hybrid oscillator-qubit state,
\begin{align}
  \ket{\psi}&=\mathrm{D}(i\sqrt{\pi}/2\sqrt{2})[a\ket{0^\prime g}+b\ket{1^\prime e}],
\end{align}
where.
\begin{align}\ket{1^\prime}&=\mathrm{D}(i\sqrt{\pi}/\sqrt{2})\ket{0},\  \ket{0^\prime}=\mathrm{D}(-i\sqrt{\pi}/\sqrt{2})\ket{0}.
  \end{align}
These states correspond to un-centered GKP states. At this point some rounds of stabilization will bring us back to the codespace where we have $[a\ket{0 g}+b\ket{1 e}]$. Now, we can use GCR to un-entangle the qubit from the oscillator yielding $[a\ket{0}_\mathrm{GKP}+b\ket{1}_\mathrm{GKP}]\otimes\ket{g}$. Note that this process uses several rounds of SBS even in the absence of errors and performs no error correction during the first and last step. In Chapter ~\ref{chapter:qec-control} we demonstrate an error-corrected gate teleportation scheme which gets read of these problems.
\section{Law-Eberly Protocol for Preparation of Fock states.}\label{law-eberly}
An arbitrary conditional displacement $e^{i(\alpha \hat x +\beta \hat p)\otimes\sigma_\phi}$ can be written in terms of ladder up and down operators $\{\hat a, \hat a^\dagger, \sigma_-,\sigma_+\}$ as follows,
\begin{align}
  \mathrm{CD}=e^{i(\alpha \hat x +\beta \hat p)\otimes\sigma_\phi}&=e^{ir(\cos{\theta} \hat x +\sin{\theta} \hat p)\otimes(\cos{\phi}\sigma_\mathrm{x}+\sin{\phi}\sigma_\mathrm{y})},
\end{align}
where $r^2=\alpha^2+\beta^2$. Now, we can express this gate in terms of sideband interactions. 
\begin{align}
  \mathrm{CD}&=e^{ir/2(e^{i\theta}\hat a+e^{-i\theta}\hat a^\dagger)\otimes(e^{i\phi}\sigma_{-}+e^{-i\phi}\sigma_{+})}\\
  &=\exp\Big[i\frac{r}{2}(\underbrace{e^{i(\theta+\phi)}\hat a\sigma_{-}+e^{-i(\theta+\phi)}\hat a^\dagger \sigma_{+}}_\text{AJC}\nonumber\\&+\underbrace{e^{i(\theta-\phi)}\hat a\sigma_{+}+e^{-i(\theta-\phi)}\hat a^\dagger \sigma_{-})}_\text{JC}\Big].
\end{align}
The last equation underlines the terms that correspond to the Jaynes-Cummings (JC) and anti-JC (AJC) Hamiltonians. Let us look at the effect of the unitary $e^{i\gamma \mathrm{AJC}}$ on $\ket{n}\ket{g}$.
\begin{align}
  \text{(Anti-JC)}^n\ket{n}\ket{g}&= e^{-i(\theta+\phi)}(2\sqrt{(n+1)})^n\ket{n+1}\ket{e},\\
  &=(2\sqrt{(n+1)})^n\ket{n}\ket{g}, \quad n\in 2\mathbb{Z}\end{align}
where $ n\in 2\mathbb{Z}+1$.
\begin{align}
  \therefore e^{i\gamma \mathrm{AJC}}\ket{n}\ket{g}&=\sum_{j=1}^\infty \frac{(i\gamma)^n}{n!}\text{(Anti-JC)}^n\ket{n}\ket{g}\\
  &=\cos{2\gamma\sqrt{(n+1)}}\ket{n}\ket{g}\nonumber\\&+e^{-i(\theta+\phi)}\sin{2\gamma\sqrt{(n+1)}}\ket{n+1}\ket{e}.
\end{align}
These calculations indicate that if we only had AJC or JC then it would be easy to prepare arbitrary Fock states using qubit rotations with $\gamma=\frac{\pi}{4\sqrt{(n+1)}}$, starting from $\ket{n=0}\ket{g}$ with single photon consumption processes. Given that the expression for $\mathrm{CD}$ also contains the JC Hamiltonian, we alternate between $\theta-\alpha=z$ and $\theta-\alpha=-z$ to collectively cancel this term. This can be easily achieved using $(e^{i\gamma x\sigma_\mathrm{y}/N}e^{i\gamma p\sigma_\mathrm{x}/N})^N$ in $N$ steps where $\theta-\phi=-\frac{\pi}{2}$ for the first gate and $\theta-\phi=\frac{\pi}{2}$ for the second. 

\begin{figure}
    \centering
    \includegraphics[width=1.2\textwidth]{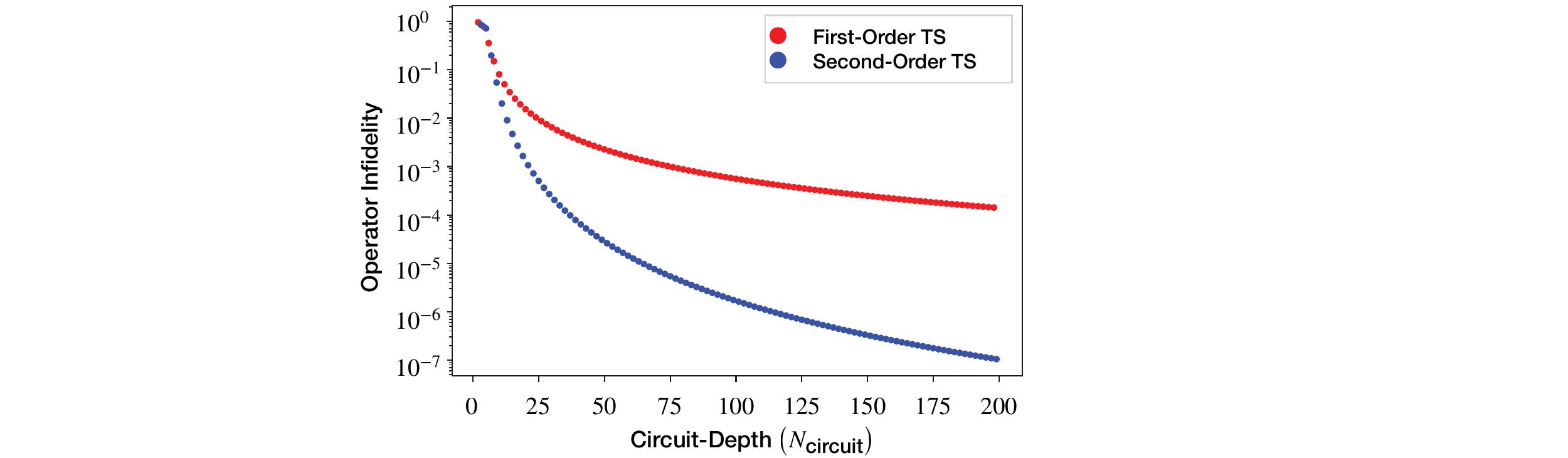}
    \caption[Fidelity of an approximate compilation of the Anti-Jaynes Cummings Interaction using the phase-space ISA.]{\textbf{Fidelity of an approximate compilation of the Anti-Jaynes Cummings Interaction using the phase-space ISA.} The $y$-axis gives the operator infidelity between the exact operator $AJC$ and its trotterized approximation obtained using $CD$s. We use the definition given in Chapter~\ref{hybrid}, with a truncated oscillator of Hilbert space dimension $d=15$. The $x$-axis corresponds to the circuit depth $N_\text{circuit}$ (number of $CD$s) of the different Trotter-Suzuki (TS) sequences used. (Red) First-order TS approximation $V=(e^{-i\frac{\theta}{r} \textrm{A}}e^{-i\frac{\theta}{r} \textrm B})^r$. Circuit depth in this case is $N_\text{circuit}=2r$. (Blue) Second-order TS approximation $V=(e^{-i\frac{\theta}{2r} A}e^{-i\frac{\theta}{r} B}e^{-i\frac{\theta}{2r} A})^r$. Circuit depth in this case is $N_\text{circuit}=2r+1$. The AJC, and JC Hamiltonians can be used to prepare arbitrary superpositions of Fock states and hence can be employed for universal oscillator state preparation via the Law-Eberly protocol~\cite{law1996arbitrary}. The value of $\theta=\frac{\pi}{{\sqrt{2}}}$ chosen for this comparison is suitable for the preparation of Fock  state $\ket{1}$. The Hilbert space of the oscillator used to compute the operators $U,V$ is $N_\text{dim}=50\gg d$ and we have checked that the results are unaffected upon a further increase in $N_\text{dim}$. Comparison with numerical optimizations discussed in Refs.~\cite{ISA,eickbusch2022fast} and state preparation of Fock state $\ket{1}$ using the Law-Eberly protocol can be found in Fig.~\ref{fig:circuit-compare}.}
    \label{fig:AJC_sim}
\end{figure}

\begin{figure}
    \centering
    \includegraphics[width=\textwidth]{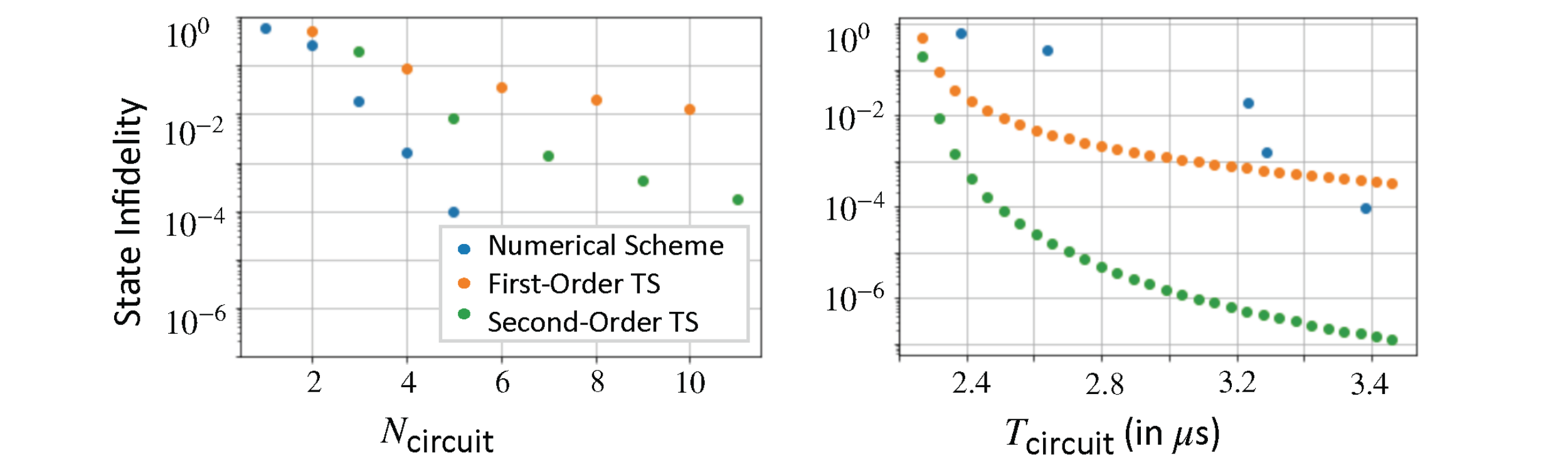}
    \caption[Comparison between numerically optimized and analytically derived circuits using techniques like Trotter-Suzuki for preparation of Fock state $\ket{1}$.]{Comparison between numerically optimized and analytically derived circuits using techniques like Trotter-Suzuki (TS). Infidelity of Fock state $\ket{1}$ preparation using the Law-Eberly protocol~\cite{law1996arbitrary} using the JC Hamiltonian synthesized via TS methods, as discussed in Fig.~\ref{fig:AJC_sim}. On the $x$-axis, we vary two metrics to improve these quantities. The left plot shows variation in the circuit depth (total count of $CD$ gates or $T_\textrm{circuit}$) and the right plot shows varying circuit duration (total amplitude of $CD$s or $T_\textrm{circuit}$) as described in App.~\ref{app:squeezing}.}
    \label{fig:circuit-compare}
\end{figure}

Alternatively, this protocol can be seen as a trotterization to achieve the sum of two $\mathrm{CD}$ Hamiltonians to achieve $H=x\sigma_X-p\sigma_Y=\mathrm{AJC}$. The operator fidelity of the resulting operation with $AJC$ Hamiltonian with respect to $N$ has been detailed in Fig.~\ref{fig:AJC_sim}, as analyzed by the author in Ref.~\cite{ISA}. In the availability of the JC or AJC Hamiltonian evolution, we can prepare arbitrary superposition of Fock states, using the protocol defined by Law and Eberly in~\cite{law1996arbitrary}, and hence universal state transfer can be achieved. The simplest of these tasks is to prepare a Fock $\ket{1}$ state. The efficiency of Fock state preparation using this scheme we have analyzed in Ref.~\cite{ISA}. This analysis is outlined below. The first row of Fig.~\ref{fig:circuit-compare} shows a comparison between the Law-Eberly protocol using the JC Hamiltonian realized in this manner and numerical optimization. 

From the plots in Fig.~\ref{fig:circuit-compare}, we can see that the numerical scheme outperforms the trotterization-based scheme in terms of circuit depth. On the other hand, the top right plot shows that the numerical scheme requires longer circuit duration for the same infidelity. We emphasize that the numerical circuits were optimized on circuit depth and not duration, which is why they do not converge to the trotterized results. In terms of circuit-depth the numerical scheme is still optimal, as expected. Our scheme described in Sec.~\ref{ssec:universal} is the only analytical scheme which presents circuit depths on par with the numerical schemes, with the same fidelities.

%%%%%%%%%%%%%%%%%%%%%%%%%%%%%%%%%%%%%%%%%%%%%%%%%%%
\chapter{Supplementary for Chapter~\ref{chapter:GKP-qec}}\label{app:GKp-qec}
Here we give the derivation dissipation engineering based stabilization scheme and its Kraus map representation, which are part of our work in Refs.~\cite{royer2020stabilization, sivak2023real}. The derivation below in Sec.~\ref{dissipation-engineering} has been adapted from our review on GKP states~\cite{brady2024advances}.
\section{Stabilization of GKP Codespace using Dissipation Engineering}\label{dissipation-engineering}
    Quantum error correction through stabilizer measurements could be thought of as \textit{dissipation engineering}. An alternative to measuring the stabilizers of the quantum codes is engineering a system-bath interaction,
    \begin{equation}
        H=\sqrt{\Gamma} (\hat{d}\hat{b}(t)^\dagger+\hat{d}^\dagger \hat{b}(t)),
    \end{equation}
    which relaxes the system to states satisfying $\hat{d}\ket{\psi}=0$, where $\hat{d}$ is known as the dissipator. 
    Any excitation in the system due to $\hat{d}^\dagger$ are transferred to the zero-temperature bath, autonomously cooling the system to the desired state $\ket{\psi}$. A Markovian model of dissipation is realized by the above Hamiltonian where the field operators (bath) obey $[\hat{b}(t),\hat{b}(t^\prime)^\dagger]=\delta(t-t^\prime)$, with $\delta(t)$ being the Dirac-delta distribution. 
    
    There are multiple ways to design dissipators into the codespace. In Ref.~\cite{royer2020stabilization}, the authors defined dissipators to the GKP codespace as the natural logarithm of the stabilizers $S$, since
    $\ln{S}\ket{\psi}=0$. An alternative definition of the dissipators was introduced in Ref.~\cite{sellem2025dissipative}. Thus, in order to find equations for GKP dissipators, we analyze the finite-energy GKP stabilizers.  As discussed in Chapter~\ref{chapter:GKP-qec}, the finite-energy GKP stabilizers can be obtained by the following code deformation of the arbitrary ideal GKP stabilizers $S$,
\begin{align}
    S_{\mathrm{x(p)},\Delta}&=\hat E S_{\mathrm{x (p)}} \hat E^{-1}=E e^{i\hat v} E=e^{i\left[\cosh(\Delta^2)\hat{v}+i\sinh(\Delta^2)\hat{v}_{\perp}\right]}
\end{align}
where $\hat{v}=\alpha\hat{q}+\beta\hat{p}$ and $\hat{v}_{\perp}=\alpha\hat{p}-\beta\hat{q}$. Here $\alpha=1,\beta=0$ for $S_\mathrm{x}$ and $\alpha=0,\beta=1$ for $S_\mathrm{p}$.
It can be easily checked that,
\begin{align}
    [E_\Delta S_iE_\Delta^{-1},E_\Delta S_jE_\Delta^{-1}]=E_\Delta[S_i,S_j]E_\Delta^{-1}= 0,
\end{align}
and thus, the new stabilizers and logical operators commute in the same way as the ideal stabilizers and logical operators, satisfying the minimum requirements for stabilizer-based error correction.\footnote{The non-hermiticity of $S_x, S_p$ or $\hat{d}_x, \hat{d}_p$ is not a problem here because we do not intend to measure these operators. Instead we want to build them into the dissipation Hamiltonian which will be Hermitian. In the next section, we discuss the engineering of this dissipation using an auxiliary qubit.} The dissipator corresponding to each stabilizer subspace becomes $\hat{d}=-\frac{i}{m\sqrt{2\cosh{\Delta^2}\sinh{\Delta^2}}}\ln S$ where $\ln S=(v_{[m/2\cosh{\Delta^2}]}/\sqrt{\tanh{\Delta^2}}+iv_\perp\sqrt{\tanh{\Delta^2}})/\sqrt{2}$. Here $v_{[l]}$ denotes symmetric version of the modular quadrature $v\quad mod\quad l$ also known as the Zak-basis~\cite{Zak_1967, Aharonov_1969, Ketterer_2016,Pantaleoni_Modular_2020,Pantaleoni_subsystem_2021,Mensen_phase_space_2021,shaw2022stabilizer, pantaleoni2023zak}. These modular quadratures are obtained from the multi-valued complex logarithm of the stabilizers $S_x,S_p$ such that $v_{[l]}\in(-l/2,l/2]$.

Focusing the discussion specifically to the single-mode square GKP states, we see that the two stabilizers generators and corresponding dissipators of square GKP code stabilizers, using the approximations $\cosh(\Delta^2)\simeq 1$ and $\sinh(\Delta^2)\simeq \Delta^2$, are given by,
\begin{align}
    S_{x}&=e^{i2\sqrt{\pi}(\hat{q}+i\Delta^2\hat{p})}\implies
    \hat{d}_x=(\hat{q}_{[\sqrt{\pi}]}/\Delta+i\hat{p}\Delta)/\sqrt{2},\\
    S_{p}&=e^{-i2\sqrt{\pi}(\hat{p}-i\Delta^2\hat{q})}\implies
    \hat{d}_p=-(\hat{p}_{[\sqrt{\pi}]}/\Delta-i\hat{q}\Delta)/\sqrt{2}.
\end{align}

\subsection{Engineered dissipation using an auxiliary qubit }\label{sec:qubit-dissipation}

One way to realize the non-local dissipators introduced just above is to use an auxiliary qubit coupled to the oscillator as a means for dissipation engineering. By preparing the auxiliary qubit in a known state, entangling the qubit and oscillator via a unitary operation, then resetting the auxiliary qubit, an effective dissipation can be realized. This method is sometimes called \textit{stroboscopic} dissipation engineering. 

As shown in Ref.~\cite{royer2020stabilization}, the continuous evolution under the Hamiltonian interaction, 
\begin{equation}
    H(t)=\sqrt{\Gamma} (\hat{d}\hat{b}^\dagger_t+\hat{d}^\dagger \hat{b}_t)
\end{equation}
can be discretized as if the system interacts with a different bath at every time step t, i.e.
\begin{align}
        U(t,t_0)&=\mathcal{T}e^{-i\int_{t_0}^t d\tau H(\tau) }\\
        &\approx\prod_{n=0}^T e^{-i\sqrt{\Gamma \delta t} (\hat{d}\hat{b}^\dagger_n+\hat{d}^\dagger \hat{b}_n) }\\
        &=\prod_{n=0}^T U_n,
\end{align}
where $t-t_0=T\delta t$ and $T\in \mathbb{Z}$. In the limit $\delta t\rightarrow 0$, we approach the continuous model. The excitation number, proportional to $\Gamma\delta t$, as shown in Ref.~\cite{royer2020stabilization}, needs to be small enough such that the $n$th bath mode contains less than one excitation. In this case, the bath mode can be realized using a qubit such that $\hat{b}_n\rightarrow \frac{\hat{\sigma}_{x,n}+i\hat{\sigma}_{y,n}}{2}$, where $\hat{\sigma}_{x,n},\hat{\sigma}_{y,n},\hat{\sigma}_{z,n}$ denote the Pauli matrices of $n$th qubit mode. The commutation relation between the bath operators $[\hat{b}_n,\hat{b}^\dagger_n]= 1$ is transformed as $\frac{1}{4}[\hat{\sigma}_{x,n}+i\hat{\sigma}_{y,n}, \hat{\sigma}_{x,n}-i\hat{\sigma}_{y,n}]=\hat{\sigma}_{z,n}$. For weakly populated qubits $\langle \hat{\sigma}_{z,n}\rangle\approx 1$, we retrieve the original commutation relation. In this qubit model, the time evolution is replaced by, 
\begin{equation}
    U(t,t_0)=\prod_{n=0}^Te^{-i\sqrt{\frac{\Gamma\delta t}{2\tanh(\Delta^2)}}\left(\hat{v}_{[\sqrt{\pi}]}\hat{\sigma}_{x,n}+\hat{v}_{\perp}\hat{\sigma}_{y,n}\tanh(\Delta^2)\right)}.
\end{equation}
Here, the qubits extract entropy from the oscillator and are left unused after. In other words, the ensemble of qubits can be replaced by a single qubit being reset after each time step, i.e.
\begin{align}
    U(t,t_0)&=\prod_{n=0}^Te^{-i\sqrt{\frac{\Gamma\delta t}{2\tanh(\Delta^2)}}\left(\hat{v}_{[\sqrt{\pi}]}\hat{\sigma}_x+\hat{v}_{\perp}\hat{\sigma}_y\tanh(\Delta^2)\right)}=\prod_{n=0}^T U_{\rm target}\label{eq:dissipation}
\end{align}

The final task is to derive oscillator-qubit circuits which realize the Hamiltonian $\hat{v}_{[m]}\hat{\sigma}_x+\hat{v}_\perp\hat{\sigma}_y\tanh(\Delta^2)$ for $\hat{v}\in\{2\sqrt{\pi}\hat{q},2\sqrt{\pi}\hat{p}\}$ via trotterization. In Ref.~\cite{royer2020stabilization}, authors specify three different circuits using first-order and second-order trotterization. Among these, the small-big-small obtained from using a first-order trotterization circuit can be made fault-tolerant under ancilla errors, and is more efficient for photon loss~\cite{royer2020stabilization}. The stabilization circuit given by unitary $U_{\rm target}$ in Eq.~(\eqref{eq:dissipation}) for $X,Z$ stabilization as, 
\begin{align}     
 \label{eq:sBs unitary}
U_{sBs}^X&=e^{i\epsilon_q\hat{q}\hat{\sigma}_y}e^{-i\sqrt{\pi}\hat{p}\hat{\sigma}_x}e^{i\epsilon_q\hat{q}\hat{\sigma}_y}, \textrm{ where } \epsilon_q=\frac{\sqrt{\pi}}{2}\Delta_q^2,\\
 \textrm{ and } U_{sBs}^Z&=e^{-i\epsilon_p\hat{p}\sigma_y}e^{-i\sqrt{\pi}\hat{q}\hat{\sigma}_x}e^{-i\epsilon_p\hat{p}\hat{\sigma}_y}, \textrm{ where } \epsilon_p=\frac{\sqrt{\pi}}{2}\Delta_p^2.
 \end{align}
 The condition on modular quadratures is replaced by conditioning the whole unitary post trotterization to remain unchanged under translation $\hat{x}\rightarrow \hat{x}+m$, permitting $\hat{x}_{[m]}\rightarrow\hat{x}$. This condition is enforced by leveraging the modularity of the qubit by choosing $\Gamma\delta t$ such that the translation $\hat{x}\rightarrow \hat{x}+m$ leads to a trivial qubit operation after time $T$.

\subsection{Ancilla Errors in GKP Small-Big-Small Circuit.}\label{ancilla_errors}
An ancilla decay during the larger conditional displacement could yield displacement errors larger than the distance of the code, and hence the logical error of the code depends linearly on ancilla decay. For example, if we define $\mathrm{CD}(\sqrt{\pi/2})=e^{i\sqrt{\pi}\hat{p}\otimes\sigma_z}$, then the effect of an ancilla decay during this conditional displacement corresponds to,
\begin{equation}
    \mathrm{CD}(\sqrt{\pi/2}-\alpha/2)(I\otimes \text{Err})\mathrm{CD}(\alpha/2)=D(\sqrt{\pi/2}-\alpha)\otimes R_Y(\pi/2),
\end{equation}
where $\text{Err}=\ket{g}\bra{e}$ corresponds to an ancilla decay event. The displacement $\alpha$ is determined by the time at which the ancilla decay happened. Thus, an ancilla error during conditional displacement can disrupt the displacement, leading to an error. Here, a displacement in position by $|x|=\sqrt{\pi}-2\alpha\in[-\sqrt{\pi},\sqrt{\pi}]$ can cause a logical error in the region where $\alpha\in[\sqrt{\pi}/4,3\sqrt{\pi}/4]$. Thus, the probability that an ancilla decay event causes a logical error is $50\%$, following this heuristic argument. 

The echoed conditional displacements used in superconducting circuits~\cite{campagne2020quantum,sivak2023real} will result in an equivalent probability of logical error rate on the GKP codewords due to ancilla decay. Ancilla dephasing on the other hand causes small displacement errors or measurement errors; the small displacement errors occur due to the dephasing errors which occur in between two conditional displacements of the $sBs$ circuit. These effects are correctable for the GKP encoding. Dependencies on ancilla errors have been demonstrated experimentally in Ref.~\cite{sivak2023real}. Thus, circuits can be modified to ensure fault-tolerance with biased-noise ancilla such as Kerr-cats, fluxonium, squeezed cats, dissipatively stabilized cats, additional flag qubits.~\cite{puri2019stabilized,Shi_2019,grimm2020stabilization}. Another approach for suppression of ancilla errors is to use a GKP ancilla for error correction as discussed in Refs.~~\cite{terhal2020towards, siegele2023robust}.                                                

\section{Similarity Transformation of the Photon Loss Operator under Gaussian Envelope.}\label{app:aE}
In this appendix we will compute the quantity $\hat E\hat a\hat E^{-1}$ required to propagate the single photon loss operator $\hat a$ through Gaussian envelope operator $\hat E=e^{-\Delta^2\hat n}$. 
\begin{align}
    \hat E\hat a\hat E^{-1}&=    e^{-\Delta^2\hat a^\dagger \hat a}\hat a\hat e^{\Delta^2\hat a^\dagger \hat a}\\
    &=\sum_{m\in \mathbb{W}}e^{-\Delta^2m}\ket{m}\bra{m} \ \hat a \ \hat \sum_{n\in \mathbb{W}}e^{\Delta^2n}\ket{n}\bra{n}\\
    &=\sum_{m\in \mathbb{W}}\sum_{n\in \mathbb{W}}e^{-\Delta^2(m-n)}\ket{m}\bra{m} \hat a \ket{n}\bra{n}\\
    &=\sum_{m\in \mathbb{W}}\sum_{n\in \mathbb{W}}e^{-\Delta^2(m-n)}\sqrt{n}\ket{m}\braket{m|n-1}\bra{n}\\
    &=\sum_{m\in \mathbb{W}}\sum_{n\in \mathbb{W}}e^{-\Delta^2(m-n)}\sqrt{n}\ket{m}\bra{n}\delta_{m,n-1}\\
    &=\sum_{n\in \mathbb{W}}e^{\Delta^2}\sqrt{n}\ket{n-1}\bra{n}\\
    &=e^{\Delta^2}\hat a
\end{align}
Similarly, $\hat E\hat a^\dagger\hat E^{-1}=e^{-\Delta^2}\hat a^\dagger, \ \hat E^{-1}\hat a\hat E=e^{-\Delta^2}\hat a, \ \hat E^{-1}\hat a^\dagger\hat E=e^{\Delta^2}\hat a^\dagger$. Here $\mathbb{W}$ is the set of whole numbers.
\section{Kraus Map Representation}\label{app:Kraus_GKP}
 The Kraus operators $K_{ij}$ for each of the four outcomes $\ket{gg},\ket{ge},\ket{eg},\ket{ee}$ are plotted in the basis of the GKP states and its error words. These error words are close to the eigenspace of $K_{gg}^\dagger K_{gg}$. This choice can be justified as follows. As we have shown before in Ref.~\cite{sivak2023real}, $K_{gg}$ applies a logical Pauli operation and the probability of outcome $\ket{gg}$ is nearly $1$. This is verified by our calculations above. In addition, for infinite-energy case, we can use $S^\mathrm{Z}=e^{i2\sqrt{\pi}\hat x},S^\mathrm{Z}=e^{-i2\sqrt{\pi}\hat p}$ to write,
\begin{align}
    K_{gg}&=\frac{\mathrm{D}(i\sqrt{\pi/2})[I+S^\mathrm{Z}]\mathrm{D}(\sqrt{\pi/2})[I+S^\mathrm{X}]}{4}\\
   K_{gg}^\dagger K_{gg} &= \frac{\mathrm{D}(-\sqrt{\pi/2})[I+{S^\mathrm{X}}^\dagger]\mathrm{D}(-i\sqrt{\pi/2})[I+{S^\mathrm{Z}}^\dagger]}{4}\\&\quad \times\frac{\mathrm{D}(i\sqrt{\pi/2})[I+S^\mathrm{Z}]\mathrm{D}(\sqrt{\pi/2})[I+S^\mathrm{X}]}{4}\\
   &= \frac{\mathrm{D}(-\sqrt{\pi/2})[I+\frac{{S^\mathrm{X}}^\dagger+{S^\mathrm{X}}}{2}]}{2}\frac{[I+\frac{{S^\mathrm{Z}}^\dagger+{S^\mathrm{Z}}}{2}]\mathrm{D}(\sqrt{\pi/2})}{2}.
\end{align}
Thus, this operator is the symmetrized stabilizer without any logical operation on the codespace. The maximum eigenvalue states of this operator is approximately close to the GKP codespace and the error spaces. The other eigenstates are close to the error space of GKP. The Kraus operators $K_{gg},K_{ge},K_{eg},K_{ee}$ were plotted in the eigen-basis of $K_{gg}^\dagger K_{gg}$. We note that $K_{ge},K_{eg}$ corrects first-order errors $f(\hat a,\hat a)=\{\hat a^\dagger,\hat a\}$ while the second-order errors $f(\hat a,\hat a)=\{\hat{a}^{\dagger^2},\hat{a}^2\}$ are corrected by $K_{ee}$. Importantly, as we predicted the second-order error $f(\hat a,\hat a)=\hat a^\dagger \hat a$ is not corrected by just one round of stabilization.

\chapter{Supplementary for Chapter~\ref{chapter:qec-control}}\label{app:qec-control}
\section{Finite-Energy SUM gate.}\label{app:finite-SUM}.
The logical gates for GKP qubits are obtained via the non-unitary gate $\hat{E}_\Delta A\hat{E}_\Delta^{-1}$ where $A$ is the gate for unrealistic infinite-energy GKP while $\hat{E}_\Delta$ is the envelope operator $e^{-\Delta^2\hat n}$ (see Chapter~\ref{chapter:GKP-qec}). The entangling gate $\mathrm{CX}_\textrm{GKP}$ such that $A=e^{i2\hat x\otimes \hat p}$ takes the following form
\begin{align}
 \hat{E}_\Delta e^{i2\hat{x}\otimes\hat{p}}\hat{E}_\Delta^{-1} &=e^{i(2\cosh{\Delta^2}\hat{x}+i2\sinh{\Delta^2}\hat{p})\otimes(2\cosh{\Delta^2}\hat{p}-i2\sinh{\Delta^2}\hat{x})} \\  
  &  \hspace*{-6ex}
  \approx e^{i(2\hat{x}_1\hat{p}_2+2\Delta^4\hat{p}_1\hat{x}_2-i2\Delta^2(\hat{x}_1\hat{x}_2-2\hat{p}_1\hat{p}_2))}\quad\text{(Finite-energy $\mathrm{CX}_\mathrm{GKP}$ gate)}\nonumber
  \end{align}
These non-unitary gates can be approximated using an auxiliary qubit, where the approximations hold in the small $\Delta$ limit such that $\cosh{\Delta^2}\approx 1$ and $\sinh{\Delta^2}\approx \Delta^2$. 
\begin{align}
  \mathrm{CX}_\mathrm{GKP}&\approx e^{-i\Delta^2(\hat{x}_1\hat{x}_2-\hat{p}_1\hat{p}_2)\sigma_\mathrm{y}}e^{i(2\hat{x}_1\hat{p}_2-2\Delta^4\hat{p}_1\hat{x}_2)\sigma_\mathrm{x}} e^{-i\Delta^2(\hat{x}_1\hat{x}_2-\hat{p}_1\hat{p}_2)\sigma_\mathrm{y}}\ket{\psi}_{\textrm{GKP}}\ket{0}
\end{align}
We can use dissipation based methods followed by trotterization~\cite{royer2020stabilization,rojkov2023two} or realize this using GCR type correction techniques where $\sigma_\mathrm{x}\ket{g}=-i\sigma_\mathrm{y}\ket{g}$ to derive this sequence. Thus, for a two-mode equivalent of the SBS type circuit we have, $\mathrm{S}\equiv e^{-i\Delta^2(\hat{x}_1\hat{x}_2-\hat{p}_1\hat{p}_2)\sigma_\mathrm{y}}$ and $\mathrm{B}\equiv e^{i2(\hat{x}_1\hat{p}_2-\Delta^4\hat{p}_1\hat{x}_2)\sigma_\mathrm{x}}$. 

We show a fast echoed conditional sequence to realize each gate in the above sequence. Let us first discuss S. Using the definition of ${\rm TMS}(r,\phi)$ in Ref.~\cite{ISA}, we first note that\footnote{Such transformations are obtained using $e^ABe^{-A}=B+[A,B]+\frac{1}{2!}[A,[A,B]]+\frac{1}{3!}[A,[A,[A,B]]]....$}, 
\begin{align}
   {\rm TMS}(\alpha,\pi)~ a~ {\rm TMS}^\dagger(\alpha,\pi) % \nonumber\\
   = a\cosh{\alpha}+b^\dagger \sinh{\alpha}\,.
\end{align}
This in turn implies
\begin{align}
{\rm TMS}(\alpha,\pi) e^{-i\chi t_\mathrm{S}a^\dagger a\sigma_\mathrm{z}} {\rm TMS}^\dagger(\alpha,\pi)&=\mathrm{exp} \left [ \rule{0pt}{2.4ex} -i\chi t_\mathrm{S} \left ( \rule{0pt}{2.4ex} \cosh^2{(\alpha)}a^\dagger a+\sinh^2{(\alpha)}bb^\dagger \right. \right. \nonumber\\
  & ~~~~~~~~~~~~~~~ \left.\left. +\frac{1}{2}\sinh{(2\alpha)}(a^\dagger b^\dagger +ab) \right) \sigma_\mathrm{z} \rule{0pt}{2.4ex} \right]\label{eq:TMS-rot}
\end{align}
The next steps of the method are inspired by the construction of echoed-conditional displacements (ECD) described above. Notice that the first two terms in Eq.~(\ref{eq:TMS-rot}) do not change signs with $\alpha$, whereas the last two will. Hence, running the pulse shown below yields an echoed two-mode squeezing, in close analogy with the echoed displacement gate previously discussed,
\begin{align}
  &{\rm TMS}(\alpha,\pi)e^{-i\chi t_\mathrm{S}a^\dagger a\sigma_\mathrm{z}} {\rm TMS}^\dagger(\alpha,\pi)\times\sigma_\mathrm{x} {\rm TMS}(-\alpha,\pi)e^{-i\chi t_\mathrm{S}a^\dagger a\sigma_\mathrm{z}} {\rm TMS}^\dagger(-\alpha,\pi)\nonumber\\
  &~~~~~ \approx e^{-i\chi t_\mathrm{S}\sinh{2\alpha}(a^\dagger b^\dagger +ab)\sigma_\mathrm{z}}= e^{-i\chi t_\mathrm{S}\sinh{2\alpha}(\hat x_1\hat x_2-\hat p_1\hat p_2)\sigma_\mathrm{z}}.
  \label{eq:Echoed-TMS}
\end{align}
Now the qubit Bloch sphere can be rotated using $\mathrm{R}_\mathrm{x}(\pi/2)$ to transform this gate to $\mathrm{S}$. In contrast with the technique using controlled parity gates to compile this unitary, the speed of the conditional two-mode squeezing gate, in this case, is decided by $t_\mathrm{S}=\frac{\Delta^2}{\chi\sinh{2\alpha}}$ instead of $\chi$. This condition helps us use the low $\chi$ regime favorable for GKP states~\cite{eickbusch2022fast}. As $\sinh{2\alpha}$ is an unbounded function, we can in principle increase it to extremely large values by varying $\alpha$. Thus, in the weak dispersive regime, we can achieve fast conditional oscillator-oscillator entangling gates by leveraging unconditional two-mode squeezing with large $\alpha$ as a resource. The two-mode squeezed frame used to actuate a large $\alpha$ can be achieved using single-mode squeezing and beam splitter operations via Bloch-Messiah decomposition. See Ref.~\cite{ISA}.

Now, for the case of $\mathrm{B}$, one could ignore $\Delta^4$ term and directly use the Bloch-Messiah decomposition for the SUM gate~\cite{ISA}. However, we can achieve this gate exactly with the $\Delta^4$ correction by going to the frame of $\mathrm{BS}(\alpha,\pi)$. Using the definition of ${\rm TMS}(r,\phi)$ in Ref.~\cite{ISA}, we first note that, 
\begin{align}
   {\rm BS}(\alpha,\pi)~ a~ {\rm BS}^\dagger(\alpha,\pi) % \nonumber\\
   = a\cos{\alpha/2}+ib^\dagger \sinh{\alpha/2}\,.
\end{align}
Thus, we have,
\begin{align}
{\rm BS}(\alpha,\pi) e^{-i\chi t_\mathrm{B}a^\dagger a\sigma_\mathrm{z}} {\rm BS}^\dagger(\alpha,\pi) &=\mathrm{exp} \left [ \rule{0pt}{2.4ex} -i\chi t_\mathrm{B} \left ( \rule{0pt}{2.4ex} \cos^2{(\alpha/2)}a^\dagger a+\sinh^2{(\alpha/2)}bb^\dagger \right. \right. \nonumber\\
  & ~~~~~~~~~~~~~~~ \left.\left. -\frac{1}{2}\sin{(\alpha)}(a^\dagger b -ab^\dagger) \right) \sigma_\mathrm{z} \rule{0pt}{2.4ex} \right].\label{eq:BS-rot}
\end{align}
The echoed conditional beam-splitter version is given by,
\begin{align}
  &{\rm BS}(\alpha,\pi)e^{-i\chi t_\mathrm{B}a^\dagger a\sigma_\mathrm{z}} {\rm BS}^\dagger(\alpha,\pi)\times\sigma_\mathrm{x} {\rm BS}(-\alpha,\pi)e^{-i\chi t_\mathrm{B}a^\dagger a\sigma_\mathrm{z}} {\rm BS}^\dagger(-\alpha,\pi)\nonumber\\
  &~~~~~ \approx e^{-\chi t_\mathrm{B}\sin{\alpha}(a^\dagger b-ab^\dagger)\sigma_\mathrm{z}}= e^{-i\chi t_\mathrm{B}\sin{\alpha}(\hat x_1\hat p_2+\hat p_1\hat x_2)\sigma_\mathrm{z}}.
  \label{eq:Echoed-BS}
\end{align}
In order to extract B from Eq.~(\ref{eq:Echoed-BS}) we need to perform single-mode squeezing of one of the modes $\mathrm{S}_a(r)$ such that $\hat x_a,\hat p_a\rightarrow e^{r}x_a,e^{-r}p_a$. Let $r>0$ and $a=1$, that is squeeze the position of the first mode. Thus, Eq.~(\ref{eq:Echoed-BS}) yields,
\begin{align}
  e^{-i\chi t_\mathrm{B}e^{r}\sin{\alpha}(\hat x_1\hat p_2+e^{-2r}\hat p_1\hat x_2)\sigma_\mathrm{z}}
\end{align}
Here $r=\frac{1}{2}\ln{\Delta^{-4}}$ is fixed. Lower the $\Delta$, larger is the value of $r$. For example, $\Delta=0.34$ requires $r\sim 2.15$. Thus, the speed of this gate is given by, 
\begin{align}
\chi t_\mathrm{B}e^{r}\sin{\alpha}=2, \ t_\mathrm{B}&=\frac{2\Delta^2}{\chi \sin{\alpha}}
\end{align}
Thus, for lower $\Delta$, this gate is much faster. Thus, for $\mathrm{CX}_\mathrm{GKP}$, we have, 
\begin{align}
  t_\mathrm{S}=\frac{\Delta^2}{\chi\sinh{2\alpha}},t_\mathrm{B}\ge \frac{2\Delta^2}{\chi}
\end{align}
Similarly, for $\mathrm{CZ}_\mathrm{GKP}$, we have, 
\begin{align}
  t_\mathrm{B}=\frac{2\Delta^2}{\chi\sinh{2\alpha}},t_\mathrm{S}\ge \frac{\Delta^2}{\chi}.
\end{align}
Note that even though the lower bound on $t_\mathrm{S}$ is half that of $t_B$ in $\mathrm{CX}_\mathrm{GKP}$, total duration of the circuit is, $t_\mathrm{CX/CZ}=2t_\mathrm{S}+t_\mathrm{B}$. Thus, both gates come down to the same speed. Thus, we have given a new circuit decomposition for fast finite-energy SUM gate sequence for logical GKP entangling operations. Our derivation also highlights a the two-mode extension of the echoed conditional displacements, which we introduced in in Ref.~\cite{ISA}.
\section{Error-Corrected Gate Teleportation}\label{app:c-pauli}
 The significance of using qubits for GKP gates is that the rotation angles on qubits preserve the periodicity at for $\mathrm{CD}(2\sqrt{\pi},\sigma_\phi)$. It emulates a torus with the GKP unit cell as proposed in~\cite{gottesman2001encoding}. Let us consider our error-corrected scheme for gate teleportation described in Sec.~\ref{ssec:piecewise-teleportation} in an architecture with two oscillators encoded in the GKP where each GKP code is stabilized by a qubit coupled to it. Now, if we entangle the GKP states and qubits using the x-entangling gadget $\mathcal{E}_{\hat x}$ on the control GKP and p-entangling gadget $\mathcal{E}_{\hat p}$ on the target GKP, perform $\mathrm{CZ}$ between the two qubits and then use the corresponding unentangling gadgets on both, we would have performed a $\mathrm{CX}_\mathrm{GKP}$ on the two GKP states with the ancilla qubits unentangled. 
 
 We can describe this teleportation, by defining  $\ket{\psi_1}_\mathrm{GKP}=a\ket{0_1}_\mathrm{GKP}+b\ket{1_1}_\mathrm{GKP},\ket{\psi_2}_\mathrm{GKP}=(c\ket{+_2}_\mathrm{GKP}+d\ket{-_2}_\mathrm{GKP})$, and writing $\mathcal{E}_{\mathrm{x}_1}\mathcal{E}_{\mathrm{p}_2}[\ket{\psi_1}_\mathrm{GKP}\otimes\ket{g_1}][\ket{\psi_2}_\mathrm{GKP}\otimes\ket{g_2}]$
\begin{align}
  &\quad=[(a\ket{0_1}_\mathrm{GKP}\ket{g_1}-b\ket{1_1}_\mathrm{GKP})\ket{e_1}][(c\ket{+_2}_\mathrm{GKP}\ket{g_2}-d\ket{-_2}_\mathrm{GKP}\ket{e_2})]\nonumber\\&\quad=\ket{GKP_{CX1}},\\
  C_1Z_2\ket{GKP_{CX1}}&\quad=[a\ket{0_1}_\mathrm{GKP}\ket{g_1}[c\ket{+_2}_\mathrm{GKP}\ket{g_2}-d\ket{-_2}_\mathrm{GKP}\ket{e_2}]-b\ket{1_1}_\mathrm{GKP}\ket{e_1}\nonumber\\&\quad\quad[c\ket{+_2}_\mathrm{GKP}\ket{g_2}+d\ket{-_2}_\mathrm{GKP}\ket{e_2}]]=\ket{GKP_{CX2}},\\
  \mathcal{E}_{\mathrm{x}_1}\mathcal{E}_{\mathrm{p}_2}\ket{GKP_{CX2}}&\quad=a\ket{0_1}_\mathrm{GKP}\ket{g_1}[c\ket{+_2}_\mathrm{GKP}\ket{g_2}-d\ket{-_2}_\mathrm{GKP}\ket{g_2}]\nonumber\\
  &\quad\quad-b\ket{1_1}_\mathrm{GKP}\ket{g_1}[c\ket{+_2}_\mathrm{GKP}\ket{g_2}+d\ket{-_2}_\mathrm{GKP}\ket{g_2}]\nonumber\\
  &=a\ket{0_1}_\mathrm{GKP}[c\ket{+_2}_\mathrm{GKP}-d\ket{-_2}_\mathrm{GKP}]\nonumber\\&\quad-b\ket{1_1}_\mathrm{GKP}[c\ket{+_2}_\mathrm{GKP}+d\ket{-_2}_\mathrm{GKP}]\ket{g_1}\ket{g_2}
\end{align}
If we start with the control GKP in $\ket{-}_\mathrm{GKP}$ state, that is, $a=1$, $b=-1$ and target GKP in $\ket{1}_\mathrm{GKP}$ state, that is, $c=1,d=-1$; the final states of the two cavities will be in the entangled GKP Bell pair, $\ket{0_10_2}_\mathrm{GKP}+\ket{1_21_2}_\mathrm{GKP}$, with both the qubits decoupled and ready for the next round of stabilization or gate operation. The success probability of this gate is $0.9987$ and the fidelity of the gate is $99.92\%$. However, these gates cannot be protected from any ancilla error using the pieceable approach. Thus, it will be a low-fidelity gate in the presence of any type of fault in ancilla. This issue can be averted using the two-qubit $P_iP_j(\theta)$ Pauli rotations, shown in Fig.~\ref{fig:two-qubit-GKP} of Chapter~\ref{chapter:qec-control}.
\chapter{Supplementary for Chapter~\ref{chapter:conc}}\label{app:conc}
\section{Constructing Hybrid Unitary for Phase Estimation}\label{app:phase-est}
The hybrid unitary $C_xU$ is constructed as follows.
\begin{equation}
e^{i\alpha\hat{x}\otimes\sigma_\mathrm{z}}S^\dagger USe^{-i\alpha\hat{x}\otimes\sigma_\mathrm{z}}SUS^\dagger =e^{i\hat{g}\hat{n}^\prime\cdot\sigma}=C_xU,
\end{equation}
where, $S=\sqrt{\sigma_\mathrm{z}}$ is the qubit phase gate. Note that, here, each expression can be presented as a quaternion where the four basis elements correspond to the Pauli vectors $\{I,\sigma_\mathrm{x},\sigma_\mathrm{y},\sigma_\mathrm{z}\}$. The Pauli vectors follow the same algebra as quaternions, and hence, we now give derivation for the exact expression of $g, \hat{n}^\prime$ using the product formulas for quaternions. 
\begin{align}
 &e^{i\alpha\hat{x}\otimes\sigma_\mathrm{z}}S^\dagger USe^{-i\alpha\hat{x}\otimes\sigma_\mathrm{z}}SUS^\dagger=e^{i\alpha\hat{x}\otimes\sigma_\mathrm{z}}e^{i\theta\sigma_\mathrm{x}}e^{-i\alpha\hat{x}\otimes\sigma_\mathrm{z}} e^{-i\theta\sigma_\mathrm{x}}.
\end{align}
For the product of quaternions (or qubit rotations)\footnote{Note that, such derivation for expressing the product of arbitrary rotations as another rotation is given in any elementary quantum information textbook.},
\begin{align}
e^{i\gamma\hat k\cdot{\sigma}}&=e^{i\alpha\hat{x}\otimes\sigma_\mathrm{z}}e^{i\theta\sigma_\mathrm{x}}, \ e^{i\gamma^\prime\hat{k}^\prime\cdot{\sigma}}=e^{-i\alpha\hat{x}\otimes\sigma_\mathrm{z}}e^{-i\theta\sigma_\mathrm{x}},
\end{align}
 Defining $\hat n\cdot\vec\sigma=\sigma_\mathrm{z}$, $\hat m\cdot\vec\sigma=\sigma_\mathrm{x}$, we can use the vector identity, $(\hat n\cdot\vec{\sigma})(\hat m\cdot\vec{\sigma})=(\hat n\cdot\hat m)I+i(\hat n\times\hat m)\cdot\vec{\sigma}$ to write,
\begin{align}
  \gamma&=\gamma^\prime=\cos^{-1}{(\cos{\alpha\hat x}\cos{\theta}-(\hat n\cdot\hat m)\sin{\alpha\hat x}\sin{\theta})}\\
  \hat k&=\frac{1}{\sin{\gamma}}(\hat n\sin{\alpha\hat x}\cos{\theta}+\hat m\sin{\theta}\cos{\alpha\hat x}-(\hat n\times\hat m)\sin{\alpha\hat x}\sin{\theta}),\\
  \hat k^\prime&=\frac{1}{\sin{\gamma}}(-\hat n\sin{\alpha\hat x}\cos{\theta}-\hat m\sin{\theta}\cos{\alpha\hat x}-(\hat n\times\hat m)\sin{\alpha\hat x}\sin{\theta}),
\end{align}
 For this case, $\hat n\cdot\hat m=0$ and $(\hat n\times\hat m)\cdot{\vec\sigma}=\sigma_\mathrm{y}$. Thus,
\begin{align}
  \cos{\gamma}&=\cos{\alpha\hat x}\cos{\theta}\implies \sin{\gamma}=\sqrt{1-\cos^2{\alpha\hat x}\cos^2{\theta}},\\
  \hat k.\vec{\sigma}&=\frac{1}{\sin{\gamma}}(\sin{\alpha\hat x}\cos{\theta}\sigma_\mathrm{z}+\sin{\theta}\cos{\alpha\hat x}\sigma_\mathrm{x}\quad-\sin{\alpha\hat x}\sin{\theta}\sigma_\mathrm{y}),\\
  \hat k^\prime.\vec{\sigma}&=-\frac{1}{\sin{\gamma}}(\sin{\alpha\hat x}\cos{\theta}\sigma_\mathrm{z}+\sin{\theta}\cos{\alpha\hat x}\sigma_\mathrm{x}\quad+\sin{\alpha\hat x}\sin{\theta}\sigma_\mathrm{y}).
\end{align}
Thus, collectively, we can write, 
\begin{align}
  \hat k_\mathrm{z}=-\hat k^\prime_\mathrm{z}&=\frac{\sin{\alpha\hat x}\cos{\theta}}{\sin{\gamma}}\\
  \hat k_\mathrm{x}=-\hat k^\prime_\mathrm{x}&=\frac{\cos{\alpha\hat x}\sin{\theta}}{\sin{\gamma}}\\
  \hat k_\mathrm{y}=\hat k^\prime_\mathrm{y}&=-\frac{\sin{\alpha\hat x}\sin{\theta}}{\sin{\gamma}},\\
  \gamma^\prime=\gamma&=\cos^{-1}(\cos{\alpha\hat x}\cos{\theta}).
\end{align}
Now, we repeat this procedure to compute the target operation $C_xU$ which is equal to,
\begin{align}
 e^{i\gamma\hat k\cdot\vec{\sigma}}e^{i\gamma\hat k^\prime\cdot\vec{\sigma}} =e^{i\hat{g}\hat{n}^\prime\cdot\vec\sigma}=\cos{g}I+i\sin{g}(\hat{n}^\prime\cdot\vec\sigma).
\end{align}
Now, we have,
\begin{align}
  \cos{g}&=\cos^2{\gamma}-(\hat k.\hat k^\prime)\sin^2\gamma=1-2\sin^2\alpha\hat x\sin^2\theta,\\
  \hat n^\prime&=\frac{1}{\sin{g}}(\hat k\sin{\gamma}\cos{\gamma}+\hat k^\prime\sin{\gamma}\cos{\gamma}-(\hat k\times\hat k^\prime)\sin{\gamma}\sin{\gamma})\\
  &=\frac{1}{\sin{g}}((\hat k+\hat k^\prime)(\sin{2\gamma})/2-(\hat k\times\hat k^\prime)\sin^2{\gamma}).\\
  \implies\hat n^\prime\cdot\vec{\sigma} &=\frac{1}{\sin{g}}\Big(\frac{\sin^2\alpha\hat x\sin 2\theta}{\sin^2\gamma}\sigma_\mathrm{x}-\frac{\sin{2\alpha\hat x}\sin{2\theta}}{2}\sigma_\mathrm{y}+\frac{\sin^2\theta\sin 2\alpha\hat x}{\sin^2\gamma}\sigma_\mathrm{z}\Big).
\end{align}
We need to choose small enough $\alpha$ such that we can ignore $\mathcal{O}(\alpha^2\hat x^2)$ terms. In this limit, $\cos{g}\rightarrow 1$, $\frac{g}{\sin{g}}\rightarrow 1$, $\sin{\gamma}\rightarrow\sin{\theta}$ and,
\begin{align}
  \hat n_\mathrm{x}^\prime&=0, \ \hat n_\mathrm{y}^\prime=-\alpha\hat x\sin{2\theta}, \  \hat n_\mathrm{z}^\prime=2\alpha\hat x.
\end{align}
If the qubit is in a particular eigenstate of $\sigma_\mathrm{y}$, after application of $C_xU$, measuring it in the $\sigma_\mathrm{y}$ will yield an average displacement of the oscillator equal to $\alpha\hat x\sin{2\theta}$.
\section{Relationship between Fock State Preparation and Quantum Random Walks}\label{app:QRW}
We will first discuss the relationship between displacements by small amplitude and Fock states. Note that the difference of displacements $[\mathrm{D}(\alpha)-\mathrm{D}(-\alpha)]\ket{n}$ yields a superposition of $\ket{n\pm 1}$. On the other hand, the sum of displacements $[\mathrm{D}(\alpha)+\mathrm{D}(-\alpha)]\ket{n}$ yields $\ket{n}$, for small $|\alpha|$. For example, ignoring normalization, we can write,
\begin{align}
    (\mathrm{D}(\alpha)-\mathrm{D}(-\alpha))\ket{n}=&(\alpha \hat p)\ket{n}\\=&-\alpha\hat a^\dagger \ket{0}=-\alpha\ket{1},\quad \text{if}\quad n=0,\\
    =&-\alpha(\ket{0}-\sqrt{2}\ket{2}),\quad \text{if}\quad n=1,\\
    =&-\alpha(\sqrt{2}\ket{1}-\sqrt{3}\ket{3}),\quad \text{if}\quad n=2...
\end{align}
 These superpositions are rare occurrences after the application of CD$(\alpha,\sigma_\theta)$ for small $|\alpha|$. However, it implies that the difference of displacements on a Fock state amounts to a superposition of adding or even removing a photon. This is the premise of quantum random walk effects captured in~\cite{aharonov1993quantum}. This calculation shows that there is a way to add/subtract photons to a system using conditional displacements. The question is how can this change in the photon number be mapped to adding exactly $m$ photons or removing $m$ photons? Or, add/remove a single photon with high probability in a measurement-based random walk (and with high fidelity in a unitary random walk).

 Somehow we should be able to relate this change in parity with the addition of exactly one photon. In this spirit, if we use a combination of conditional displacements $CD(\alpha,\sigma_\mathrm{y})=e^{-i2\alpha\hat p\otimes\sigma_\mathrm{y}}$ and momentum boost $CD(i\alpha,\sigma_\mathrm{x})=e^{i2\alpha\hat x\otimes\sigma_\mathrm{x}}$, controlled on the orthogonal qubit axes, 
\begin{align}
\mathrm{CD}(\alpha,\sigma_\mathrm{y})\mathrm{CD}(i\alpha,\sigma_\mathrm{x})&=  [\mathrm{D}(\alpha)+\mathrm{D}(-\alpha)][\mathrm{D}(i\alpha)+\mathrm{D}(-i\alpha)]\ket{g}\bra{g}\\
  &\quad+ [\mathrm{D}(\alpha)+\mathrm{D}(-\alpha)][\mathrm{D}(i\alpha)-\mathrm{D}(-i\alpha)]\ket{g}\bra{e}\\
  &\quad+ i [\mathrm{D}(\alpha)-\mathrm{D}(-\alpha)][\mathrm{D}(i\alpha)+\mathrm{D}(-i\alpha)]\ket{e}\bra{g}\\
  &\quad+ i [\mathrm{D}(\alpha)-\mathrm{D}(-\alpha)][\mathrm{D}(i\alpha)-\mathrm{D}(-i\alpha)]\ket{e}\bra{e},
\end{align}
we compute (to first-order in $\alpha$), 
\begin{align}
\mathrm{CD}(\alpha,\sigma_\mathrm{y})\mathrm{CD}(i\alpha,\sigma_\mathrm{x})\ket{n}\ket{g}&=i2\alpha (\hat x\sigma_\mathrm{x}-\hat p\sigma_\mathrm{y})\ket{n}\ket{g}\\&=2\alpha\hat a^\dagger\sigma_{+} \ket{0}=2\alpha\ket{1}\ket{e},\quad \text{if}\quad n=0,\\
    &=2\sqrt{2}\alpha\ket{2}\ket{e},\quad \text{if}\quad n=1,\\
    &=2\sqrt{3}\alpha\ket{3}\ket{e},\quad \text{if}\quad n=2....
\end{align}
 We direct the readers to the discussion on the composition of conditional displacements in our work~\cite{ISA} to understand the various trajectories the qubit could end up in, after each successive conditional displacement. Thus, combining conditional displacements controlled on different axes of the qubit basis yields a way to regulate the amplitude of each term in a superposition of Fock states. For the case of orthogonal qubit axes as used here, the protocol for Fock state generation corresponds to the Law Eberly gadget using JC or AJC. The next question is, what if we were allowed to use non-orthogonal phase space and qubit Bloch sphere vectors? And, how can we increase the probability of this rare occurrence (the `minus' superposition)?

\begin{myframe}
\singlespacing
 \begin{quote}
     In Ref.~\cite{aharonov1993quantum}, the author points out that ``the important displacement of the distribution (which is the rare occurrence) after only ten steps, by an amount larger than the original width, and much larger than the maximum classically allowed one, is quite apparent." This statement was made for a random walk where the qubit was measured after every conditional displacement. It would be interesting to see if this random walk strategy could be used without any measurements to now increase the fidelity of the final state with the required oscillator-qubit state.
\end{quote}
\end{myframe}
 
 \doublespacing
 
 To formalize the problem of engineering Fock states, we analyze the effect of the two conditional displacements about arbitrary axes on the $X-Y$ plane on the Bloch sphere ($\sigma_\phi=\cos{\phi}\sigma_\mathrm{x}+\sin{\phi}\sigma_\mathrm{y}$) to rotate the qubit from $\ket{g}$ to $\ket{e}$. This picture enhances the idea that any rotation of the qubit on the Bloch sphere is emerging from the phase space dynamics of the oscillator under conditional displacements. Remember that the eigenstates of $\sigma_\theta$ are $\ket{\pm \phi}=\ket{g}+e^{i\theta}\ket{e}$; for composing displacements we can use,
\begin{align}   e^{i\vec{\beta_1}}e^{i\vec{\beta_2}}&=e^{i\alpha}e^{i(\vec{\beta_1}+\vec{\beta_2})},\\\quad\text{where}\quad\alpha&=\frac{|\beta_1\beta_2|}{2}\mathrm{Im}[e^{i(\theta_1-\theta_2)}]=\frac{|\beta_1\beta_2|}{2}\sin{(\theta_1-\theta_2)},
\end{align}
where $\beta_i=|\beta_i|(\cos{\theta_i}\hat x+\sin{\theta_i}\hat p)$. The following analysis is targeted towards three goals, (i) achieving the correct rotational symmetry in phase space using $\theta_i$s (ii) populating the oscillator with the required superposition of photons using the amplitudes of CD, $|\beta_i|$s (iii) unentangling the qubit from the oscillator using $\phi_i$s. While all parameters contribute to the fidelity of the target state, the above classification highlights the main goal for each parameter space $\{\beta_i,\phi_i,\theta_i\}$. For brevity, we will often use $\vec{\beta}_{a+b}=\vec{\beta}_a+\vec{\beta}_b$ and ignore the overall normalization constant.  
\allowdisplaybreaks{
\begin{align}
\ket{\psi_1}&= e^{i\vec{\beta_1}\sigma_{{\phi}_1}}\ket{0}\ket{g}=(\ket{\beta_1}+\ket{-\beta_1})\ket{g}\nonumber\\&~~~~~~~~~~~~~~~~~~~~~~~~~~~~+e^{i\phi_1}(\ket{\beta_1}-\ket{-\beta_1})\ket{e}\\
\braket{\sigma_\mathrm{z}}_1&=\braket{\beta_1|-\beta_1}=e^{-|\beta_1|^2}\\
\ket{\psi_2}&=e^{i\vec{\beta}_2\sigma_{{\phi}_2}}\ket{\psi_1}\\&=\Big[\cos{\frac{\phi_2-\phi_1}{2}}\ket{\vec{\beta}_{1+2}}+ie^{-i\alpha}\sin{\frac{\phi_2-\phi_1}{2}}\ket{\vec{\beta}_{2-1}}\Big]\ket{+\phi_2}\\
&+\Big[ie^{-i\alpha}\sin{\frac{\phi_2-\phi_1}{2}}\ket{\vec{\beta}_{1-2}}+\cos{\frac{\phi_2-\phi_1}{2}}\ket{\vec{\beta}_{-2-1}}\Big]\ket{-\phi_2}\\
&=\Big[\cos{\frac{\phi_2-\phi_1}{2}}(\ket{\vec{\beta}_{1+2}}+\ket{\vec{\beta}_{-1-2}})+e^{i(\frac{\pi}{2}-\alpha)}\sin{\frac{\phi_2-\phi_1}{2}}(\ket{\vec{\beta}_{1-2}}+\ket{\vec{\beta}_{-1+2}})\Big]\ket{g}\\
&+\Big[\cos{\frac{\phi_2-\phi_1}{2}}(\ket{\vec{\beta}_{1+2}}-\ket{\vec{\beta}_{-1-2}})+e^{i(\frac{\pi}{2}-\alpha)}\sin{\frac{\phi_2-\phi_1}{2}}(\ket{\vec{\beta}_{-1+2}}-\ket{\vec{\beta}_{1-2}})\Big]\ket{e}\\
\langle\sigma_\mathrm{z}\rangle_2&=\braket{\vec{\beta}_{1+2}|-\vec{\beta}_{1+2}}\cos^2{\frac{\phi_2-\phi_1}{2}}+\braket{\vec{\beta}_{1-2}|-\vec{\beta}_{1-2}}\sin^2{\frac{\phi_2-\phi_1}{2}}\nonumber\\&~~~~+\sin{\alpha}\sin{(\phi_2-\phi_1)}\braket{\beta_2|-\beta_2}
\end{align}
}
From the expression for $\braket{\sigma_\mathrm{z}}_1$, it is evident that we need overlapping Gaussian wave functions; in other words, small $|\beta_{1+2}|,|\beta_{1-2}|,|\beta_{2}|$ to achieve $\braket{\sigma_\mathrm{z}}_2\neq 0$. After two conditional displacements, it can be seen that $\braket{\sigma_\mathrm{z}}_2<0$ is possible if $\phi_1\neq\phi_2$. Thus, from this analysis, it is clear that non-commuting vectors in the Bloch sphere enable the desired un-entanglement, while non-commuting vectors in phase space enable rotational symmetry. The magnitudes of conditional displacements can be optimized for each pair of circuit depth and target Fock state $(N,\ket{n})$, independently. 

These insights could aid in developing a constructive algorithm for the deterministic preparation of Fock states. More importantly, one could borrow techniques from quantum random walks to increase the probability of the rare events where $m$ photons are added using circuits composed of very small conditional displacements.

%%%%%%%%%%%%%%%%%%%%%%%%%%%%%%%%%%%%%%%%%%%%%%%%%%%%%%%%%%%%%%%%%%%%%%%%%%%%%%%%%%%%%%%%%%%%%%%%%%%%%%%%%%%%%%%%%%%%%%%%%%%%%%%%%%%%%%%%%%%%%%%%%%%

   \addcontentsline{toc}{chapter}{Bibliography} % To get an entry in the TOC.
    \setstretch{1.0}
    
\bibliographystyle{elsarticle-num}     %\bibliographystyle{plain} 
    \bibliography{thesis.bib}

\end{document}